Juhani Risku

# Improving the Performance of Early-Stage Software Startups

## Design and Creativity Viewpoints

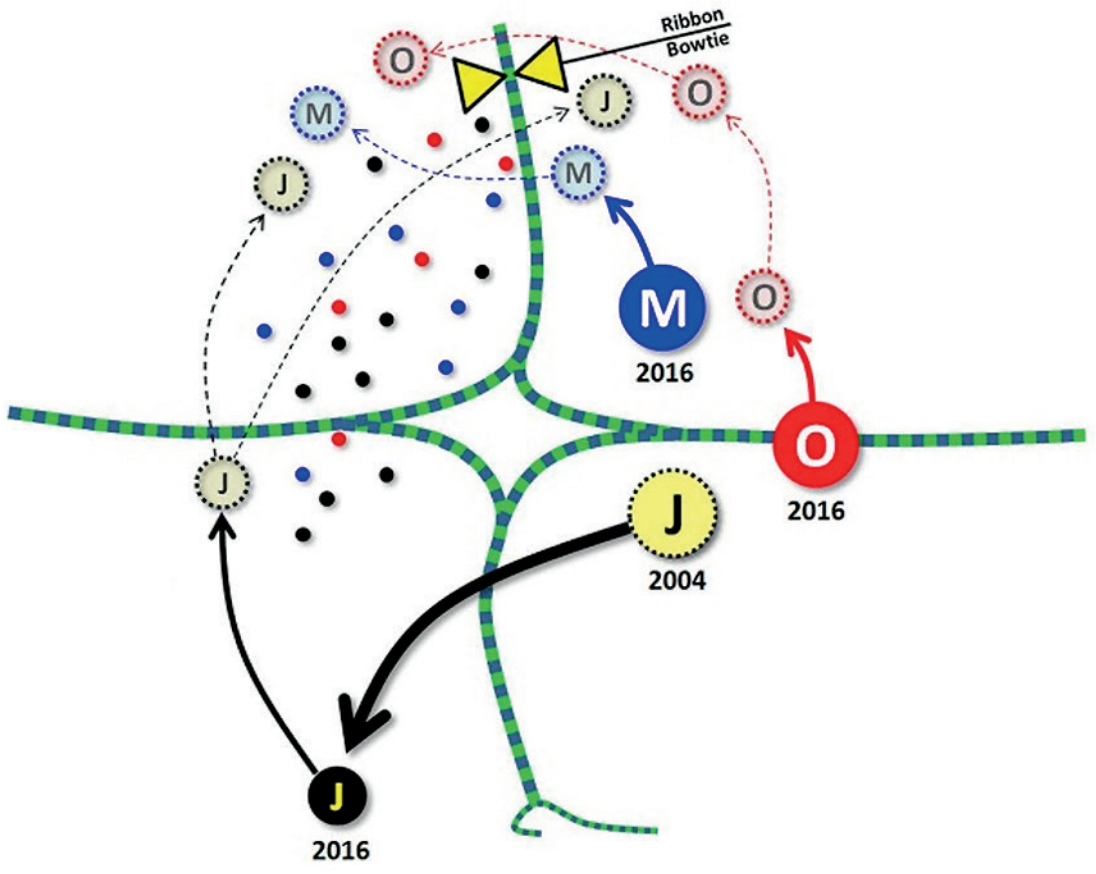

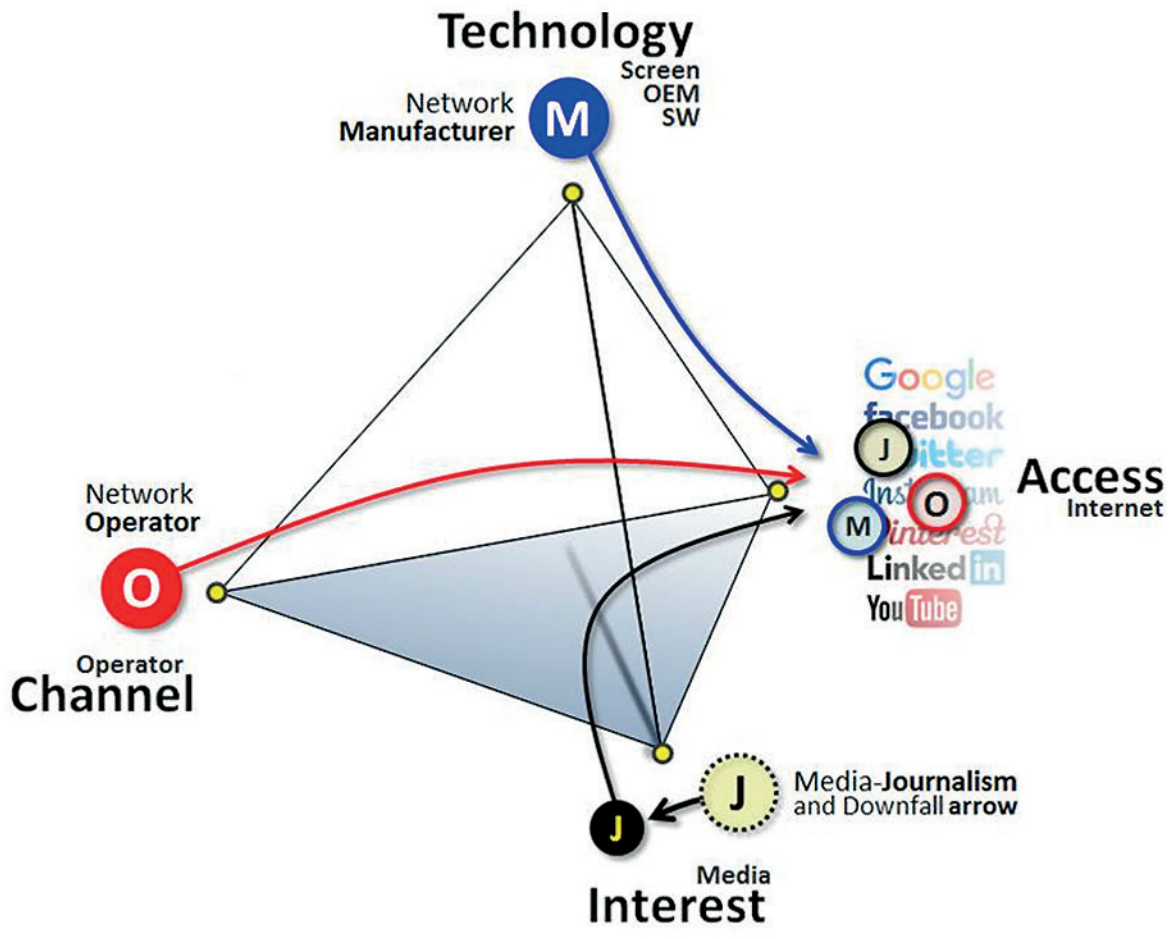

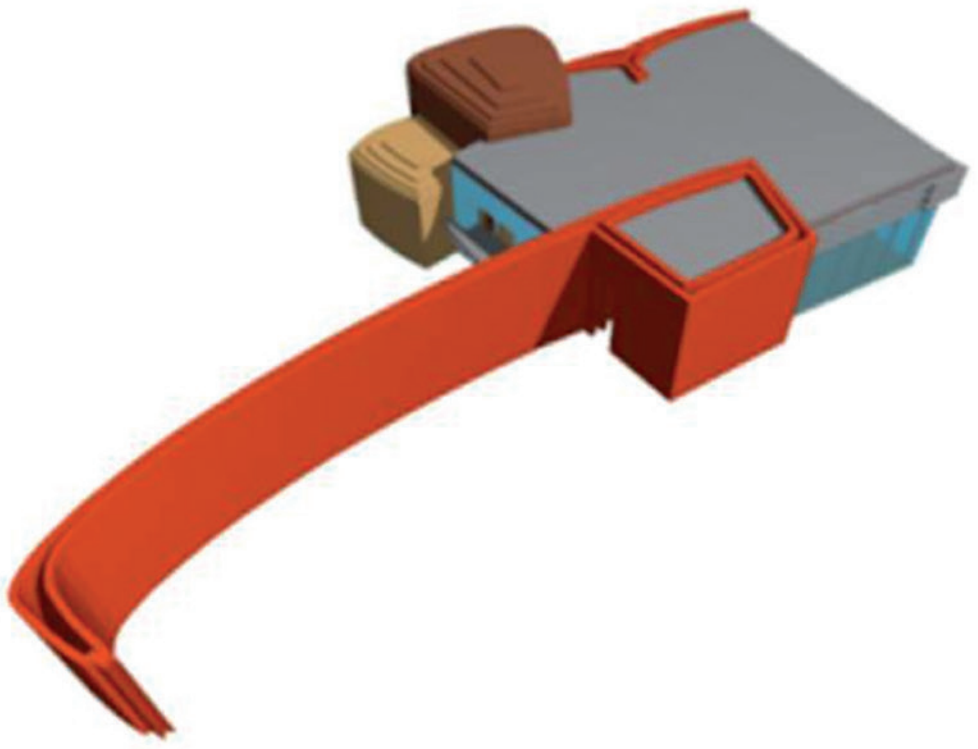

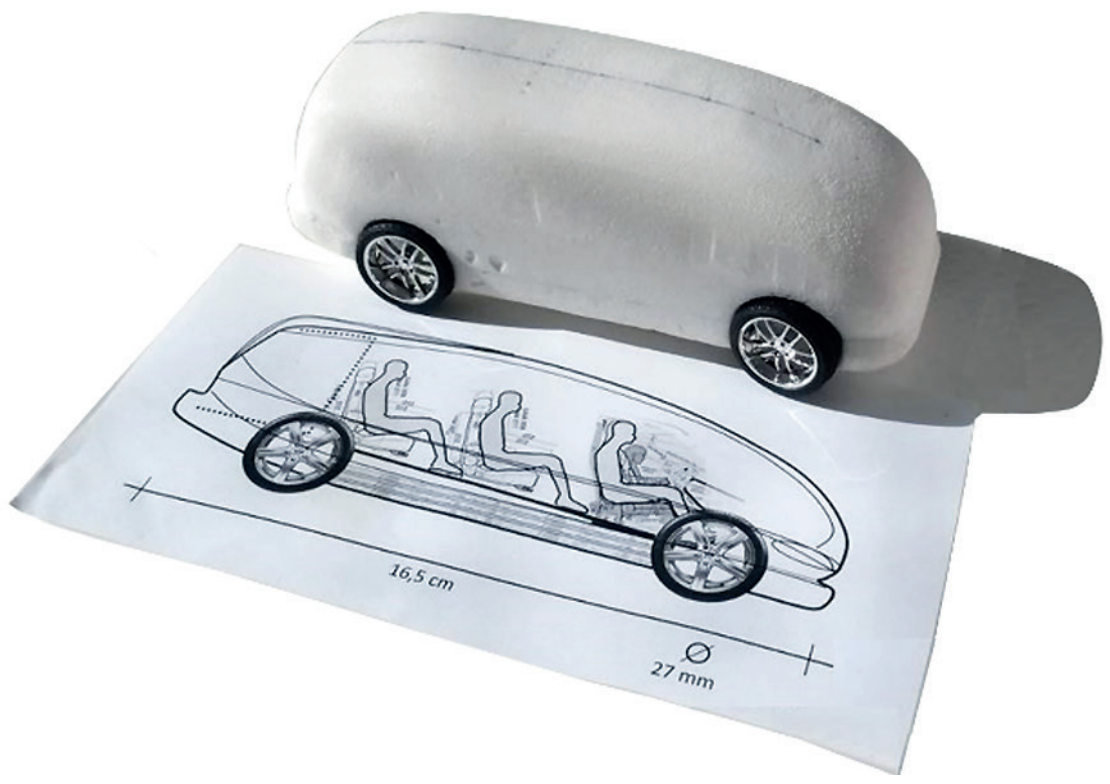

**Juhani Risku**

# Improving the Performance of Early-Stage Software Startups

## Design and Creativity Viewpoints

Editors
Marja-Leena Rantalainen
Faculty of Information Technology, University of Jyväskylä
Ville Korkiakangas
Open Science Centre, University of Jyväskylä

Cover picture: Cynefin framework applied in Article II; Tetra framework applied in Article II; Rankkasauna, Harsh Sauna, house concept 1992; RodMobile, electric car concept 2016-2021.

# ABSTRACT

Risku, Juhani

Over the last 20 years, a very large number of startups have been launched, ranging from mobile application and game providers to enormous corporations that have started as tiny startups. Startups are an important topic for research and development. The fundamentals of success are the characteristics of individuals and teams, partner investors, the market, and the speed at which everything evolves. Startup's business environment is fraught with uncertainty, as actors tend to be young and inexperienced, technologies either new or rapidly evolving, and team-combined skills and knowledge either key or fatal. As over 90% of software startups fail, having a capable and reliable team is crucial to survival and success.

Many aspects of this topic have been extensively studied, and the results of the study on human capital are particularly important. Regarding human capital abilities, such as knowledge, experience, skills, and other cognitive abilities, this dissertation focuses on design skills and their deployment in startups. Design is widely studied in artistic and industrial contexts, but its application to startup culture and software startups follows its own method prison. In the method prison, old and conventional means are chosen instead of new techniques and demanding design studies. This means that when a software startup considers design as a foundation for creativity and generating better offerings, they can grab any industry with a disruptive agenda, making anything software-intensive. The concept of design can be expanded and deepened to a new level. Business can escape the method prison if it adopts artistic design to help stagnant industries and uses disruptive methods with realistic self-efficacy.

Through five partially overlapping articles with varying details, this dissertation clarifies the daily themes and interests of startups required to survive and succeed. This dissertation is a reflective practitioner's investigation of startup practices using a mixed-methods approach. With design-based creativity, startups will be stronger and more successful in the future. They can cause or protect themselves from disruption. Startup can retain customers and its self-efficacy strengthen.

# TIIVISTELMÄ (ABSTRACT IN FINNISH)

20 viime vuoden aikana startup-yrityksiä on perustettu erittäin suuri määrä. Tässä joukossa on mobiilisovellusten ja pelien tekijöitä mutta myös suuryrityksiä, jotka ovat aloittaneet pieninä startupeina. Startupit ovat tärkeä aihe tutkimuksessa ja kehitystyössä. Menestyksen perustekijöitä ovat yksilöiden ja tiimien ominaisuudet, kumppanina toimivat sijoittajat, markkinat sekä yleinen nopea kehitys. Startup-yrityksen liiketoimintaympäristö on täynnä epävarmuutta, sillä toimijat ovat yleensä nuoria ja kokemattomia, tekniikat uusia tai nopeasti kehittyviä, ja tiimin yhdistetyt taidot ja tiedot menestyksekkäitä tai tuhoisia. Koska yli 90% ohjelmistostartupeista epäonnistuu, kykenevä ja luotettava tiimi on ratkaisevan tärkeä menestymisen kannalta.

Aihetta on tutkittu paljon ja varsinkin inhimillistä pääomaa koskevan tutkimuksen tulokset ovat vakuuttavia. Tässä väitöskirjassa keskitytään tietoon, kokemuksiin, taitoihin ja muihin kognitiivisiin kykyihin, joihin design-taidot ja niiden käyttö liittyvät. Suunnittelua tutkitaan laajalti taiteellisessa ja teollisessa kontekstissa, mutta sen soveltaminen startup-kulttuuriin ja ohjelmistojen startup-yrityksiin tapahtuu omassa menetelmävankilassa (method prison). Menetelmävankilassa otetaan käyttöön vanhoja ja perinteisiä keinoja uusien tekniikoiden ja vaativien suunnittelututkimusten sijaan. Jos ohjelmistoyritykset ottavat designin perustaksi luovuudelleen, ne voivat tarjota paremman tuotevalikoiman. Ne voivat tarttua mihin tahansa teollisuuden alaan disruptiivisella agendalla ja tehdä kaikesta ohjelmistointensiivistä. Samalla designin käsitettä voidaan laajentaa ja syventää uudelle luovuuden tasolle. Näin voidaan myös paeta menetelmävankilasta.

Tämä väitöskirja selventää monimenetelmällisen tutkimuksen keinoin viidessä artikkelissa, miten startupit voivat selviytyä ja menestyä paremmin päivittäisissä toimissaan. Tutkimus tuottaa vahvempia ja menestyvämpiä startup-yrityksiä design-keskeisen luovuuden avulla. Samalla ne voivat aiheuttaa disruptiota tai suojautua siltä. Startup-yritykset voivat pitää asiakkaansa ja niiden minäpystyvyys vahvistuu.

**Author** Juhani Risku
Faculty of Information Technology
University of Jyväskylä
Finland
juhani.risku@jyu.fi
ORCID 0000-0002-0587-4431

**Supervisors** Pekka Abrahamsson
Faculty of Information Technology
University of Jyväskylä
Finland

Tuure Tuunanen
Faculty of Information Technology
University of Jyväskylä
Finland

**Reviewers** Kari Systä
Faculty of Information Technology and Communication Sciences
Tampere University
Finland

Pertti Seppänen
Faculty of Information Technology and Electrical Engineering
University of Oulu
Finland

**Opponent** Sami Hyrynsalmi
School of Engineering Science
Lappeenranta University of Technology
Finland

# FOREWORD

It is all about lifelong learning. For me, my dissertation is a milestone in an extensive round. I began my academic studies at the University of Jyväskylä at the Department of Mathematics and Physics in 1976. A year later, I moved to architecture at Tampere University of Technology, where I graduated as an architect. I had already started an apprenticeship as a turbine mechanic, which laid the foundation for my craftsmanship and Arts & Crafts skills.

Later I studied at Nokia Corporation and its Learning Center industrial design and brand management to take part in usability and user experience design. Just before starting my doctoral studies in Jyväskylä, I had been an apprentice for 15 masters' of machinery, acoustics, photography, ceramics, carpentry, stone masonry, forestry, stained glass, architecture and city planning. The University of Jyväskylä finally took me to science. I systematically and scientifically caught up with everything I had done in practice, such as industrial and graphic design, user interface design, and design science. In fact, I was able to understand retrospectively the rationale, background, and future to deepen my practical work.

I was lucky to get Professor Pekka Abrahamsson as my supervisor for my doctoral studies. He is a creative, prolific and visionary scientist and leader. Simultaneously with his scientific input, he leads the Startuplab of the University of Jyväskylä, where new products, services and solutions are created. While the work of the startup lab is at the crest of disruption and scientifically justified, we have fun in the lab. Professor Abrahamsson is known for taking on all challenges, while at the same time developing new ways of working and leading by example. He always answers yes if you have to tackle a wicked problem, and solves it. I want to thank Professor Pekka Abrahamsson for his tireless support and inspiration for both study and design. Through him, civil courage is transmitted to both science and the development of services, products and new futures.

My other supervisor, Professor Tuure Tuunanen, is a special case of the university: he is a multidisciplinary entrepreneur and scientist. In addition, he is a well-known expert in a discipline close to myself, design science. The origins of design science lie in the architects' quest for a systematic approach to design. Under the leadership of Professor Tuunanen, I was able to taste design science. I want to thank Professor Tuure Tuunanen for his always happy and sympathetic support.

I want to thank all the researchers and students of the University of Jyväskylä startup laboratory. There could be no better community around shared enthusiasm. There is always a huge bustle in the Startup Lab. It hurts and happens all the time, each of which gets stimulated and the shared joy grows. I found myself young in the same team where the others are 35 years younger. There is something wonderful and peculiar going on all the time that we are researching and making science into. Special thanks to Johannes Impiö, who finds everything and even more on the internet and makes it a company, video or creative concept. He is a graphic master of notes. I want to thank Johannes for his daily determination to either do business, videos or music. Everything goes. Doctoral student Kai-Kristian Kemell is a person that everyone would need to cope with visions, practice and diligence in science. He saves his scientific partner with his superior diligence. He is also the most credible performer. I want to express my utmost gratitude to my multitalented colleague Kai-Kristian for his example in thorough and diligent research.

I have had the luxury of joining the multidisciplinary research and development community of the University of Jyväskylä. Special thanks to Dr. Elina Jokinen, the most inspiring and cheerful teacher for both writing and communicating about science. During Elina's courses, I met a wide range of people from different faculties and interest circles. An entire university can be built around her. I want to thank research assistant Marja-Leena Rantalainen. Because of her, our university stands stable, and processes run. I also want to thank political scientist Ms. Outi Alapekkala, Sciences Po Paris, our native French from the Arctic Lapland. We wrote together an impressive article about restructuring the media. It changes everything in the media business.

In the end, I wrote my dissertation mainly in the art center Järvilinna, Laukaa, in Central Finland. Its generous heart and inspiring atmosphere, as well as working with Kauko Sorjonen and the artists of the art center, made my work a celebration. Thank you!

I have received strength and courage late from my father, who graduated with a high school diploma at the age of 69 and a master's degree in economics at the age of 75. My mother is still my best supporter, and she always finds an understanding perspective on situations. My warmest thanks to them for the best starting points for studying and science!

My daughter Sade Risku has encouraged me in my last year of saving so that science got "our universal dancer" on its occasion. One day, she will surely dance in the realm of science. My son Pasi Risku and his extended family have always supported my even strange actions with understanding. I want especially like to thank my loved one, Mirja Nylander, who has been following my work throughout the scientific research trip, supporting and facilitating it. Being both a forester and an agricultural and forestry scientist, Mirja, as a farmer's daughter, knows what hard work is. She is also one of the masters in my field of study during my apprenticeships. Mirja has taught me how to sharpen and use a chainsaw, and identify tree species, not to mention forest biology. I got more than most of it. She earths and encourages people to make ever-nobler endeavors. A new order is soon coming to forestry. Thank you, Mirja!

Jyväskylä 25.5.2021
Juhani Risku

## LIST OF INCLUDED ARTICLES

I Risku, J. & Abrahamsson, P. (2015). What can software startuppers learn from the artistic design flow? Experiences, reflections and future avenues. In Abrahamsson, P., Corral, L., Oivo, M. & Russo, B. (eds.), *Product-Focused Software Process Improvement: Proceedings of the 16th International Conference PROFES 2015* (pp. 584-599). Lecture Notes in Computer Science, 9459. Springer, Cham.

II Risku, J. & Alapekkala, O. (2016). Software startuppers took the media's paycheck: Media's fightback happens through startup culture and abstraction shifts. In *2016 International Conference on Engineering, Technology and Innovation/IEEE lnternational Technology Management Conference (ICE/ITMC)* (pp. 1-7). IEEE.

III Kemell, K-K., Risku, J., Evensen, A., Abrahamsson, P., Dahl, A. M., Grytten, L. H., Jedryszek, A., Rostrup, P. & Nguyen-Duc, A. (2018). Gamifying the escape from the engineering method prison. In *2018 IEEE International Conference on Engineering, Technology and Innovation (ICE/ITMC)* (pp. 1-9). IEEE.

IV Risku, J., Kemell, K-K., Schweizer, S., Nguyen-Duc, A., Suoranta, M., Wang, X. & Abrahamsson, P. (2020). What makes a digital game addictive? A player viewpoint on player retention. Accepted to be presented at 2020 IEEE International Conference on Engineering, Technology and Innovation (ICE/ITMC). [1]

V Risku, J., Kemell, K-K., Kultanen, J., Feshchenko, P., Carelse, J., Korpikoski, K., Suoranta, M. & Abrahamsson, P. (2020). Exploring the relationship between self-efficacy and creativity: Case IT & business education. To be submitted.

In Articles I, II and V the author was responsible of designing the research, gathering the research data, analyzing the data, and drawing the conclusions. In Article III, the author was in charge of board game design and realization, user experience and usability. In Article IV, the author was in charge of the game design specific objects, the influence of self-efficacy and creativity in gaming context when playing and designing games, and the impacts of retention in game design.

[1] The paper was peer reviewed and accepted, but not included to the proceedings due to an obstacle to participate the presentation caused by the corona pandemic. Currently, the work is an unpublished manuscript that will be submitted to a conference.

## FIGURES

## TABLES

## CONTENTS

# 1 INTRODUCTION

This chapter describes the research area of this dissertation, and introduces the main concepts of the focus areas. An outline of the structure of this dissertation is also presented.

## 1.1 Motivation

Software startups have significant implications for innovation and economies. These implications have an extensive impact on people, specifically users who adopt their products. Multiple corporations have a history as startups: Apple, Google, Facebook and Microsoft originated as startups. Now they are global actors in the software industry, developing various technologies, including smartphones, computers, applications, search engines and social media platforms (Dolata 2017). The dynamics of software startups can be seen in the cases of Google and Facebook, which are also advertising and marketing companies. Their platforms are interesting for advertisers because of the enormous number of users. Equally, Amazon, Booking.com and Airbnb, being commercial or mediation platforms, have grown to have a large user base (Rochet et al. 2003).

A software startup is a company that creates high-tech software products and services. The company's nature is to be a temporary organization. The nature of the company is that it has little or no operational experience and tries to grow fast and scale its business in extremely scalable market areas (Giardino et al. 2014). Through software startups, the release of new markets, approachable technologies and venture capital globally is actualized (Blank 2005). Startup companies that become successful businesses, like Facebook, LinkedIn, Spotify, Instagram, Groupon and Dropbox, began as fresh technological adventures. Still, many startups fail before achieving their business prospects (Crowne 2002). More than 90% of startups fail, mainly because of self-destructiveness, not competition (Marmer et al. 2011).

One of the main challenges for a startup is to find the right people for all roles, including the core team, management and investors (Seppänen 2018). In the initial core team, there are three roles: the founder, the expert and the implementation team member. In different startups, founders may have the role of expert and the implementation team member simultaneously. The founder's personal capabilities lead to role plurality, carrying out multiple roles in a top-down direction from the founder to the expert and the implementation team member. This is obvious to founders with deeper and broader capabilities (Seppänen 2018). Marvel et al. (2007) divide the founders' praxis into experience depth and experience breadth (Marvel et al. 2007).

The core tasks of a software startup are to execute and manage the creation and delivery of software-intensive products of high value. Product and software engineering practices are key components of the performance and success of the startup. The notions "engineering performance" and "success" need to be specified in the software startup context (Klotins 2019).

In this dissertation, startups have their origin in creativity, innovation and design, the attitude and spirit to find something new and to quickly realize the idea and launch it broadly for public use. The general idea of initial concepts and active ways of working scales from software startups to startups in different industries and trades, moving to a software-intensive mode. All these startups orchestrate their work through software in the form of applications, Web-based solutions and system-level platforms.

The author's personal motivation to research creativity, innovation and design as a reflective practitioner is based on their background in various professional activities. Thirty years of sketching, drawing, conceiving, sculpting, and prototyping and building houses, furniture, prototypes, pieces of art and societal systems has taught the author that systems, objects and details slide into one another and scale in their structure. The only common factor seems to be the qualities of the creators and designers: eagerness to ideate and conceptualize, diligence to craft and iterate, and determination to finish the functionality, form and beauty of the artifact.

Curiously, the winterization of a Segway in Lapland, and riding with it in Murmansk, coroneted the author's career as a designer (Figure 1). The project proved that anything is possible, even when there is a meter of snow, and it is –32°C in the shade.

The qualities above are those of the startuppers. JYU Startuplab as a home base for the author, is a fruitful environment for further research of creation and design. Research is the evidence and assistance for the development to combine practical design and research of startups.

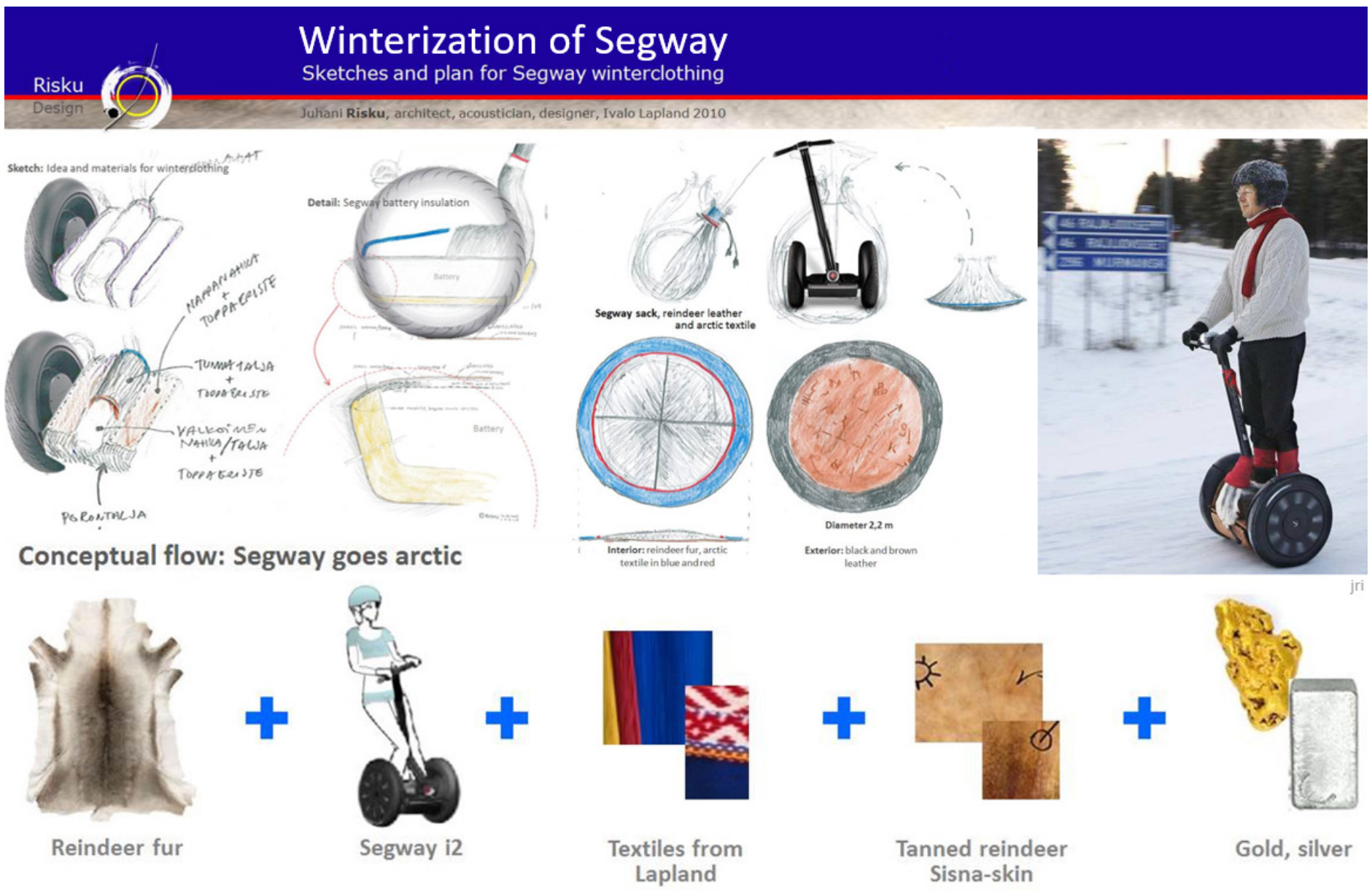

FIGURE 1. Winterization of Segway (generated by the author, 2010)

This dissertation is a horizontal and a detail–system and creation–research product of startuppers' actions in various contexts. Despite the scales or approaches of the research subjects, visions, creativity, innovation and research can be built into the workflow. As designers approach the next assignment, they act as both a researcher and a startupper.

Outcomes from Articles' I-V initiatives with different staring points and processes come to a common and compatible conclusion: startup spirit, as an approach and attitude, scales and motivates to creativity and design to any given task. This dissertation uses a mixed-methods approach with a reflective practitioner's viewpoint. Thus, qualitative and quantitative techniques alternate be-tween the articles and are reflected in the summary (Wu 2011).

Creativity is extremely complex and manifold scientific subject with connections to any matter (Mumford et al. 1988). Creativity has also been ignored as a research problem because of its mystical dimension and spiritual nature, which are not easily adopted in the academic environment (Blumenthal 1980). When Haeberlin (1916) expresses, that "all psychic phenomena are, therefore, creative products of synthesis: they are, when seen from this point of view, acts of will" (Haeberlin 1916), the educated and experienced reflective practitioner can feel the meaning of will, but the notion needs research to be applied to wider creation contexts. One fundamental motivation arose from the expression ~~of Smith et al. (1979)~~ "noncreative people tend to maintain one single interpretation. In order to be creative, regardless of the subject area, the individual must be able to go beyond the conventional interpretations of reality" (Smith et al. 1979). The word noncreative triggers the scientific motivation in form "am I creative or noncreative", can creativity be defined both by practical outcomes and spiritual existence, simultaneously. The fuzziness of creativity motivates to understand creativity as a human capability, a verb to generate something, and to research the meaning of the notion.

## 1.2 Research questions

The general purpose and scope of this dissertation are to clarify the nature of software startups creating, surviving and succeeding in different phases, cases and industries. Design as an umbrella term during all phases of the startup's product development flow, is a factor enabling companies reach their own exceptional competitive factor compared to competitors (Čirjevskis 2016). As critical success factors for design startups, Kim et al. (2018), lists 1) Entrepreneur's ability, philosophy, and leadership to lead the design startup business success, 2) Idea and innovative capability to make or lead the new market for the success of the design startup business, 3) Technology, product, and business modeling capability of the design startup business success, and 4) Funding and financial ability to lead the design startup business success (Kim et al. 2018). This approach offers a set of conditions, environments, requirements and progression patterns from early-stage startups in the game industry to successful ones that made an enormous exit, either they go public or are acquired by another firm. The startups, in case of success, can continue with superb service and solution offerings and grow independently.

A startup requires ideas, skills to develop concepts and financial and business support from experienced parties in order to succeed. The human factors—both the team and their personal qualities—are crucial to startups. A startup is in an ambivalent arrangement: it has a contradictory setting of a highly demanding business environment and technological challenges but has a temporary nature, little or no experience, a lack of resources, possibly technical debts and an incomplete team (Giardino et al., 2016, Klotins 2018).

The main research question is:

RQ. How to improve the performance of software startups?

The research questions are defined as follows:

RQ1. How can design and creativity viewpoints support early-stage startups to develop their ideas to concepts better? (Articles I, II, and V)

RQ2. How can software startups develop their processes for better customer retention? (Articles III and IV)

RQ3. How can creativity be enhanced in software startups? (Articles III and V)

In the following section the articles and their connection to each other is explained.

## 1.3 Structure of the dissertation

As stated earlier, scientific motivation for this dissertation Creativity as a notion, characteristic and concept is a praised and required entirety in the arts, design, education, at the Academia and industries. This is motivating in practice and scientific circumstances. The research consists of five articles that horizontally cover software startups' actions in different domains, circumstances and ways of working. The cases present a broad granularity of businesses, roles of the startups, their internal capabilities and various sets of customers, users, markets and futures. This is what startups face during their life-cycle.

Article I presents a design approach borrowed from art and design cultures, which have developed practices and skillsets that are still in wide use throughout industries. Even an ancient design culture, with its methods, practices and people, can affect startups' survival and success when it is implemented in the culture. A set of actions is presented as an alternative avenue for change. Article I answers to the RQ1 by emphasizing the importance of design skills and utilization of design processes. These are largely used in the Arts, architecture and various design domains, and civil engineering. The importance of creativity and design for startups is crucial, when they have to propose new and unique concepts to survive on the markets. The skillset of various design areas is advisable, because the development speed is fast. Here all communicative means help to ensure the startups' credibility.

Article II describes how software startups grow and cause disruption in global businesses. Later, a startup may face threats from disruptive actions targeted by the underdog parties. The article envisions and builds a strategy to challenge the present-day market leaders using the Cynefin framework to determine actions, their order and consequences for the future. This challenge is seen as a medium for a single startup or internal startup. Article II answers to the RQ1 by encouraging the early-stage startups, on the contrary to the three industrial conglomerates from the mobile network manufacturers, network operators and media houses, the startups bravely challenge the industries by bypassing the problems of three conglomerates. Then the startups can establish something what industries have not done before. Young early-stage startups have several advantages on their side: nobody believes that a novice can penetrate the billion dollar industries with a brilliant concept.

In Article III, the Essence theory included in SEMAT is strengthened through an educational game. SEMAT is a theory of restructuring software engineering to have common ground and punctual rigorous discipline. SEMAT represents a library of practices that form a software development kit with drawings and set of modules to fit any software development cases. To create educational offerings for software engineering students, it is important to enable studies on and dynamic access to the development of structured software. Article III describes a gamified development initiative to create a board game for educating software development and its project management practices. The Essence theory of software engineering as part of the SEMAT initiative, with the ever-present threat of method prison, was also recognized. The board game showed that software development processes can be treated dynamically and situationally rather than monolithically and exclusively. The game flow proceeded through the team's interactive aspirations, ending with a common understanding of dynamic and varying paths while fulfilling the robust target with various alternatives in the team process. For this thesis, Essence-theory provides the mechanism, language and notation to Essentialize practices for startups to use. Article III answers to RQ2 and RQ3 by process development factors: startups in the context of game development, profit from comprehensive design of the balance principle, rewards and how different playstyles and ingame activities effects, and advance social interactions and feelings of achievement as aspects of the game. The question of escaping the method prison is twofold: to free oneself from the method prison, is a function of skills and creativity, and how professionally to utilize the skills in design, and creatively to use development processes.

Article IV describes user and customer retention, specifically player [2] retention in the digital game industry. User retention can be seen as a fundamental factor in any business where a product or service is offered. It presents a threat to the achieved status of a business in a highly competitive market. The article studied the startups' role in ensuring that game elements keep players playing the game. In the article, motivation was studied through rewarding players. These two factors, customer retention and motivation through rewards, are generally important for any business and are highlighted in startups' product creation. Article IV answers to RQ2 by findings that help startup design processes in game design, where multiple factors must be exploited: 1) fine-tuning the rewards, 2) test emotions through achievements by in-house play, and 3) try in advance the game styles, in-game activities, and exploration. By taking these factors into account in the design process, startups can ensure better customer retention as players continue to play and eagerly return to the game.

Article V describes how creativity is correlated with self-efficacy among students according to self-efficacy questionnaire results and design work evaluations. According to the results, self-efficacy and creativity were not clearly correlated. In addition, against the presumptions, seasoned and skillful students in design achieved only medium scores in self-efficacy, which may demonstrate realism and self-critique when comparing oneself to designers in general. Article V answers to RQ1 and RQ3 by evaluating the design skills reflected to creativity and self-efficacy. The design skills were in focus when evaluating the creativity of the students through practical design projects, where visualization and conceptual ideas were emphasized. This combined both RQs by the importance of a design skillset that is useful in planning concepts and visualizing the design.

The summary part of the thesis is structured as follows. Section 1describes the idea of software startups and startups in general, through the common aspects of innovation, creativity and design. In Section 1, the author's motivation is explicated with a brief personal description, where the approach and attitude in the workflow and the memorable highlights of the author's professional career emphasize the role and nature of a reflective practitioner within scientific research. In this section, the research scope and objective are explained, and the research questions are structured.

[2] In this thesis player is a key stakeholder in the digital game industry and referred to as a customer or a player, depending on the given context.

Section 2 defines the software startup as a phenomenon and presents an overview of the challenges that startups face over their lifetime. The means of success are analyzed as meaningful motivations for startups. Because startups are on their way to an established state of entrepreneurship or either going public or been acquired, their life-cycles are studied. An overview of corporative internal startups is presented because they are a medium in which large systemic constructs can be built. Large, complex solutions are highly disruptive and barely possible for small, early-stage startups. Antipatterns, the unfortunate products without customers, are explained in this section.

Section 3 presents the scope of the research and the research methodology. In the five articles of the dissertation, the multifaceted nature of startups and their subjects of creation and development are described. The idea of a reflective practitioner is analyzed and seen as a practically oriented scientific researcher. Typically, surgeons base their professionalism in science and practice. In the best-case scenario, they derive new scientific findings from their practical experience of drilling inside people.

Section 4 summarizes the original publications. Findings and connections to the objectives of the dissertation are presented. Creativity can be seen as a holistic process throughout a workflow, and product creation and design can be considered practical actions. Creativity can also be seen as a disruptive counter-action for friendly revenge, challenging competitors. This happens across businesses and industries all the time. Soft-ware development is filled with processes and the needs of management. Using various self-created processes instead of one rigid procedure is important for startups. In addition, continuously attracting and winning back customers is crucial. Startups require strategies for customer retention. Finally, creativity and self-efficacy are significant qualities for startuppers. Still, it is not self-evident that they are linked in a realistic way.

Section 5 collects the results of the dissertation to describe the future of startups in ever-demanding businesses. Design practices help startups to act more professionally in the turbulence of a competitive environment. Theoretical and practical contributions are described, limitations explained and future research envisioned.

## 1.4 Other scientific contributions by the author

This section lists other scientific articles by the author related to this dissertation. These articles contributes directly to the software startup culture and software engineering practices:

# 2 THEORETICAL BACKGROUND

The theoretical background is based on the topical and situational literature on software startups, Essence, creativity and self-efficacy. The processing method used in the dissertation varies between theoretical studies, practical design, reflection, analyses and visionary outlook, on-site and in-class studies. Systematic literature studies can be run in different ways. There are various guidelines of different first steps, e.g. start with a search strings in different databases or start with reference lists a set of papers. Snowballing is recommended as a first step in information systems. The method used to identify the relevant literature is called backward-snowballing, which refers to a technique to find applicable hits from the reference lists (Jalali et al. 2012). In this study, the references in publications Seppänen (2018) and Klotins (2019) have performed a significant role in software startup research.

## 2.1 Software startups

In this section, the nature of software startups is determined. Large technology companies often have a strategy of innovation and expansion into new business areas in the form of "acquisition instead of cooperation." This means that the corporations expand their expertise through direct acquisitions of startups, not through cooperating with them. This ensures that the startup's resources and competencies are integrated directly into the design and development of the organization (Rothaermel 2001; Roijakkers et al. 2006; Hagedoorn et al. 2000).

Exit strategy is a fundamental target of startups. Startups have four optional strategies for exit: First option, being become public is through an initial public offering IPO. Second option, to be acquired by an industry player. Third option, to be acquired by a financial investor. Fourth option, to apply leverage strategy, possibility to preserve full ownership control with lenders' funds (Deenitchin et al. 2005).

The optimal exit pattern depends on various elements, such as the anticipated lucrativeness of the startup, degree of uncertainty of the product and markets and the imbalance of information between potential buyers and new investors (in-siders, outsiders). In addition, potential conflicts of interest among the buyers and venture capital qualities can create ambiguity for the startup (Akerlof 1970; Basu et al. 2011).

It is optimal for technology startups to be specifically designed to be acquired by larger corporations. These exits are mainly executed a few years after the beginning of the startup. Early exit has also turned out to be more probable than waiting years for an initial public offering (IPO) (Peters 2009). Timing and management of different parties' interests are crucial for startups' exit strategies; otherwise, the dream does not come true.

### 2.1.1 Definition of software startups

Software startups have been defined with different term depending on the source. By adjectives, a software startup is described to be either quick, young, immature, beggarly, resourceless, temporary, creative, robust, agile, fresh, dynamic, unorganized, informal, vagabond, aimless, decentralized, weathercocking (by pivots), of triumph, of monetarization, horny, adolescent, software-intensive, nerdy, unstable, lucky, lottery-winners, tiny, eager, cowboyish, non- engineering, believing, trying, addictive, illusory or concoctive. Many of the terms can be found in the news of startup success or failure, and in the miserable investors' minds.

Here we concentrate on certain important factors of software startups. Carmel (1994) defines software startups as early as year 1994, including their product development characteristics as follows: minimize time-to-completion, increase innovation/features, maximize quality, minimize product cost, and minimize development cost (Carmel 1994). Klotins (2019) defines the product development process:

> Product development [in start-ups] is driven by the following five goals: (1) minimize time-to-completion, (2) increase innovation/features, (3) maximize quality, (4) minimize product cost, and (5) minimize development cost. It is impossible to pursue all five goals at once. Developers must make trade-off decisions - implicitly or explicitly. Any such decision will, by definition, affect time-to-completion. (Klotins 2019, p. 4, xref).

Ries (2011) proposes a broader definition of a startup, which also scales outside the software and technology industries to other lines of business: "A startup is a human institution designed to create a new product or service under conditions of extreme uncertainty" (Ries 2011, p. 17). Blank (2005, et al. 2012) defines a startup as a temporary organization that creates innovative high-tech products and has no earlier operating history. This separates startups from established organizations with more resources and positions in mature markets. Blank (2005, et al. 2012) also states that a startup seeks a scalable, repeatable and profitable business model because of its willingness to grow. Here, the definition differentiates between startups and small business companies that do not essentially strive to grow. Therefore, small business companies lack a scalable business interest and model.

From these definitions Klotins (2018) proposes a compressed definition of software startups. There the notion 'software-intensive product or service' include e.g. software as code and usability, application as an artefact or embedded to physical products, service as a systemic solution from front-end view to all-inclusive and widely spread structure on the Web. As an interpretation of Klotins (2019), a software startup company can be understood as a recently created institution with a focus of launching an innovative software-intensive product or service to market.

As a common conclusion, software startups share many similarities, such as dealing with uncertain conditions, being willing to grow fast, intending to develop innovative products and aiming for scalability (Unterkalmsteiner et al. 2016). Like Sutton (2000) states, there are factors that differentiate software startups from other types of startups, such as changes in the software industry, new computing and network technologies, and new sorts of computing devices. Software startups also need to use frontline equipment and techniques when developing innovative software products and services. Sutton (2000) also describes software startups through the challenges they encounter. First, startups have little to no experience in managing development processes and organizations. Second, they have little to no resources. They try to launch a product, advertise it and acquire strategic coalitions. Third, startups experience several stimuli, including pressure from financiers, clients, associates and contestants to make decisions. Although each of these actors is important, they may cause confusion in decision-making. Fourth, startups face challenges from vibrant technologies and businesses. The seminal nature of software startups urges them to regenerate or perform development with distracting technologies to reach highly prospective and target markets (Sutton 2000). These requirements reflect a temporary, young, eager and indigent phenomenon as a software startup must act in a difficult environment that can shift from tranquil (getting funding) to chaotic (failure and aftershocks).

Startups and small and medium size enterprises resemble each other by characteristics like small number of employees and limited resources (Kamsties et al. 1998, Laporte et al. 2008). Table 1 lists a wide set of characteristics that are not collectively agreed, and they vary by definitions of different authors. Therefore it is challenging to apply these characteristics direct to startups (Paternoster et al., 2014).

TABLE 1. Characteristics of software startups (Paternoster et al., 2014, Seppänen 2018, p. 31)

| **Characteristic** | **Description** |
|---|---|
| Lack of resources | Economical, human, and physical resources are extremely limited |
| Highly reactive | Startups are able to quickly react to changes of the underlying market, technologies, and products (compared to more established companies) |
| Innovation | Given the highly competitive ecosystem, startups need to focus on innovative segments of the market |
| Uncertainty | Startups deal with highly uncertain ecosystems from many perspectives: market, product, competition, people, and finance |
| Rapidly evolving | Successful startups aim to grow rapidly |
| Time pressure | The environment often forces startups to release fast and to work under constant pressure (terms sheets, demo days, investors' requests) |
| Third-party dependency | Due to lack of resources, to build their product, startups rely heavily on external solutions: External APIs, open source software, outsourcing, commercial off-the-shelf solutions, etc. |
| Small team | Startups start with a small number of individuals |
| One product | A company's activities gravitate around one product/service only |
| Low-experienced team | A good part of the development team is formed by people with less than five years of experience and often recent graduates |
| New company | The company has been recently created |
| Flat organization | Startups are usually founder-centric and everyone in the company has big organizational responsibilities, with no need of high-management |
| Highly risky | The failure rate of startups is extremely high |
| Not self-sustaining | Especially in the early stages, startups need external funding to sustain their activities (venture capitalist, angel investments, personal funds, etc.) |
| Little work history | The basis of an organizational culture is not present initially |

A scale-up, scaleup, post startup, or a scaled startup is based on the philosophical statement by Reid Hoffman (2015), the co-founder of LinkedIn: "First mover advantage doesn't go to the first company that launches, it goes to the first company that scales" (Hoffman 2015). Also Markides (et al. 2004) express, that often in case of radical innovations later entrants reign the markets over the early explorers. This may happen in industries of cars, tires, plastics and Web searches (Markides et al. 2004). When an early-phase startup survives its first 2 to 3 years, in fortunate circumstances, it can strive for the growth phase. The expansion of marketing and sales means scaling up the business (Zajko 2017).

When Isenberg (2012) says, "Extraordinary value creation cannot occur without growth, and entrepreneurial growth post startup has numerous challenges which can be an order of magnitude more difficult than simply starting a venture," he recommends various possibilities to re-orient the strategy:

First, a re-start can happen through acquisition, redefinition, spinning-off or arrangement of underused or underestimated resources. Second, you can end the treatment of venture capital as an indicator of policy success and start promoting growth efforts. Third, instead of long-term support from several low-value startups, you can accelerate their extinction and recycling to enrich the local workforce—an integral part of an efficient ecosystem (Isenberg 2012, Zajko 2017). In practice, all startups aim to scale up their business by improving their activities.

### 2.1.2 Challenges that software startups face

The environment and epoch where software startups are working, consists of severe requirements, when building an own role in existing entrepreneurial ecosystems. As Wang et al. (2016) express, based on empirical data of 4100 startups, challenges that software startups face, varies throughout their different life-cycle phases. Startup companies work in broad business domains, like "travel, art and gifts, fashion, e-commerce, social network, idea management, event management, social advertising, project and task management, mobile and social games, luxury hobbies, real estate, e-learning, financial services, health care, etc.". (Wang et al. 2016, p 173). Altogether, the challenges of software startups vary by emphasis of individual cases in uncertain markets, ever changing technologies, uncertainty to get funding, rapid evolution, time pressure, third-party dependency, high risk, and not being materially independent, self-contained.

One specific challenge, concerning creativity and design, are the product related matters. When combining the startup's endeavor to build its first product to the later *minimum viable product* MVP, the product specific challenge is the biggest. When adding *product market fit* as a design issue to the two aforementioned MVP and the first product, product related matters form the most challenging topic of startups (Wang et al. 2016, Table 2, p. 176). In this dissertation creativity and design are important to create artefacts, and design skills enhance to create products. The key challenge, product and design are interrelated, and can be exceeded by required design practices.

These factors of uncertainty and working with the unknown can be clarified into four comprehensive measures: product, finance, market, and team (Giardino et. al. 2015, MacMillan et al.). Here two necessities, succeeding in technological uncertainty and gaining the first paying customer (Giardino et.al. 2015). In addition, contradictions in leadership, like dichotomies between managerial strategies and implementation can lead to startup failure (Giardino et.al. 2014). In Table 2, the top 10 challenges that a software startup faces are organized by the challenge category, its detailed description and to which dimension of the activities of the startups they include.

TABLE 2. Top 10 challenges structured according to MacMillan et al. (1987) classification, here applied to software startups (Giardino et.al. 2015)

| Challenge | Description | Dimension |
|---|---|---|
| Thriving in technology uncertainty | Developing technologically innovative products, which require cutting-edge development tools and techniques | Product |
| Acquiring first paying customers | Persuading a costumer to purchase the product, e.g. converting traffic into paying accounts | Market |
| Acquiring initial funding | Acquiring the needed financial resources, e.g. from angel investors or entrepreneurs' family and friends | Financial |
| Building entrepreneurial teams | Building and motivating a team with entrepreneurial characteristics, such as the ability to evaluate and react to unforeseen events | Team |
| Delivering customer value | Defining an appropriate business strategy to deliver value* | Market |
| Managing multiple tasks | Doing too much work in a relatively short time, e.g. duties from business to technical concerns | Team |
| Defining minimum viable product | Capturing and evaluating the riskiest assumptions that might fail the business concept | Product |
| Targeting a niche market | Focusing on specific needs of users willing to take risks on a new product, such as early-adopters and innovators | Market |
| Staying focused and disciplined | Not being particularly sensitive to influences from different stakeholders, such as customers, partners, investors and competitors (both actual and potential) | Team |
| Reaching the Break-even Point | Balancing losses with enough profits to continue working on the project | Financial |

As seen in Table 2, three dimensions describe how challenges influence the various development and learning phases of startups. The product is typically a software product, an application, digital service or a physical, software-intensive product. The financial dimension is crucial for the progression of the startup when trying to grow and position itself in the market. Knowledge of the market is fundamental to understanding customer needs. When the market and product-specific challenges are similarly significant in the early problem evaluation stage, the market comes to the fore of the startuppers' perception in the mature stage of the product development. The product and detail-level factors are the focus of early-stage concerns; the strategic market-specific factors are emphasized later during the mature stage (Giardino et al. 2015).

The founders' roles in startups as leading persons are significant (Seppänen 2018). Therefore, it is crucial that the founder can bear the central challenge of leadership and cross-functional operations. When recruiting members to the initial startup team, the founder must broaden their human capital. As Seppänen (2018) states, a balanced startup team is balanced both externally (the problem) and internally (the startup's mission to deliver its tasks successfully). A balanced startup team is achieved when the common structure of the generic human capital model can be identified. Three factors affect the ability to create a balanced startup team: the founders' prior human capital, team growth and learning. These factors are dependent on several conditional elements: the problem and the product idea, the founder's earlier human capital (Marmer et al. 2011), the customer segment and customer contacts, the human capital available and the financial situation of the startup (Seppänen 2018).

Human capital theory claims that individual employees have a range of skills and talents that they can enhance or acquire through training and schooling (Becker 2009). The theory of human capital can be applied to human activity at different levels, from individuals to all of humanity. In the context of business, enterprise and entrepreneurship, human capital is defined as an individual's cognitive abilities, experiences, knowledge and crafts; these have been expanded to include the groups in which these individuals operate (Becker 2009). In this context, the founder of the startup has a significant role in business performance: "To be more specific, former experience of the business founder in the industry in which he starts his business appears to improve all performance measures" (Bosma et al. 2004, p 12).

Human capital shortages are a challenge during the early phases of a startup's evolution. Iterative processes have been proposed for startups (Ries 2011; Bosch et al. 2013). Both linear and iterative processes are recognized during a startup's evolution path (Seppänen et al. 2018). During iteration sessions, the business value is validated, and solid customer cases are created to consider different ideas. The iteration cycles can be semi-controlled or uncontrolled. The purpose of this semi-controlled iteration round is to decrease the uncertainty and risk-level typical of early-stage startups (Paternoster et al. 2014). The uncontrolled iterations lead to learning for the initial teams, as seen in the semi-controlled iterations (Ries 2011; Bosch et al. 2013; Seppänen 2018).

Interestingly, Unger et al. (2011) found a stronger dependence between success and dominant human capital (original skills and knowledge) than between success and education (investments in human capital). This presumably returns to the origin of startup culture, where young, eager people used scarce experience and brilliant ideas to surpass all expectations. Startups that can balance challenges that require tenuous skills and an extreme will be successful and make a difference.

In this dissertation, the definition of a software startup is interpreted in the compressed form given by Seppänen (2018), "that a startup is a small company exploiting under high uncertainty a new business opportunity with software-intensive products, services and/or solutions that are not well known."

The notion of a software-intensive system allows a startup to participate in industries, branches of businesses or domains other than software. A software-intensive system is defined as "any system where software contributes essential influences to the design, construction, deployment, and evolution of the system as a whole" to encompass "individual applications, systems in the traditional sense, subsystems, systems of systems, product lines, product families, whole enterprises, and other aggregations of interest" (IEEE 1471, 2011).

Software-intensive startups in the farming sector, “AgTech,” were successful in collecting venture investments in 2014–2015 (WEF 2016). In the retail and finance industries, Kickstarter offers a global crowdfunding platform that enables creative projects. In the music delivery and purchase industry, Spotify has developed a novel way to consume music online. Airbnb, an online holiday rental trading service, has created a disruptive service that competes with traditional hotels by renting out rooms in private homes globally (Shontell 2012). These examples illustrate the diversity of software-intensive startups. Giardino et al. (2016) point out that from an engineering perspective, startups must constantly apply predominant knowledge in order to open up disruptive routes to better their business branch. They have to organize training for full-stack engineers, develop techniques for constant challenges in engineering and develop methods to manage technical debt (TD) (Unterkalmsteiner et al. 2016).

### 2.1.3 How a software startup succeeds?

The definition of success, according to Merriam-Webster (2020), is “favorable or desired outcome.” Lexico (2020) defines it as “the accomplishment of an aim or purpose.” In the startup context, a favorable result is to be acquired by another company or to reach an IPO on the stock markets. For a startup, it may be more profitable to aim for acquisition and an IPO if the risks from present operations are higher than the estimated profits of the exit (Wennberg et al. 2014).

**Performance of the team**

Performance is the ability to gain results for targets defined in advance (Laitinen 2002). Performance can be identified in detail by measuring turnover, market share, product quality and internal efficiency by their changes (Reijonen et al. 2007; Egorova et al. 2009). Good engineering performance, according to Egorova et al. (2009), means that 1) customers are involved in the engineering process, 2) there is a good under-standing of customers’ problems, 3) the team follows relevant best engineering practices, 4) business and engineering objectives are aligned, and 5) tasks are completed on time and within. In addition, a good performing product features 6) good quality in relation to customer expectations, 7) customer satisfaction with features and quality, and 8) reliability and suitability for the intended purpose budget (Egorova et al. 2009).

The performance and success of a startup depend on the diversity of factors behind software engineering, such as the idea, proficiency of the original team, structure of the organization, economic situation, repertory of marketing and sales strategies, business models and other relevant factors (Chorev et al. 2006). Regardless, product engineering is one of the central actions of startups and has a significant impact on the inclusive performance of the startup.

**Success factors**

Success factors for software startups include business model elements like unique selling points, the market, resources, collaborators, cost and revenue frameworks (Huarng 2013; Dubosson-Torbay et al. 2002; Osterwalder et al.

Pigneur 2010). In addition, components of success can include the startup's attitude and methods during the development cycle regarding the target, product itself, customer needs, financial advantages, operating model, earnings, the core team and the culture of ways-of-working (Churchill et al. 2000; Kumbhat et al. 2017). Still, neither of these success factor classifications self-evidently leads to success in practice. Kumbhat et al. (2017) also proposed a list of categorized success factors of software startups.

The success factors for software startups in structured form are as follows:

- Factors related to Product Offering
  - Value Drivers
  - Innovation
  - First Mover Advantage & Scope for Late Entrants
  - Quality and Time to market
- Factors related to Startup Team
  - Personality
  - Human and social capital and social skills
  - Entrepreneurial Experiences
  - Gender
  - Attitude towards Radical Innovation
  - Effectuation
  - Founding Team
- Factors related to Other Environmental Circumstances
  - Market
  - Investors
  - Investment Sizes and Timing
  - Stock Options to Employees
  - Location
  - Culture
  - Others
    - Age of Entrepreneur
    - Pre-startup planning
    - Capacity to stay afloat
    - Right mentors
    - Internationalization
    - Cost optimized operations
    - Balanced scaling

The list is clarified as follows.

**Product offering**

Product offerings, whether a physical product, service, application or a large solution, are the central outcomes for software startups. The product itself is crucial for the startup's success. The most common ways for startup companies use to develop their products are copying existing products, prototyping, leveraging expert assistance, and working together with customers (Seppänen et al. 2017). Software startups can affect the range of offering under the notion of software-intensive products. When main part of the software-intensive innovations are dependent on Open Innovation (OI), and often are carried out with Open Source Software (OSS) (Wnuk et al. 2016), the software startups can benefit from these. Value proposition enabled with business deal performance is important for the buyer and seller. Here, rapid decision-making is crucial, as is offering a wide scale of product or service alternatives. In addition, by creating attractive offerings for potential customers, including vertical and horizontal complementary products and services, a bundled set of products or service extensions can be delivered simultaneously. These extended offerings on top of the initial product can include horizontal add-ons (e.g., and online travel service that suggests topics that may be of interest for the traveler, such as weather reports, currency exchange or clinical service information). An example of vertical complementarities would be a software startup that offers produce customization, personalization or individualization services to their market, users or brand (Amit et al. 2000). Complementary items increase customer retention through positive lock-in so that the customers remain motivated purchasers of the products. This is a significant factor for an early-stage startup's business because it is very expensive to attract new customers.

**Value drivers**

Amit and Zott (2001) present four dimensions for value creation potential in e-businesses: efficiency, complementarities, lock-in, and novelty. Efficiency reduces costs, complementarities are bundled to the product to provide more value than the product alone. The lock-in effect binds customers around the company's offering and thus prevents strategic customers from shifting to competitors, creating value for the company. Novelty relates to innovating and pioneering with products and services. These dimensions also act in combinations with each other (Amit et al. 2001).

**Innovation**

A startup gains success through the innovation culture that it builds beside intellectual property rights (IPRs). Independence from routines and resilience in ways of working strengthens this innovation attitude. An early-stage startup should focus on mass markets instead of experimental products because innovations of this kind may surpass the capabilities of the startup and endanger its subsistence. This advice for young startups is twofold: Externally generated innovations developed with a partner are usually produced gradually. In this situation, the established partner company has a better negotiating position. In contrast, innovations with a disruptive character are easier to produce internally in the startup (Rosenbusch et al. 2011). This can be contradictory to the mass market–oriented solution. A risk-centric startup may find the disruptive development pattern more tempting and lucrative than the conventional method.

Choia et al. (2007) observed that unlawful or only partly legal organizations are pioneers in applying new technologies and creating business opportunities. These organizations also provide invaluable market insights and help the emergence of new legal and canonical business models for legitimate industries (Choia et al. 2007). That is why startups should always keep an eye for inspiration from such sources. Pirate communities and diverse developers of revolutionary innovations are on the frontline of disruptive innovations. Startups should monitor these parties when seeking opportunities to re-engineer and transform their own business practices and emergent technologies for legitimate affairs (Kumbhat et al. 2017).

Christensen (2013) expresses that companies can be both competitive and strong by following new technologies while still improving existing technologies. In addition, he proposes that companies should studiously look for forerunner status in businesses only when the new technologies are disruptive. Emerging markets have generous first-mover benefits and earnings for groundbreaking companies. Examples include the hard disk, motorcycle and personal desktop computer manufacturing.

Christensen (2013, p. 9) defines technology as "the processes by which an organization transforms labor, capital, materials, and information into products and services of greater value. All firms have technologies." Christensen (2013) states that disruptive products are more delicate and simpler when compared with the conventional products available. Moreover, disruptive products are likewise considered inexpensive, smaller, simpler and handier to use than the dominant products on the market. Disruptive technologies, generated by innovation, emerge first in small markets, before reaching the status of mainstream products and producing remarkable business profits (Christensen 2013).

Christensen's (1997, p. 18) list of established and disruptive technologies lists 24 pairs of established and disruptive technologies, including Silver halide film vs. digital photography, notebook computers vs. hand-held digital appliances, printed greeting cards vs. free digital cards on the Web, and offset printing vs. digital printing. This comparison produces several interesting observations: Christensen identifies early the evolution from old to new technologies, e.g. the emergence of the electric car. Also, some predictions now have a new form. As Christensen (1997) argues, medical doctors seek online medical advice rather than the advice of peer practitioners in-house, as Ma et al. 2018 continues (Ma et al. 2018). A unique and timely development occurred due to the Covid-19 pandemic: classroom and camp-based instruction transitioned into distance education over the Internet (Khalil et al. 2020).

Today, when startups follow new technologies and want to operate within them in either a conventional or disruptive way, they have to find sources that contain analyses and structural patterns of the future technologies and business opportunities visualized in a consistent way. In addition, the startup has to decide what alignment to follow: disruptive or conventional product creation. Several of the outcomes predicted in Christensen's (1997) comparisons are mainstream industries today and are business-as-usual. Surprisingly, the electric cars predicted are not yet a realistic threat to gasoline cars; Christensen (1997) still mentions that the electric car is a potentially disruptive technology and a future threat to existing automotive companies.

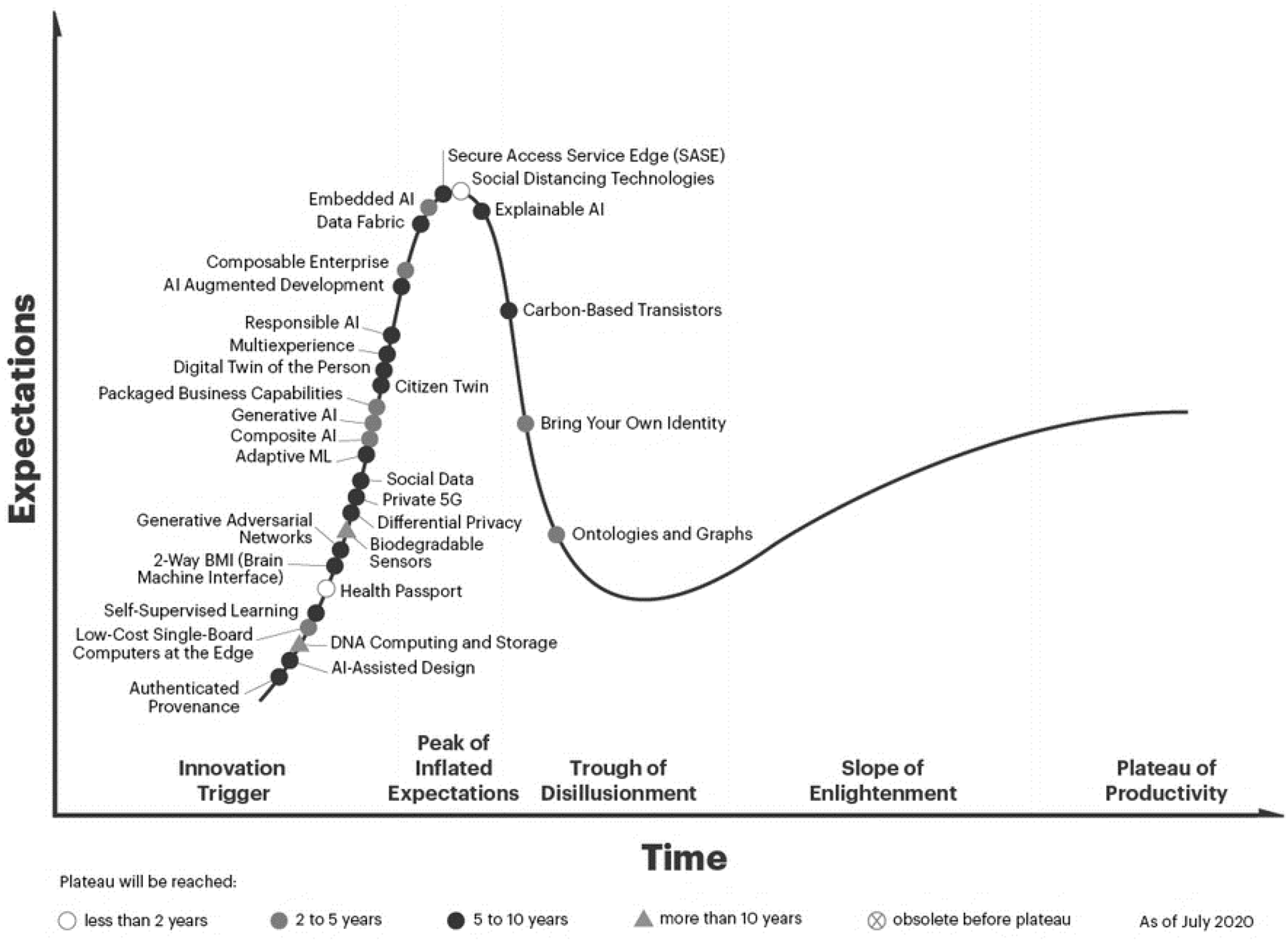

FIGURE 2. Hype Cycle for emerging technologies (Gartner 2020)

As a present-day reference to Christensen's (1997) comparisons, the Gartner Hype Cycle (Gartner 2020) is a collection of embryonic technologies that need a vast amount of research, development and resources. Gartner's hype cycle is one of the most significant and the most influential consulting models for large companies counseling on their future technology strategies (Figure 2). It is used by both practitioners and academic researchers to explain and confirm whether to invest or not to certain technologies (Steinert et al. 2010). The Hype Cycle opens opportunities for internal startups in large corporations or their joint ventures rather than in small early-stage startups.

A software startup has a strong advantage when entering a market early as a first-mover. This results in better operating income, a better share of the market and opportunities to survive in the business (Lieberman et al. 2013). However, the first-mover may face both advantages and disadvantages. When the winner takes it all, it is an advantage, but if the industry or technology is immature, the first-mover may encounter disadvantages when trying to access the market (Markides et al. 2013).

A startup can demonstrate technological leadership through a steep learning curve, IPRs and patents, or by developing products in secrecy (Lieberman et al. 1988). The learning curve and first-mover advantages and effects happen mostly simultaneously, as "the learning curve creates entry barriers and protection from competition by conferring cost advantages on early entrants and those who achieve large market shares" (Spence 1981, p. 21).

Amazon profited from its first-mover position in e-commerce when it entered the book e-retailing market. Amazon was able to generate greater profits than its competitors did 20 months earlier. With this business offensive, the company had a head start in understanding customer preferences. It patented its processes and had a better opportunity to bargain with publishers about its new channel of market entry and the pricing of the service (Mellahi et al. 2000).

When a startup develops products in secrecy, it avoids public attention. A strict disclosure level fosters privacy. This means reducing information exchange between other risk-oriented ventures by emphasizing the protection of secret information like behavior and performance of the company itself. Fostering transparency means that plans that highlight collaborative interaction more likely lead to the disclosure of partners' manner and execution (Cohen et al. 2019). The lean startup tries to get customer feedback when preparing a minimum viable product to match the outcome to the customer's needs.

Blank (2013, p. 6) positions two opposite approaches, the lean startup and the stealthy startup:

> During the dot-com bubble, startups often operated in stealth mode (to avoid alerting potential competitors to a market opportunity), exposing prototypes to customers only during highly orchestrated beta tests. The lean startup methodology makes those concepts obsolete because it holds that in most industries' customer feedback matters more than secrecy and that constant feedback yields better results than cadenced unveilings Blank (2013, p. 6).

**Team**

A fresh startup team is at the center of balancing the survival and success of their intention because the performance of entrepreneurial intention, as a driver for success, differs from the struggle for survival when they are alternatives for daily actions (Luca et al. 2012). An openness to experiment and a tendency to take risks are favorable factors when starting a business but are negatively correlated to startup survival and success because the average level of risk is related to sustainable survival, while the extreme (low and high) levels of risk-taking lead to the abandonment of the endeavor (Zhao et al. 2010).

These factors depend on the personalities of the team. Individuals with stronger self-efficacy and an achievement-oriented attitude not only favor challenging activities but also have a higher resilience in these activities, leading them to become successful entrepreneurs. Qualities of personality are described by capabilities (e.g., verbal, numerical, spatial and emotional intelligence), motivations (e.g., desire for affiliation and achievement), values, attitudes and actions (e.g., openness to experience, conscientiousness, extraversion, agreeableness and neuroticism [Brandstätter 2011]). Attention to detail, high energy, a proactive stance and a positive mindset are personality aspects of entrepreneurs involved in successful companies.

The ability to identify great opportunities and obstacles in a timely manner also leads to success for businesses. Successful entrepreneurs also have significantly greater adversity quotient scores (Stoltz et al. 2000) and higher perceived management adversities, are more capable of resisting and quickly coping with setbacks and failures than others, and take greater responsibility for the outcome in adversity regardless of their origin (Markman et al. 2003). Preston (2001) proposed that the chances of a startup company succeeding significantly increase according to team size, as long as the size reaches four or five founders. A founding team of two to three people is considered the most ideal. When the skills of the founding team members complement one another, the company's chances of success are greater than with a single technology (Vidal et al. 2013). Similarly, when skills and abilities are combined, team members with complementary skills contribute to the realization of the first-mover's impact (Preston 2001).

**Environmental conditions**

Environmental conditions form a significant potential success factor for the working context of early-stage startups. The startup environment consists of business actions, business interests, financial factors, physical environment, culture, people and other operative factors. The markets dictate the nature of environmental attributes. The products that the startup is developing might be an everyday thing or uninteresting addition to existing markets. When directing a new market with new products and services, startups need a long-standing engagement to educate customers and to institutionalize the offering. These actions demand wide-reaching efforts, which jeopardize the continuation of the startup. If the target market is developing by itself, it invites investors with its glamour and appeal. A large consumer market with a high potential for growth allows a startup to scale its assortment and captivate new investors. The function of investors is crucial for startups because they are able to provide significant leverage, which opens new doors and saves time-consuming efforts (Kumbhat et al. 2017).

In addition to enabling competitors to enter the market aggressively by killing startups in small batches, appropriately timed smaller investments are important for the founding team of a startup, as they require the team to spend time growing the business rather than only raising money. This is important because the amount and timing of fundraising are critical to a startup's survival and success (Kumbhat et al. 2017). When ownership is distributed among employees, they develop a different attitude and participate industriously in the growth and survival of the startup. This ability to provide stock options to employees has a positive impact on startup success (Preston 2001).

Location and culture are important contextual surroundings for a startup's survival and success. The location of a company is a key factor for success, which is why the business should be located close to the main competitors or its most arduous customers: financial clusters. Locating a startup in a cluster with complementary or competitive skills makes finding the necessary resources, support and infrastructure easy (Porter 2001; Pe'er et al. 2013). It is important that the operative culture is passionate without fear. Failures and drawbacks should be celebrated as they are very important to advancing the startup's business and internal spirit. The influence of innovation on firm performance is much stronger in cultures with a nature of collectivism (typically in Asia) than in individualistic cultures (like the U.S.). Generally, a higher level of individualism hinders the synergies of teamwork and social interplay (internal and external). These are essential for the success of innovation and its influence on the firm's stability (Rosenbusch et al. 2011).

As another success factor for an early-stage startup, mentors play a remarkable role in supporting the team, especially for a startup based on exceptionally high-tech solutions. It is important that startups have clear insight and vision to choose the most appropriate mentor to facilitate the development of the forefront technologies (Jain et al. 2017).

The differences between startups and established companies are fundamental. Despite this, a large and experienced company can have, in certain departments and even divisions, a fresh and innovative spirit that does not reflect the whole organization.

### 2.1.4 Life-cycle of a software startup

The life-cycles of diverse small and medium enterprises have been studied intensively. The focus has been the phases of the business development process and the respective conditions. A critical event, crisis, pivotal impetus or momentum, when the conversion from one stage to another happens, is an important moment in the evolution of a small or medium enterprise or a new company. There are commonly recognized challenges and crises that happen during these phases, and instructions are provided to conquer them. Some of them propose organizational and managerial changes and upgrades like renewed ownership structure or complete direct control in management (Kroeger 1974; Galbraith 1982; Quinn et al. 1983).

Because life-cycle models are not universal and specifically self-evident, the number and content of life-cycle stages are questionable (Phelps et al. 2007; Salamzadeh et al. 2015). Therefore, it is important to identify the steps that startups follow during their growth. A life-cycle model identifies the functions, organizational resources, entrepreneurial qualities, milestones and players through which startups can acquire the necessary resources, develop entrepreneurial capabilities, achieve results and overcome potential emerging crises and problems. Reaching a milestone is a crucial moment for any startup, as it shows the transition to the second phase (Table 3).

One way for startups to survive is to build partnerships with different parties in the business. These include venture capitalists and business angels, researchers and educators, innovators, science and technology clusters and incubators, other partners, companies and suppliers. In this way, startups are able to fill their resource gaps and respond to various challenges effectively. Thus, the study of the life-cycle of startup companies and the challenges and factors of the different stages is topical. Development paths need to be identified in order to reduce the number of failed startups. In addition, the weakness of the global political, economic and social system, the lack of dynamism and the identification of small and medium-sized actors need to be developed.

TABLE 3. Startup's four-stage life-cycle model by Passaro et al. (2016), based on interviewing startups and the literature review

| | **Ideation** | **Intention** | **Start-up** | **Expansion** |
|---|---|---|---|---|
| **Definition** | Potential idea generation | Entrepreneurial intention readiness, opportunity validation, and pre-start-up activities | New venture creation | Consolidation, scalability and self-sustainability |
| **Required resources and capabilities** (organizational level) | Technical resources and entrepreneurial culture | Financial, technical and managerial resources | Financial, technical, physical and managerial resources | Financial, technological, physical and managerial resources |
| **Key factors** (individual start-upper level) | Creativity, Intuition, prior experience | Entrepreneurial and risk-taking orientation, self -confidence, motivation | Entrepreneurial and risk-taking orientation, self-confidence, leadership | Leadership, coordination ability, strategic orientation |
| **Key activities** | • Discovering idea<br>• Market opportunity intuition<br>• Resources needs and availability | • Market opportunity validation<br>• Engagement/commitment<br>• Team building<br>• Resource searching/validation | • Business planning<br>• Product and commercial development<br>• Searching for additional funding resources | Massive customer acquisition, back-end scalability improvements, new personnel and first executive hiring, internationalization |
| **Milestones** | Idea viability | Prototype | 1st invoice | Scale-up |
| **Ecosystem's actors** (mainly) | Higher education systems, Governor and local agencies (entrepreneurial culture developers) | Higher education systems, Start-ups initiatives, Family, Friends, Business Angels, Fablab, Business centres, TTO, Incubators | Incubators, Accelerators, co-working spaces, crowdfunding platforms, Venture capitalists, partners (suppliers, customers, SMEs, large firms, start-ups) | Accelerators, venture capitalists, partners (suppliers, customers, SMEs, large firms, start-ups) |

When startups themselves know how to manage and utilize optional functions and resources, they can progress through their life-cycle, from idea to business arrangements. In this way, startups can develop their own strategies from market to market. The startups can also create their own business and socio-economic position and value for the future (Passaro et al. 2016).

The four stages of the startup life-cycle model (Passaro et al. 2016), as seen in Table 3, are defined as follows:

**Ideation:** The startup creates a prospective idea and focuses on developing it. Emphasis is placed on the significance of the potential to fulfill the problem (Marmer et al. 2011). The startup concentrates on ensuring a realistic market opportunity (Keating et al. 2010).

**Intention:** The startup highlights converting the idea to fit the business. The market opportunity has to be validated to occupy the customers' interests. In this phase, the important qualities of the startup are self-esteem, motivation and risk management.

**Start-up, starting up:** The fresh business based on the recent feasibility studies of the developed idea for the markets is launched (Davidsson et al. 2003). At this step, the startup evaluates the actualized business by its idea relevance and success and recognizes the required positive or negative assets by their relevance. During this period, a startupper can be considered a serious entrepreneur who invests their time and effort in running a viable and independent business (Reynolds et al. 1997).

**Expansion:** The startup has scaled the business to respond to the markets, and it is now self-sustaining. The startup can begin to grow its expertise and capabilities to increase revenues, encourage personnel to stay motivated and engage and sympathize with the customers and providers. An operation to plan future market actions abroad and recruit potential business partners are effective strategic procedures. At this stage, startups should begin to develop their leadership, strategic orientation, and coordination skills (Ensley et al. 2000; Brännback et al. 2008).

Salamzadeh et al.'s (2015) startup life-cycle model (Figure 3) provides a conclusive view to the flow of startups' evolution. The bootstrapping stage is the early phase, where the idea is turned into a business. During this period, investments are collected from the classic setup called "fools, family and friends" (Khayesi et al. 2011; Vissa 2010, 2011). This is valid because of the closeness of the first investors. These near individuals are helpers from the social relationships of households, diverse friendships and business connections. In these cases, an agreement of financial compensation is usually arranged (Biais et al. 2008; Hellmann et al. 2002). In general, startups and close investors rely on individuals with whom they have worked in the past because information about the reliability of these individuals is readily available (Granovetter 1985; Gulati 1995). Therefore, the primary investors for startups are people they know and can trust (Kotha 2012).

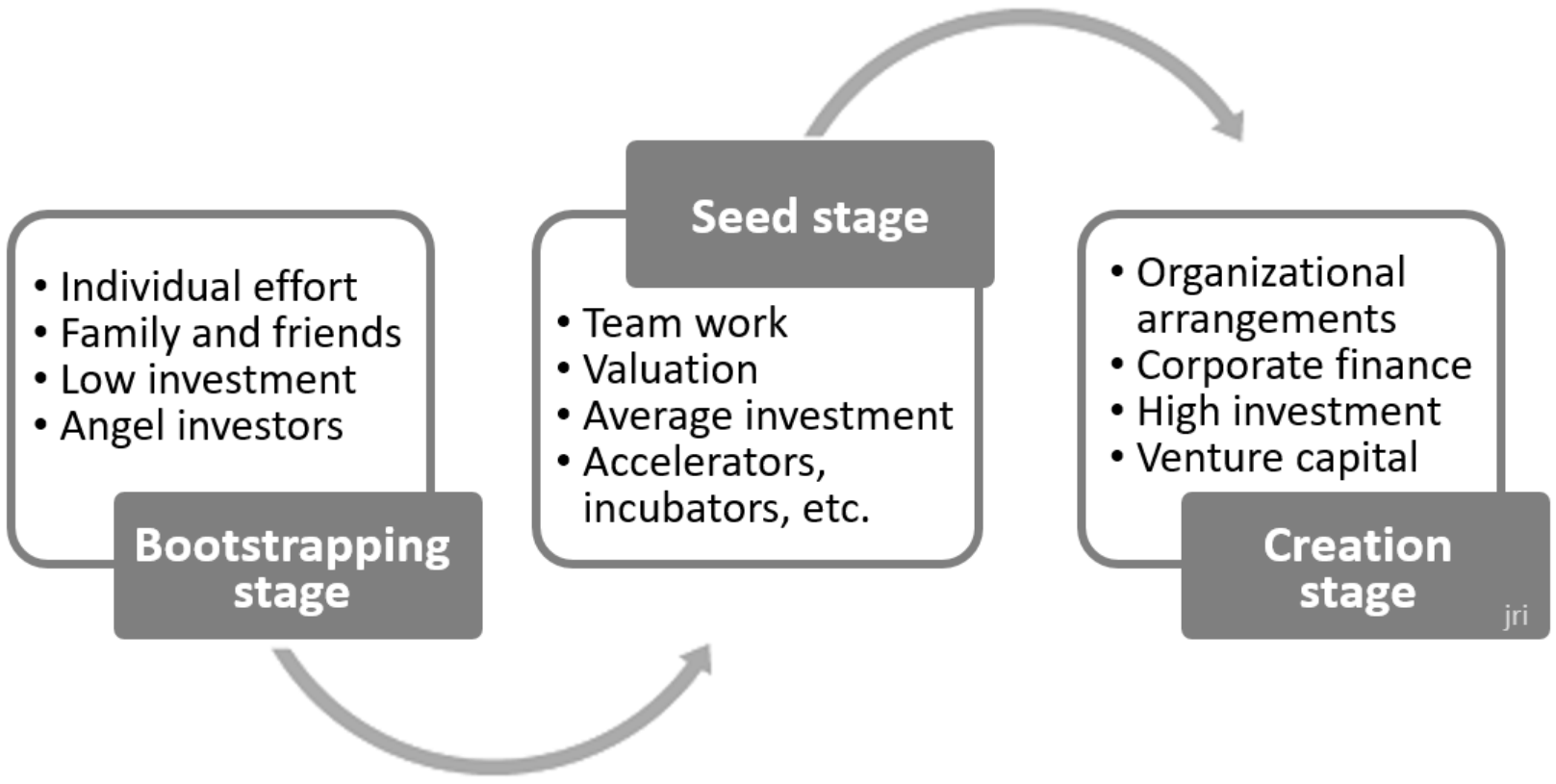

FIGURE 3. The life-cycle of startups (Salamzadeh et al. 2015); the sequence of activities and stages might vary between startups

### 2.1.5 Internal startup

Today, internal startups in conventional and large corporations are innovation accelerators for companies. An internal startup is a separate organization with agreed-upon independence, new ideas and innovations as a target. The internal startup may have a manifold role in the product creation process, either as the creator and innovator or as the testing team to iterate and reshape existing products or to test and plan the go-to-market actions and product launch. A testing team may even reinforce the current product portfolio.

In this dissertation, startups are discussed both in the early-stage and innovative sense, and in the latter form, as an innovative, established company with disruptive solutions, services and products. Both of these definitions are discussed in Article II, about how software startuppers have taken the media's paycheck. Companies that originated as startups found innovations and practices (e.g., Google) to take the livelihood and financial foundations from print media houses. These companies conquered the advertisement business by redirecting the advertisers' payments for daily ads in newspapers and Web news platforms to a new host through search engines and social media algorithms.

### 2.1.6 Antipatterns

When developing innovative technology-centric products and services under extreme conditions and ever-changing markets, startups face different challenges according to their stage of product development. Three phases stand out: constructing and publishing the initial release of a product, appealing to customers to choose the product over early adopters and beta testers, and expanding the product to new markets. Klotins (2019) argues, that antipatterns express challenges related to the release of the first version of the product, captivating customers, and broadening the product to new markets. Antipatterns indicate that the challenges and failure scenarios that seem to be connected to a business or market, are based, at least in part, on deficiencies in technology. Klotins (2019) presents three separate antipatterns: (1) failing in getting the first product launched, (2) not engaging consumers to the product, and (3) challenges to scale the product for new markets (Klotins 2019). These three phases describe the antipatterns that a software startup encounters during the progression process.

In Figure 4, the plotted line represents the expected growth of a company. The arrows inclining downwards indicate alternate paths, antipatterns that hinder the development of the startup. Under the figure, in the text boxes, the lists express potential reasons for the progression. The first antipattern, I, embodies the notion of (not) releasing a market-worthy product. This first version of the product, called the minimum viable product, is good enough for basic use and is a test version to find out if the customers want it. Simultaneously, the minimum viable product reveals if the concept is valid for further investments.

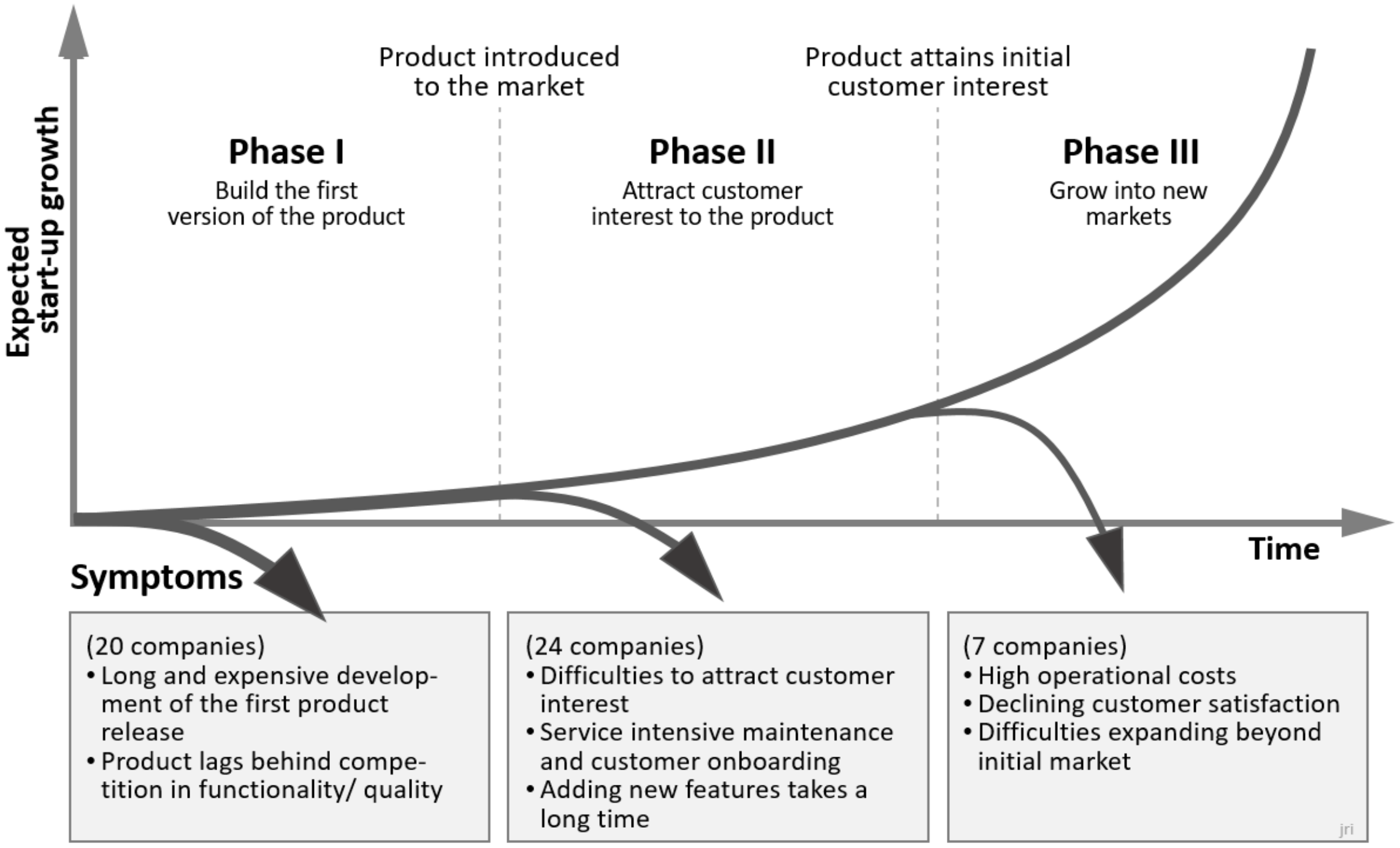

FIGURE 4. Startup milestones and symptoms of antipatterns (Klotins 2019)

Because of the importance of the rapid launch of the first product, startups emphasize simple and quick evolution processes for the first release. Still, startups report that building the first variant of a product is too slow of a process, requires too many resources and decreases motivation. The market potential is worn out, if not lost. The first version of the product is an indicator of team coherence and coordination of the work. At the same time, it depicts whether the team has the appropriate skills and expertise to produce an adequate product (Klotins 2019).

Software business practitioners claim that time-to-market is shorter when integrating existing open source or third-party components. These components generally provide a number of interrelated functions straight out-of-the-box, thus placing emphasis on the main features exclusively (Munir et al. 2016). What a startup decides to build or buy and what components it decides to use depends on the desired features of the product. Therefore, it is important to know what kind of quality aspects are important and what level of quality is expected (Regnell et al. 2008). The focus is placed on the conventions of how the customer uses the product rather than creating new lists of feature ideas (Cloyd 2001). Development time and resources are saved when fewer features are developed. Quick processes and attitudes during the development of the first product version also save effort, especially if features late prove to be unsuccessful. In other words, minimizing efforts within the product scope should involve reducing work by lowering the quality of design. The target should be to launch fewer features but features that are good quality (Klotins 2019).

The second antipattern II, (not) attracting customers, occurs when the startup has difficulty finding potential customers and transforming them into paying customers. One reason for this is that customers are not involved in the presentation and validation of requirements, resulting in a product that is irrelevant to the market. The second reason for poor results in the market is that the product does not stand out from the competing products, is clumsy to use, misses essential functionality or is unreliable. The primary reasons for these issues include the unfulfilled requirements of quality, absence of quality testing, and optimization of the incorrect features of the product using false metrics (Klotins 2019).

The principal benefit of startups over large companies is velocity in the early phase of product development. However, the startup may be driven off course in the process by piling up TD. This delays the advancement of new features, raises difficult quality issues and undermines team morale (Tom et al. 2013). Common complaints from customers concern functional mistakes in mobile applications, slowness in normal use, and unexpected crashing of applications (Khalid et al. 2014).

## 2.2 Essence in software engineering

In this section, Essence is introduced, starting with its background and motivation. After the model itself and the criticism of it are discussed.

### 2.2.1 Background and motivation for Essence

The community behind Software Engineering Methods and Theory (SEMAT) designed the Essence framework. In their opinion, it was time to change the way that software engineering methods were understood. The aim was to redefine software engineering fundamentals according to the problems that were recognized in the existing situation of software development. In addition, a rigorous discipline was developed (Jacobson et al. 2012b).

As a foundation, the SEMAT Call for Action Statement was published (Jacobson et al. 2009). In this publication, solutions for actual problems and topics were described. The problems, immature practice, observed in software engineering by the SEMAT community included (Jacobson et al. 2009, p. 7):

- prevalence of fads being more typical in the fashion industry than in engineering;
- lack of a sound and widely accepted theoretical basis;
- huge number of methods and method variants, with differences that were little understood and artificially magnified;
- lack of credible experimental evaluation and validation;
- split between industry practice and academic research.

The SEMAT Call for Action Statement (Jacobson et al. 2012a) described key principles to redefine the process of software engineering based on a uniform theory, substantial principles and best practices. The principles stated that the process 1) include a kernel of widely agreed elements, extensible for specific uses; 2) address both technology and people issues; 3) be supported by industry, academia, researchers and users; and 4) support extension in the face of changing requirements and technology.

The solution published to redefine software engineering principles was the Essence framework. It consists of general foundations for the creation, use and generation of software development procedures. The main parts of Essence are the kernel and language. Essence enables the comparison, evaluation, customization and exploration of the use of essential elements of current and future methods and practices. The framework also allows for the continuous specification of the progress and status of software development processes (Jacobson et al. 2012b; OMG 2014).

### 2.2.2 The core framework model

The core of the Essence framework is the kernel, which describes the similarity of software development conventions. It consists of the central elements that are dominant in all software development efforts, including the team, requirements and stakeholders. These elements have positions and properties that represent the advancement and condition of the element. Distinct methods are evaluated by the qualities of the kernel, which allows software engineers to make new judgments about the practices they use (OMG 2014; Elvesæter et al. 2012). The kernel defines the general basis for software development methods that enable the comparison and tailoring of different methods.

The meaning of the language is to specify the abstract syntax, dynamic semantics, graphic syntax and textual syntax of the kernel. New practices are formed according to the language (OMG 2014). Several methods have been developed for software development, so Essence and other methods can be utilized simultaneously (Dwolatzky 2012). The power of Essence is that it enables the comparison of different practices and their feasibility with the present function. The flexible Essence framework can also be combined with other methods to fit the particular development situation. One advantage of the kernel is that it enables engineers to begin software development with a minimal set of methods and later add routines as needed in the future (OMG 2014).

Three main aspects of the kernel—customer, solution and endeavor—focus on a specific aspect of software development. The actual use and exploitation of the software system are handled in the customer area. The solution field addresses everything to do with the software system specifications and development. Subjects of the team and its way of working are treated in the endeavor area.

The areas of concern contain a small number of alphas, activity spaces and competencies. The independence of the Essence framework from other software development methods means that it does not need to include other elements that are necessary in the context of a specific procedure. Alphas are things that are needed to manage, produce and use during the developing process to maintain and support software. Alphas are required to determine the progress and condition of a software effort. Alphas are also anchors for the auxiliary sub-alphas needed to link the other software development methods used during the development process. Alphas are substantial factors in the development of software (OMG 2014; Jacobson et al. 2019; Jacobson et al. 2012b).

### 2.2.3 Activity spaces and competencies

Activity spaces are substantial matters in software development. They represent the challenges of developing, maintaining and supporting software systems that occur during the development process. Activity spaces also point to the objectives of the team, see Figure 5 (OMG 2014).

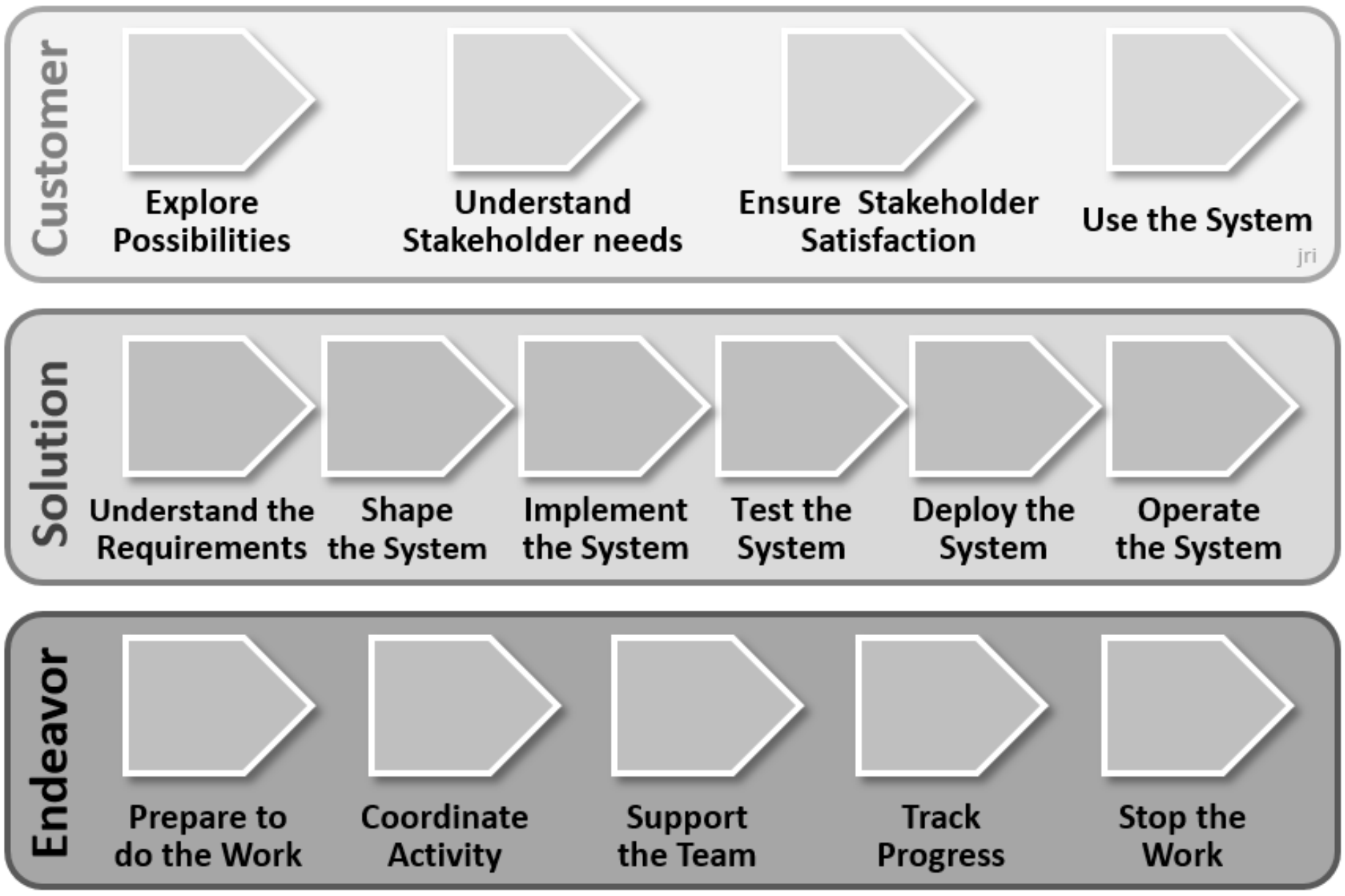

FIGURE 5. Activity spaces (OMG 2014, Ravaska 2020)

Competencies describe the necessary capabilities for software development aspirations. The competencies finish the alphas and the activity spaces with the needed capabilities so that the development process can be performed by following the alphas and the activity spaces, see Figure 6 (OMG 2014; Jacobson et al. 2012b).

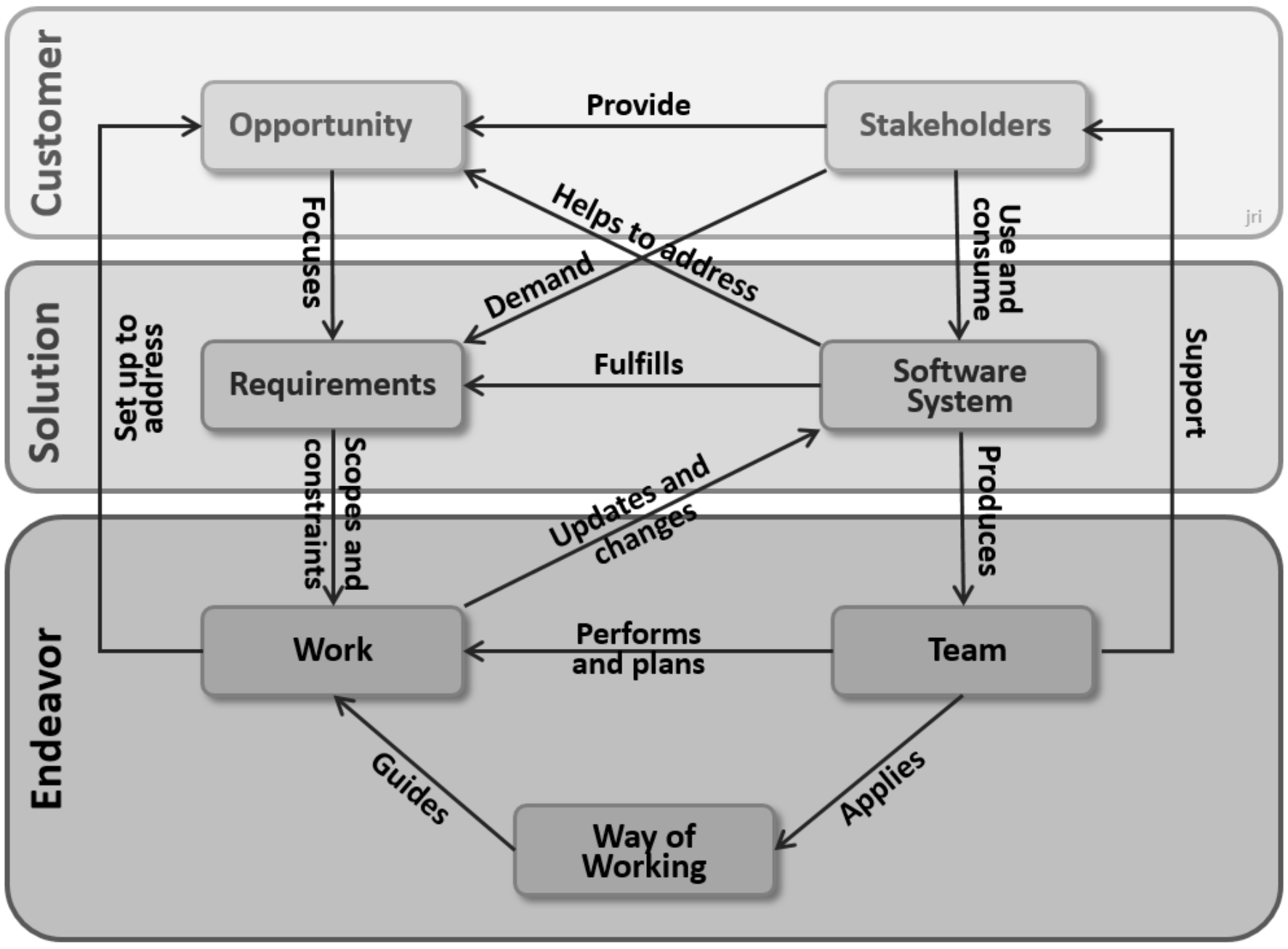

FIGURE 6. Alphas (OMG 2014, Ravaska 2020)

### 2.2.4 Alphas and their relations

The central constructs in software development are the alphas in the kernel. The alphas have pre-defined states to monitor their progress and condition. States are used to analyze and handle the risks and challenges for the alpha. Pre-defined checklists of states are used to define the state of the alpha. State are a tool for the continuous review of the progress of that alpha. This process can be run several times in iteration cycles. The alphas are indicators, not just physical partitioning or abstract work products. The team members can represent both the team alpha and the stakeholders' alpha (OMG 2014; Ng et al. 2013; Jacobson et al. 2012).

In practice, Essence is an abstract model of the things that require the most control in software system development. Essence can also be too abstract method of conducting the software development process. Therefore, it can be used simultaneously with other selected procedures and practices. In every software development endeavor, requirements may vary by project. User stories can be used in one project to handle use cases (Jacobson et al. 2012b; OMG 2014).

First, Essence focuses on project management. Teams can monitor their work and design the next steps in spite of the chosen methods and practices in the project's entirety. All actions can be planned with Essence practices, including all the most important aspects of the development process. This allows the team to progress further, keeping the alphas in balance. The team can work at a stable pace with the necessary factors of the project without omitting anything important. Essence is a dynamic method that allows practices to be added or removed from the project whenever needed. If the team finds better practices during the workflow, they can continuously fine-tune the progress according to the encountered situation. In a software development process, organized according to the Essence framework, continuous iteration is possible throughout the process from the original idea to the completed product (Jacobson et al. 2012b).

Second, Essence describes various software development methods and practices. Because of the abstract nature of Essence, it is possible to add practices from the ongoing development methods while mixing different procedures. Practices are presented as separate modular units with the kernel so that they can be used or discarded, or changed on the fly. Usually, software development methods are considered to be an inseparable set of practices that only work together as a whole (Jacobson et al. 2012b; Jacobson et al. 2019).

Third, Essence can be applied to teaching. Universities are adopting the Essence method to teach students about the fundamental principles of software development. Positive experiences have been noticed after adopting the agile side-by-side methods of Essence in teaching undergraduate students the principles of software development (Gil et al. 2014). Essence can also be used in software engineering research because of the nature of the common framework used to report empirical findings (Huang et al. 2014).

### 2.2.5 Essentialization of a practice

The kernel and language of Essence can be used to clarify different practices and present them explicitly. The procedure of describing practices with the Essence framework is called essentialization (Jacobson et al. 2019). Essentialization makes practices from different origins compatible and helps the group adopt new practices. Practices can come from a variety of sources that have different ways of describing them, or practices may not have been described before. This makes it difficult to combine them, which can be difficult for new team members to understand. Essentialization supports these aspirations in a consistent way by characterizing all the practices used by the group in a unified language and symbols.

In addition, with essentialization, researchers can describe practices for different sources in an explicitly coherent way. For a practice to be accurate and reusable by means of the essentialization method means that the practice is proven, and its elements are necessary elements only. In general, teams need many structurally proven practices that are meant to be used together in practice architecture. Although practices can be seen as separate elements, they are generally not independent. The practice architecture may have a layered form with more general practices at the bottom and more specialized practices at the top of the structure (Jacobson et al. 2019; Ravaska 2020).

The central idea of the essentialization method is to start with a chosen practice and assemble the essential elements of that practice, describing them with the kernel and the Essence language. A practice consists of things that need to be done (activity), the abilities and capabilities needed to conduct the practice (competency), the actual things generated using that practice (work product), and things that need to be monitored, taking into account their advancement (alpha).

Certain practices include other essential elements (patterns). In the Essence framework, the notion of pattern points to the essential elements of a practice other than activities, competencies, work products and alphas. Patterns can either play a role in a team or be control points that synchronize alpha state development flow.

Figure 7 shows how a scrum team, product owner and scrum master roles perform as patterns. Sprint planning, daily scrum, sprint review and sprint retrospective are the activities. Product backlog, sprint backlog and increment are work products. Sprint and product backlog item are the alphas (Jacobson et al. 2019; Elvesæter 2013).

Figure 7 also shows that a certain method can be attached to Essence to expand the framework in accordance with the needs of the team. Added alphas or sub-alphas have separate states, such as alpha elements in the kernel. For practical use of the kernel, the states of the added alphas and sub-alphas are necessary to monitor the advancement of added practices. The language is designed to assist in the definition of different perspectives that suit different practitioners. For example, some teams may choose to use alpha state checkpoints to guide their work, while others may choose to use a list of activities to track progress from one state to another (Elvesæter et al., 2012).

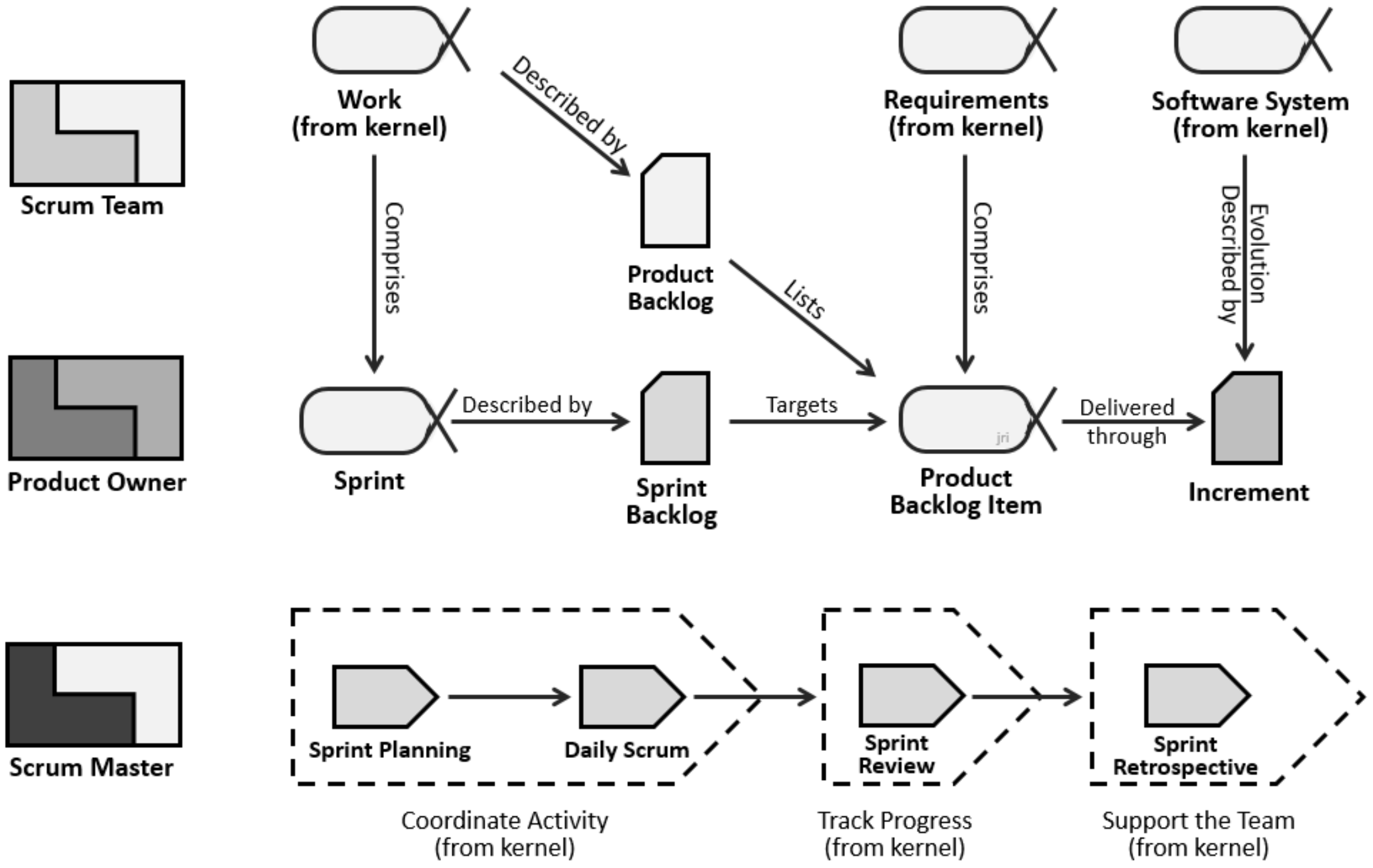

FIGURE 7. Example essentialization of the Scrum method (Jacobson et al. 2019, Ravaska 2020)

Essencery is a Web application and tool developed by Evensen et al. (2018) that allows professionals and academics to simulate various software development practices and methods using the Essence language's graphical grammar. Essencery creates graphical structures according to Essence graphical syntax. The creators call it the Essence Practice Editor (Evensen et al. 2018). For software startups, the Essencery tool can also support the use of practices by presenting them in a conventional form following the Essence language. This enables us to add casual software startup practices to agile approaches and reshape those methods to fit the practical efforts of startup ventures. This tool can be used to combine different formal and informal practices. In their study, Evensen et al. (2018) explain that Essence has not been adopted in practice; therefore, their intention was to create a tool that could be used to essentialize practices and methods. According to their research, the Essencery tool is easy to learn and use and can be used for practical use in the framework.

### 2.2.6 Known issues in the Essence model

The SEMAT initiative and the Essence framework have faced criticism from the academic community. Smolander et al. (2013) proposed criticism and complementary elements to the Essence framework by publishing their own model: the "coat hanger" model. This model theorizes development practices in software development and focuses on practical work and the consequences of using certain practices in different situations. The coat hanger model may expose shortcomings in present hands-on research and suggest some enhancements to practical theory. The model makes the components of software development practices more explicit and presents structured views on theorizing practices. Smolander et al. (2013) argue that their theorizing of SEMAT improves practices by choosing and evaluating best practices in a specific context. Learning and reflection are the means to improve understanding and theories. They express that the ontology of Essence is fixed and based on earlier knowledge, ordinary opinion and the most probable speculations. They propose that the coat hanger model, with its ontological difference, gives more freedom to the creation of theories (Smolander et al. 2013; Päivärinta et al. 2015).

In the criticism of the Essence framework's one-time standpoint, Smolander et al. (2013) and Päivärinta et al. (2015) say that software engineering theory has to be an ongoing and mutual process from experience to theory. They also point out that SEMAT is merely standardization in which concepts are based on consensus. Likewise, they argue that SEMAT does not take into account the need for theories to evolve. Despite the authors' criticism of the Essence framework, they see it as a justified proposal, and the core of their critique is targeted theorizing in general and the concepts of theories.

Cockburn (2010) sees software development methods as a trend or fixed policy when they are right. However, it is difficult to change methods because people have a tendency to be attracted to new popular trends. Therefore, people who are not very familiar with the software engineering frameworks should use alternative methods and stay critical of SEMAT. His critique in detail is:

1. The call for action was inflammatory, poorly researched and logically broken; it uses the very hype-for-fashion language that it decries; it is internally contradictory, the problems it deplores, it cannot fix, and its proposed solutions do not address the problems named. It sets a direction in tone and content that does not do the topic justice. It is a red herring, intended to generate support through appeal-to-authority, hype and ambition.
2. They are unlikely to discuss either engineering or engineering theory—a more accurate name for the initiative would be the Meta-Process-Kernel initiative.
3. Whatever they produce is unlikely to affect topics that matter to the industry. (Cockburn 2013)

Fowler's (2010) main point of criticism is that although people are a key part of software engineering and are naturally non-linear and unpredictable in their efforts, the core of the software meta-method is doomed to failure. This is based on the idea that people are not predictable actors that can be characterized by adaptive mathematics. As further criticism shows, there is a lack of consensus regarding Essence as an accepted model. Respected practitioners and researchers of agile methodologies have expressed negative comments on the SEMAT initiative (Cockburn 2013; Fowler 2010; Aranda 2009). One reason is the opinion that, in general development, the Essence framework is hard to utilize because of its lack of implementation tools. Because software engineering is highly connected to empirical research, experimental data is crucial when using Essence. This leads to a better understanding of its theoretical and practical significance and scientific evaluation (Wohlin et al. 2000; Graziotin et al. 2013).

Implementation tools and practices linking the Essence model and framework to appropriate software development are needed to combat criticisms. Thus, the SEMAT Accelerator (SematAcc), as a Web-positioning environment for software engineering needs, is a practical scheme. SematAcc implements SEMAT's Essence kernel, enables the use of the Essence framework and learns to understand the Essence kernel in practice. The SematAcc tool is a medium for teaching, adoption and research of the Essence framework in controlled experiments and case studies (Graziotin et al. 2013). Similarly, the Essencery tool, discussed above, is a Web-based tool like SematAcc that can simulate software development practices and methods with the Essence methods (Evensen et al. 2018).

## 2.3 Design viewpoints

In this section, the term design is described in the context of this study. Firstly, the definition of design is handled as a practical action, as a verb. Secondly, design is software startups as a practice is described through technical debts, that are specifically design related in in practice to avoid the debts. Thirdly, design in Software Engineering is described by combining both artistic design conventions and software engineering practices. Fourthly, the use of Cynefin framework in this study is described by its nature of sense making model.

### 2.3.1 Definition of design

Design as a term with manifold meanings and contextual usage cause communicative problems. Design, as in this study, is understood as a verb when creating and making and artefact differs from design as a subject and object. Design as a creative process has different meaning than design as an artefact.

Sketching by hand is a crucial technique in engineering, science, and innovation, when generating, communicating, and evolving ideas (Ullman et al. 1990; Goldschmidt 1991; Schütze et al. 2003; van der Lugt 2005; Ainsworth et al. 2011). Early sketches of design are often used to assess the quality of an idea, and creativity is often an important aspect. The sketches that influence the perceived creativity of an idea are based on a recent study that found the clarity of the draft concept, affect the perceived creativity of the idea that the draft describes (Kudrowitz et al. 2012). In this study *the notion design as a verb* addresses to the product/service creation process which is more common in the Arts, architecture and e.g. automotive design. Sketches are made to refine ideas, share ideas, and evaluate the qualities and value of concepts. The evaluation of sketches is based on several characteristics, but in case of innovation in the form of new products, many researchers agree that creativity is a crucial criterion (Amabile 1982; Besemer 1998; Christiaans 2002; Horn et al. 2009). Various studies examined the ways to generate more creative ideas and simultaneously found, that the more you had ideas, the more creative ideas were in the outcome (Diehl et al. 1987; Rietzschel et al. 2006; Kudrowitz et al. 2012).

Sketching as an idea generation and development technique, as in this study, is commonly used in traditional design professions in Arts, architecture and e.g. automotive design. The product gets its form during the sketching and concepting process by individuals and teams. In brainstorming and team meetings the sketches are mainly quick line drawings to describe ideas to other persons than professional designers (Goldschmidt 1991, 2006; Lipson et al. 2000; van der Lugt 2005).

The notion *design* as a flexible construct has the same obscurity in meaning and definition as the notion *architecture*. When in software development, technical debt focuses design on detailed design, code smells, design tools and processes (Li et al. 2014), and uses practical design techniques in graphic design usability and designing industrial and physical artefacts, design has diverse abstractions and meanings. Here, the building blocks and how they are organized to a higher-order constructs in software design, have a base in both software and design. The validity threat appears when two different cultures, traditional design and software design join their forces or stay separate.

The term of design in this study is understood as a combination of the creation process from idea to artefact. The idea is first crafted (to sketch, develop, refine, to perfect), then communicated (to visualize, interpret, induce, and reflect) and finalized (to iterate, complete, modify, build, and to productize).

### 2.3.2 Design in software startups

This study approaches the obscurity of the notion *design* through design practices in the arts and design in form of artistic design flow. As a proposition in Article I, novel ways to interact between traditional art creation techniques and software development methods allows new combinations in the practices. When the software graphic designers and usability specialists are more like artist-engineers with hands-on skills of sketching, drawing and visualizing, a startupper's skillset may lack possibilities to give form, create structure, visualize and communicate in an understandable way. A fruitful level of interaction between the two cultures of art and software is reached when the exchange is mutual, both parties are receiving parties. One possibility is to create processes, ontologies, taxonomies, classifications, structures and work-flows for the creative processes. The outcome in this *art & software combination* can be an act of opening to improve design and creativity in software startups.

Technical debt (TD), in the case of a rapidly growing software startup, is a harsh challenge when trying to overcome the simultaneous complexity of the need to grow fast, deliver products or services on time and survive the ever-changing technical environment. Here, the different dimensions of TD might require the startup personnel's skills in the development procedures used and their diligence to follow each step of the development process. The *term design in technical debt meaning* addresses to technical shortcuts that are taken in detailed design, software architecture, code smells, design patterns, code, code evolution, user interface design, design tools (e.g., modelling tools), design as planning process (detail design) and design as a verb. As a conclusion, technical shortcuts that are taken in detailed design are called Design TD (Li et al. 2014).

As Kruchten et al. (2012) and Nord et al. (2012) express, TD is analogical to describe incorrect engineering solutions in a deliverable, which adds friction to its improvement and support. The additional effort related to this resistance (i.e., reimbursement) needs to be refunded when an incomplete solution requires modification (Kruchten et al. 2012; Nord et al. 2012). McConnell (2013) advanced the definition of TD to be "a design or construction approach that's expedient in the short term but that creates a technical context in which the same work will cost more to do later than it would cost to do now (including increased cost over time)" (McConnell 2013, p. 3).

Startups may transform the TD into an investment in which the payback time never arrives. Tom et al. (2013) call this "debt amnesty," which happens when the feature or product does not succeed. Further, Avgeriou et al. determined that TD is:

> [A] collection of design or implementation constructs that are expedient in the short term, but set up a technical context that can make future changes more costly or im-possible. TD presents an actual or contingent liability whose impact is limited to internal system qualities, primarily maintainability and evolvability (Avgeriou et al. 2016, p. 112).

As seen in Table 4, the definition of TD varies by practitioner and re-searcher, which leads to unclear and rough interpretations of the concept (Li et al. 2015). Debt, as a term, also points to any software features and matters like code, usability and design. Because several available tools support the identification, measurement and repayment of code, most of the attention has focused on code TD. The code, on the other hand, is also explicit and effortless to perceive. Coders, due to their experience as everyday practitioners, are familiar with code analysis.

TABLE 4. Classification of 10 types of Technical Debt (TD) at different levels (Li et al. 2015)

| TD type | Definition |
|---|---|
| Requirements TD | Refers to the distance between the optimal requirements specification and the actual system implementation, under domain assumptions and constraints |
| Architectural TD | Is caused by architecture decisions that make compromises in some internal quality aspects, such as maintainability |
| Design TD | Refers to technical shortcuts that are taken in detailed design |
| CodeTD | Is the poorly written code that violates best coding practices or coding rules. Examples include code duplication and over- complex code |
| Test TD | Refers to shortcuts taken in testing. An example is lack of tests (e.g., unit tests, integration tests, and acceptance tests) |
| Build TD | Refers to flaws in a software system, in its build system, or in its build process that make the build overly complex and difficult |
| Documentation TD | Refers to insufficient, incomplete, or outdated documentation in any aspect of software development. Examples include out-of-date architecture documentation and lack of code comments |
| Infrastructure TD | Refers to a sub-optimal configuration of development-related processes, technologies, supporting tools, etc. Such a sub-optimal configuration negatively affects the team's ability to produce a quality product |
| Versioning TD | Refers to the problems in source code versioning, such as unnecessary code forks |
| Defect TD | Refers to defects, bugs, or failures found in software systems |

Design TD consists mostly of code smell, which points more at philosophical assumptions, like offenses against design principles and attempts against the quality of code (Li et al. 2015). Code smells affect design quality and can be indicators of TD. Code smells are usually not defects; they are not technically flawed and do not interfere with the operation of the software. Rather, the smells point to design deficiencies that can slow development or increase the risk of errors or failures in the future. Martin (2009) presents a list of code smells based on values that construct behavior as a value system of the skills and expertise of a coder as a professional craftsmanship in software development.

It is challenging for startups to innovate and iterate products and simultaneously manage technical design as the company grows, balancing flexibility and speed. When it succeeds in creating growth as the number of customers, employees and product operations increases, the startup is forced to manage and control its chaotic software development environment. The most significant challenge for a startup is to find a comfortable point between entering the market quickly and managing the amount of accumulated TD (Giardino et al. 2015). When a startup from a broader industry or branch of business originally outside the software domain conquers a fundamental confrontation to access digital solutions in the form of software and applications, this allows for possibilities of similar success (e.g., Spotify and Airbnb).

### 2.3.3 Design in software engineering

A potential conflict is that two design cultures, the traditional art and design cultures and software creation cultures do not understand and use the term design in a consistent, correlative and mutual way. In this study, the author represents several design and creativity areas as a practitioner: Arts & Crafts, architecture, acoustics, city planning, industrial design, graphic design, mobile communication device creation, and visionary and imaginary theories (Appendices). Design as a fuzzy and all-inclusive term is a relative to the blurred term *architecture* used in (traditional and constructed) *architecture* and '*software architecture*'. The notion *architecture* is based on several definitions, but also on the words *technē* and the prefix *archi*. *Archi* refers to being first and being a ruler as in professions *archiereus* or *architheōros* (a ruler priest or a high priest). The word *technē* of the highest degree demands an absolute wisdom of the entire working process and the masterly professionalism to create a well-wrought ultimate artefact (Holst 2017).

Software architecture, highlighted by Baragry et al. (2001), has qualities like 1) identify and theorize about the large-scale structures of software systems, 2) the large-scale structures are observed as the 'architecture' of the software system, 3) research as a medium to improve the development process at the software architecture level of design, 4) various system representations are required to describe the architecture of a software system, 5) those representations are seen as analogous to the diverse representations of traditionally built artefacts, 6) the state of software architecture and how it is represented is still unclear (Baragry et al. 2001).

When describing software architecture by Reed's (2001) highlights, software architecture is a large-scale structure under construction of its development processes and system representations. Traditionally built artefacts and research can lead the development. Confusion still exists in the idea of software architecture, therefore we call it only "architecture". (Reed 2001).

When comparing architecture with software "architecture" (with parentheses), software "architecture" has no intentions to be *archi* and *technē* in the same meaning as in architecture. Structure is a term already used by Baragry et al. (2001) and in case of *large-scale structures of software system* a clear and understandable notion for is *large-scale system*. Luckily Baragry et al. (2001) determines software "architecture" to be under construction. This allows further research and development without the demanding notion of *architecture* and concentrate on reaching the meanings of *archi* and *technē*.

### 2.3.4 Design dynamics in Cynefin framework

Cynefin sense-making framework is used in this study, in Article II, to clarify the structure of three industries with common possibilities to create a new business model.

Cynefin is a medium to sense-making in complex and complicated environments. Cynefin challenges to make sense in organizational support and strategy for decision making assumptions of order, of rational choice, and of intent. The value of Cynefin is on the sense-making and decision-making capabilities of decision makers of a wide range of unspecified problems, helping them to break out of conventional ways of thinking and to consider unmanageable problems in fresh ways. The Cynefin framework is used principally to consider the dynamism of situations, decisions, perspectives, conflicts, and changes in order to reach an agreement for decision-making under uncertainty. The framework has five domains, ordered domains 1) *known* (simple) causes and effects, and 2) *knowable* (complicated) causes and effects, and unordered domains 3) *complex* relationships and 4) chaos. The fifth domain is disorder, which indicates the different views of decision makers looking at the same setting from different perspectives. Cynefin can be used to obtain new insights on argumentative topics. Here contextualization allows a dynamic way to handle diverse items, like "communities, products, actions, motivations, forces, events, points of view, beliefs, traditions, rituals, books, metaphors, anecdotes, myths, and so on" Kurtz et al. 2003 (p. 471). The narrative database in Cynefin describes the situations, actors, events, and forces of the topic in question. Convergence methods allow to build different future views based on the narrative database or the participants' own experiences. Use of alternative history allows to describe the history of the item, where turning points of periods when small events generated big changes (Kurtz et al. 2003).

According to the dynamism of Cynefin framework, it was used in Article II in the context of three different industries with probable mutual interests. The narrative was constructed to combine their strengths in technology and customer relations to create a new and disruptive combination, a new business model to gain back the lost or weakened business models.

Cynefin framework has dynamics to move strategically between the four domains *simple*, *complicated*, *complex* and *chaos*. When moving between *simple* and *complicated*, the result is incremental improvement. When moving from *simple* to *chaos*, the item has collapsed. Movement between *chaos* and *complex* cause divergence or convergence depending on the direction of the move. The movement from *complicated* to *complex* means exploration, and backwards from complex to complicated means just-in-time transfer (Kurtz et al. 2003). Most influential movement and interaction, in the case of three industries in Article II, was to take a strong position in the proximity on the borderline between complex and complicated domains. This position allows to move in an agile way from exploration to just-in-time transfer. This means, that the three industries find new ideas and disruptive forces to bring to their own, strong and existing businesses.

When applying the Cynefin sense-making framework by putting the three industries (media industry, network manufacturers and network operators) into the five domains (simple, complicated, complex, chaotic and disorder), the timely evolution of each industry is clarified. Network operators largely abandoned offering content, entertainment and mobile phone applications because of the pressure and content-sharing platforms of newcomers. These newcomers were Web service providers like Yahoo, which launched its Web portal with a search engine, free email, news, music and other early Web content and mobile offerings. The network operators stepped down from the Web content industry but continued the growing operational practices of the mobile network.

The operators remained one of the closest vendors for the end-user: the mobile device owner and proprietor of the SIM card. The users' main interest was not to own a chip; they wanted entertainment, news and user-generated content. Thus, the network operators lost significant mobile and Web business and became an actor in a zero-sum game of managing mobile networks and selling SIM cards. Network operators still stay between simple and complicated on the Cynefin diagram.

Mobile network manufacturers like Ericsson and Nokia have been technologically in the complicated sector of the Cynefin diagram. The systems have an extremely complicated structure, modularity, functionality, reliability, usability and customization for diverse operator needs and requirements. Today, the position on the diagram is stable because of the upcoming 5G network systems and the potential of IoT (Internet of Things). In addition, the world political situation in 2020 has helped Western manufacturers acquire new customers from network operators (e.g., by banning the Chinese company Huawei from delivering 5G).

Media houses, especially newspaper and Web news companies, mostly in journalism, have a miserable position in the Cynefin diagram. Both businesswise and in their social media content, media houses are partly in the chaotic sector, or even in the disorder domain. Money no longer flows from news and journalistic production, and the modern solution of Web advertising is owned by conglomerates like Google and Facebook. The present situation of the media business is mired in chaos.

System-level creativity indicates the dynamic use of the Cynefin framework as a sense-making device and uses it to propose decisions. Cynefin can be used in complex and varied dynamic contexts. When organizing the three industries into the Cynefin framework, the companies, industries and markets have a causal dependence on one another, and the sense-making approach can clarify the dependence. This means that there is a clear assumption of order. The assumption of rational choice allows industries to realize when a framework logically shows the possibilities of choosing between alternatives and selects a decision that optimizes the cooperative action.

## 2.4 Creativity in focal contexts

### 2.4.1 Definition of creativity

Creativity has been defined for decades from different perspectives. According to the standard definition, originality and effectiveness are required factors of creativity (Runco et al. 2012). Originality is not enough to explain creativity because it may happen randomly, and its result can be useless. Therefore, other attributes are needed to specify the notion of originality. The result has to be useful, adequate and suitable, aspects that can be referred to as effectiveness in this context (Runco et al. 2012).

Effectiveness is described by Marcus Vitruvius Pollio in The Ten Books on Architecture (Pollio 1914, originally c. 27 BC). Vitruvius describes the aesthetic principles for good design as utilitas, firmitas and venustas, or commodity, firmness and delight, as well as strength, utility, and beauty. These attributes are used as cornerstones for usability and user experience in software engineering (Van der Voordt 2009). Compared with the standard definition of creativity, utilitas and firmitas, or utility and strength, are synonymous with effectiveness and adaptiveness. Likewise, the notion of venustas translates to beauty, delight and authenticity, which are synonyms for genuine, real and original (Reich 2007). The standard definition of creativity is supported by this ancient definition of originality and effectiveness as factors of creativity, in the form of utilitas, firmitas and venustas.

Amabile (1988, p. 126) suggests that creativity is a "production of novel and useful ideas by an individual or small group of individuals working together." Creativity is a phenomenon that happens under conditions in which both criteria—novelty and usefulness—are bound together. This is commonly agreed upon in software engineering. Products' creativity can be rated by factor generation, reformulation, originality, relevancy, hedonics, complexity and condensation (Taylor 1975).

The Creative Product Semantic Scale (Besemer et al. 1987) defines creativity by three factors: novelty, resolution and a combination of elaboration and synthesis. Novelty means that the product is original, surprising and germinal, and these factors relate to notions of originality, beauty, delight and authenticity (whether the product is genuine, real and original). Resolution addresses how valuable, logical, useful and understandable the product is. These relate to notions of effectiveness, commodity, firmness, delight and solidity, as well as to utilitas and firmitas. Elaboration and synthesis refer to the attributes of a product being organic, elegant, complex and well-crafted. Elaboration echoes the ancient definition of Vitruvius' venustas, beauty, as a more objective and measurable dimension of an artifact.

Creative productivity is the quantity of ideas generated that are original and adaptive. Quantity measures numerical productivity, but only quality may determine genuine creativity. Therefore, as in Article V mentioned, the judges (senior designers and those who are professionally productive) evaluate productivity based on quality as a priority. Here the creative productivity depends on career age, not chronological age. Commonly, productivity starts when an individual is approximately in their 20s and peaks in the late 30s or early 40s. Then there is a steady decline in creative productivity starts (Simonton 1988). The students at the design class in Article V were between their 20s and 40s, so high potential for creative productivity was assumed.

Cropley (2000) looked at various creativity tests in a multifaceted way that defines creativity in terms of products, processes, motivation, and personal factors. Table 5 provides an overview of these (Cropley 2000).

TABLE 5. Collection of creativity elements applied to multilateral attributes (products, processes and personal factors) (Cropley 2000)

| Product | Process | Motivation | Personality/abilities |
|---|---|---|---|
| • Originality<br>• Relevance<br>• Usefulness<br>• Complexity<br>• Understandability<br>• Pleasingness<br>• Elegance/well-craftedness<br>• Germinality | • "Uncensored" perception and encoding of information<br>• Fluency of ideas (large number of ideas)<br>• Problem recognition and construction<br>• Unusual combinations of ideas (remote associates, category combination, boundary breaking)<br>• Construction of broad categories (accommodating)<br>• Recognizing solutions (category selection)<br>• Transformation and restructuring of ideas<br>• Seeing implications<br>• Elaborating and expanding ideas<br>• Self-directed evaluation of ideas | • Goal-directedness<br>• Fascination for a task or area<br>• Resistance to premature closure<br>• Risk-taking<br>• Preference for asymmetry<br>• Preference for complexity<br>• Willingness to ask many (unusual) questions<br>• Willingness to display results<br>• Willingness to consult other people (but not simply to carry out orders)<br>• Desire to go beyond the conventional | • Active imagination<br>• Flexibility<br>• Curiosity<br>• Independence<br>• Acceptance of own differentness<br>• Tolerance for ambiguity<br>• Trust in own senses<br>• Openness to subconscious material<br>• Ability to work on several ideas simultaneously<br>• Ability to restructure problems<br>• Ability to abstract from the concrete |

Creativity, as well as innovation ability, can be used in any kind of action, business or industry. Everyone is capable of creativity in artistic professions like art, music and architecture. Creativity and innovativeness are comprehensively considered in relation to entrepreneurial circumstances (Ahlin et al. 2014; Khedhaouria et al. 2015), where innovative ideas change markets and create new ones. Graziotin et al. (2014) emphasize creativity in problem-solving among programmers in information technology. Similarly, Carberry et al. (2018) underline innovativeness among engineering students.

Creativity is an important element for university students because they have to face new and unfamiliar duties and assignments and constantly postpone changing their attitude to their future work, vocation and circumstances (Jackson 2013). The problem of creativity in the academic world is not that it is missing but that an analytic culture penetrates the academic domain (Jackson 2008). Five factors influence education at universities: 1) Creativity lays the foundation for human existence. It is based on human skills, and it affects self-awareness and how we achieve our goals and live our lives. 2) Both creative and academic advancement are fundamental if we believe that higher education is crucial to students' ability to enable progress in a normal life. 3) Teachers can encourage students to grow creatively by organizing demanding and fascinating courses. 4) The subjects taught and future profession pursued promote insight for creative advancement. The creativity, mastery, crafts and abilities of the profession are related to the creative progress of the students. 5) The creativity developed in students is scalable and suitable for all topics in life, not just in the academic environment. Higher education can promote students' lifelong goal of creativity and progression in their studies (Jackson 2011a, 2011b; Barnett 2011).

Several disciplines and curricula in academia emphasize taking notes on creativity. In art, design and teaching logbooks, research records and field books are used as tools for recording notes and producing new knowledge when working in real-world environments (Schön et al. 1986; Newbury 2001). Students of architecture also commonly use study diaries to make notes and sketch their ideas. Practicing architects use drawings and sketchbooks to enhance their practical and operational planning (Heynickx 2013).

According to these examples, the importance of note-taking and the use of the implementation task as a tool for creation are emphasized by both notes and their enhanced development, note-making, and using them as a natural tool in design courses, if not in the entire academic environment.

### 2.4.2 Creativity in different contexts

Design and creativity are closely bound. Design is an umbrella notion understood through historically developed meanings and today's significantly broader, fuzzier meanings. Historically, design is related to vision, skills, art, creativity and artifacts. Today's ever-growing openness to develop processes and artifacts allows the application of any relevant, and even at first irrelevant, means to improve the quality and essence of the result. Members of the classical design community, with their abilities, are ready to apply their proven creativity, skills and outcomes to modern use in new environments like software development.

A certain set of skills and capabilities enable and enhance the fulfillment of creative acts. In Feldhusen's (1986) study of the early life of creatively productive people, a set of signs were found. These included 1) high-level intellect, memory and deduction capacity; 2) early mastery of technics and methods and praxis in a specific field; 3) aim to generate high energy levels and engagement or dedication to studies or to work; 4) internal belief in controlling the results of happenings in their lives (locus) and a belief in their own creative power; 5) emphasized sensitivity to detail; and 6) working alone and extreme independence (Feldhusen 1986).

There are two different categories of tools in the context of creativity: 1) personal skills and capabilities and 2) external utensils and instruments to be used. The utensils and instruments that enable and enhance creative acts, extending thinking to construct artifacts, are practical and physical utilities, as well as digital systems and applications.

Gamification and creativity are implemented by the developers who create new games and the players who creatively play the games. Digital games are a relatively new game category compared to historical board games. The first digital games were launched in 1979 by Commodore 20 (Vic 20). Since then, the game industry has grown to include naturalistic video games and mobile devices.

The game industry constantly has to renew itself by inventing new game categories and features and capturing qualities for growth and player retention. Innovation and creativity are the driving forces for game developers, who constantly improve their existing games and try to scale the game mechanics in new variations.

Business and creativity are linked to success in business and industry. In their constant changes in context, competition and life-cycle, uncertainty dominates. Different means of surviving and growing in this everlasting ambivalence can be exploited.

### 2.4.3 Measuring creativity

Creativity as a human quality has been studied in numerous circumstances and across disciplines in art and science. Creativity can be seen as generating new and useful ideas in all areas (Amabile et al. 1996). Although the definition of creativity is generally accepted, creativity needs to be defined more precisely on a case-by-case basis if it is to be measured appropriately for the purpose of research.

Creativity has typically been measured by examining the results of the creative process and the work leading to the creation of the creative results (Amabile 1982; Davis 2009). Concretely, this often means that participants, 96 undergraduate students in Foregeard's (2011) study, create creative solutions to rare problems (Foregeard 2011; Kaufman et al. 2007). These participants should be unfamiliar with the problem being addressed so that they do not have an opportunity to use solutions that they know are well suited to solving the problem in a creative way. Judges, one of whom may be the author of the study, can score these solutions in order to assess the creativity of the solutions and the participants (Graziotin et al. 2014).

Measuring creativity also seeks to understand the factors that influence creativity. To achieve this goal, human factors, such as personality (Wolfradt et al. 2001) and cognitive style (Hayes et al. 1998), as "the way people perceive stimuli and how they use this information to guide their behavior," are linked with creative performance (Beeftink et al. 2012, p. 72). These factors involve a psycho-social work environment (Shalley et al. 2000), which is associated with creative performance in many situations. In the context of information technology and software development, Graziotin et al. (2014) studied developer happiness and creativity. Their findings further supported Forgeard's (2011) observations that the mood of the moment influences a person's creative thinking.

### 2.4.4 Creativity and self-efficacy

This section explores self-efficacy as a factor of correlation between design and creativity among students. Self-efficacy relates to the studies conducted and reviewed in this dissertation. Self-efficacy describes how well a person can execute a given task required to deal with potential situations and the perception the person has of their own capabilities in carrying out the mission. For example, self-efficacy would include an individual's perception (or belief) of whether their C++ programming expertise is sufficient to execute a work task (Carberry 2018).

Self-efficacy is important for performing diverse tasks successfully. If a person believes in their abilities, they are more likely to cope with challenges, to pursue primarily related tasks and to be more internally motivated (Bandura 1994). Likewise, someone may feel less willing to begin a task if their self-efficacy is poor, even if they have the necessary skills or knowledge to perform the task. Therefore, self-efficacy can lead to productivity (Gist et al. 1992).

According to Bandura (1994), self-efficacy is influenced by four factors: (1) mastery experiences, (2) vicarious experiences, (3) social persuasion, and (4) physiological states. Mastery experiences are past task completions (or failures) that hold a notable impact on one's self-efficacy. In comparison, vicarious experiences can weigh on one's self-efficacy. Vicarious experiences are gained by observing others with similar abilities perform tasks. Social persuasion refers to support from prestigious individuals or individuals we respect. This can positively affect self-efficacy. Finally, current physiological states, such as stress or simply being tired, can influence self-efficacy (Bandura 1994).

Although self-esteem relates to one's personal feelings, where low self-esteem indicates negative feelings about oneself, self-efficacy relates only to assignments. Even if an individual's self-efficacy in relation to a task ahead, such as winning a baseball game, is weak, it does not matter, as the individual still believes in himself (Gist et al. 1992). In addition, self-efficacy is associated with certain task skills (Carberry et al. 2018). Although an individual might consider their programming skills to be good, the skill attached usually pertains to a specific task.

Self-efficacy changes over time, as training and skill development are likely to have a positive impact on self-efficacy, especially in relation to tasks related to specific skills (Carberry et al. 2018). This aspect is studied in Article V as we look at the self-efficacy of students before and after taking a design course intended to support self-efficacy in relation to various skills required in design work. However, the aim is also to understand the potential link between self-efficacy in these skills and creativity in practice. We thus further discuss self-efficacy in the specific context of creativity in the following section.

Studies related to creativity also include self-efficacy and creative self-efficacy in particular, which Tierney et al. define as the belief that a person has the ability to produce creative results (Tierney et al. 2002), meaning that trust in your own creativity improves your creative performance. In addition, Mathisen et al. (2009) argue that creative self-efficacy can be improved through training, but note that further research is needed on whether this also improved creative performance.

Schack (1989) explored the self-efficacy and creative productivity of talented children. She found that a (creative) self-efficacy exercise did not necessarily produce creativity in students and had little effect on their later initiation of autonomous (creative) projects. In fact, it was found that self-efficacy training can lead to lower self-efficacy (e.g., if participants find the tasks challenging), leading to negative management or vicarious experiences. This poses challenges for those who want to design self-efficacy courses.

Much effort has been made to explore the connection between creativity and creative self-efficacy, but less research has been done on the self-efficacy of creative skills. Ahlin et al. (2014) studied the self-efficacy of entrepreneurship in relation to creativity and innovation in the context of entrepreneurship, while Khedhaouria et al. (2015) examined overall self-efficacy in the same context. Beeftink et al. (2012) examined the self-efficacy of design in relation to the performance of creative professions and also claimed that self-regulation firmly maintains self-efficacy in the design work context.

In this dissertation, an investigative technique is applied to studying self-efficacy in a large number of skills, using the mechanism developed by Carberry et al. (2018) to measure self-efficacy. In this way, we attempt to expand our understanding of how various types of self-efficacy affect (or do not affect) creativity.

There are four dimensions that develop people's beliefs about their efficacy. These include experiences of mastery, experiencing people like themselves succeeding in meeting the requirements of a task, relating to social persuasion so that a person is able to succeed in a particular activity, and conclusions about somatic and emotional conditions that are proportional to personal strengths and vulnerabilities. Everyday reality is full of obstacles, adversity, setbacks, frustration and inequality. Thus, people need to have a solid sense of efficiency to withstand the relentless efforts needed for success. Successful life-cycles highlight new types of competency requirements that require further development of personal efficacy to ensure successful operation. The nature and extent of perceived self-efficacy change throughout an individual's lifetime (Bandura 1994).

Self-efficacy terms that are meaningful in Article V are listed by Bandura (1994):

- Mastery experiences
  - Successful experiences boost self-efficacy, while failures erode it
- Vicarious experience

- Observing a peer succeed at a task can strengthen beliefs in one's own abilities
- Verbal persuasion
  - Describes the positive impact that our words can have on someone's self-efficacy i.e. encourage and motivate
- Somatic and emotional states
  - A positive mood can boost one's beliefs in self-efficacy, while anxiety can undermine it.
- Belief in one's capabilities to achieve a goal
  - The stronger self-efficacy, the more you challenge yourself
  - The stronger self-efficacy, you likely achieve your personal goals
  - Self-efficacious people also recover quickly from setbacks
- Poor self-efficacy
  - Poor self-efficacy reflects low aspirations
  - May result disappointing academic performances
- Is there a correlation between self-efficacy and creativity?
  - Self-efficacy is a belief
  - Creativity is a more like a dynamic capability (provocation and movement), or a static capability (creative attitude) (Azadegan 2008)
  - Creativity is a more like a mindset and attitude related (Choi 2019)
  - Creative mindset as the decision to be creative (definition by Choi 2019)
  - Important to clarify correlations
- Self-efficacy vs. motivation vs. design skills
  - Potential to find educational means for self-efficacy and creativity.

Despite that the findings in Article V did not verify straight correlation between creativity and self-efficacy in the design class with non-design students, the hypothesis of clear correlation of creativity and self-efficacy stays as a future research subject.

A person's decision to be creative is related to his or her beliefs about their psychological traits and abilities, which are important for the development of his or her motivation and behavior. According to Dweck's definition, people have different beliefs that represent their state of mind and explain their subsequent behavior (Dweck 2000). People can experience their abilities in two ways: some feel their abilities are fixed (fixed mindset: unable to change), whereas others consider that their abilities are modifiable (growth mindset: capable of change) (Dweck et al. 1988).

Innovation Self-Efficacy (ISE) survey by Carberry et al. (2018) lists the activities in the test as follows:

1. Understand the needs of people by listening to their stories.
2. Find connections between different fields of knowledge.
3. Seek out information from other disciplines to inform my own.
4. Identify opportunities for new products and/or processes.
5. Question practices that others think are satisfactory.
6. Come up with imaginative solutions.

7. Make risky choices to explore a new idea.
8. Consider the viewpoints of others/ stakeholders.
9. Evaluate the success of a new idea.
10. Apply lessons from similar situations to a current problem of interest.
11. Envision how things can be better.
12. Do things in an original way.
13. Set clear goals for a project.
14. Troubleshoot problems.
15. Keep informed about new ideas (products, services, processes, etc.) in my field.
16. Communicate ideas clearly to others.
17. Provide compelling stories to share ideas.
18. Learn by observing how things in the world work.
19. Solve most problems if I invest the necessary effort.
20. Be resourceful when handling an unforeseen situation.
21. Suggest new ways to achieve goals or objectives.
22. Test new ideas and approaches to a problem.
23. Share what I have learned in an engaging and realistic way.
24. Make a decision based on available evidence and opinions.
25. Relate seemingly unrelated ideas to each other.
26. Think of new and creative ideas.
27. Model a new idea or solution.
28. Find new uses for existing methods or tools.
29. Explore and visualize how things work.

Directions are: Rate your degree of confidence that you can do each of the activities listed below on a scale from 0 to 100: 0 = not at all confident, 100 = extremely confident (Carberry et. al. 2018).

### 2.4.5 Creativity in software engineering

Software engineering as such is a wide domain with multiple variables and motives. In the context of creating software, there is a specific target like application, and requirements describing the need and functionality of the outcome. The process in managed by a team, time pressure is apparent, and methods used vary from traditional waterfall to agile and scrum practices. On top of these factors, the competitive situation on the markets require creativity. Creativity in software production has been identified as a potential market incentive for maintenance. However, related research remains limited to compensate for the challenges of an ever-evolving competitive market (Amin et al. 2011). Industries are investing more in creativity research to explore new approaches to software development, which can lead to the creation of unique and higher quality software products (Gu et al. 2004). Knowledge and skills in the field are required for successful software development; software designers should therefore strive to own solid domain skills and creativity (Graziotin et al. 2014). Developing high-quality and efficient software to achieve the desired cost, quality, and time is becoming increasingly expensive and challenging (Elberzhager et al. 2012). Software development is complex and knowledge-intensive in terms of operations and expertise (Hedge et al. 2014). Software development requires a number of skills and competencies, such as developer communication, design, and making strategies to generate creative and innovative ideas. This is because no developer has absolute knowledge of how to complete a task (Crawford et al. 2012). A team of software designers with an intercultural dimension is seen as a potential for creativity because of differences in people's perspectives and knowledge that can be combined into knowledge synergy (Dalberg et al. 2006). The traditional software development procedure, which is based on a repetitive and mechanical approach, prevents the developer from exploring and collaborating, unlike the second-generation approach, which does not show a precise models, patterns, processes, or frameworks for addressing development, fostering designer creativity (Fallman, 2003).

Because creativity is an ambiguous phenomenon, the assessment of creativity is complex and difficult to achieve in a shorter time span (Weiley et al. 2011). As a measure of how to evaluate creativity in a software project can be used by observing the number of features or functions added per release (Paulson et al. 2004). Teamwork promotes greater creativity in a software development project rather than an individual developer, because team performance has improved compared to the individual person's impact (Fagerholm et al. 2014).

Creativity plays a fundamental role in artistic design flow (Article I), when embedding practical design skills to improve creativity in software creation and development. By applying the artistic design flow, skills and tools and especially in large and complicated design processes, software development gets one credible and tested method for creation from the Arts. In this study, creativity is seen as a quality that scales from system-level planning via organizing businesses to the creation of products, services and solutions for customers. Creativity and disruption are bound together in this section. (Article II).

## 2.5 Research gaps by RQs

Research gaps faced in this dissertation are related to startups in design and creativity (RQ1), processes and customer retention (RQ2), and enhancing creativity (RQ3).

The creativity gap has a cultural and fundamental background. As Heartfield (2005) says, "creativity and design have become the conventional wisdom" (Heartfield 2005). Here Heartfield points to Landry that design, multimedia and internet companies that provided 'the buzzing atmosphere on which cities thrive, experimenting with new products and services' (Landry 2000). The idea was that the creative sector could revitalize cities, which has been replicated in all major cities in the UK. If one was not creative in his/her work, it was considered a sin. Well-known advertising agencies thought that creativity was the exclusive prerogative of creative companies. However, they argued that the commercial success of any business depends on creativity (Heartfield 2005). Now the creative industries (Figure 8) are dynamically proposing new products, services, business models, or organizational structures in any business areas. They also utilize innovative solutions from other technology areas, particularly from the information and communication technology (Müller et al. 2009, Europea 2006). Creative industries can have a significant impact to economic growth, either directly and indirectly (Dapp et al. 2011). The submarkets of creative industries, classified by Söndermann et al. (2009), are the music industry, book market, art market, film industry, radio and TV, performing arts, design, architecture, press, advertising industry, software and games industry, and a residual category of other creative activities (Söndermann et al. 2009). Within these submarkets, software and games industry and the advertising business are largest from the startup perspective (Kohn et al. 2018).

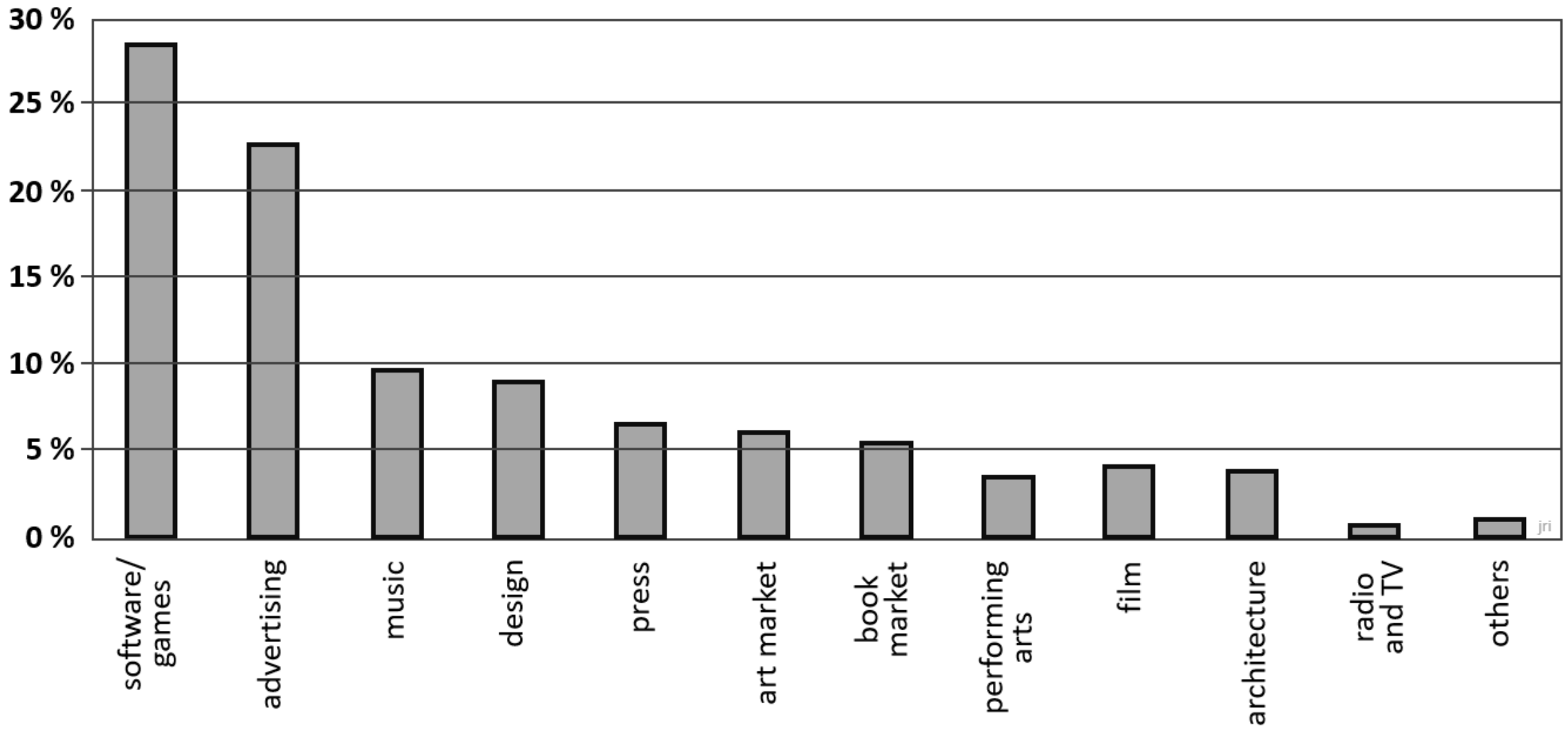

FIGURE 8. Submarket shares among creative industries, calculation using population weights; data source: KfW Start-up Monitor, waves 2003-2010 (Kohn et al. 2018)

When considering the startups creative and being a strong actor in software and game submarkets of creative industries, two gaps can be found: the eleven other submarkets could be environment for startups in general, but especially for software startup to find software-intensive opportunities there. The other gap relates to further research needs of design and creativity. When Seaman (2003) argues, that an overall higher education level is found in knowledge-based creative industries (Seaman 2003), it would be important to research the possible skills and competences that improve startups' performance in the submarkets other than software and games. Here the art and design skills as drawing, sketching, visualizing and handcrafting may have a positive impact to startups, also in their most embraced software and games. Here, additional measures are needed to support startups to navigate in creative industries to reach better sales growth and employment creation (Kohn et al. 2018). In the future research, the measures are more sophisticated when the design skills are also studied.

The research gap in startup processes for better customer retention (RQ2), is a fundamental factor in the business. The most common measures for software startups, emphasized by the practitioners, are user retention and user churn, active users and user engagement, short-term focused financial metrics (month-on-month growth and cash burn rate), and User-focused financial metrics such as User Acquisition Cost. This was selected from a software startup metrics collection of over 100 measuring subjects (Kemell et al. 2019). When the most of software startups fail (Unterkalmsteiner et al. 2016), it is important to find solutions to the known metrics, especially to the decisive ones, that are considered important. In this dissertation self-efficacy relates to Primary Empirical Conclusions 1-8 in Article IV, which points to the players' characteristics as motivation to play, enjoyment through active role and achieving something in the game, and balancing between challenges. These factors form an important unity between self-efficacy among the players, and a game design criteria set for game developers to consciously make more compelling games. This future challenge is a gap for the present customer retention practices by knowledge and understanding the link between self-efficacy, design, creativity, and player retention.

The research gap in RQ3 relates to the in software startups can enhance creativity in their work. When Parkman et al. (2012) argue, that the balance between creativity and business objectives leads to successful the creative startup (Parkman et al. 2012). This is extremely important case, when creativity and lucrativeness are often hard to link for creative businesspeople (Swedberg 2006, Mills 2011). To ensure the startups' capabilities not to fall in the method prison, locking one into certain methods, even unsuitable ones, is an important reason (Article III). Also, the weak correlation of creativity and self-efficacy found in Article V, needs further research to clarify their relationship. When entrepreneurial orientation (EO), has been seen as a result business owners' individual creativity and self-efficacy (Fillis et al. 2010), it could scale to students and startups. Simultaneously, creativity is connected with self-efficacy in general, and it is determined as a belief of entrepreneur's capability to fulfill named grades of achievement (Markman et al. 2002). These two nominators, the correlation of creativity and self-efficacy, and the general confidence that creativity is connected with self-efficacy, need more research in startup community to clarify the needs of skills and personal characteristics for success.

# 3 RESEARCH METHODOLOGY

This chapter presents an overview of the research approach and methodology.

## 3.1 Research approach

In this dissertation, five overlapping themes of startups' practices and businesses are bound together by a mixed-methods approach, combining both qualitative and quantitative techniques. These two techniques vary depending on the current or sequential time ordering of the two methods or the balancing between the degrees of dominance of these methods (Wu 2011). The research methods used in the dissertation vary by article and within articles (Table 6).

TABLE 6. Methods used in the articles

| Article | Method |
|---|---|
| I | Reflective practitioner, creative analysis |
| II | Reflective practitioner, creative analysis |
| III | Quasi experimentation |
| IV | Observation |
| V | Quasi experimentation, creative analysis |

Articles I and II are qualitative, discussing the factors of startups and future possibilities in various businesses with better design skills in general and as through a decisive partner between three stagnating industries. Articles III, IV and V concentrate on quantitative techniques in the form of surveys, questionnaires and interviews. However, in Articles III, IV and V, qualitative aspects were present in the form of player interviews, observations of board game players and an evaluation of creativity in students' sketchbooks. Self-reflection by the reflexive practitioner is also used as a dimension of the mixed methods approach (Schön et al. 1986). In this dissertation, the author alternated between reflection and planning and between observing and reflecting. After finishing a full cycle, an-other iterative cycle followed and enabled the author to improve the achievements (Johnson et al. 2019).

Articles I, II, III, IV and V represent a horizontal overview to a reflective practitioner's multifaceted approach to design and creativity. These articles reflect, on the one hand, the variety of a designers daily and weekly work, and, on the other hand, intentional trial to find structural conformity in creativity and design in diverse contexts, and in different granularity of depths in focus. Article I states the ideas and practices for the use of artistic design practices in software startups. The article proposes four alternative avenues for the coexistence of design and software.

Article II tackles a large-scale problem, the distress of three industries, media industry, network manufacturers and network operators. Each of them have their own journey to the present circumstances, and the article proposes them a common and shared future with disruptive qualities. Article III dives into possibilities to create openness in development processes by gamification and escaping the method prison. Here the method agnostic attitude and traditional board game surprises and offers opportunities to assimilate various practices for software engineering in a sound way.

Article IV interacts with a user group, video game players, where the focus is player-centric view on retention. Retention in any business is a crucial phenomenon, otherwise one does not have customers or developer community in the future. The article finds factors of an attractive product, a game, and builds design objectives to retain customers, players. Article V addresses to a challenging theme, searching for correlation between creativity and self-efficacy, through design. The setup stays demanding, because the focus group, students designing in a non-design faculty, understand the notion self-efficacy in respect of creativity, and own design performance incoherently.

Articles I, II, III, IV and V punch creativity and design from detail to entirety, from individuals to large corporations. The conclusive consequence is determined by a reflective practitioner in architecture, acoustics, industrial design, and usability. The view to creativity and design is a mixture of "reflective practitioner and design professional in normal creativity and design related circumstances".

## 3.2 Reflective practitioner

The author considers himself as a reflective practitioner. A reflective practitioner uses reflection to obtain understanding and experience, gaining the ability to choose suitable operations for handling a task (Schön et al., 1986). Being one of the first architects to start modeling in 3D with ArchiCAD 3.1 on a Macintosh LC (20 Mb hard disk and 2.5 Mb RAM) in 1990 was a groundbreaking experience that went beyond normal pencil and paper sketching and drawing. Advancing in the digital industry, the author moved to Nokia Corporation in 2001, which was a shocking transition: the company had procedures, guidelines, standards, leadership models, its own design approach and amazingly good products. The market share grew up to 40% globally. Participating in the diverse forms of design opened the author's eyes to the digital product industry as a type of rocket science. It was a feeling of privilege to participate in the most complex system design under human control. This mobile network management system, called NetActTM, facilitated effective fault diagnosis and disaster avoidance (Lin et al. 2002). The system needed a solid and stable user interface, which had to be modular, scalable and compatible with the brand. It had to be easy to use when monitoring over 20 million cars moving simultaneously on the German motorways. When the warning traffic lights, red–yellow–green, would suddenly blink, you had to let the system analyze the severity of the issue. Our system followed the traffic light sequences, and the notification "stop—prepare-to-stop—go" was applied to the form "severe—instant-check—OK" for each instance (Sivarao et al. 2010; Marzuki et al. 2015). The author was impressed by two exceptional things: the system was always fixed remotely from the control room, while the competitors used a manned van and a screwdriver to travel to the base station.

Schön et al. (1986) argue that reflection-in-action is subject to a discipline that is both similar to and different from the exactness of scientific research and controlled experimentation. A reflective practitioner is more than an expert of techniques. The role instead involves an inclusive approach to practice that requires the practitioner to engage in professional and personal progression (Anderson et al. 2004, Knowles et al. 2001). This, keeping a journal or writing a book about the experiences, equals with what Scanlan and Chernomas (1997) express: Students at clinical courses must keep a reflective diary documenting their retrospective thinking in practice. As they express, students who keep journals will become reflective practitioners. Their assumption is, that journaling develops the students to be reflective practitioners (Scanlan et al. 1997). As Kelly and Young (1996) say: "For reflection to be most successful, educational experiences should actively prepare students to use reflective techniques that can be transferred to practice" (Kelly et al. 1996). The author of this dissertation approached the role of a reflective practitioner by writing a journal during the period of nine years, and wrote a book and future forecasting report after working at that specific communication and mobile technology corporation for 9 years. These texts and structural analyses of over 200 pages, can be seen as the author's exit reports from the corporation, or as efforts to help the corporation through a painful and disastrous situation caused by its outdated technologies and a collapsing global business.

Schön et al. (1986) see the use of knowledge-in-action as a central part of the skill and expertise of professional practice. Knowledge-in-action has also been divided into a twofold phenomenon of craft expertise (Knowles et al. 2001) and tacit knowledge (Martens 1987). By reflecting on what we actually do, our own experience improves our practice and professional development. This reflection generates an in-depth knowledge grounded in practice (Ghaye et al. 2014). This knowledge is used by practitioners in their work, and in this way, it evolves into knowing-in-action. Much of this insight and knowledge can be difficult to verbalize, but it is realized in the practitioner's behavior (Schön et al. 1986).

Kolb (1984) defines experiential learning as knowledge is created through the transformation of experience. This means that experiential learning is a holistic process that includes all of our life experiences (Kolb 1984; Houle 1987; Pamungkas et al. 2019). Kolb (1984) expresses that effective learning must undergo four processes: concrete experience, reflective observation, abstract conceptualization and active experimentation abilities. The concrete experience of learning requires apprentices to be receptive and able to adjust themselves systematically to access problem circumstances. In the reflective observation process, apprentices monitor simple presentations by expressing figurative imaginations and outlooks on why and how they happen. Reflective observations transform experience into learning (Carson et al. 2006; Scott 2010; Sheehan et al. 2009). This helps apprentices to understand concepts generally based on real-life experiences and reflective observations. Logical thinking is needed to understand problems during the abstract conception phase.

The active experiment stage completes the learning journey. At this stage, apprentices are able to use the theory achieved during the abstract conception stage to forecast various situations in practice. The apprentices are able to construct reasonable constructs, and their self-confidence grows to the level of capability to solve problems and make relevant decisions (Pamungkas et al. 2019).

Kolb (1984) also divides learning styles into four types. First, in a diverging style (feeling and watching), apprentices watch rather than act and appear to collect information and use imagination in problem-solving. Second, in an assimilating style (thinking and watching), apprentices favor forming idealistic notions. As Kolb (1984) proposes, one of the strongest abilities of assimilators is to build a theoretical model that combines different observations into comprehensive descriptions. Third, in a converging style (thinking and doing), apprentices have the ability to manage practical problems and advance in attitude and decision-making. Fourth, in an accommodating style (feeling and doing), apprentices prefer to encounter occasions directly by dealing with problems and exploiting their own or other people's praxes, knowledge, expertise and analyses. After this, it is possible for the apprentices to perform simple experiments because of their growing courage to take risks.

In this dissertation, the reflective practice is used as method described by Amulya (2004), where reflection is a dynamic process that proves our own experience so that we can look at it more closely (Amulya 2004). The reflection has been situationally twofold, either during an activity or separately later as a reflective examination. As Amulya (2004) argues, the objective of reflective practice is to produce learning from experience. The reflection can be practiced at different *frequencies*, vary in *depth*, and serve various *purposes* (Amulya 2004). The reflection by the author of this dissertation has practiced reflection in context of each process that addresses to creativity, design and their practical outcome like created artefacts, concepts, products, services and systemic solutions. In these contexts, the reflection frequency has been every day when the contextual need occurs, like when lecturing. The depth of reflection has been professionally controlled by making notes during the sessions and leaning on the course material based on literature and own practice. Reflection has been done with the purpose to help the students' team to be more effective and innovative in their designs, the Trio or startups (Article II) to be more aware of their business opportunities. Also the author has used reflection to reinforce common thinking or to open the structures behind complex and systemic endeavors like design skills (Article I) and disruptive actions (Article II). The reflective practice procedure has continued as an ongoing process during the dissertation project.

The reflective practice process can be either individual or collective (Amulya 2004). During lecturing and writing articles, the reflection of the author has been individual, except when writing with colleagues. When preparing lectures, lecturing and reading scientific literature, the reflecting has been collective. Here collective means to reflect the content and meaning of literature to own experience and arisen consciousness of respective topics.

Intuition is significant factor for reflective practitioners. Like Schön (1984) express, that some practitioners bring practical epistemology to uncertainty, instability, uniqueness by artistic and intuitive processes. Schön (1984) continues by saying, that in a spontaneous execution of day-to-day life activities, we feel aware in a special way. Often we cannot express what we know and we may us confused, or we develop explanations that are apparently unsuitable. In these cases, our expertise is usually tacit, indirect in our models of operation and in our sense of the things with which we are working. Schön (1984) concludes, that here it appears correct to state that our knowledge emerges in our actions (Schön 1984, p 49). The author of this study, as a practicing architect, artist and industrial designer, finds himself to ideate and concept(ualize) in a way, where the mind, hand, pencil and sketch paper are in notionless, indescribable and nonverbal connection at the present moment, "Now". Stern (2004) argues, that the instant while we are living, is psychologically and consciously the hidden but obvious reality right now, in the present moment. According to Stern (2004), the present moment is not a moment as any other; it appears to comprise a punctual but continuously rich mediator that has a great transmuting prospect (Stern, 2004). Stern (2004) emphasizes the innovative power of the present moment. Describing the innovative power of the now-moment, Stern says that while they seem so ordinary, every moment is by nature a feeling of something new. Often reflection-in-action depends on the perception of surprise to the extent, that when intuitive behavior causes surprises, pleasing and promising or unwanted, our reaction is possibly reflecting-in-action (Schön 1984, p 56).

Kautz (2005) separate "true" intuition from other types of "general intuition". Kautz declares that some practitioners feel that

> Intuition is the mental process of acquiring information and knowledge directly into the mind, without the use of reasoning, sensing or even memory (in the usual sense of that word). This definition implies that, if one is to show that a piece of new information is truly intuitive, he must demonstrate that it could not have been obtained by one of these other three means. (Kautz 2005, p. 8)

The practitioner and intuition can be understood by the idea, that intuitive skills can be cultivated with practice or strengthened with constructive circumstances (Hogarth 2001; Kautz 2005). Hogarth (2001) expresses, that emotions and sensations are an important factors of intuition, but questions if intuition can be taught (Hogarth, 2001). An interesting viewpoint is, that several practitioners' opinion to intuition is, that if intuition is reliable, all emotions are outside, and if emotions exist, many intuitive persons characterize intuition as unreliable and in some way ''contaminated' (Myss 2005).

The author of this dissertation has as a reflective practitioner, according to the author's self-monitoring, performed intuitive judgment, design, structuring and form-giving. Intuitive judgment was present in Article V, when the assignments during the design course were evaluated. In that case the intuition of that time focused both on holistic view of the design works, and on the details in the visual touch and how the design communicated the given task. In Article V the students' design works were evaluated according to the standard definition of creativity. There originality and effectiveness form the basis of creative factors (Runco et al. 2012; Corazza 2016). For the judges' use in the evaluation process, the contents of Effectiveness was characterized as useful, solid, adequate and suitable; commodity and firmness, and utilitas and firmitas as Vitruvius wrote 27 BC (Pollio 1914, 27 BC). Similarly, the contents of Originality was characterized as authenticity, genuine, real and original, beauty and delight, and venustas as Vitruvius wrote 27 BC (Pollio 1914, 27 BC). Throughout the evaluation operation of the students' design works, the judges had to estimate the qualities of the works against various synchronous factors that the constructs *effectiveness* and *originality* include. Author of this study became aware of, that intuition as an action explain accurately the character of the evaluation process.

As a designer's and form-giver's practice, the author of this study, prolongs the lifespan of the drawn imaginary sketches and abstract figures, to later get a meaning and scale and position in a realized idea. The present author feels, that the statement of architect Louis I. Kahn (1961), originally an Estonian from Kuressaare in Saarenmaa, later US citizen, pronounced, that a great building "must begin with the unmeasurable, must go through measurable means when it is being designed and in the end must be unmeasurable" (Kahn 1961, Delpino 2017). In practice this means for the present author, that the sketching process is "an intimate moment without connection to time and space but connecting only with the magical instant of form-giving. The present author finds this form-giving process similar now as it was over 30 years ago, and strongly sees it as a reflective practitioner's intuitive initial design process. By the same time when the drawn sketches seem to be purposeless compositions, they are scalable and reusable universal lines with more meaning than the artist can right now become aware of. This, in the present author's case, means that the sketches are surprising and promising, valuable for later use (examples in the Appendix).

As an individual finding, the author of this study has had the opportunity to use several sources of a practitioner's professional backgrounds and notes. When facing a new notion, concept or construct, they usually have similarities, taxonomy, processes or systemic nature with the earlier experiences and knowledge that the author has undergone. For example the notions of *design* and *architecture* have been borrowed from the history, and applied in more superficial ways to software development and computer science, meaning more the word *structure* and a large and fuzzy and context dependent set of definitions, descriptions, processes, plan and/or outcomes. Here the architect and industrial designer can easily see the connection to historical notions of *design* and *architecture* that often is more precisely defined (more in section 5.2.1). Still, the use of historical words with strong meanings is laborious for an old school professional to accept and use without particularizing the essence. An example of systemic nature is comparing the design and manufacturing of mobile phones and an individual apartment building. The phone has to be prepared to mass production, but the house has to be individual. The scalability of the phone platform has to be easily and cheaply multiplied, but the individual house should be individual from beginning to end. These comparisons between the present author's professional histories makes the reflective practice a meaningful and inventive method.

The present author also reflects the process of this dissertation approach to his backgrounds and experience. Outcomes of this experience have been a centric part of the author's work practice during the years 1986-2021 (examples in the Appendix). The Appendix presents a collection of reflective practitioner's notes, designs and sketches drawn over time illustrating the work process, intermediate results and the reflection process.

## 3.3 Observation

Observation as a research method is complex because it requires the researcher to take on multiple roles and use multiple techniques, such as his/her own senses, to gather data. Furthermore, to his/her degree of participation in the research team, the researcher's primary task is to remain sufficiently independent in collecting and analyzing information relevant to the problem under study (Baker 2006). Despite that there are few definitions of observation, Chatman defined ethnography as a method by which a researcher can gain insider insight by observing and participating in social conditions that reveal reality as the members of these ordinances live (Chatman 1992, p. 3). Gorman and Clayton (2005) argue that observation define observation studies as those that "involve the systematic recording of observable phenomena or behavior in a natural setting" (Gorman et al. 2005, p. 40). Spradley wrote that participatory observation "leads to ethnographic description" (1980, p. vi). He defined ethnography as the depiction of culture as a work that aims to understand another way of life from an original perspective (Spradley 1980). Becker et al. (1970) defined observation as either a hidden or overt contribution in which the observer participates in the daily lives of the people being studied, observing events, listening to what is being said, and questioning people for some time (Becker et al. 1970).

The researcher may have various roles in relationship with the research team. According to Gold (1957), the typology of Buford Junker with four roles in participation with the research team, researchers can have roles of a complete observer, observer-as-participant, participant-as-observer, and complete participant. Later, Adler et al. (1994) argued that the roles of complete observer and observer-as- participant had transformed to greater involvement in qualitative observation. It meant that the researchers adopted a more active membership role in the research team (Adler et al. 1994).

In this study, in Article IV, the participants were observed when playing an online videogame. The observation was carried out in the players' natural habitat, in which the players acted as they normally did in their own familiar settings. During the observation sessions, thinking-out-loud, think-aloud, method was used in data collection. Participants shared their experiences, and they typically did it cautiously, creating a dialogue explaining what and why they were doing during the play. After the observation period, a semi-structured interview was performed with each participant. As in the preliminary study, the pre-prepared questions focused on the structures of the research framework. In addition to the pre-planned questions, each participant was asked questions based on both the observation and their answers to the pre-planned questions. The interview and observation were recorded and later transcribed. Transcripts were analyzed for this study. This applies to data from both the preliminary study and the main study.

## 3.4 Experimentation

In the experiment, intervention is introduced on purpose monitor its effects. This is control that essentially allows detection of processing and outcome in experiments. Controlled experiment is a method of studying causal processes through cause-and-effect analysis. As a result, predictive findings and descriptive results are produced. Controlled experiments can be used to test theoretical ideas, even with small groups of people. In addition, the results can be extrapolated based more on logical reasoning than on statistical judgment. Controlled experiments can test the cause-and-effect relationship in a setting that avoids irrelevant variations in the phenomenon. This means that the variables are controlled so rigorously that irrelevant variables cannot produce the same effect. In a controlled experiment, there is also a control group against which the experimental results are compared (Grabe et al. 2003). The strength of the experimentation is its ability to explain the causal rationalization. The weakness of the experimentation is the question about the amount to which the causal relationship can be generalized (Cook et al. 2002).

Experiments in which research units are assigned to experimental groups nonrandomly are called quasi-experiments. They can be used to investigate causation in situations where randomization is inappropriate, impractical, or too costly (Kampenes 2008). Both experimental and quasi-experimental research plans investigate if a causal relationship exists between independent and dependent variables. As such, an independent variable is a variable of effect and a dependent variable is a variable that is affected (Loewen & Plonsky, 2016). Because of this, the independent variable is anticipated to cause about certain variation or change in the dependent variable.

In software engineering, experiments are often quasi-experiments in which it has not been possible to randomly divide the participants in the experiments (Wohlin 2012). In Article III, the experiment was arranged by following quasi-experimental way to get sufficient amount of students to participate the experiment. For practical reasons, the IT students from the University of Jyväskylä were invited to the experiment. The experiment was performed during two consecutive evenings. Participants had to attend either on the second night or both nights. The objectives of the experiment was to create an educational board game that had to meet the objectives 1) students should learn the basic concepts of Essence and SE in a fun way, 2) the board game should teach a method agnostic view of software engineering, and 3) the board game should teach the importance of teamwork and communication in software engineering project work. Here the students were participating in different teams, but any of the teams were control groups to each other.

The gathered data was qualitative, because the focus was on the subjective experiences of the individuals playing the board game. After the experiment, a quantitative post-game survey was held. The survey consisted of a wide range of criteria of game specific factors, usability, User Experience, and education usability (Scholtz 2016). The findings of this quasi-experimentation was, that that the game 1) managed to teach first-year software engineering students the basic concepts of Essence and software engineering in a playful manner, 2) teaches the method agnostic perspective of software production, and that software engineering methods are modular, and 3) teaches the importance of teamwork and communication in software engineering project work.

In the university setting, where students participate suddenly organized experiments, quasi-experimentation allows to build practical tests and experiments supplemented with surveys. The practical session like gaming together, added with the data collected with a survey, gives material for further analyses. A step to follow the true experimental patterns allows comparisons between the control group and experimental group, which requires larger arrangements and practical setup for the experiment session.

## 3.5 Creative analysis

Creative analysis is a multifaceted experimental method that uses illustrative and verbal phrases of communication (Zierer 1976). Originally, the testable and theoretical structure of creative analysis, developed from Dr. Ernest Zierer's (1976) definition, was designed to set different standards for "normal" and "ab-normal" art. The initial doctrine arose from the idea that mere verbalization will not produce results. Nevertheless, concurrent oral and visual juxtaposition can affect the experience and insight of emotional difficulties (Zierer et al. 1952; Zierer 1976). This notion of creative analysis was developed in the context of creative therapy, which "attempts more than mere description," as defined by Fenichel (1948).

Creative analysis is used in Article I as an experimental and reflective method describing the use of artistic design flow in the arts. From this description, the flow is applied to software engineering and software startups' use. Partly the artistic design flow and its conventions already exist in software engineering through visual and graphic design, usability and User Experience, as well as the physical and industrial design of respective devices, accessories and mechanics.

In Article II, creative analysis goes through the analysis of three technology and media industries and their possible joint venture, the Trio. The three industries, the network operators, network manufacturers and media houses are more as customers for each other, not close cooperative partners. The three stagnating industries, according to their interests to gain back their days of success and incomes, may rise from same interests. Here creative analysis is in the midst of solving their common trouble, refreshing their businesses with a mutual venture, the Trio.

The original concept of creative analysis was applied in Article V as a judgment method to analyze the students' creativity in their visual design assignments and sketchbooks completed during the course. As the study setting was a design course, not a therapy session, the method was applied because of the congruence between creation during assignments and the evaluation by design professionals. The assignments steps of ideating, sketching, drawing, note-making and visualization were prototyped during preceding courses, and these activities were present at every lesson throughout the course in practical work. In addition, the judges used evaluation practices that had been traditionally used throughout their education and their own teaching careers. The agreed-upon evaluation criteria for the assignments were:

1. Effectiveness: useful, solid, adequate and suitable; commodity and firmness (Vitruvius's idea of *utilitas* and *firmitas*).
2. Originality: authenticity, genuine, real and original; beauty and delight (Vitruvius's idea of *venustas*).

When comparing Zierer's (1976) original methods and criteria with Article V procedure, the similarities are obvious. The significant color interrelation experiences in Zierer (1976) equate to design and conceptual maturity in Article V assignment qualities. Similarly, the problem-solving approaches in Zierer (1976) equate to students' design skills and progress.

In the case of Article V, creative analysis revealed how individual progress within skills and concentration led to an awareness of the student's functional growth, problem-solving ability, and growth in design skills and conceptual design. The assignments required students to delve into design and diligence to make notes and ideate given topics. Zierer (1976) suggests that therapeutic testing can be seen as a broad tool for self-understanding and self-realization, and it also serves as an effective healing and social educational tool. In the context of students at a design class, this proposition by Zierer (1976) scales that students get a glimpse of for self-understanding and a growing conditional self-efficacy. These two thoughts and beliefs can be seen as enlightenment through hard work and a continuous pursuit of design. Creative analysis can be seen as a matchmaking method between two different counterparts, finding connections and interfaces for their coexistence.

# 4 OVERVIEW OF THE ARTICLES

## 4.1 Article I: What can software startuppers learn from the artistic design flow? Experiences, reflections and future avenues

Risku, J., Abrahamsson, P., (2015). What can software startuppers learn from the artistic design flow? Experiences, reflections and future avenues. In *Product-Focused Software Process Improvement: Proceedings of the 16th International Conference PROFES 2015* (pp. 584-599). Springer, Cham.

**Research questions**

In this paper, creativity is seen through a holistic process and a set of operations. The process can be described as an artistic creation process, which starts from an idea and ends as a real-world artifact. When the target is an artifact, the process follows a waterfall-type path to an endpoint or a realization. During the process, several iterative loops are executed until a viable version is accepted.

The creation process has been described as an individual artist working alone as a first-mover. The artist's creation is in the form of abiogenesis, coming into being for the first time. The history of art and design is an interminable source for startups' product and service creation. In addition, besides thinking about design (design thinking), product creation and design are considered a practical action that requires both designers' way of thinking and a full skillset of design principles and practices.

A software startup company begins with an idea, and it should be realized and launched. Startups have to face several obstacles in the process of get-ting to the market, and more than 90% of startups fail (Giardino et al. 2014). When four crucial dimensions of a startup company—product, team, business and market—are viable, a thorough design process is needed (MacMillan et al. 1987). This paper explains and applies artistic design flow–related factors to the creation process of startups. Future steps to overcome the difficulties that may lead to the failure of a startup and a proposal to embed design principles from the arts into software development are presented.

**Findings**

This study identified several design factors using traditional art and design forms and processes. In software development, there are several art and de-sign–related entities, such as usability, user experience design and user-centric design, which have been in the core of architecture and industrial design for ages. Graphic design has an important status in Web design because the look and feel of sites and their pages attract wider audiences, especially in branches where competition is harsh.

Other widely used artistic design methods in software development include paper prototyping, sketching, structuring by drawing, flow charts, simulations and even scale models when the product is physical. Usually, the startupper does not have organized education on these topics

As a popular topic, design thinking takes possession throughout industries and academic circumstances (Cooper et al. 2009). Design thinking, as a process of conceptual and practical creation of artifacts with a collection of methods and procedures, fits the software development process. A startupper who is a first-mover, a lonely innovator, can become a qualified designer. The startupper becomes a designer who does more than think: they craft.

The paper presents the idea of embedding designers' ideals as an integral part of software creation and development. Four alternative avenues to combine design and software include a radical, conventional, arrogant and ignorant direction. In the first alternative, the software designer is seen as an idealistic figure, a skillful, enthusiastic and educated designer. In the second alternative, conventional, the prevailing condition of today, to outsource design, would continue. Third, the arrogant alternative serves to improve the existing design habits of software development. Lastly, the ignorant direction allows design to have a peripheral status, where code and its technical paradigm are enough.

The proposal to embed artistic design principles into software development involves organizing practical educational subjects, starting research programs looking into how students at universities could study design, and preparing the universities' teaching staff to provide up-to-date courses. The research and piloting of design courses would be joint actions to find suitable curricula for software development, design and fit for appropriate skillsets in software creation.

**Connection to the objectives of the dissertation**

This study aimed to answer RQ1 by understanding how to embed practical design skills to improve creativity in software creation and development by applying the artistic design flow, skills and tools. The focus groups were startuppers, university students and some university teaching staff. The focus groups formed a horizontal and comprehensive group that participated in innovation and product creation from different perspectives, periodically adopting and exchanging their roles.

## 4.2 Article II: Software startuppers took the media's paycheck: Media's fightback happens through startup culture and abstraction shifts

Risku, J. & Alapekkala, O. (2016). Software startuppers took the media's paycheck: Media's fightback happens through startup culture and abstraction shifts. In *2016 International Conference on Engineering, Technology and Innovation/IEEE lnternational Technology Management Conference (ICE/ITMC)* (pp. 1-7). IEEE.

**Research objectives**

In this paper, creativity is seen through disruption and counter-creativity as friendly revenge; joint-endeavors are undertaken by industries that were not originally in the same boat, so to speak, but have encountered the same difficult circumstances. The paper describes the changes in communication and technology industries as having enormous effects on financial and monetary flows. It also discusses the possibility of repatriating funds from their own businesses through joint ventures and counter-disruption. This happens when three partly related industries (in the communication field) share common interests (to survive, to recover business and to make a disruptive change in the technology and communication ecosystem).

Interests have separated the three industries from each other. Journalism and media houses are in the news and entertainment business. Network operators have a direct relationship with the end-user, the smartphone user, through their SIM card. The network manufacturer is an outsider without a loyal network carrier because the latter has abandoned the relationship. The network manufacturer has to balance between prior relationships and the general atmosphere and evolution in the communication business.

These three separate businesses, with their media content, delivery system and wires, are poised for change. This change has been made automatically, gradually and silently with the transition of the Internet. This transition has brought us browsers, Web pages, search engines and algorithms. Today, algorithms are extremely complicated, secret and hidden. So are the cookies, browsing histories and attacks from advertisers. There is no visibility into the extremely complex reality of what happens to data, media and communication. You do not even own your digital self (your physical self is luckily still your own).

The paper proposes a disruptive start or renaissance for the Trio, the strategic joint venture of the media industry, network manufacturers and network operators, as payback against the newborn vampires, like search engines and social media companies. In addition, companies making money off of others' content are enemies.

**Findings**

In this paper, creativity is seen as a correlative process, where the evolution of businesses goes through incremental steps, pivoting and responding to competitors' challenges, or through disruption. Disruption may happen accidentally or intentionally.

An interesting finding of the study was the possibility of using the Cynefin framework like clockwork. The study shows that all three industries are moving toward the border of complex and complicated. Mobile network manufacturers and operators have the shortest distance to travel. Journalism has the longest journey through chaotic phenomena like the sensation of inability, poverty to invest, and the despair of losing centric business strengths and audience.

Following the success stories of startup culture, the study found that disruption is the main notion. Among the three industries, disruptive actions can be taken by joining forces. Here, the solution is the Trio, a joint-venture internal startup. The structure to combine three industries is based on the TAIC-SIMO model, where the four vertices of the tetrahedron from a union of technology, channel, interest and access. These dimensions describe the components that would be organized and managed by the Trio. Each of the three industries has specific strengths and focus points by dimension and comprehension. Three common factors form a bond between the industries: they are all in trouble, they live on content, and they understand one another. If all three move toward the border of complex and complicated domains, their businesses will be hard for competitors to overtake.

The threats of the complex domain can be avoided because each industry has spent time in the simple and chaotic domains at some point in their histories. The only challenge is finding the means to balance complex and complicated domains. Therefore, a stable position between Cynefin's Complicated and Complex is proposed, called Ribbon-Bowtie. It is a strategic endeavor that requires balancing creative (managing complexity) and stable (creating complicated enough products and services) mindsets. Here, the leadership model is crucial, which gives more authority for creative and disruptive forces and talents. In other words, the leadership model follows the freedom of ideal startup companies.

**Connection to the objectives of the dissertation**

This paper contributed to RQ1 by exploring how three global technology and communication industries can regain power when they have lost their dominance in their own industries through disruptive actions and counter-creativity. The three industries—media houses (a.k.a. journalism), network manufacturers and network operators—all are experiencing industry-specific death struggles. Journalism has lost the superiority of newspaper-ads as moneymakers, and advertisers have shifted from printed newspapers to Web services. Network manufacturers are in a zero-sum game, struggling with 5G, IoT and network operators' unwillingness to invest in newer technologies. Network operators are in a poisonous battle of footprint, growth and existence. Uncertainty follows any actions because of the response by competitors and the external conditions.

The study proposes that an internal startup, the Trio, which is a collaborative joint-venture between operator, manufacturer and media house, acts as a disruptive startup. It would have massive funding because the three partners already have large innovation investments. The Trio could act independently and stealthily, with reasonable properties. The intention is to create unprecedented business and industrial solutions, boosting two high-tech industries into a new era. Startup sensitivity can also be applied by offering competitors, who have neglected to develop similar solutions, parts of the developed solutions and services. In this way, every competitor could receive a customized solution to compete within the specific industrial context, but with a different composition, price and support. The Trio could also be an early-stage startup with a strong attitude towards disruptive action in the three industries.

The proposal of three different industries to join their forces may sound impossible and dangerous in their established business areas. Here, the early-stage either meet an insurmountable challenge because of the magnitude of the three industries and their incoherence. The three industries serve each other, are dependent on each parties' businesses. Also the industries represent corporative policies and are not easily approachable. But, on the other hand, the adolescence of early-stage startups and the startuppers' inexperience can turn to a cutting edge, when taking a stance of disruptive actions. When Article II propose that a joint internal startup the Trio would be the answer, an early-stage startup could bypass the problems of three conglomerates to establish something what they have not done before. The young startup has several advantages on their side: nobody believes that a novice can penetrate the billion dollar industries with a brilliant concept.

In this paper, creativity is seen as a quality that scales from system-level planning via organizing businesses to the creation of products, services and solutions for customers, whether they are consumers or business partners.

As an application of this concept, the Trio would be established, the journeys toward the line of complex and complicated would start, and the common and overlapping products would be created. The assumption of intentional capability means that each partner of the Trio would concentrate on its talents and strengths and rely on them but would also accept accidental acts and ways to proceed. At its most obvious dimension, the Trio would involve a division of work, roles and responsibilities, but along the disruptive journey, new roles and actions would come up (Snowden 2000; Kurz et al. 2003).

Creativity in organizing businesses in the case of the Trio would mean that the three representatives from different industries would have to adopt new practices and agendas simultaneously when their own business confronts adversities. The companies would have to find new ways of solving problems and change their focus to new and viable trades. Xerox and Kodak are an example of when the innovation culture turned to aversion in building technologically new solutions. The reason was that the businesses ran well, and replacing their long-standing success was too troublesome. It would have meant taking risks and changing their business model (Cuthbertson et al. 2015).

The Trio could mainly continue with their strongest expertise and business models, but the joint venture would introduce mutual overlaps in interests and capabilities to organize content creation and distribution, prepare plat-forms and prepare developer communities. This approach would widen and deepen the Trio's partners' readiness to serve customers, from individual users even to competitors.

The Trio and its area of operation, the business environment of the media industry, network manufacturers and network operators, is an extremely complex, gigantic and futuristic. However, the industries struggle with slow growth, uncertainties of the breakthrough in new technologies, and the incomes based on advertising. The idea of the Trio is constructed on the problems each of these industries face: Network operators are struggling with zero sum game, where acquisitions seems to be the only solution. Network manufacturers are few, and they have not yet benefitted from IoT and 5G. Media industry, especially the print and journalistic media, has lost their incomes because of the dynamic and easy Web based advertising services by search engine and social media companies.

The Trio as either a joint internal startup of three different industries, or a strong early-stage startup with a disruptive agenda to congregate three parties from the industries.

The early-phase startup perspective to the Trio relates to RQ1 and RQ2, how design and creativity can support the development of their ideas to concept better. When the combination of three different businesses, network operator, network manufacturer and media house has to join their forces, creativity, design and concept creation are needed. Especially the contextual factors are harsh: the three industries have been well established, they have an impact on international level, and the companies are conglomerates. If an early-phase startup wants to propose limited and detailed solutions for these industries and companies, together or separately, it is difficult even to contact the parties. When the startups propose systemic and groundbreaking solutions to restructure whole industries and ways-of-working, it may sound impossible. But, here the startup culture has an advantage: a startup can come like an ulterior and uncharted actor, which can be invited as a potential rainmaker. This possibility is built in the startup culture, because startups can disrupt stable business models (Märijärvi et al. 2016). The early-stage startups, in case of proposing disruptive and systemic solutions for business conglomerates, have to prepare themselves with equal qualities as a large technology research and development companies. As an example, the company PricewaterhouseCoopers PwC, has a wide set of solutions and services for the telecom & media industries and companies (Gustavsen et al. 2018). In the startups, this means to both study design and industry specific subjects, and recruit investors and people who are experienced in these industries. This means in practice, that the startup has to concentrate on topics like the culture and processes of the respective businesses, understanding complex and systemic businesses and their present strategies, and propose new disruptive and viable solutions, recruit senior professionals to guide and contact the parties, base the actions on research, and be aware of the means to ensure the retention of the parties' customers.

The Trio relates to RQ3 in the sense that method prison can be avoided by the Essence practices, which leads to method agnostic routines and allows to differ from established procedures and even ideas. By the same time, creativity and professional skills in recruitment are crucial, especially when it is debatable that do they correlate with desirable self-efficacy qualities.

TAIC-SIMO tetra(hedron) model organizes the three industries of the Trio to their common joint venture. The vertices are named with the industry specific factors that represent their relation to the consumers and users in form: Technology refers to network manufacturers, Channel refers to network operators, and Interest refers to the media industry. Their common endeavor refers to Access. The TAIC-SIMO model makes it possible to crystallize the industries by nature and create a new joint program with new offering. In the model, the three industries maintain their clear competence area and expertise, but with the Trio they have a common product and service portfolio for themselves and through licensing for their competitors. As a head start, the three industries have a significant footprint in continental and even global magnitude. Also these three industries can profit from each other, and support in a new way the other's businesses.

Designing products, services and solutions for consumers and competitors is a challenge to creativity. Consumer products are already or have been in the network manufacturers and network operators' portfolios for more than a decade. Journalistic media houses write articles and news for media users, but broadcasting also has a platform dimension. The publication is available to readers on a publishing channel for a few days only. This manner changes when a developer community is scaled to a global reporter community. Creativity and disruption are bound together in this section.

## 4.3 Article III: Gamifying the escape from the engineering method prison

Kemell, K.-K., Risku, J., Evensen, A., Abrahamsson, P., Dahl, A.M., Grytten, L.H., Jcdryszek, A., Rostrup, P. & Nguyen-Duc, A. (2018). Gamifying the escape from the engineering method prison. In *2018 IEEE International Conference on Engineering, Technology and Innovation (ICE/ITMC)* (pp. 1-9). IEEE.

**Research objectives**

This paper examined how game-based education helped software engineering students to escape from the software engineering method prison. To fulfill this objective, an educational board game was created with three characteristics: 1) provide an entertaining way of learning Essence concepts and software engineering in general, 2) teach software engineering in an agnostic and comprehensive way by explaining the modularity of the methods and 3) emphasize teamwork and communication in software engineering.

During the research and development process, the theory behind SEMAT and Essence was analyzed. These entities naturally characterize the game and influence the game and ways of learning. When SEMAT is represented as a modular library of practices, like a software development "Lego kit," it is familiar and understandable for first-year software engineering students. The idea of the board game was to allow practical studies of and dynamic access to the development of structured software for software engineering students. These initiatives included project management practices as an umbrella to fit software development practices into the application creation process. Depending on the development environment, the processes can be described as wide, large, slow, fuzzy, vulnerable, pressing, distressing, controllable and uncontrollable, rewarding and meaningful. For first-year students, the touch to software creation and development process can be confusing. The board game as a medium was rationalized based on the characteristics and their stated outcomes for the study and development project.

Essence is an abstract model of the most important and monitored matters in software system development. It can be too abstract in leading the software development process, especially during the first end-to-end development projects executed by a first-year student team. Therefore, when Essence was featured as a project management window into the development process, the game had to clarify its topics, team, roles and responsibilities. All diverse soft-ware development methods and practices needed to be parts of the game in practice. When universities are assimilating Essence to teach the students about the primary principles of software development, a game fits this objective. The importance of Essence is that it can be used to explain various practices and present them clearly to students.

**Findings**

The developed game taught software engineering students about the basic concepts of Essence and software engineering in an active way and an encouraging atmosphere. The late evening case sessions at a remote campus building were communal and lively, and the startup culture was at its best. Many young people participated in teambuilding, which led to coherent groups of people who had been strangers. There was a colorful board game with well-explained instructions and pizza. The game walk-through lasted several hours, and everybody had the energy to play and give feedback. As a research and development process, the board game was an encouraging technique to find just-in-time and live practices and flow on the fly.

The gamified model, a game-based learning tool, was an enjoyable experience for the participants. As the board game fulfilled the set objectives, it was deemed a useful means of teaching the Essence method in the future. The notion of method prison in the context of startups and student may have a twofold meaning: either, as a novice, you cannot be in the prison because of studies on methods, processes and practices, or precisely this is the reason. Like Article I emphasizes to learn a meaningful design skillset for software development, Article V evaluates the skills reflected with creativity. Here in Article III, to free oneself from the method prison, is a function of skills and creativity to free oneself professionally to use the skills in design, and creatively to use development processes. During the play of the board game the students found pleasant, which led to play the game all the way. Apparently, one has to use all available means to escape the method prison.

**Connection to the objectives of the dissertation**

This study addressed RQ2 and RQ3 by providing a theoretical explanation of Essence and by developing a board game that illuminated how game-based education can help the software engineering students to escape from the software engineering method prison. The method prison was processed from the fundamentals of the respective prison. A situation in which an organization is locked into using one or several specific method(s) is called a method prison. This happens regardless of whether students fit the current software engineering setting of the organization because they think this is the standard condition of an information technology organization.

When RQ2 emphasizes the processes and to escape the method prison, startups, beside technical debts in detail, have debts of processes. Typically there is lack of development process among startups. The lack of processes for managing, identifying, and prioritizing TD means that TD decisions are often made in individual cases and the team does not know how to make consistent decisions. This is important as the team grows to ensure compliance (Besker et al. 2018). When bypassing the method prison, the startup team uses creatively the conventional methods and can tackle given tasks rapidly. By the same time, the startups in large-scale initiatives like the Trio, can let their creativity to be the counterforce against the corporative culture. This would be a highlight of creativity by startups that RQ3 calls for. The escape from the method prison is enabled both by the board game and Essence theory. The game with its attractiveness for the players smoothens the way to play with software development methods and practices, and by the same time, the Essence and its method agnostic nature is presented.

In the board game context, creativity as a factor in developing something new and exhilarating was actualized in the form of the researchers' enthusiasm to create something extraordinary. When this happened at an academic startup lab in which practices of "concoct—concept—prototype—test—deliver" were combined and where academic research standards were fulfilled, two milestones were achieved: expectations were set and realized by a good company. This is the basic setting for startup cultures: try in practice and follow with academic research.

## 4.4 Article IV: What makes a digital game addictive? A player viewpoint on player retention

Risku, J., Kemell, K.-K., Schweizer, S., Nguyen-Duc, A., Suoranta, M., Wang, X. & Abrahamsson, P. (2020). What makes a digital game addictive? A player viewpoint on player retention. Accepted to be presented at 2020 IEEE International Conference on Engineering, Technology and Innovation (ICE/ITMC).

**Research objectives**

This paper examined what online game elements and rewards influence players' motivation to continue playing digital games.

The acquisition and retention of users are a constant topic of digital innovations. Digital games are progressively moving toward free-to-play schemes, which rely on various in-game interactions to produce benefit, so retaining active players becomes increasingly essential for the developers of digital games. Online games have faced this challenge over the last 10 years, particularly those in the mobile game industry. In contrast, player retention as a component of loyalty has not been a notable area of academic study. Attracting new users and retaining them is a common interest of all digital products, services and innovations.

The research framework consisted of four elements: social, rewards, game mechanics and self-efficacy. The social dimension represented the player's connection to collaboration with the team of players, the relationship between players in general and socializing. Rewards were connected to in-game rewards that the player received and to the feelings associated with getting rewards. Game mechanics explain the structure of the game and how it is built. Mechanics also explain how the progression of the game changes during the play.

**Findings**

Our findings highlighted the importance of some existing good practices in the area and the potential negative impacts of others. We hypothesized that higher self-efficacy has an impact on player retention.

Self-efficacy belief relates to the impression that a player has of their own expertise in performing a mission in the game. The belief is strong when the player succeeds during play. The creativity dimension is built into the definition of self-efficacy in its eight categories: exploration, iteration, implementation, communication, resourcefulness, synthesis, vision and creativity. Self-efficacy is related to player retention. Self-efficacy's category implementation (by making risky choices or suggesting new ways of achieving goals) relates to achieving challenges, overcoming difficulty spikes and gaining rewards. These increase players' motivation to continue playing.

Difficulty spikes also relate to resourcefulness when challenges increase. Balancing between active or passive roles as a player relates to exploration in self-efficacy categories. This means that the player is observing people, considering viewpoints and understanding people's needs. From the self-efficacy clusters, creativity relates to the player's approach to challenges and gaining a feeling of achievement and survival when playing with random elements. The player's social aspect and relation to the character also relate to self-efficacy's exploration. Based on these factors, specifically the connection between player retention and self-efficacy, and the hypothesis that higher self-efficacy impacts player retention was validated. Motivation, enjoyment and success were all indicators of the player's higher self-efficacy.

One of the practical findings in answer to RQ2 was that the player participants considered the balance principle to be the most important for choosing to continue playing a game. In addition, when the game became gradually more difficult during the play session, players found it more challenging. This happened even when the players advanced in skills, and their characters got stronger. The players found that the feelings of achievement originating from different factors were exceedingly motivating.

In response to RQ2, rewards were found to be satisfying for the players, especially when overcoming challenges. In addition, positive feelings through achievement were essential. These can be enhanced through in-game rewards. Rewards were also emphasized by different actions, such as being rewarded for diverse playstyles, in-game activity, exploration and completion of extra tasks in the game. These findings lead to startups' design processes in case of game design, where several factors has to realize: 1) rewards has to be fine-tuned, 2) feelings through achievement can be tested in-house by playing, and 3) playstyles, in-game activity and exploration can be experimented beforehand. When taking these factors into account in the design process, the startups can ensure better customer retention, because the players continue to play the game, and they avidly return back to the game.

Further development and research into more specific game elements can reveal player retention factors in detail. Players' self-efficacy belief, which explains the connection between player retention and gaming, can be organized in a more customized questionnaire. For game developers, the results of detailed surveys and future research based on these are important sources that could produce more creative games.

**Connection to the objectives of the dissertation**

This study aimed to answer RQ2 by providing a focused view of the important genre of software startups to game developers. The factors, like retention, motivation, rewards and feelings of achievement, are common to all startup endeavors and even to all industries, offering consumer products, services and solutions.

The results found that online game elements keep players playing the game by strengthening the game players' relationship to their activities, motivation and feelings during the play. The study presents a formula for game developers to understand the player, their paying customer. Retention of the customer indicates loyalty to use and choose the supplier in the future. Therefore, player or consumer retention is a prominent factor to be considered in all stages of product and service design among startups.

## 4.5 Article V: Exploring the relationship between self-efficacy and creativity: Case IT & business education

Risku, J., Kemell, K-K., Kultanen, J., Feshchenko, P., Carelse, J., Korpikoski, K., Suoranta, M. & Abrahamsson, P. (2020). Exploring the relationship between self-efficacy and creativity: Case IT & business education. To be submitted.

**Research objectives**

This paper explored the question of whether there is a correlation between students' creativity and the results of self-efficacy questionnaires and design work evaluations.

To evaluate creativity, a course was begun, focusing on designers' ways of working. It included note-making, drawing, designing and visualizing ideas. The design work of the students comprised a sketchbook with notes made during the lessons and when working on assignments. Note-making included writing notes, outlining ideas and concepts related to the assignment topic, sketching, drawing and structuring with different methods, like mind-mapping, charts, tables, bullet-point lists and other visual techniques.

The fundamental difference between note-making and note-taking was explained: note-taking is the passive action of capturing information mainly from dictation and recording this information by hand on paper. This kind of note-taking is linear and records the lecturer's presentation with summaries, sentences, abbreviations and centric matters after listening or seeing. Note-taking was explained as a feeble and defective method when the intention is to ideate and create new ideas and concepts. In comparison, note-making is a medium of explaining thoughts and plans, organizing content, thinking creatively, building relationships and achieving views. Note-making was understood as a designer's tool for gathering their own ideas and designs in a structured and visual arrangement to be used later. The students were told that the notebooks were mandatory for the course.

For the creativity evaluation, students completed two separate design tasks. The first task was to design appealing services and entertainment for new visitors at an event. The outcome was a one-page concept with drawings, text and event-related factors. The second task was to design new city-centric attractions to invite young people to enjoy and even move to the city. Students had to create a two-page proposal with one rough ideation page and one conceptual design with overall and detailed concepts. The whole course was a hands-on design course, which handled design as a verb, which meant that the assignments followed a practice flow: envision–ideate–concept(ualize)–design–prototype-report–communicate in an iterative way. The students' design works were evaluated according to the standard definition of creativity, where originality and effectiveness form the fundamental notions (Runco et al. 2012; Corazza 2016). For the judges' use, a detailed guideline of the contents of Effectiveness was: useful, solid, adequate and suitable; commodity and firmness, and utilitas and firmitas as Vitruvius wrote 27 BC (Pollio 1914, 27 BC). A detailed guideline of the contents of Originality was: authenticity, genuine, real and original, beauty and delight, and venustas as Vitruvius wrote 27 BC (Pollio 1914, 27 BC). These additions were interpreted from the original articles' various set of synonyms of the notions effectiveness and originality. The notions firmitas, utilitas, and venustas by Vitruvius were important to consider, because these concepts are known among educated and professional designers and architects. During the evaluation process of the students' visual design works, the judges had to weigh the qualities of the works against multiple simultaneous factors that the notions effectiveness and originality consist of. Author of this dissertation understands, that intuition as an action explain attentively the nature of the evaluation activity.

Measuring the self-efficacy rate was the second part of the study. Self-efficacy was measured with the innovation self-efficacy measure (ISE) method (Carberry et al. 2018). The ISE survey is based on 29 activities related to innovation and creativity. There are eight categorized of activities: creativity, exploration, iteration, implementation, communication, resourcefulness, synthesis and vision. The ISE questionnaire was distributed during the assignment. The students filled out the ISE questionnaire on a Web survey service twice: at the beginning and at the end of the course.

**Findings**

Three experienced designers formed a panel of judges to evaluate the creativity of each student on a 7-point Likert scale. The score of creativity was then computed against the rates of self-efficacy. The results were analyzed and compared to the self-efficacy questionnaire results to evaluate the impact of the course on students' progress in design skills and creativity. The results revealed that self-efficacy and creativity were not strongly correlated. This leads to several findings, suggesting the relevance of self-efficacy as a related notion in a design-related course in a non-designer educational establishment. This would suggest that a professional designer who is satisfied with a design artifact and sees themselves reflected in other designers would achieve a naturally high score in self-efficacy.

In the study, the spread of correlation was wide, and no straight dependence between self-efficacy and creativity was indicated. One specific result appeared: students who were more experienced and skilled in design also got average scores in self-efficacy, which may indicate realism and self-critique when comparing oneself to design as general. A possible reason for the results was that the ISE questionnaire itself is long, and the questions are partly generic, wishful and difficult to direct to practice. Therefore, respondents may answer effortlessly or diffusively. In addition, the scale of the rate to input varies between 0 and 100, which may lead to general confusion to find meaningful differences between, for example, 57 and 63. When completing 29 questions, some inaccuracy can be accepted.

The findings suggest that new ways to bring design to academia are needed to improve students' design skills in general, especially outside of professional design universities and faculties. Because creativity and design are emphasized widely across industries and startups and in everyday work, it is appropriate to consider implementing design as a study module or even a curriculum at the Academia and Humboldtian universities.

Self-efficacy belief as a practical indicator (e.g., when objectively trying to evaluate creativity grades and judging recruitments or awards based on the results) need further research and development. In the research, it was proposed that embedding the designer's ideal in computer science and software development would enhance the students' readiness in their studies and later workplace. The Academia could, with a thorough program of research and development of design principles, integrate the ideal of creative and design capable students into trades, industries and workplaces. This could be realized by design education according to traditional design methods and new design methods from computer science and software development research. In addition, academic staff, researchers and teaching persons could manage better with new design skills. In this way, a new design culture can be created at the Academia and Humboldtian universities.

**Connection to the objectives of the dissertation**

This article answers to RQ1 and RQ3. Creativity and high self-efficacy are essential parts of the personal qualities of startuppers, impacting product perspective, finding a market fit, establishing a solid and productive team, and working in ever-changing, extremely uncertain conditions. Creativity can blossom when a person has practical design skills. Simultaneously, one's self-efficacy grows, increasing their courage and productivity. With a creative mindset and skillset, suitable means can be generated for survival.

The article evaluates the design skills reflected to creativity and self-efficacy by their correlation. The correlation was weak, which led to a proposal to concentrate on groups with various skills, and compare them.

The design skills were in focus when evaluating the creativity of the students through practical design projects, where visualization and conceptual ideas were emphasized. This combined both RQs by the importance of a design skillset that is useful in planning concepts and visualizing the design.

# 5 RESULTS AND CONTRIBUTIONS

This section presents the results of the research and the contributions to the literature and practices. Threats to the validity of the study are discussed in this section and limitations and ideas for future research are presented.

## 5.1 Results

This section introduces the findings of this dissertation on design practices in startup work, startups winning back lost resources (media, mobile technology), familiarization with software engineering through game-based education, customer retention in a gaming context, and a perspective on creativity and self-efficacy.

### 5.1.1 Design practices in startups

Software startups apply an array of artistic design practices in their daily work. Typical design-centric topics for startups are usability, user experience and user-centric design. These are used in application development to ensure the suitability of applications and physical products for the end-users.

Design thinking is an innovative approach with a set of three factors that define the innovativeness of an artifact: viability, desirability and feasibility (Chasanidou et al. 2015). Design thinking involves applying design principles in interdisciplinary teams to a wide range of innovation challenges. When innovation is a central requirement for companies and organizations to compete in the ever-changing circumstances of markets, it is important that the whole workforce have a common base to participate in their design objectives. In this case, design thinking is a widely applicable collection of tools for professional designers and non-designers (Seidel et al. 2013).

Visualization is an important aspect of Web design distribution. Startups use graphic design, icons and visuals to attract users. Visualization is important when considering the complexity of a system or reflecting a converging or different view of the design depending on the design phase (Chasanidou et al. 2015).

Other design procedures used by startups in product and service creation include the use of flow charts, paper prototypes, simulations, scale models, drawings, storyboards, collages and iterative improvisation. With these practices, the five phases of the design thinking—flowing, empathizing, defining, ideating, prototyping and testing—can be planned and ensured (Brown 2009).

To utilize these practices, software startups need a deeper and more thorough drawing and handcrafting skillset. Article I describes this finding, which is based on the idea that professional designers have an elaborate education and their working experience becomes gradually wider. Startuppers could also benefit from these professional skills and attitudes. This could also improve the early-phase development of startups' products, services or applications. When the number of team members is limited or the timeframe to produce the first version of an artifact is compressed, it is better to have all the skills and competencies required in the team. More advanced design skills also improve team performance. This lightens the obvious mental pressure achieved through self-efficacy and self-confidence.

### 5.1.2 Startup winning back lost resources

Companies in any industry may face a moment when the competition is stiff, and the evolution of technologies has passed the company's capabilities. Historically, there have been some technological revolutions in which an entire industry has lost its *raison d'etre*. The car, for instance, marginalized the use of the horse in locomotion. The horse had been the wheels of the social machine, the operative factor whose disappearance would harm all social classes and damage trades, agriculture and social life. Horses industrialized cities and mechanized agriculture. *Horseless Age* (published from 1895 to 1918), a magazine for car owners, declared that getting rid of the horse would reduce noise and save money. Steam engines replaced the horse in long-distance freighting and agricultural work. However, the automobile did not clean up cities because the horse dung was merely replaced with invisible gas. In addition, public transportation was abandoned by the rich and the middle class. The new car scaled the benefit that the horse created (Nikiforuk 2013).

This has also happened to print media houses, newspapers and journalism. The new technology, or the car, in this case, is the Internet, search engines and social media. Old media is the horse. The biggest losers in this situation are journalism and print media, but not free speech. Everybody can express their opinions on social media, but its footprint in publicity is disorganized and chaotic. We certainly need an organized, educated and critical public news media. However, resources like salary are missing. Search engines and social media were created by innovative startuppers, so there is a place for a counter-revival.

Article II describes the conquest to redirect advertisement money held by startup-originated search engines and social media companies back to media houses and journalism. The proposed vehicle is a joint-venture internal startup with three actors: one mobile network manufacturer, one mobile network carrier (operator) and one journalistic media house. They would form the Trio, which would organize a full group of platforms, applications, procedures, developer communities and training, encouraging people to join the business.

### 5.1.3 Game-based education

Software engineering as a notion and convention is under rapid development but is bound to the heritage of old and functional practices. To create applications and systemic solutions in this situation, the startupper must find appropriate methods and tools for creation and development.

Article III describes the environment of a supposed startup studying software engineering. An early-stage startup with a good and novel idea may lack the skills, knowledge and capabilities to build a full set of methods in days or weeks. Here, the SEMAT initiative, with its Essence theory, offers a solution. By studying the fundamentals of software engineering and SEMAT through a gamified board game, players can easily familiarize themselves with the procedures. The game developed for software engineering students teaches the basic concepts of Essence and software engineering in an entertaining way. The game teaches engineering processes, management and the team's roles and responsibilities during the play. The game is available to software engineering communities from academic surroundings as well as startups to learn the complicated ways of working to develop applications and software in teams.

### 5.1.4 Customer retention

Player retention, like customer retention in general, is a crucial dimension of product and service creation. Player retention can be improved by allowing virtual avatar customization, adjusting the game difficulty, organizing social relationships between players, providing in-game rewards and improving game mechanisms. Players' self-efficacy improves their enthusiasm to keep playing and makes the bond between the game and other players stronger.

The findings of Article IV can help game developers to enhance games by using the balance principle to scale the difficulty of the gain and by fostering feelings of achievement when playing. The players can strengthen their social interaction with formerly unknown people and still play with well-known players. Introducing various playstyles and rewards for players' activity, exploration and completion of extra tasks can also be motivating.

### 5.1.5 Correlation between self-efficacy and creativity

In a design course with students from non-designer faculties, self-efficacy and creativity are not strongly correlated. As an observation, skilled and experienced students in design received average scores in self-efficacy, which indicates realism and self-critique in their design assignments.

Article V envisions the Academia playing an important role in enabling creativity and self-efficacy among students and personnel. By improving the students' design expertise and routines, the Academia can convey design as a medium for creativity and overall quality in studies and assignments. Establishing a complete program to research and develop design principles in university curricula would enable the model of creative and design-oriented students to be integrated into businesses, industries and workplaces.

## 5.2 Validity threats

In this section the validity threats of the study is discussed. The term validity indicates to the approximate truth of a knowledge claim (Shadish et al. 2002). In this thesis validity definition of Runeson et al. (2009) is used. The validity categorization comprises of construct validity, internal validity, external validity, and reliability (Runeson et al. 2009). This classification is accepted in the software engineering research community and its use and applicability for case studies.

Reflecting on experience is an increasingly important part of professional development and lifelong learning. However, with regard to evaluation in particular, there is still uncertainty as to how the principle can best be applied in practice (Koole et al. 2011). This section examines these uncertainties to find practical ways to assess reflection from the mixed-methods and reflective practitioner's perspective of validity threats and reliability.

### 5.2.1 Construct and internal validity

The validity of construct describes the degree to which research methods for collecting research data and how drawing conclusions describe the desired answers to research questions (Runeson et al. 2009). Construct validity concerns in particular the validity of the building blocks and the way in which these blocks are assembled and abstracted into higher-level structures (Maxwell 1992; Shadish et al. 2002).

RQ1 addresses to startups' possibilities to develop their ideas to better concepts. This is a combination of relevant design skills and creativity that enables concepts and offering that lead to successful business. On a general level, a person's creative personality has tendency in courage to chance taking. By the same time, product creation involves teamwork to avoid individual disadvantages (Li et al. 2007).

Internal validity refers to how much erroneous causal relationships between the constructs are reduced (Yin 2003). In this study, causal relationships are handled in Article V. A research result may argue that Theme 1 influences Theme 2, also the Theme 2 may be influenced by Theme 3 (Runeson et al. 2009). In this thesis, the constructs self-efficacy and creativity are studied in Articles I, II, and V. When examining the relation between self-efficacy and creativity through design practices (RQ1), the practices of designers and artists are both analyzed as techniques, but also as methods to produce concepts and designs. When Tierney et al. (2002) define creative self-efficacy as the belief one has the talent to produce creative results, there is a mutual connection between the terms self-efficacy and creativity. The connection of self-efficacy and creativity is described by Mathisen et al. (2009) that a person's confidence in his/her own creativity makes him/her more creative (Mathisen et al. 2009). In this study creativity was evaluated by the produced concepts during design classes, judged by professional designers, and reflected to Innovation Self-Efficacy (ISE) Measure instrument developed by Carberry et al. (2018).

The results showed, that there is not a strong statistical correlation between self-efficacy and creativity in the investigated students' designs versus judged creativity, the internal validity can be questioned. But, as Schack (1989) found, that self-efficacy training may actually be driven to lower self-efficacy, if the participants find the tasks challenging, resulting in negative mastery and/or vicarious experiences efficacy (Schack 1989). This means, that there are possibilities that awareness of one's expectations cause a threshold in managing to fulfil a challenging task. In the study, internal validity from the research perspective, using the ISE instrument was adequate, but the setting of diverse skills and experience in design skills and creativity among the students spread the correlation of self-efficacy and creativity apart. The heterogeneous skills and experience of the students, and the possibilities to negative mastery and/or vicarious experiences efficacy, led to a conclusion, that when making a comparative study by two groups of persons with different design skills and experience, the correlation between self-efficacy and creativity can assumingly give a consistent result of the relationship. The comparative study by two groups would be ideal when the other group of students is diverse in design skills, and the other professionally oriented.

### 5.2.2 External validity

External validity refers to the generalizability of research findings (Shadish et al. 2002). It also suggests how much the findings are of interest to other parties outside the study (Runeson et al. 2009). In this thesis *design* and *design skills* are found crucial in the context of the Arts, architecture and e.g. automotive design (RQ1). In these classical design based and also handcrafted industries, skills to draw is a technique for form-giving of artefacts. Competent designers use the methods of art, craft, innovation and research to create form and artefacts (Keinonen et al. 2013). By the same time design is used in a similar way in software design as in the classical design industries. As Vermeeren et al. (2015) argue, that design in User Experience demands empathy and design skills from the designer (Vermeeren et al. 2015). Graphic design as an essential part of Web design, is a medium to generate ideas, use it as a problem-solving process, communicate messages, and interpret meanings to the audience. The evolution process of sketching has three stages: explorative, explanative, and persuasive (Olofsson et al. 2005) which relates to a taxonomy by Ferguson (1992): thinking sketch, talking sketch and prescriptive sketch (Ferguson 1992). These three phases represent the genesis of a product concept, numerous variances of a concept with different details, features or uses, and a finalized visualization with colors, views and contexts of a selected concept (Kudrowitz et al. 2012). As Kansrirat et al. (2016) argue, by the same time graphic design produces drawings and plans as results e.g. for architecture, fashion and industrial design. When generating graphics in form of visualizing data, Kansrirat et al. (2016) found need for skills in four topics: 1) data analysis; 2) data interpretation; 3) designing ideas; and 4) display by sketch design (Kansrirat et al. 2016). In this setting graphic design skillset includes software related topics like data analysis, data interpretation and designing ideas.

When combining design as a notion including both the *process to create* and the *result as an artefact*, traditional art and software development are using them both. Also graphic design being an essential part in creating, sketching and communicating ideas, data, messages and physical products, the generalizability of graphic design can be seen dynamic, eligible and essential both for the Arts and software development.

The generalizability of player retention (RQ2) scales to the importance of customer retention in other industries. As Kotler (1994) expresses, "the key to customer retention is customer satisfaction" (Kotler 1997, p.20). This phrase scales to satisfaction, which the customer gets through *compensatory* (rewards) and *non-compensatory* (penalties) qualities that influence on the formation of satisfaction (Hennig-Thurau et al. 1997). Player retention has relation to similar phenomena, like technology acceptance and technology use and continuance of use services (Davis 1985; Bhattacherjee et.al. 2008; Soliman et al. 2015). These areas of research belong as parts of in Information Systems research. Player retention in Article IV was related to social aspects, rewards and game mechanics. Sicart (2008) informs, that the traditional game mechanics describe how players interact with rules, and with formal properties of a game such as game goals, player actions and strategies, and game states (Sicart 2008). These factors relate to the findings in Article IV, that the players are satisfied when the interaction is challenging (rewards and feeling of achievement), applying different playstyles, rewards for activity and exploration (game goals), (also) different playstyles and exploration (player actions and strategies), and the balance principle (difficulty, game state). What is missing in the definition by Sicart (2008) is the social interaction of the players, where they contribute with the player society. Still, the findings in Article IV are convergent with the player and customer loyalty and player retention that follows from the game construct. The study also proposes that now the game developers and designers have more structured basis to create more appealing games and products with proven possibilities for customer retention.

The generalizability of notions *creativity*, *design* and *startups* vary from the viewpoints of different industries and research areas. The threat diminishes when defining and structuring taxonomies and meanings of the notions in context. Collaboration between these industries and research areas in joint research programs can lead to mutual understanding, especially when the Academia and (reflective) practitioners join their forces. Also the method portfolio, being dynamic and multifaceted, can generate a competitive forms and paradigms for research practices.

### 5.2.3 Reliability

The reliability of qualitative research is based on the consistency of methods. In other words, another researcher must be able to study the work and draw similar conclusions (Ryan et al. 2002, p. 155; Koskinen et al. 2005, 258). As Tashakkori et al. (2003) state, "Did we accurately capture/represent the phenomenon or attribute under investigation?" (Tashakkori et al. 2003, p. 694). The reader should be able to evaluate, based on careful documentation and reporting, how the researcher has collected, produced, and interpreted the data. Admittedly, there are threats to reliability at all stages of the qualitative research process (Lillis 2006, p. 472).

Data collection in Articles III, IV and V (RQ2 and RQ3) was based on experimental gaming and development, interviews, questionnaires, practical design works, and their evaluations by design professionals. Uncertainties in the responses to interviews and questionnaires include potential threats because of several reasons: The answers may depend on the mood, understanding, willingness or readiness of the participant, or contextual reasons like form of the interview or questionnaire and timely matters, or language and conceptual reasons. This, as described in the articles, can be avoided with having larger participant groups as actors, and comparison between the new data gathered together with the prevailing data. As an assumption concerning the relation between creativity and self-efficacy, differences between experienced and novice designers should be researched by dividing groups in respective categories, and reflect them to control groups.

In Articles I, II and V (RQ1), there are dynamic constructs as design, media and their future among startups. Design as a notion is described in Construct validity section. Therefore, a reflective practitioner can see design as a verb more than a theoretical construct. This perspective is converging with the attitude of startuppers, who have the appetite for rapid design, realization and profit. Here the differences of design, realization and profit between industries may vary by more specifically defined notions. When a ceramist cultivates her product portfolio according to, e.g., Japanese traditional Raku pottery, a very time-consuming practice and technique (Pitelka 2005, pp. 26, 27), that happens slowly and with respect for the culture, a software startupper has an entire opposite standpoint: she or he wants to develop quickly, and even uses disruption as a rainmaker's magic touch. Media as an industry facing fundamental changes, especially its journalistic and print media section, has ever changing futures depending on who expresses them. The newest trends and episodes are communicated in the daily media like newspapers and on the Web. When the research is based on latest news, the facts vary and cannot easily to be verified. If the research is based on older data and opinions, some trends and personal visions can be structured and futures envisioned by future forecasting methods.

Here the reliability is relative and depends on the hit accuracy in large contextual matters and in details.

In this study, in Article II, three industries was researched and combined. These three areas, network operator businesses, network manufacturing and media houses, include two more stable fields, the network operator businesses, network manufacturing. Some hypotheses can be speculated, like the growing business trend of network manufacturers because of the 5G solutions and IoT technologies. Along these trends, network operators can improve their businesses through growing bandwidth needed to deliver more videos and augmented reality content. Therefore, when combining the three industries together allows practical reasoning of their mutual and shared future. Still, threats to reliability stay, and certainty can be reached through continuous balancing with old and new data, future forecasting methods, and practitioners' enlightened opinions.

### 5.2.4 Triangulation

Triangulation refers to the combination of different methods, researchers, data sources, or theories in a study that combines multiple methods and approaches. Different research methods or perspectives can simultaneously produce conflicting research results on the same phenomena (Tuomi et al. 2002, 2009). Triangulation relates to the use of several different approaches to achieve a better understanding of a particular theory or phenomenon (Burton et al. 2011). Research according to mixed methods is built on the idea that understanding is increased through methodological triangulation (Molina-Azorin, 2007; Torrance, 2012). And as Weick (1969) states, several methods or techniques are needed, each of which is incomplete in different ways. When several methods are used, the imperfections of the methods tend to cancel each other out (Weick 1969, p. 21).

In this study mixed methods, different manners of approach to creativity and design, and the role of reflective practitioner are reasons to approach the topics from different angles. RQ1 is handled in Articles I, II and V, where the themes are studied from the practical possibilities of startups to develop their creation skills 1) through artistic design flow, 2) through getting on with large systemic approaches (media, mobile network operators and manufacturers) even in a disruptive way, and 3) strengthening own self-efficacy and creativity through design and note-making skills. This practice allows to evaluate the differences of design, creativity and self-efficacy from the perspective of personal qualities (Article V), from the viewpoint of a structured proposal to improve one's design skills (Article I), and proposing a strategy of disruptive innovation opportunity in a large scale initiative. These three approaches focus on the same target posed in the RQ1. To avoid threats and missing angles in the triangulation, the articles propose to use larger groups of participants with various and wider scale of skills in creativity and design.

The possibilities for startups to develop their processes for better customer retention (RQ2) was handled through SEMAT processes with Essence theory (Article III). As findings, the board game that was developed for the study, taught well the basic concepts of Software Engineering and Essence concepts, taught method agnostic view to Software Engineering, and expressed the importance of teamwork and communication in Software Engineering projects. Beside Software Engineering processes, successful product qualities were studied (Article IV) by game elements like the balance principle, role of social interactions, the meaning of rewards, and the importance of motivation. These two approaches triangulate the RQ2 from different perspectives, and form a parallel view to combine both Software Engineering processes and the qualities of certain application (game) in application design.

The question of enhancing creativity in software startups (RQ3) has a twofold approach: study creatively the Software Engineering practices and combine the skills in design projects (Articles III and V). When the board game to study the processes and practices, is creative product and pleasant to play, the practical creation works at a design class are funny when the creativity improves after several rounds of exercise. Despite that the self-efficacy and creativity scores did not have clear correlation, the course produced a wide set of visual concepts and plans. This is exceptional during a class in a non-designer faculty, and encourages to continue separate hands-on design lectures.

## 5.3 Contributions

This research provides insight and foresight into the startup culture in software-intensive product, service and solution creation. The research has a cross-sectional and aggregative nature: it is a reflective practitioner's view and a scientific study. The notions of creativity, design, startups in a disruptive context, self-efficacy, and customer retention and method prison are highly cross-sectional. The concepts are dynamic and flexible; they scale from practices to industries and from discussions to science. In this dissertation, the contributions stem from the reflective practitioner's perspective in a scientific environment. This viewpoint includes knowledge-in-action with craft expertise and tacit knowledge. This motivates the practitioner to widen the visual angle of ways of working outside of stabilized assumptions. This could be explained by a startupper's attitude.

### 5.3.1 Theoretical contributions

The outcomes of this dissertation contribute to diverse areas of research. Design and creativity have been thoroughly investigated in the literature, but combining them in the practical environment of startups, especially taking into account their impact on industries, is an emerging area of research and development. In addition, what applies to early-stage startups and their young personnel applies to university students, as they share similar ages and attitudes toward entrepreneurship. A study of 193 law, management and finance students at Kozminski University in Poland showed that students have the potential for entrepreneurship (Olszewska 2015). The participants emphasized creativity, leadership skills and diversity management skills as the most desirable characteristics for an entrepreneur, but they also expressed their wish for opportunities to start a business. These findings are echoed by Krueger et al. (1994, p. 2): "Before there can be entrepreneurship, there must be the potential for entrepreneurship." A strong will supporting these intentions was expressed by the European Commission in 2012 (EU 2012). The report declared that an investment in Europe's human and cultural capital was needed. The report described people who would create, constitute and continue Europe's cultural traditions, including future philosophers, artists, writers, entrepreneurs, craftsmen and women.

Education and apprenticeships organized for startups and students impact their self-efficacy positively. This factor points to self-efficacy, the belief in being able to fulfill a given task successfully, and follows both groups throughout their work history. Articles I, II and V describe various conditions in which startups and students fulfill assignments: in practical studios in a small team, in the environment of digital industry conglomerates, and in an academic setting when studying design.

When bridging software design and computer science with design, new ways of working can be developed on top of present practices in design of software and applications. When a large set of design practices are already embedded in software engineering, such as usability, user experience, user-centric design and graphic design, there are still practices and aspects of design professionalism that can be added. Ferré et al. (2001) suggest that a wide set of methods and design areas are involved in usability and user interface design in software engineering. These design species include conceptual design, visual design, prototyping, usability evaluation and leadership. The specialist requires qualities like system-level understanding and awareness of psychology and sociology (Ferré et al. 2001). This means that design is understood as a mindset and culture.

This is the doctrine of design thinking, with its philosophical dimensions fostering an encompassing idea of creating a design culture and mindset in schools, public organizations and companies (Diefenthaler et al. 2017). This attitude requires incorporating artistic design principles into the design approach of software. Design as culture is mutual: the customer or user identifies and understands the artifact through design; and simultaneously, the software designer gets a wider set of design skills and tools to understand the artifact and human. Article I opens the research area of design to be more reciprocal and developmental. Artistic design can have a significant impact on software engineering, and software engineering can have help to advance the arts and its design principles. These approaches require programs of simultaneous research and development.

This dissertation contributes to the research area through a broad and prolonged series of experimental studies. The findings reveal that creativity can be measured during design classes, but the spread of results between the correlation of creativity and self-efficacy is a complex phenomenon (Article V). This means that a design class in which creativity is studied in correlation with self-efficacy could be organized simultaneously with comparison groups of design students and non-designer students. Also, a pilot course, Design at the Academia, could be used as a platform for research and development to find adequate means and methods for design education and the creativity that follows when individuals are encouraged to express it.

### 5.3.2 Practical contributions

Established global corporations face periodic disruptions in business. Some of them either pivot immediately, struggle financially or go bankrupt. To avoid the curse of digital transformation, three possible strategies are available: 1) develop new customer segments, 2) create new business models, or 3) restructure the value chain (Bughin 2017). This also presents an opportunity for startups or internal startups to act on the side of the chosen party, the winner or the loser. Article II describes a case of the digital industry disrupting print media and the turnover of journalism. This has had an impact on the three industries: network operator businesses, network manufacturing and media houses. If these parties form an internal startup, the Trio, resources could be collected for research and development. This endeavor could lead to the reorganization of the Fourth Estate, a societal watchdog of press, media and journalists, expanding on the traditional European concept of the three estates of the realm: the clergy, the nobility and the commoners. If the Fourth Estate were disrupted, it would essentially be left without pay. As van Aelst et al. (2008) express, if, or when, the relationship between journalists and politicians turns into a *marriage de raison*, determined by mutual mistrust, the situation will become undesirable (van Aelst et al. 2008).

With the establishment of the Trio, changes in the industries would become self-evident. Large philosophical, societal and technological research programs are needed within the media, journalism and free speech communication. For example, in-car and car-to-car communication could be transformed into a car-as-journalist concept with sensors, online news and artificial intelligence combined in one connected broadcast system. Article II describes the future research needed on the ethics, societal impacts and business models of such a car, system, people and data

Article III describes how game-based education helps students of software engineering to escape from the software engineering method prison. A practical software development class used a board game to study the basic concepts of SEMAT/Essence and software engineering in a pleasant way. The results showed that the game-based method of teaching was an enjoyable experience for the students. A research agenda was proposed based on practical results for an adequate fit in learning software engineering practices, combined with extant literature. The agenda included research on board game suitability and the effect of the adoption of Essence and software engineering practices and methods.

Article IV contributes to the existing literature by describing the main game elements that players found most important. The use of the balance principle to scale the difficulty and feelings of achievement when playing was a factor in the decision to continue playing. Feelings of achievement and social interactions during play were regarded as important. Being rewarded for overcoming challenges resulted in a positive feeling of achievement, which was essential. The contributions of this phase improve the existing literature and can be used to plan further research programs to deepen our knowledge of the topics.

This dissertation proposes further practical contributions in the form of research topics on the themes presented in this study, pragmatic operations and strategies for startups and corporations, as well as educational curricula for the Academia. Practical concepts for courses in design at the Academia have been a strategic contribution of Articles I, II and V.

Article III presents further educational means for adopting Essence and software engineering practices and methods in software engineering classes. Article IV identifies targets for game developers to emphasize game design. This could lead to more stable player retention during the life-cycle of the game. The results obtained in Article V emphasize the importance of note-taking. The use of the implementation task as a tool for creation, emphasized by both notes and note-making, should be applied in design courses, if not in the entire academic environment (Figure 9).

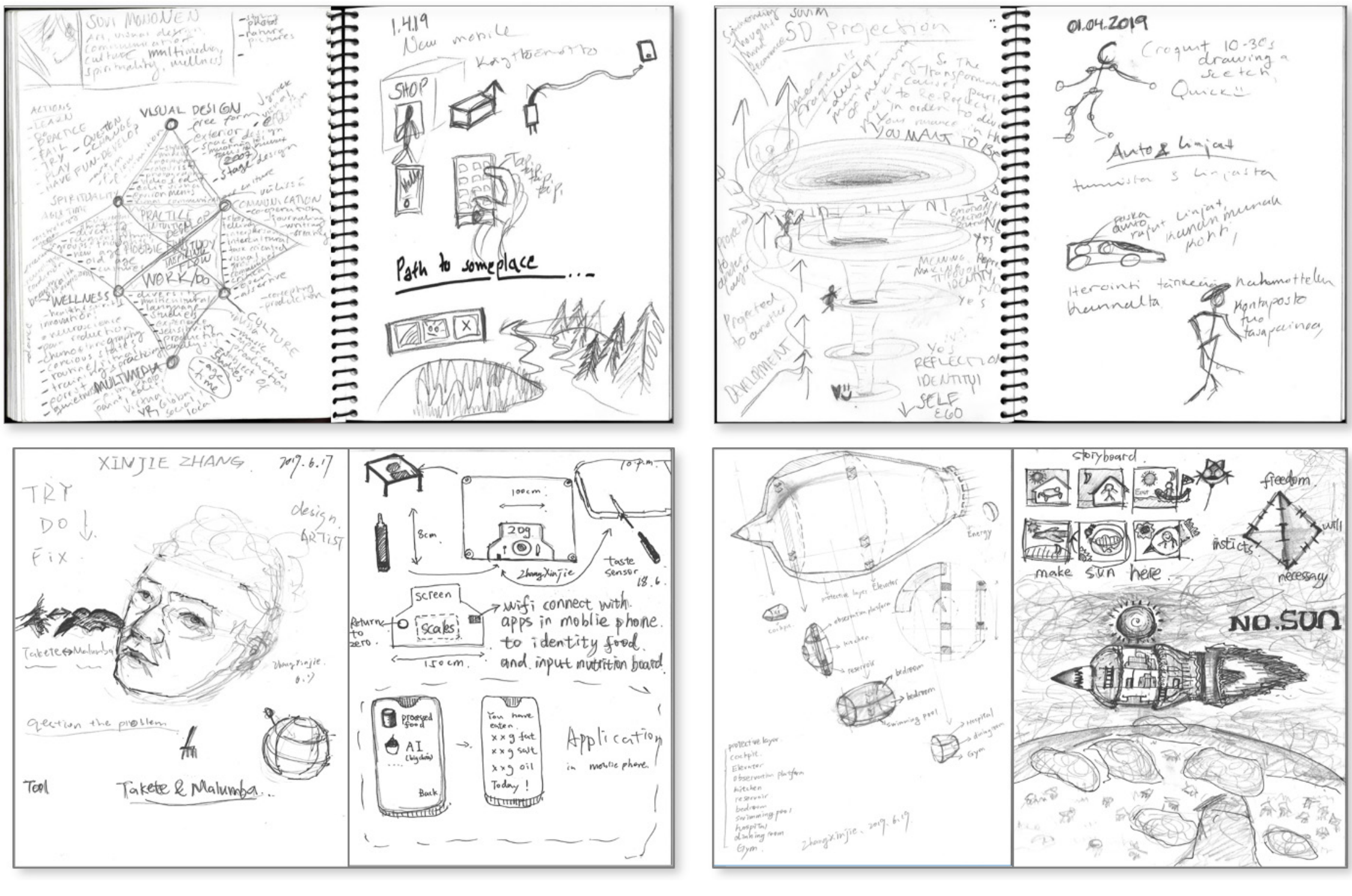

FIGURE 9. Students' sketchbook material from the Fuzzy Front End Design course (2019), handled in Article V[3]

[3] Sketchbooks reproduced with permission from the students.

As an overall comprehension of the combination of research and development in the form of a reflective practitioner at work, crucial for the author. A practical understanding of the topics, professionalism and skills produced within the research topic artifacts will provide an advantage to investigating acute research themes and their research questions.

### 5.3.3 Limitations

Creativity is linked to nearly all actions and demands that an individual encounters at home, at work and during leisure. In addition, society and industries emphasize creativity in the form of planning, designing, ideating, innovating and acting artistically. New methods, practices and offerings, like design thinking, service design, video classes, free art creation tools and Web applications, allow everyone to access creative acts without boundaries. Design classes within different professions and areas of study allow individuals to develop their skills and readiness, even reaching a level of creative professionalism.

Everyone is a designer and, therefore, a creator. Individuals design their daily schedules, organize their workflow and choose what to buy at the grocery for dinner. Creativity is needed when thinking of a birthday present for a person who has everything. A folklorist or museologist may be driven to pursue a new and innovative museum project. Hobbies are at the center of creativity. When knitting a pullover, for instance, patterns can be found in a knitting book than strengthen the forms and structure when the knitting goes wrong. Serendipity captivates us to create our own patterns, giving us confidence.

One limitation of this study is that the notion of design is a practical, forecasting and active verb, as well as an elastic and fuzzy construct. At the Academia, design is treated as a highly professional topic with extremely demanding entrance examinations, fundamental curricula and long study time. Still, design with its components and elements, such as vision, ideas, plans, sketches, mind-maps, prototypes and actualization, belongs to everyone. The students and the academics' day begins with ideating and planning the day, constructing the presentation, figuring out the homework, and writing a menu and the seating order for the night's dinner. These actions are all about design, but we design plans for the future without preparatory courses and skills. This limitation applies to Articles I, II and V.

Cognitive biases can be seen as a limitation in Articles I and V. These articles discuss design, creativity and self-efficacy, which are flexible and dynamic qualities from person to person. Cognitive biases have a central impact on decision-making, and therefore a long-lasting effect on one's future actions, and the cognitive biases are interlinked to each other (Mohanani et al. 2018, Table 11). Cognitive biases need a thorough research plan to recognize and find dependencies in the practical design of software, whether they are systemic or at the detail level within a startup or a digital conglomerate. Cognitive biases result from presupposed beliefs and prejudices that cause systematic deviations and errors in optimal reasoning (Ralph 2011; Stanovich et al. 2009).

By researching cognitive biases' impact on creativity and design among startuppers and students, the interrelation between self-efficacy and creativity can be achieved in Article V. It is hypothesized that strong self-efficacy belief and creativity are positively correlated. A research program on cognitive biases in the context of students, startups and entrepreneurs versus professional designers and artists would explicate the dichotomy of illusionary and realistic self-efficacy beliefs. Simultaneously, a set of criteria of biases and their dependencies should be present. Educational action to decrease both illusionary self-efficacy and cognitive biases could be tested in design classes, practical startup hackathons and respective occasions. Also, when planning design classes for the Academia, the design curricula for diverse faculties could recognize the threat of cognitive biases described in Table 7 (Mohanani et al. 2018).

The concept of addiction presents limitations in Article IV. Although addiction can have a positive connotation in an optimistic dimension, it can also reflect a pathologically addictive nature and the qualities of a gambling disorder. When video game players were investigated in clinical conditions, it was found that video game addiction was dominant among people with gambling disorders (Jiménez et al. 2014). Gambling problems were linked to the use of and participation in video games.

Method prison as an obstacle to studying and using software engineering methods and principles presents limitations in Article III. The notion of a method prison itself is clear, but the question of how to avoid being locked into one opens new tracks for research. Startups must prepare themselves in their daily practice of their own methods and principles. This is obvious because of startups' habit of sticking to concepts that are innovative solutions to unknown problems under constant uncertainty. Although the solution may be disruptive, no handbook or guidelines, including common instructions, provides advice as to how to choose the right methods. One of the worst-case scenarios would be that a startup finds a solution outside of the method prison that is not good enough. Even worse would be if the startup did not realize the inadequacy of the method they had chosen or created. Here, the structure of the method prison needs clarification and further preparation.

TABLE 7. Cognitive biases investigated in software engineering knowledge areas (Mohanani et al. 2018)

| Cognitive Bias \ Knowledge Area | Computing foundations | Configuration | Construction | Design | Economics | Engineering foundations | General | Maintenance | Management | Mathematical foundations | Models and methods | Process | Professionalism | Quality | Requirements | Testing | Total |
|---|---|---|---|---|---|---|---|---|---|---|---|---|---|---|---|---|---|
| Anchoring/adjustment | | | 5 | 8 | | | | | 9 | | | 1 | | 1 | 2 | | 26 |
| Attentional | | | 1 | | | | | | | | | | | | | | 1 |
| Availability | | | 5 | 4 | | | | | 3 | | 1 | | | 1 | 1 | | 15 |
| Bandwagon effect | | | | 1 | | | | 1 | | | | | | | | | 2 |
| Belief perseverance | | | | 1 | | | | | 1 | | | | | | | | 2 |
| (Over)-confidence | | | | 1 | | | | | 15 | | | | | | | | 16 |
| Confirmation | | | 6 | 4 | | | 1 | 1 | | | | | | 1 | 3 | 7 | 23 |
| Contrast effect | | | | 1 | | | | | | | | | | | | | 1 |
| Default | | | | 1 | | | | | | | | | | | | | 1 |
| Endowment effect | | | | | | | | | | | | | | | 1 | | 1 |
| Fixation | | | | 1 | | | | | | | | | | | 1 | | 2 |
| Framing effect | | | | 1 | | | | | | | | | | | 2 | | 3 |
| Halo effect | | | 1 | | | | | | 1 | | | | | | | | 2 |
| Hindsight | | | | | | | | | 3 | | | | | | | | 3 |
| Hyperbolic discounting | | | | 1 | | | | | | | | | | | | | 1 |
| IKEA effect | | | | | | | | | | | | | | | 1 | | 1 |
| Impact | | | | | | | | | 2 | | | | | | | | 2 |
| Information | | | 2 | 1 | | | 1 | | 1 | | | | | | | | 5 |
| Infrastructure | | | | 1 | | | | | | | | | | | | | 1 |
| Invincibility | | | | 1 | | | | | | | | | | | | | 1 |
| Mere exposure effect | | | | 1 | | | | | 2 | | | | | | | | 3 |
| Miserly info. processing | | | | | | | 1 | | | | | | | | | | 1 |
| Misleading information | | | | | | | 1 | | 2 | | | | | | | | 3 |
| Neglect of probability | | | | 1 | | | | | | | | | | | | | 1 |
| Normalcy effect | | | | 1 | | | | | | | | | | | | | 1 |
| (Over)-optimism | | | 1 | 1 | | | | | 1 | | | | | | 3 | | 6 |
| Primacy/recency | | | | 1 | | | | | | | | | | | | | 1 |
| Representativeness | | | | | | | | | | | 1 | 1 | | 1 | 1 | | 4 |
| Selective perception | | | | | | | | | 1 | | | | | | | | 1 |
| Semantic fallacy | | | 1 | | | | | | | | | | | | | | 1 |
| Semmelweis effect | | | | 1 | | | | | | | | | | | | | 1 |
| Status quo | | | | 2 | | | 1 | | | | | | | | | | 3 |
| Sunk cost | | | | | | | | | 1 | | | | | | | | 1 |
| Time-based | | | | 1 | | | | | | | | | | | | | 1 |
| Valence effect | | | | 1 | | | | | | | | | | | | | 1 |
| Validity effect | | | | 1 | | | | | | | | | | | | | 1 |
| Wishful thinking | | | | 1 | | | | | 2 | | | | | | | | 3 |
| Total | 0 | 0 | 22 | 39 | 0 | 0 | 5 | 2 | 44 | 0 | 2 | 2 | 0 | 4 | 15 | 7 | jri |

### 5.3.4 Future research opportunities

Future research on self-efficacy and its relation to creativity and innovation is important for present-day companies and universities. A wide range of methods, tools, practices, curricula and research can help students to encourage themselves, inspiring themselves to create and innovate privately and as part of a team. This development often happens in companies on the fly, but the Academia can develop these skills in advance, supported by research. With a theoretical, practical and educational procedure, new ways of working motivate students to pursue meaningful studies and exploration. At the Academia, the spirit and eagerness of students from diverse faculties and backgrounds form an ideal setup for teams. This is important both for the industries the students will later work in and for the teams that the students will join after graduation.

Future studies of creativity and self-efficacy should examine a thorough organization of courses embedded with design topics or even a curriculum of design studies for non-design-related faculties. This could lead to a broad portfolio of tailored study modules on design to respond to the requirements of creative and skillful graduates. An inclusive movement of creativity and design skills at universities could even be called Design at the Academia.

In Table 8, key findings and contributions are listed by articles, research questions, research objectives, key findings, and contributions.

TABLE 8. Key findings and contributions

| Article | Research question(s) | Research objective | Key findings | Contribution |
|---|---|---|---|---|
| I. | RQ1. How do design and creativity viewpoints encourage early-stage startups to develop their ideas into better concepts? | To support early-stage startups in efficiently developing their ideas into concepts. To determine what tools, methods, skills and education provide appropriate design. | Software development already includes several artistic design principles and practices. Usability, user experience and user-centric design are used daily in application development. On Web pages, graphic design, icons and visuals attract users. Other practices from design are common in startups' product and service creation, like the use of flow charts, paper prototypes, simulations, scale models, drawings, storyboards, collages and iterative improvisation. For these practices, drawing and handcrafting skills are needed. | This presents a radical proposal for software startups to adopt more appropriate design by embedding designers' ideals as an integral part of software creation and development. This would require new programs to research and develop design principles in computer science and software development. Design education should be based on traditional design and newfound design methods. Artistic design principles strengthen software development in general, especially among first-movers: startuppers. The designer-developer-startupper becomes better equipped to address future challenges. |

| Article | Research question(s) | Research objective | Key findings | Contribution |
|---|---|---|---|---|
| II. | RQ1. How do design and creativity viewpoints encourage early-stage startups to develop their ideas into better concepts? | To build a successful countermove with disruptive actions and joint-venture startup practices to win back the lost market position from new global actors. | Large, established global industries and corporations may face disruptive attacks from innovative companies intruding on their business. These companies may also have a history of their own startup era.<br>By combining three partners, including a set of companies from mobile network manufacturers and operators and media houses, a consortium known as the Trio could be founded. It would be an internal startup fighting back to revitalize their businesses and starting a new business to gain back their lost market leadership. | In the future, software startuppers, with their applications and communities, will be the drivers for this abstraction shift in media and journalism. The Trio consortium would build new systemic products to enable new tools, formats and platforms for their developer community to create and share content. This content would be broadcasted in the form of text, photos, videos, film, social media content and new appealing formats. This movement would create opportunities for extreme changes in advertisement channels and incomes, reporter-developer communities, and even the structure and nature of the Fourth Estate. |
| III | RQ2: How can software startups develop their processes to improve customer retention?<br>RQ3: How can creativity be enhanced in software startups? | How does game-based education helps students of software engineering to escape from the software engineering method prison? | The presented game successfully taught first-year software engineering students the basic concepts of Essence and software engineering in a fun way. It taught a method-agnostic view of software engineering and that software engineering methods are modular. It taught the importance of teamwork and communication in software engineering project work. | The game is a gamified model of teaching the Essence method because the method still suffers from a lack of interest among practitioners. The game was determined to be a useful method for teaching Essence because it was an enjoyable experience for the participants, and the board game fulfilled the set objectives. As process development factors, startups in the context of game development, profit from thorough design of the balance principle, rewards and how different playstyles and ingame activities effects, and advance social interactions and feelings of achievement as aspects of the game. |

| Article | Research question(s) | Research objective | Key findings | Contribution |
|---|---|---|---|---|
| VI | RQ2: How can software startups develop their processes to improve customer retention? | What online game elements keep players playing the game? How do rewards affect the player's motivation to continue playing video games? | For player retention, several factors are crucial, including virtual avatar customization, game difficulty, social relationships between other players, in-game rewards, game mechanisms and self-efficacy. | The practical findings can help game developers to improve games by using the balance principle, scaling the difficulty and providing a feeling of achievement when playing. Social interactions can be improved by playing with previously unfamiliar people as well as well-known players. Different playstyles and rewards for activity, exploration and task completion can be motivating. |
| V | RQ1. How do design and creativity viewpoints encourage early-stage startups to develop their ideas into better concepts? RQ3: How can creativity be enhanced in software startups? | How does creativity correlate with self-efficacy and design work? | The results show that self-efficacy and creativity are not strongly correlated. Surprisingly, experienced and skilled students in design got average scores in self-efficacy, which may indicate realism and self-critique when comparing oneself to design as general. | To bring design as a medium for creativity and overall quality to the Academia, methods of improving the students' design skills and practices are needed generally. Establishing a research and development program for design principles in the curricula will allow creative and design-skilled students to integrate into present trades, industries and workplaces. |

# YHTEENVETO (SUMMARY IN FINNISH)

Startup-yritykset ovat merkittäviä innovaatioiden ja uuden liiketoiminnan synnyttäjiä. Startup-yritykset ovat luoneet keksintöjä, niistä sovellettuja innovaatioita ja vaurautta jo usean vuosikymmenen ajan. Startup-kulttuuri on laajentunut useille toiminnan ja teknologian alueille, jotka ovat mullistaneet maailmaa. Startupeilla on myös haasteita ja vajavaisuuksia, joihin pyritään löytämään ratkaisuja niin tutkimuksella kuin käytännön kehitystyöllä.

Aiemmassa tutkimuksessa startup-yrityksiä ja niiden toimintaa on tutkittu laajasti, etenkin on selvitetty niiden toiminnan perusteita, kuten ohjelmistojen hyödyntämistä tuote- ja palvelukehityksessä. Startupeilla on merkittävä uudistava vaikutus muuttaa jokin vanha toimiala tuottavammaksi ja globaaliksi. Merkittävää on myös startupien kyky eri toimialojen mullistamiseen, disruptioon. Tällaisia häiriöitä toimialalleen aiheuttaneita yhtiöitä ovat muun muassa maailman suurin taksiyhtiö Uber, ilman ainuttakaan autoa, maailman suurin hotelliketju Airbnb, ilman ainuttakaan huonetta, ja maailman suurin vähittäiskauppa Alibaba, ilman ainuttakaan varastoa. Näitä yhtiöitä yhdistää globaalin alustan, platformin, rakentaminen. Ohjelmistoteknologialla tehokkaasti toteutettu alusta mahdollistaa maailmanlaajuisen toiminnan aivan käytännöllisille toimille, että se kuulostaa uskomattomalta. Usein menestyvät startupit ovat eri alojen tavallisiakin liiketoimintoja digitalisoinnilla kehittäneitä yhtiöitä.

Samalla kun startupit luovat uutta innokkaasti, on niiden liiketoimintaympäristö täynnä epävarmuuksia. Startupin toimijat ovat yleensä nuoria ja kokemattomia ja käytettävät tekniikat joko uusia tai nopeasti kehittyviä. Startupien tiimit ja niissä yhdistyvät taidot ja tiedot ovat joko korkeatasoisia tai laadultaan vaihtelevia. Kun yli 90% ohjelmistostartupeista epäonnistuu, on edelleen paneuduttava kaikkiin jo tutkittuihin vaikuttajiin ja löydettävä uusia mahdollisuuksia edistää startupien liiketoimintaa.

Startupin tiimi on usein todettu elintärkeäksi, joten juuri ihmisten ja tiimien ominaisuuksia ja taitoja kannattaa kehittää. Tässä tutkimuksessa keskitytään tietoon, kokemuksiin, taitoihin ja muihin kognitiivisiin kykyihin, joihin design-taidot ja niiden käyttö liittyvät. Suunnittelua tutkitaan laajalti taiteellisessa ja teollisessa kontekstissa, mutta sen soveltaminen startup-kulttuuriin ja ohjelmistojen startup-yrityksiin tapahtuu omassa menetelmävankilassaan. Tutkimuksessa keskitytään muotoilun ja luovuuden merkitykseen startupeissa. Samalla esitetään kolmen laajan globaalin liiketoiminnan yhteistyötä, jossa startup-asenne voisi tuottaa menestystä, startupille itselleen mutta myös näille kolmelle teollisuusalalle. Tätä tukemaan, tutkimuksessa käsitellään ohjelmistosuunnittelun uusia opetusmahdollisuuksia ja sitä, miten asiakas saadaan pidettyä ja miten startup-tiimin henkilöiden minäpystyvyyttä voidaan kehittää. Menetelminä tutkimuksessa on käytetty takautuvasti päättelevän ammattilaisen (reflective practitioner) työtapaa, luovaa analyysiä (creative analysis), kvasikoetta (quasi experimentation) ja havainnointia (observation).

Tutkimuksen tuloksena ehdotetaan startupien suunnittelu- ja muotoilutaitojen lisäämistä taiteen ja designin keinoin. Tällöin piirtämisellä, sommittelulla, konseptoinnilla ja käsityötaidoilla katsotaan olevan laadullista merkitystä startupien luodessa uusia tuotteita ja palveluita. Startupeilla on mahdollisuus kehittää laajoja ja globaaleja liiketoimintoja, joista esimerkkeinä ovat muun muassa Uber ja Airbnb. Tutkimuksessa esitetään malli, jolla voidaan organisoida medialiiketoiminta uudella tavalla niin, että journalismi saa takaisin Internetin hakukoneiden ja sosiaalisen median keräämät mainosrahat. Mallissa kootaan yhteen kolme joko taantuvaa tai nollasummapelin teollisuutta, mobiiliverkonvalmistus, mobiilioperointi ja mediateollisuus, erityisesti journalismi. Näiden kolmen teollisuushaaran liitto mahdollistaa uusien palveluiden rakentamisen uusin ideoin ja uudella kokoonpanolla. Samalla luodaan liiketoiminnalle tärkeät kaksi toimintoa, alusta ja kehittäjäyhteisö.

Tutkimuksessa esitellään myös ohjelmistokehityksen ja sen prosessien menetelmä, jossa startup-tiimi voi lautapelin avulla, pelillisen kokemuksen kautta, oppia työssä tarvittavat käytännöt ja prosessit. Tavoitteena oli opettaa opiskelijat valitsemaan vapaammin menetelmiä ja työtapoja ohjelmistokehitykseen, jotta ei ajauduta menetelmävankilaan (method prison). Välttäessään menetelmävankilan startup-tiimi käyttää luovasti perinteisiä menetelmiä ja pystyy käsittelemään annetut tehtävät nopeasti.

Asiakkaan säilyttäminen on kaikelle liiketoiminnalle tärkeätä, sillä se osoittaa asiakkaan uskollisuutta ja halua valita sama toimittaja tulevaisuudessa. Peliteollisuudesta tutkimuksessa löydettiin pelaajia erityisesti kiinnostavia ominaisuuksia, jotka herättivät myönteisiä tuntemuksia pelattaessa. Tulosten mukaan online-pelielementit pitävät pelaajia pelaamassa, kun pelaajien suhdetta, motivaatiota ja tunteita vahvistetaan pelin aikana. Tutkimuksessa esitetään kaava, jonka avulla pelinkehittäjät ymmärtävät pelaajaa paremmin. Siksi pelaajien tai kuluttajien säilyttäminen on merkittävä tekijä, joka on otettava huomioon tuotteen ja palvelun suunnittelun kaikissa vaiheissa startupien keskuudessa. Näihin panostaessaan pelinkehittäjä voi varmistaa tuotteensa haluttavuutta.

Minäpystyvyys (self-efficacy) kuvaa ihmisen uskomuksia tai arvioita omiin kykyihin suoriutua erilaisista tehtävistä. Luovuus on ennen kaikkea uusien ennenäkemättömien, toimivien ja hyödyllisten asioiden tuottamista. Luovuus tuottaa alkuperäisiä, hyödyllisiä, toimivia, merkityksellisiä ja kauniita asioita. Tutkimuksessa selvitettiin minäpystyvyyden ja luovuuden suhdetta opiskelijoiden harjoitustöiden yhteydessä, jossa piti suunnitella palveluita piirroksin, tekstein ja konseptuaalisin suunnitelmin. Sen jälkeen opiskelijat vastasivat minäpystyvyyskyselyyn. Näitä kahta tutkittiin tilastollisin menetelmin eikä luovuudelle ja minäpystyvyydelle löydetty vastaavuussuhdetta. Tutkimuksessa ehdotetaan, että koe suoritetaan eri ryhmin siten, että vertailuun voidaan ottaa ammattimaiset designerit ja vähemmän muotoilua harrastaneet. Tutkimuksessa arvioitiin myös, että ihmisten käsitys minäpystyvyydestä voi vaihdella suhteessa ammatillisiin kykyihin ja että designin kohdalla ihmisillä on laadullisia ja sisällöllisiä taitoeroja.

Muotoilu ja luovuus startup-kulttuurissa tarvitsevat lisää tutkimusta. Jo startupien 90-prosenttinen epäonnistuminen on riittävä peruste tutkimukselle, mutta paljon muutakin voidaan löytää. Esimerkiksi startupien disruptiivista rohkeutta tarvitaan kaikilla aloilla, digitaalisuutta ja ohjelmistoja voidaan soveltaa laajemmin yhteiskunnassa ja liiketoiminnassa ja nuoret tulevat työelämään innokkaina uudistajina.

Startupit, kuten muutkin työelämän toimijat, tarvitsevat yhä laajempaa valikoimaa menetelmiä, työkaluja, käytäntöjä, opetussuunnitelmia ja tutkimusta menestyäkseen. Uudet taidot ja tiedot auttavat ihmisiä ja yrittäjiä rohkaistumaan ja inspiroimaan muita ja itseään luomaan ja innovoimaan niin yksin kuin osana tiimiä. Kun teoria ja käytäntö yhdistyvät opiskelussa, uusilla opetusmenetelmillä opitut työskentelytavat motivoivat opiskelijoita jatkamaan mielekästä tutkimusta ja kehitystä. Yliopistoissa eri tiedekuntien ja aineiden opiskelijat muodostavat yhdessä ihanteellisia tiimejä. Tämä on tärkeää sekä teollisuudenaloille, joissa opiskelijat työskentelevät myöhemmin, että ryhmille, joihin opiskelijat liittyvät valmistumisensa jälkeen.

Tutkimuksessa nähdään tarpeellisena monipuolisten suunnitteluteemojen sisältämistä opetussuunnitelmiin ja kursseihin. Lähes kaikissa tiedekunnissa tulisi olla laaja suunnittelutieteen ja -tutkimuksen (Design Science) opetussuunnitelma. Kaikkihan suunnittelevat päivittäin arkisia asioita, antavat muodon esimerkiksi juhlalle ja sen kattaukselle. Luovuus ja muotoilu yhdistettynä antavat perustan tutkimuksen ja kehityksen rohkeudelle.

# APPENDIX: REFLECTIVE PRACTITIONER'S NOTES

This appendix describes the author's contribution to design, creativity and startup culture. Appendix shows the author's comprehension as a designer in various species of art and design. The personal stance has led to concepts, artefacts, products and buildings mainly alone or in a small team, by own hands.

The author's journey on his way to be a reflective practitioner is illustrated with examples from his professional career as an architect, industrial designer, acoustician, cabinet maker apprentice, and artist. *Design* in these examples has been perceived as a *verb*. The collection is a set of artefacts created between years 1986 and 2020. Copyrights of the following pictures, concepts and plans belong to the author and are free to be reproduced in scientific circumstances.

### Sketches and notes

Monday, Tuesday, Thursday and Saturday are self-made typefaces before the first Macintoshes' WYSIWYG (What You See Is What You Get) era (Figure 10). The author was 18 years old. Later, the author studied architecture and created own typology and architectural morphology to design a series of houses, the 47/47 series (Figure 11).

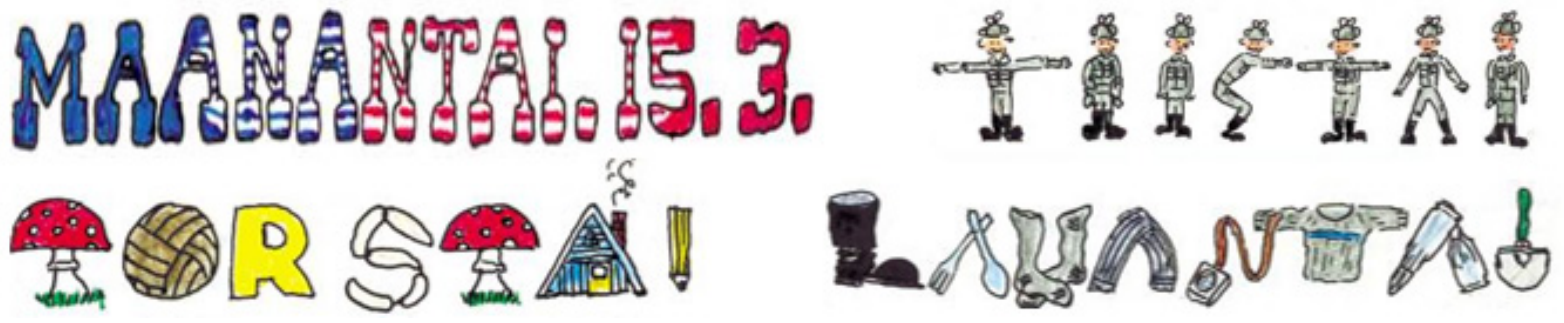

FIGURE 10. Typefaces and fonts: Early drawings in the author's notebook during his service in the Finnish army

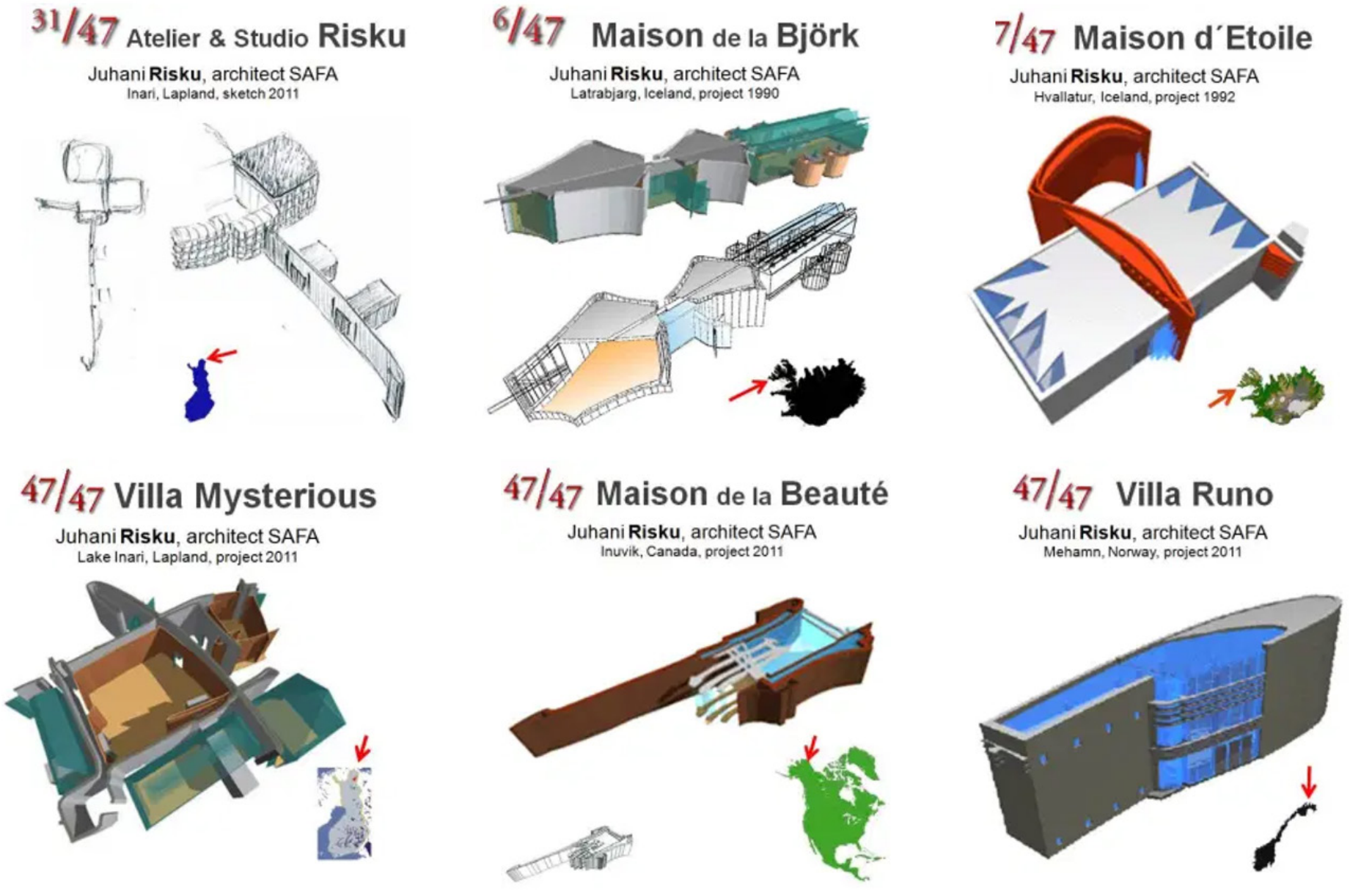

FIGURE 11. Typology and architectural morphology to design a series of houses, the 47/47 series

The vision, idea and meaning of an innovation lab and makerspace was concepted in a very early phase to be base for further ideation and development. It is easier to communicate to a team with draft plans and pictorial papers. Visualizing the idea enables better communication and possibilities to continue planning. Here the concept is a proposal including keywords and structured progression to build a lab (Figure 12).

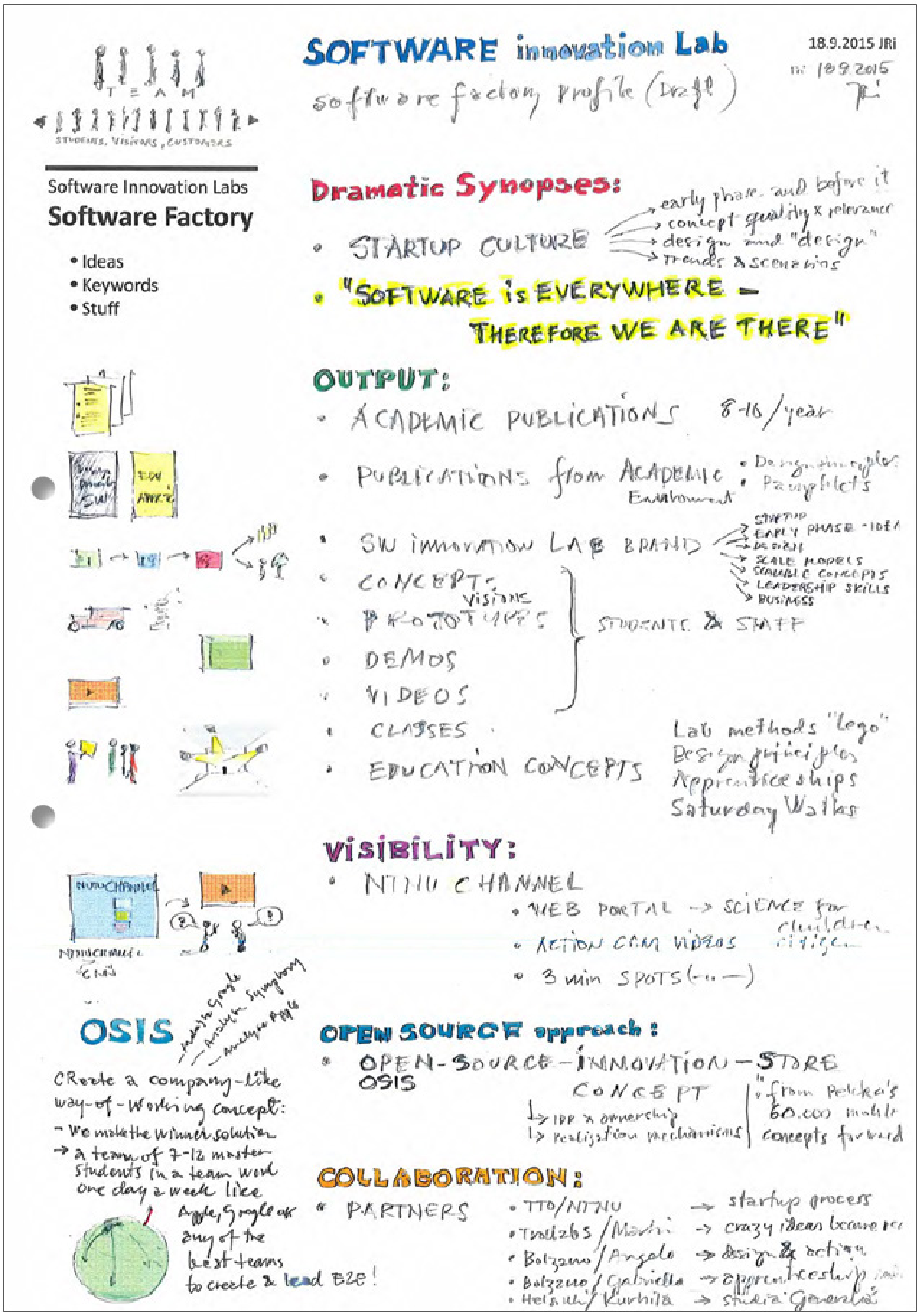

FIGURE 12. Conceptual sketch about NTNU IDI Software Innovation Labs (A4 paper, pencil and colors, the author, 2015)

The mind and hand get a sensitive touch to the paper and pencil. The drawing begins to approach the mission with right forms. This interaction leads to an initial plan, which starts the whole design process (Figure 13).

The fragile sketching session lasted 45 seconds giving the form for a chapel. Later a 3D model was done, 200 drawings and a scale model 1:50. The granite wall found its place with a variation of +/- 5 cm on the bedrock. Further structures like roofing, details and finishing required an ongoing sketch-to-realization process (Figure 14).

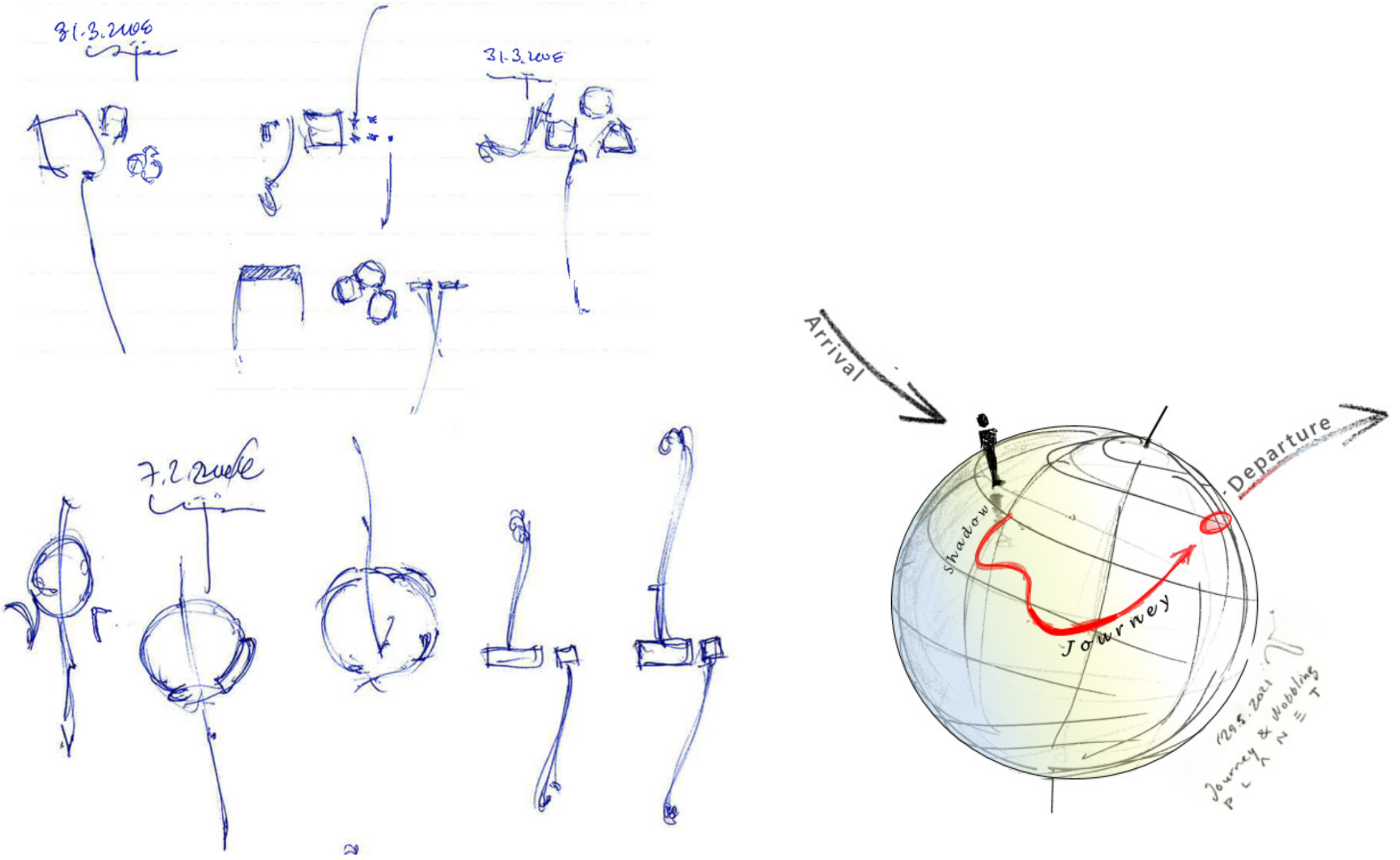

FIGURE 13. Calibrating the hand with free sketching before actual design

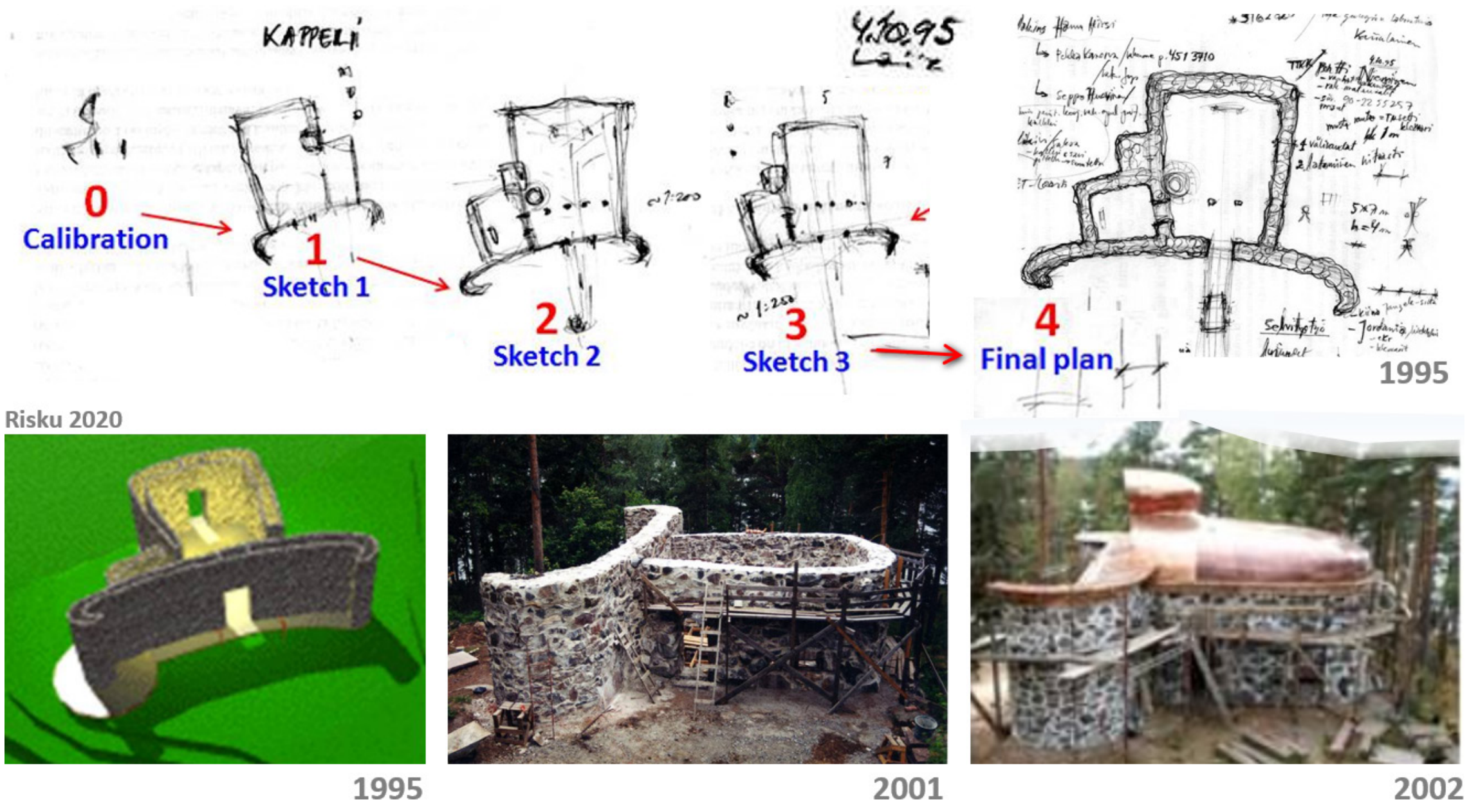

FIGURE 14. From sketch to stone: Stone Chapel in Vivamo by the author, 1995-2045

## Artefacts and conceptual models

Handicraft is a form-giving medium for human-sized artefacts (Figures 15 and 16). Hands and fingers give a direct and sensual touch to the coincident idea, meaning and form in the material and its structure. The form-giver is in the midst of form and its origin.

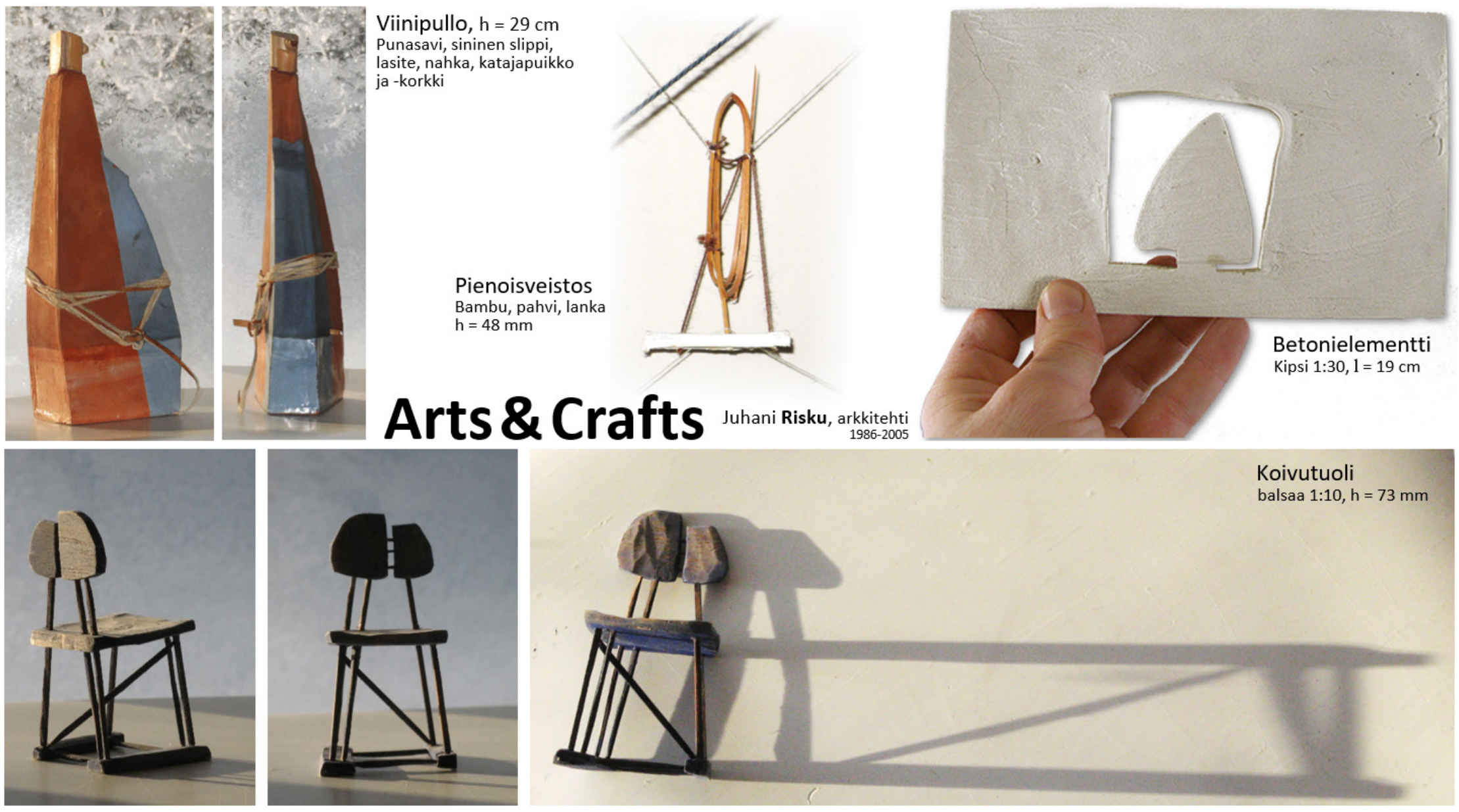

FIGURE 15. Top left: hole in a wine bottle, 0.5 liters, red clay, blue slip (engobe), glazing, leather, and juniper stick and bottle stopper; Top middle: micro-sculpture, height 48 mm, bamboo, cardboard, yarn; Top right: concrete element, scale 1:30, plaster, casted in a mold; Bottom: chair concept, scale 1:10, balsa

| **Mercy,** gift | **Art,** manmade | **Genesis,** elements | **Technique/phase** |
|---|---|---|---|
| 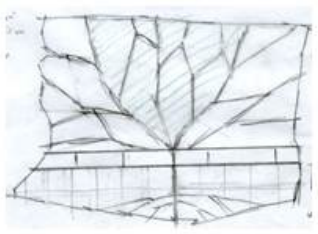 | 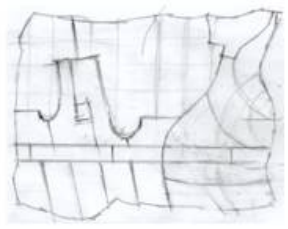 | 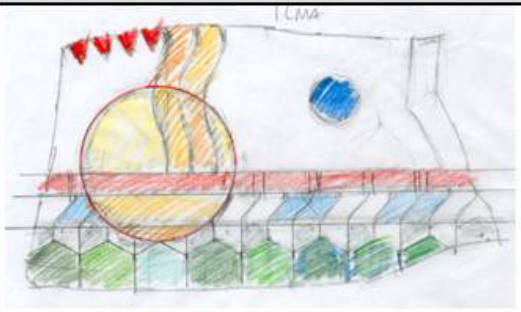 | Sketches on sketchpaper, scale 1:1<br>All forms and colours are defined. |
| 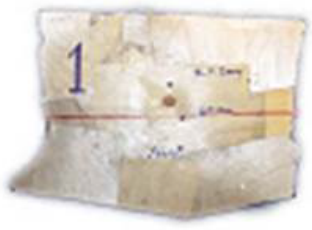 | 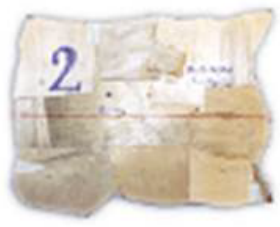 | 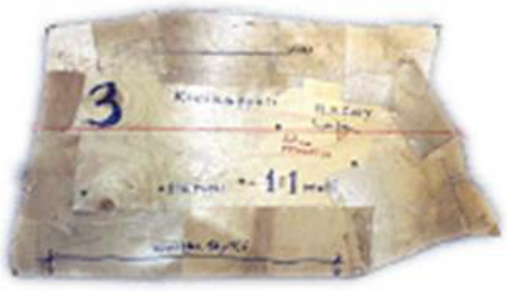 | Plywood model 1:1, exact measures to the granite wall window hole. |
| 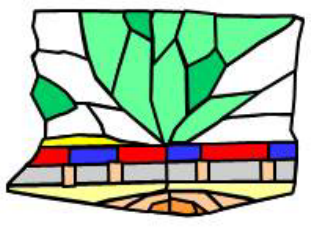 | 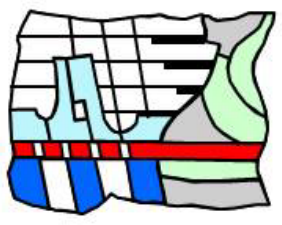 | 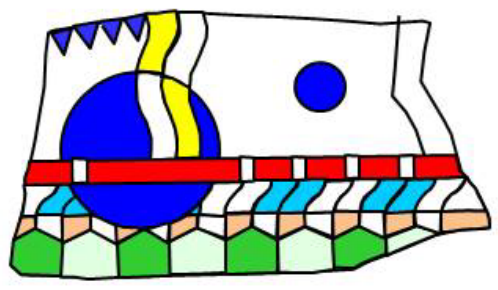 | Colour model simulation on computer.<br>Colours equal to real coloured glass products |
| 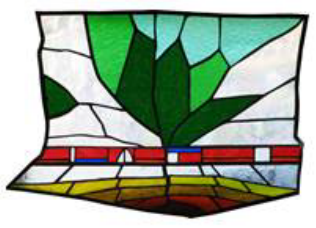 | 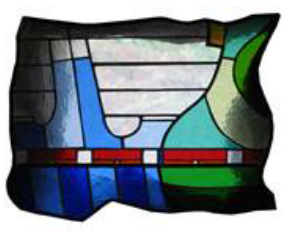 | 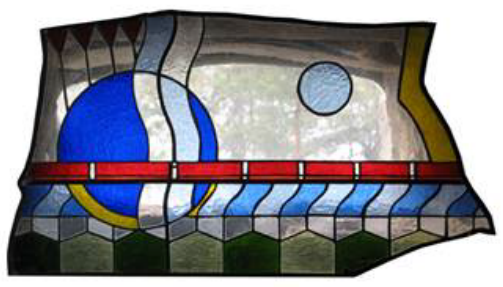 | Stained Tiffany glass windows assembled to the window holes.<br>Lead came on the outer borders. |

FIGURE 16. Stained glass windows at the Stone Chapel in Vivamo, Finland, 1995-2045 (sketches, plywood models, and form-giving realized by the author)

### Industrial design

Cast iron joint systems have been designed for concrete and wooden columns. As a component system, a light mass customization method has been used. These systems are intended for public buildings, railway stations, concert halls and outdoor use (Figure 17).

The Interest Machine™ (2012) is a system of screens and devices connected by liquid software to form a unified environment of a communication. The phone and tablet are stretchable so that the screen scales from small to large size, even from a phone to a tablet. The liquid software allow the user to use all his/her devices smoothly and securely in all environments. The Interest Machine is a reality, knowledge and media machine, as well as a wisdom, drama and beauty machine (Figure 18).

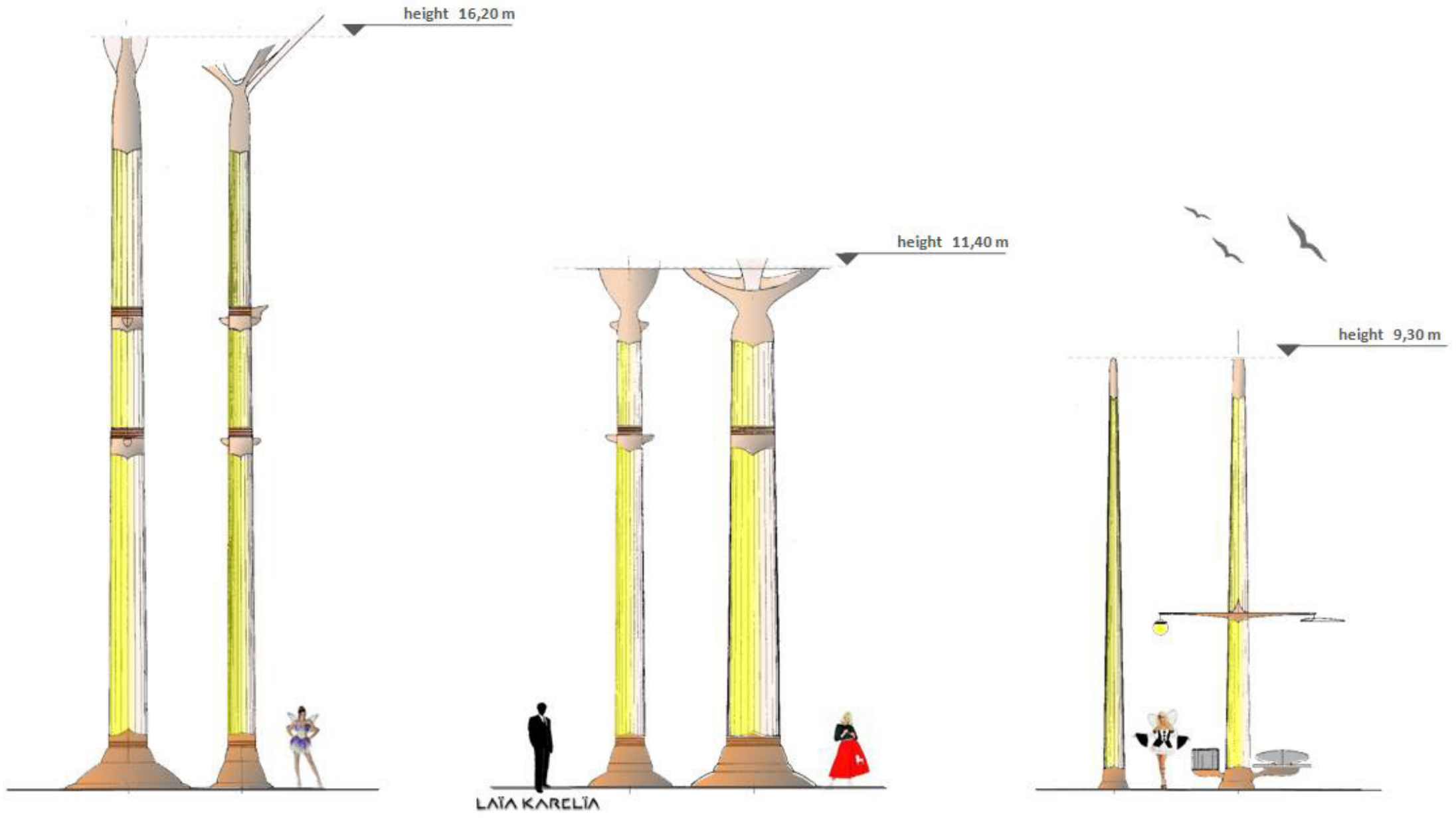

FIGURE 17. Ledoux system: Cast iron joint system for laminated wood as a bearing structure

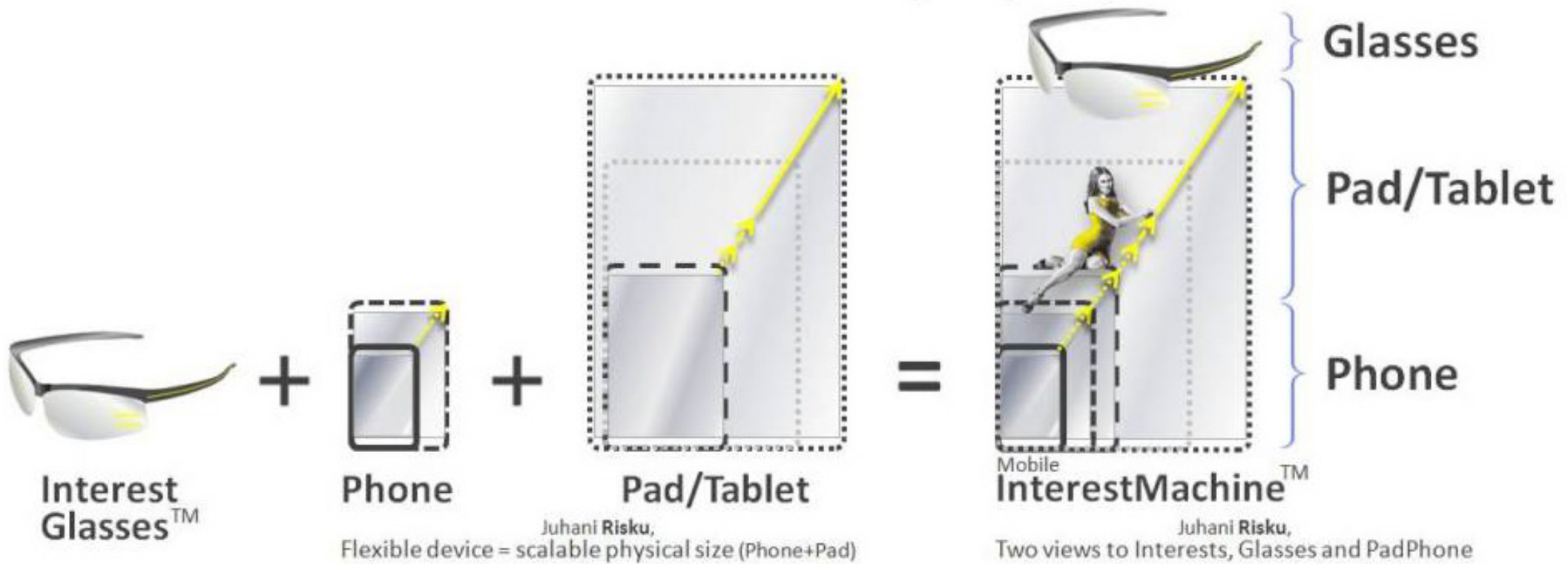

FIGURE 18. Interest Machine™: personal devices from stretchable handhelds and liquid solutions to a new device category (the author, 2012)

## Realization of the building

The design process of the stone chapel from the original sketch (1) to the finished drawing (4) took five minutes (Figure 19). The sketching was done on A4 paper. 3D modeling (5) was performed with ArchiCAD 4.2 and the designed graffiti was sprayed on the washed stone. Note the 8 degree error correction from yellow to red spray paint. A quick sketch for the parish (7) and a sketch of the stained glass windows (8) were made with the Nokia Digital Pen SU-1B. The 2 m high windows (16) were made by the author and master using Tiffany technique with lead strips and rigid frames. The largest window (15) was made with the same technique. The materials used were granite, bent wood, and copper (13) and (14). Work has been ongoing since 1995 and details are being finalized.

The granite wall (thickness 0.7-1.1 m) was completed in September 2001. The copper roof provided the final touch to the exterior in August 2002. The concave front wall is an acoustic reflector that concentrates and amplifies the sound of the choir through the forest to a nearby recreation center (Figure 20).

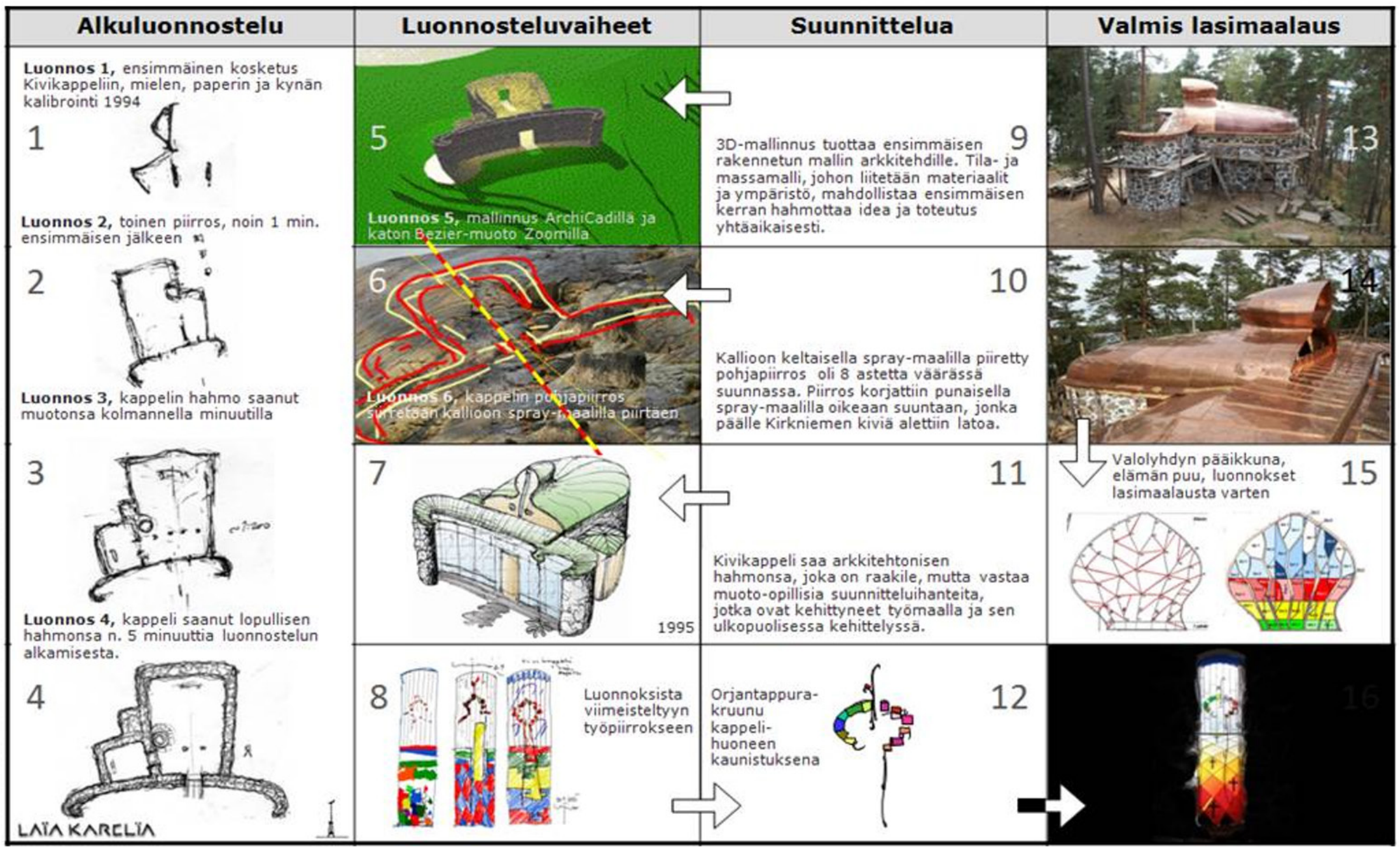

FIGURE 19. Design process of a stone chapel in Vivamo, Finland

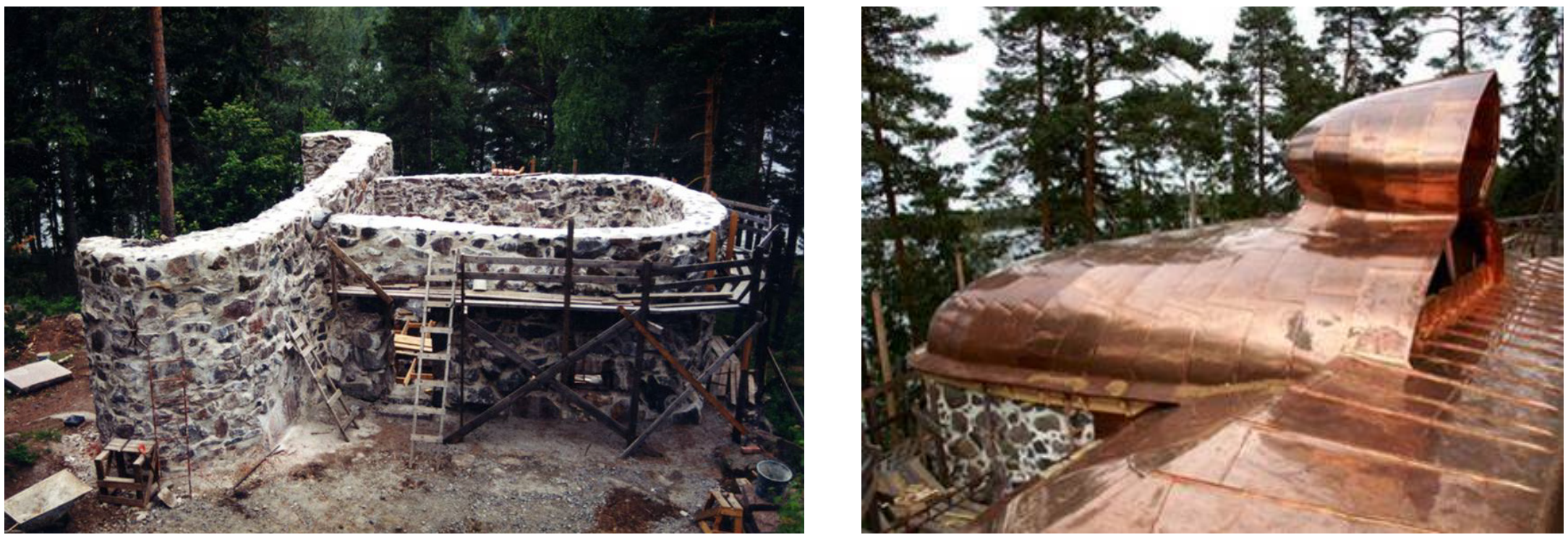

FIGURE 20. Granite wall and copper roof of a stone chapel in Vivamo, Finland

## Systemic solutions: Interest Machine, TAIC-SIMO, media

Interest Machine™ is a future device solution that consists of mobile devices, computers and TV sets (Figure 21). All its machines are interconnected and dynamically managed. Interest Machine applies new abstractions of computing and technologies, like Virtual Reality (VR), Artificial Intelligence (AI), Augmented Reality (AR), and media, like dynamic visualization, knowledge and meaning based search, sorting and cross matrix computing. Usability, understandability and relevance are drivers for next generation devices and solutions, where data converts to information, information to knowledge, knowledge to truth and wisdom, truth and wisdom to drama and beauty, drama and beauty to meaning. Interest Machine replaces the smartphones and tablets as we know them today (Figure 22).

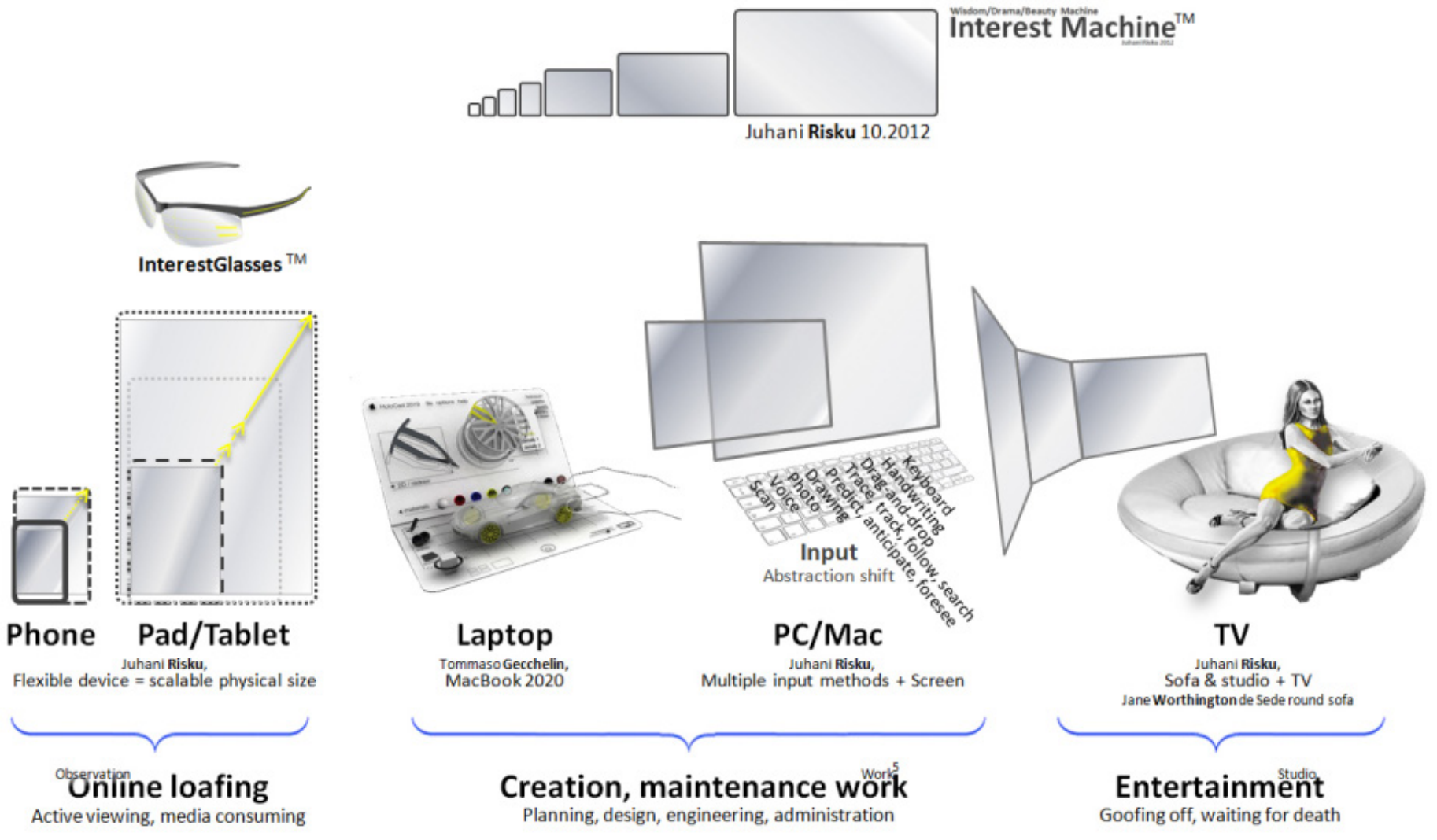

FIGURE 21. Future liquid screens

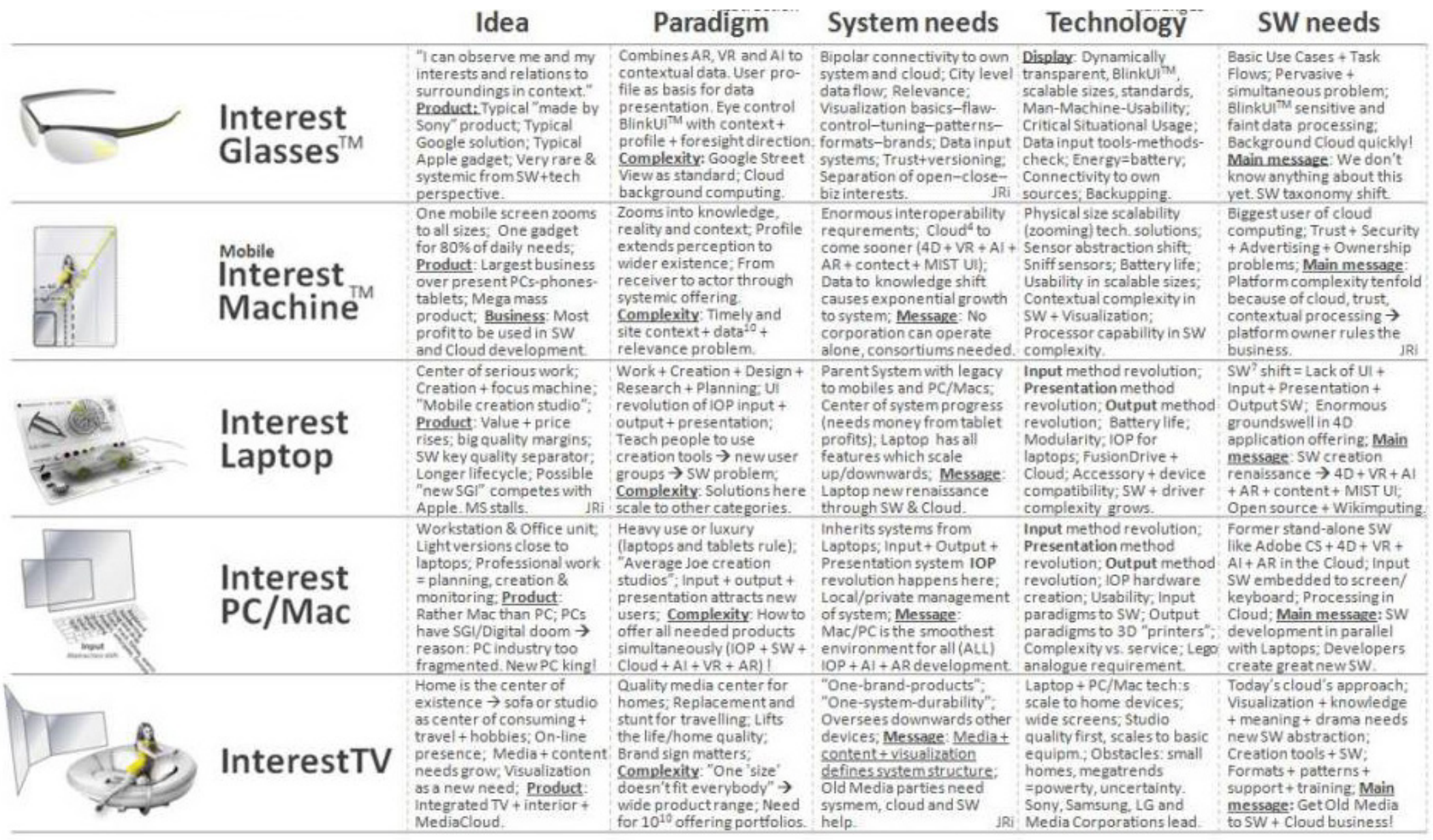

| | Idea | Paradigm | System needs | Technology | SW needs |
|---|---|---|---|---|---|
| Interest Glasses™ | "I can observe me and my interests and relations to surroundings in context." **Product:** Typical "made by Sony" product; Typical Google solution; Typical Apple gadget; Very rare & systemic from SW+tech perspective. | Combines AR, VR and AI to contextual data. User profile as basis for data presentation. Eye control BlinkUI™ with context + profile + foresight direction. **Complexity:** Google Street View as standard; Cloud background computing. | Bipolar connectivity to own system and cloud; City level data flow; Relevance; Visualization basics–flaw–control–tuning–patterns–formats–brands; Data input systems; Trust+versioning; Separation of open–close–biz interests. JRi | **Display**: Dynamically transparent, BlinkUI™, scalable sizes, standards, Man-Machine-Usability; Critical Situational Usage; Data input tools-methods-check; Energy=battery; Connectivity to own sources; Backupping. | Basic Use Cases + Task Flows; Pervasive + simultaneous problem; BlinkUI™ sensitive and faint data processing; Background Cloud quickly! **Main message**: We don't know anything about this yet. SW taxonomy shift. |
| Mobile Interest Machine™ | One mobile screen zooms to all sizes; One gadget for 80% of daily needs; **Product**: Largest business over present PCs-phones-tablets; Mega mass product; **Business**: Most profit to be used in SW and Cloud development. | Zooms into knowledge, reality and context; Profile extends perception to wider existence; From receiver to actor through systemic offering. **Complexity**: Timely and site context + data[10] + relevance problem. | Enormous interoperability requrements; Cloud[4] to come sooner (4D + VR + AI + AR + contect + MIST UI); Data to knowledge shift causes exponential growth to system; **Message**: No corporation can operate alone, consortiums needed. | Physical size scalability (zooming) tech. solutions; Sensor abstraction shift; Sniff sensors; Battery life; Usability in scalable sizes; Contextual complexity in SW + Visualization; Processor capability in SW complexity. | Biggest user of cloud computing; Trust + Security + Advertising + Ownership problems; **Main message**: Platform complexity tenfold because of cloud, trust, contextual processing → platform owner rules the business. JRi |
| Interest Laptop | Center of serious work; Creation + focus machine; "Mobile creation studio"; **Product**: Value + price rises; big quality margins; SW key quality separator; Longer lifecycle; Possible "new SGI" competes with Apple. MS stalls. JRi | Work + Creation + Design + Research + Planning; UI revolution of IOP input + output + presentation; Teach people to use creation tools → new user groups → SW problem; **Complexity**: Solutions here scale to other categories. | Parent System with legacy to mobiles and PC/Macs; Center of system progress (needs money from tablet profits); Laptop has all features which scale up/downwards; **Message**: Laptop new renaissance through SW & Cloud. | **Input** method revolution; **Presentation** method revolution; **Output** method revolution; Battery life; Modularity; IOP for laptops; FusionDrive + Cloud; Accessory + device compatibility; SW + driver complexity grows. | SW[7] shift = Lack of UI + Input + Presentation + Output SW; Enormous groundswell in 4D application offering; **Main message**: SW creation renaissance → 4D + VR + AI + AR + content + MIST UI; Open source + Wikimputing. |
| Interest PC/Mac | Workstation & Office unit; Light versions close to laptops; Professional work = planning, creation & monitoring; **Product**: Rather Mac than PC; PCs have SGI/Digital doom → reason: PC industry too fragmented. New PC king! | Heavy use or luxury (laptops and tablets rule); "Average Joe creation studios"; Input + output + presentation attracts new users; **Complexity**: How to offer all needed products simultaneously (IOP + SW + Cloud + AI + VR + AR) ! | Inherits systems from Laptops; Input + Output + Presentation system **IOP** revolution happens here; Local/private management of system; **Message**: Mac/PC is the smoothest environment for all (ALL) IOP + AI + AR development. | **Input** method revolution; **Presentation** method revolution; **Output** method revolution; IOP hardware creation; Usability; Input paradigms to SW; Output paradigms to 3D "printers"; Complexity vs. service; Lego analogue requirement. | Former stand-alone SW like Adobe CS + 4D + VR + AI + AR in the Cloud; Input SW embedded to screen/keyboard; Processing in Cloud; **Main message:** SW development in parallel with Laptops; Developers create great new SW. |
| InterestTV | Home is the center of existence → sofa or studio as center of consuming + travel + hobbies; On-line presence; Media + content needs grow; Visualization as a new need; **Product**: Integrated TV + interior + MediaCloud. | Quality media center for homes; Replacement and stunt for travelling; Lifts the life/home quality; Brand sign matters; **Complexity**: "One 'size' doesn't fit everybody" → wide product range; Need for $10^{10}$ offering portfolios. | "One-brand-products"; "One-system-durability"; Oversees downwards other devices; **Message**: Media + content + visualization defines system structure; Old Media parties need sysmem, cloud and SW help. JRi | Laptop + PC/Mac tech:s scale to home devices; wide screens; Studio quality first, scales to basic equipm.; Obstacles: small homes, megatrends =powerty, uncertainty. Sony, Samsung, LG and Media Corporations lead. | Today's cloud's approach; Visualization + knowledge + meaning + drama needs new SW abstraction; Creation tools + SW; Formats + patterns + support + training; **Main message:** Get Old Media to SW + Cloud business! |

FIGURE 22. Interest Machine™ chart providing embedded AR solutions

Ecosystem ONE™ (TAIC-SIMO) is a platform of a mobile operator, network manufacturer and media house (Figure 23). New solutions are created for devices and background systems. They form an internal startup to gain back the incomes that the search engine and social media corporations took by the digitalization of the media (Article II).

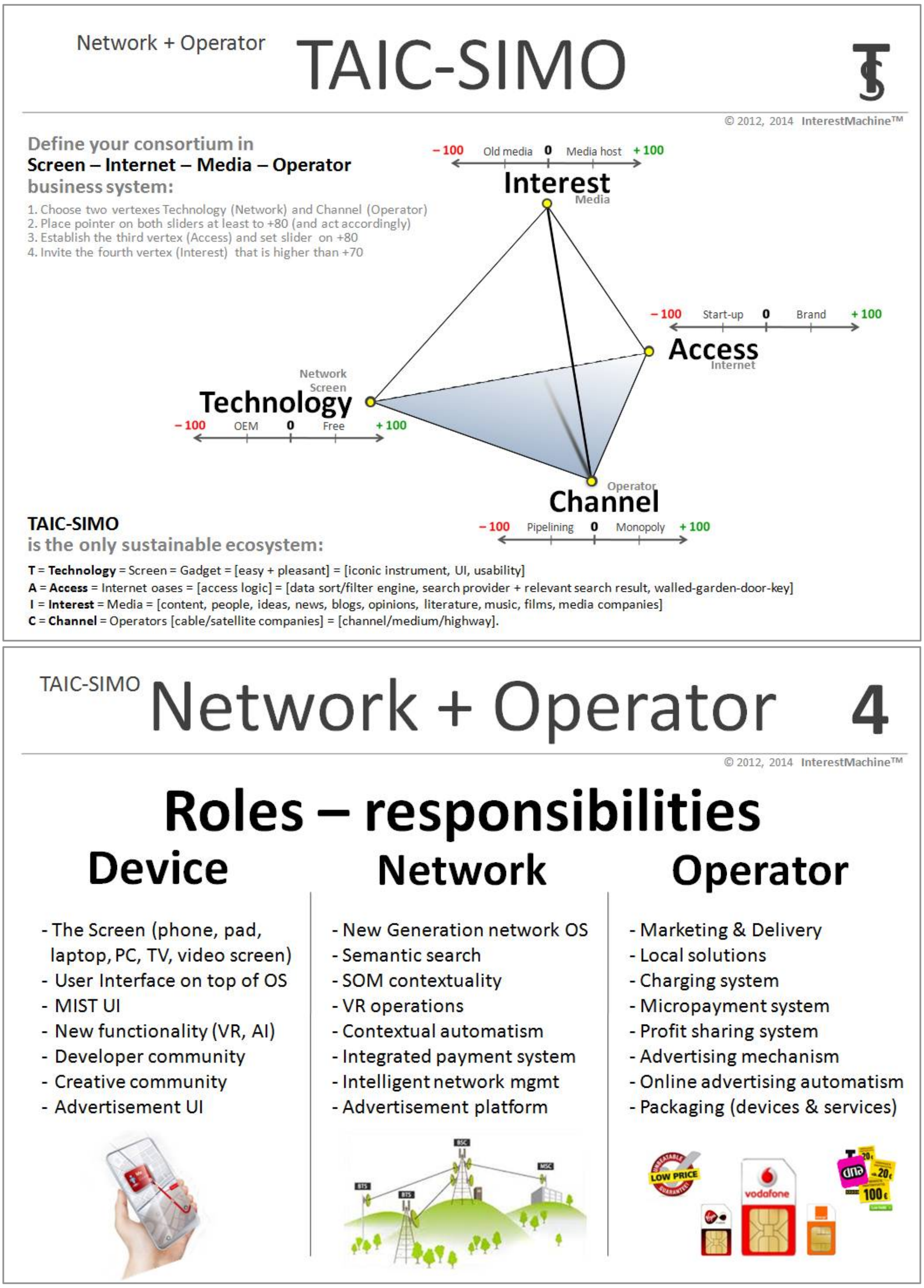

FIGURE 23. Ecosystem ONE™ roles with the Tetra model from both a device and technology perspective

Ecosystem ONE™ invites media houses to take part in the restructure of media with the network operator and network operator (carrier) (Figure 24). Media houses join the solution by providing their own and user-created content. These three parties form an actor named the Trio (Article II).

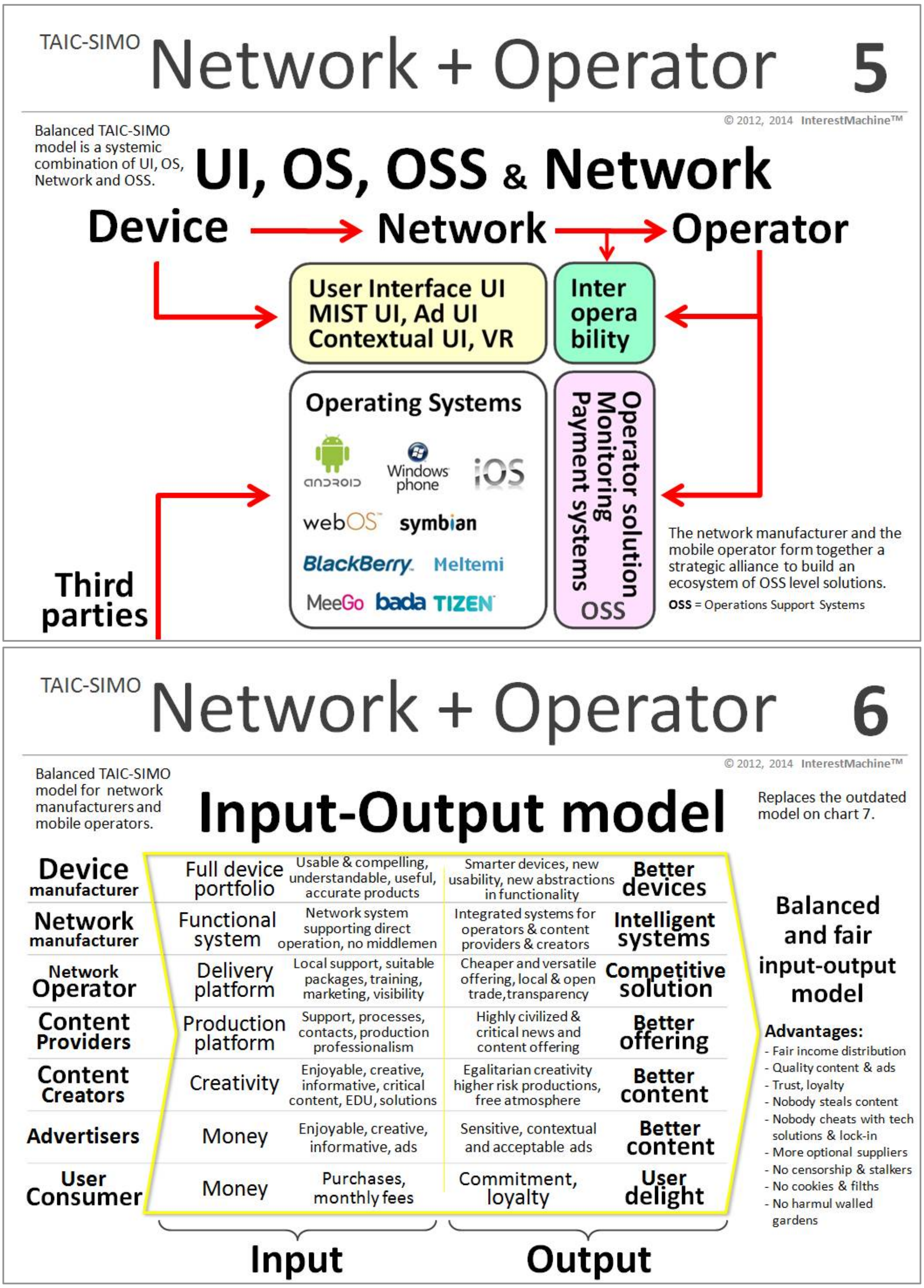

FIGURE 24. Conceptual and structural descriptions for Ecosystem ONE™

Restructure the media, conference poster, invites media houses to take part in the restructure of media with the network operator and network operator (carrier). These three parties form a startup named the Trio (Article II), based on TAIC-SIMO model (Figure 25).

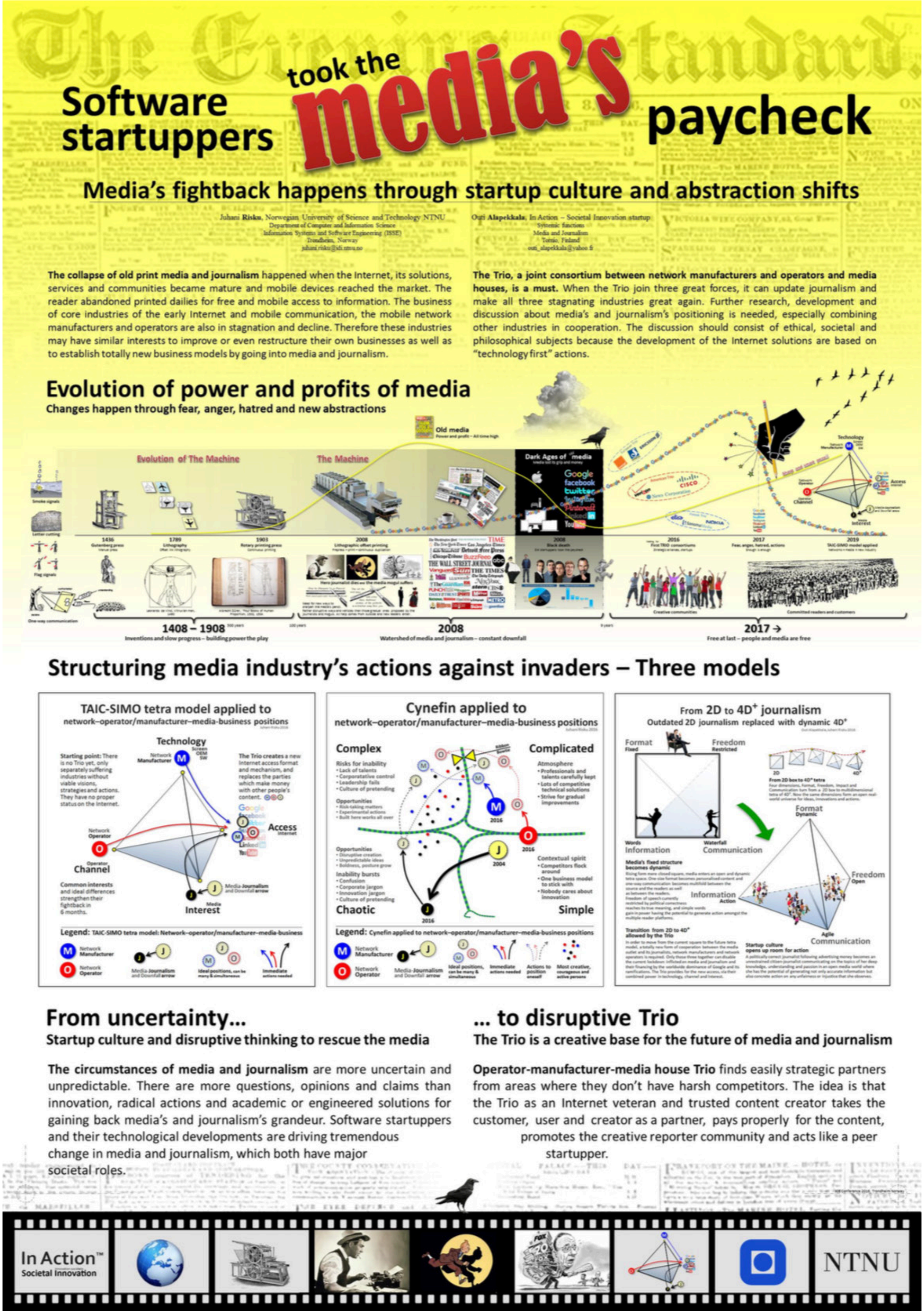

FIGURE 25. Media restructuring: holistic visualization of the timeline

## Forestry and wood construction

The post-pulp industry has a future in wood construction (Figure 26).

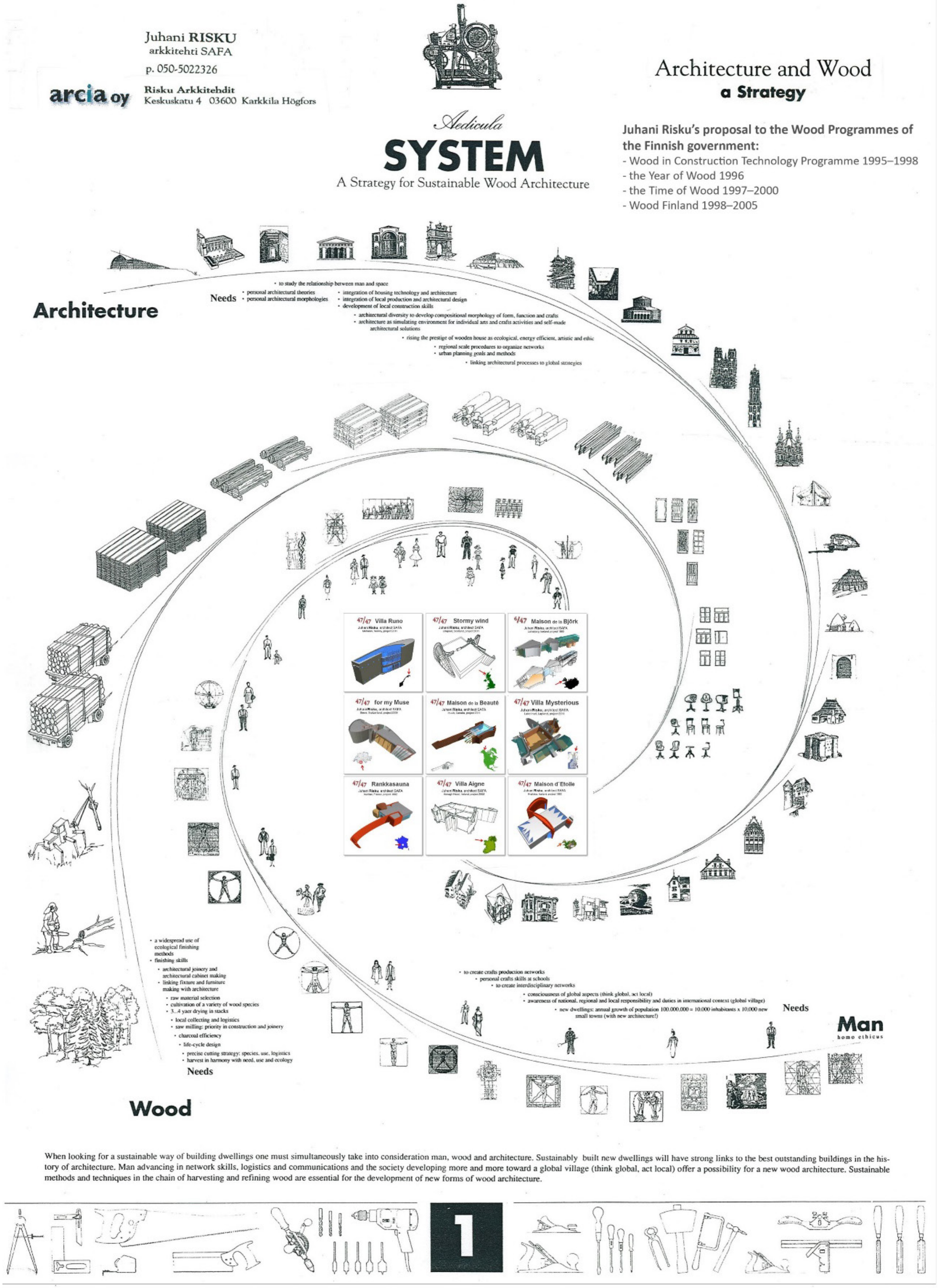

FIGURE 26. System for organizing the Finnish forestry and wood construction industry (A0 poster by the author, 1996)

## Imaginary theories, MIST dimensions

The author's model combines four dimensions Materia-Idea-Space-Time in one construct (Figure 27). A conceptual sketch is given in Figure 29. This opposes Einstein's 2D universe of three spatial components and time.

Erosion and weathering hits a temple through simultaneous MIST evolution. Matter, idea, space and time are interlinked and the transformation happens in all MIST dimensions. Each MIST dimension has a range of –100,…,0,…,+100 (nothingness … zero … perfection) showing its state of existence (Figure 28).

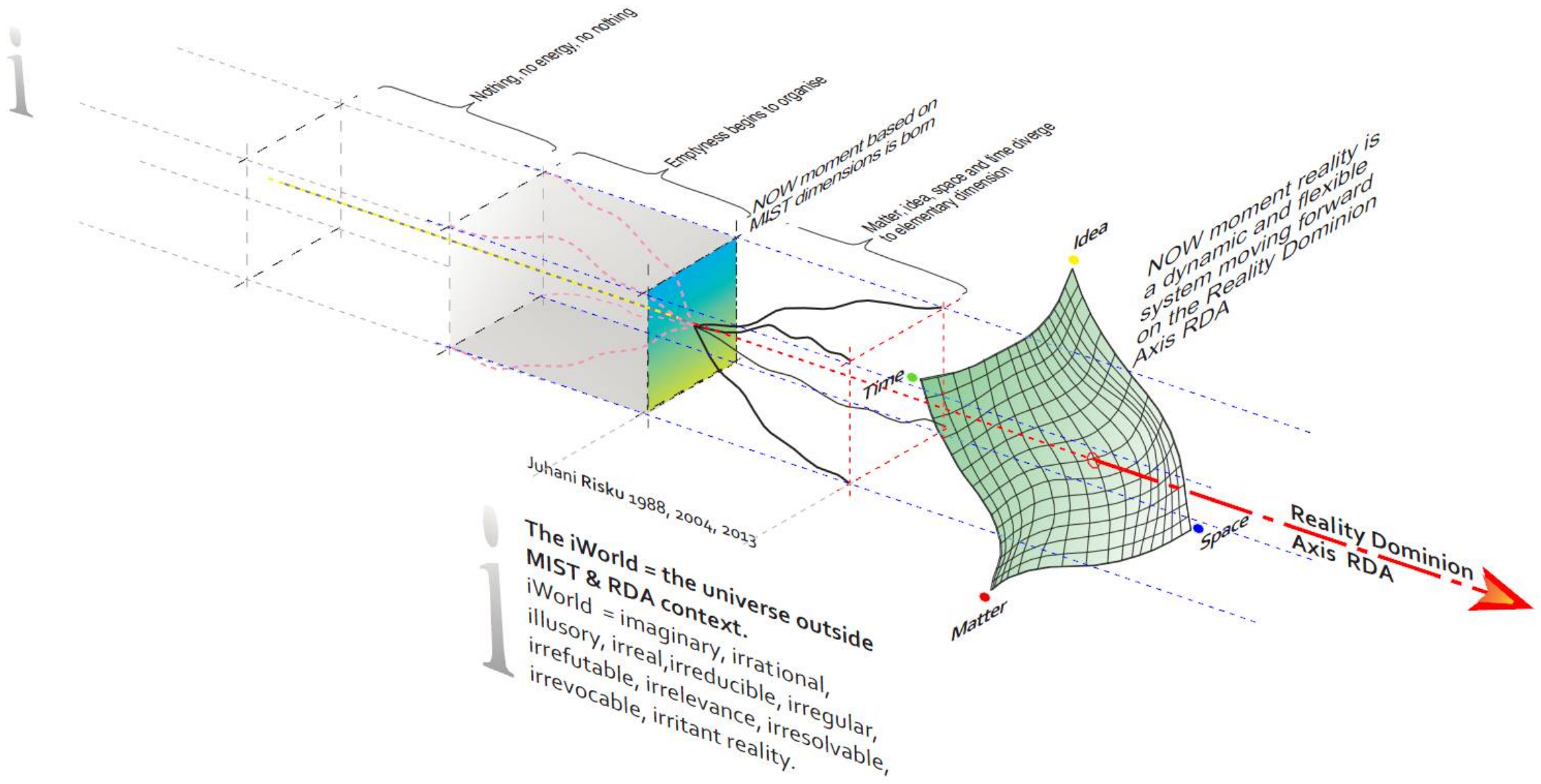

FIGURE 27. Now Moment Reality (NMR) grid moving forward on the Reality Dominion Axis (RDA)

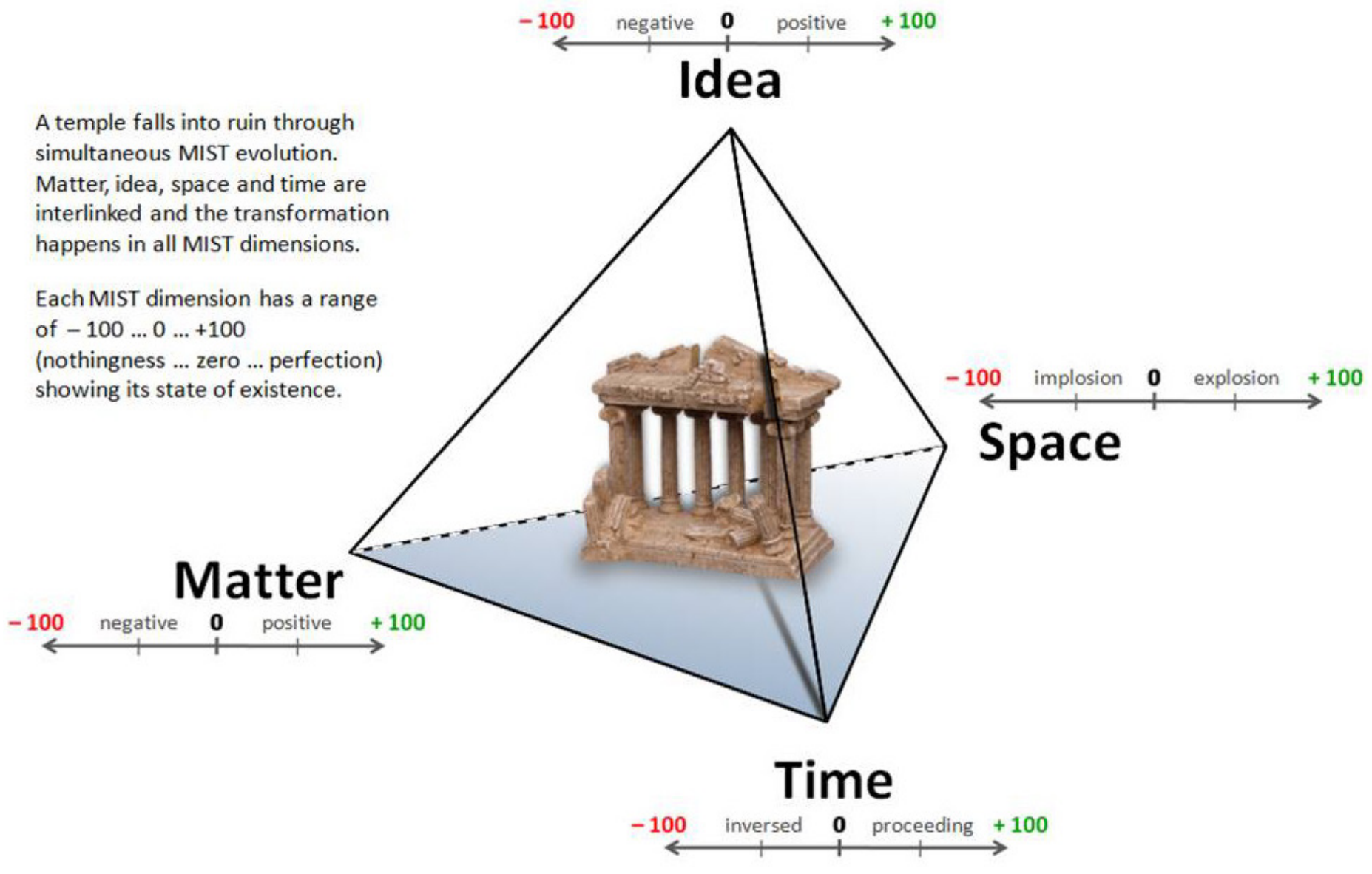

FIGURE 28. Tetrahedron geometry of MIST dimensions, temple-test of MIST dimensions: Destruction of a temple (vision by the author, 2013)

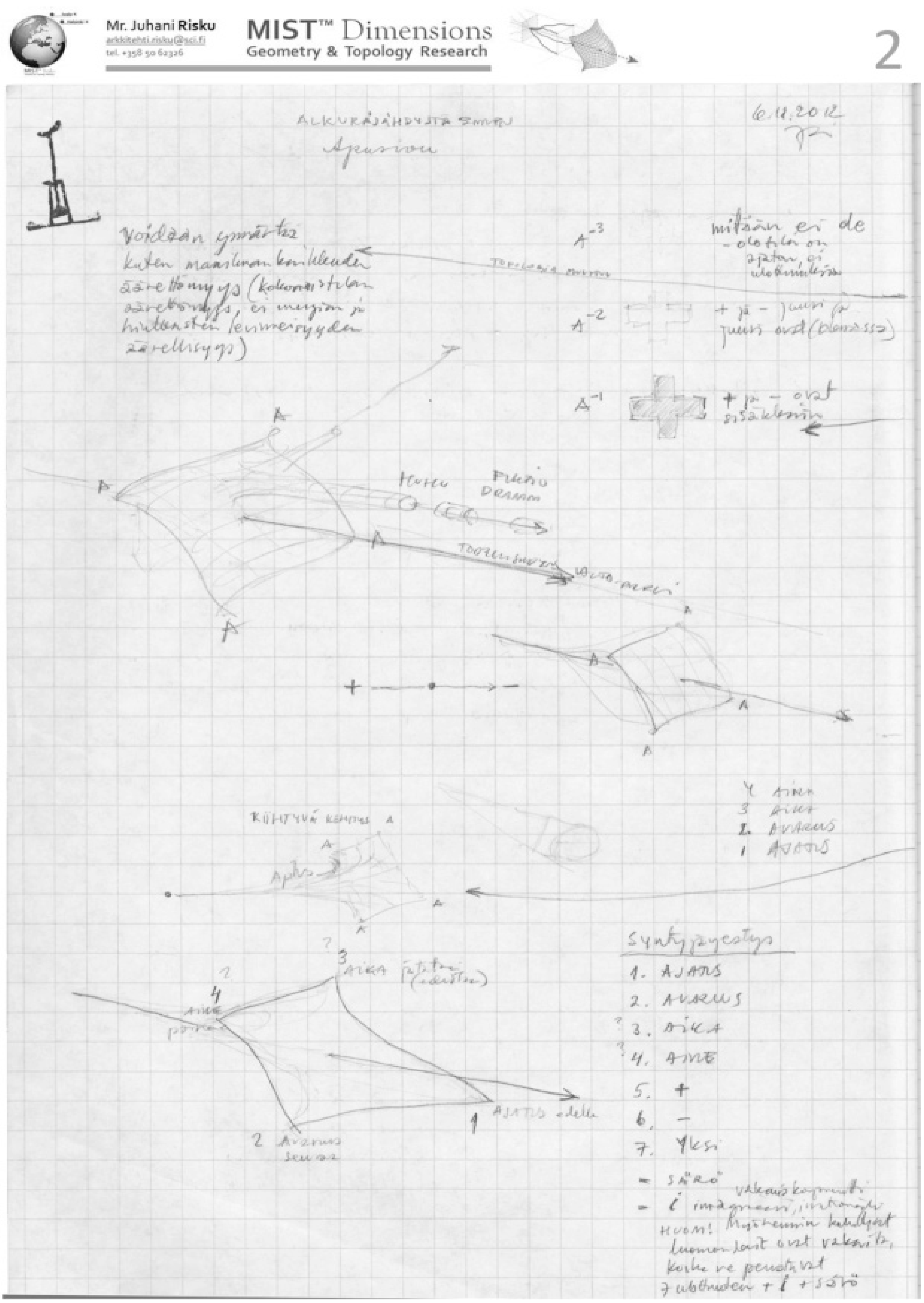

FIGURE 29. Conceptual sketch of Now Moment Reality (NMR) grid moving forward on the Reality Dominion Axis (RDA)

Corrosion is a 4D phenomenon. Corrosion and rust hits a knife through simultaneous MIST evolution. Matter, idea, space and time are interlinked and the transformation happens in all MIST dimensions (Figures 30 and 31).

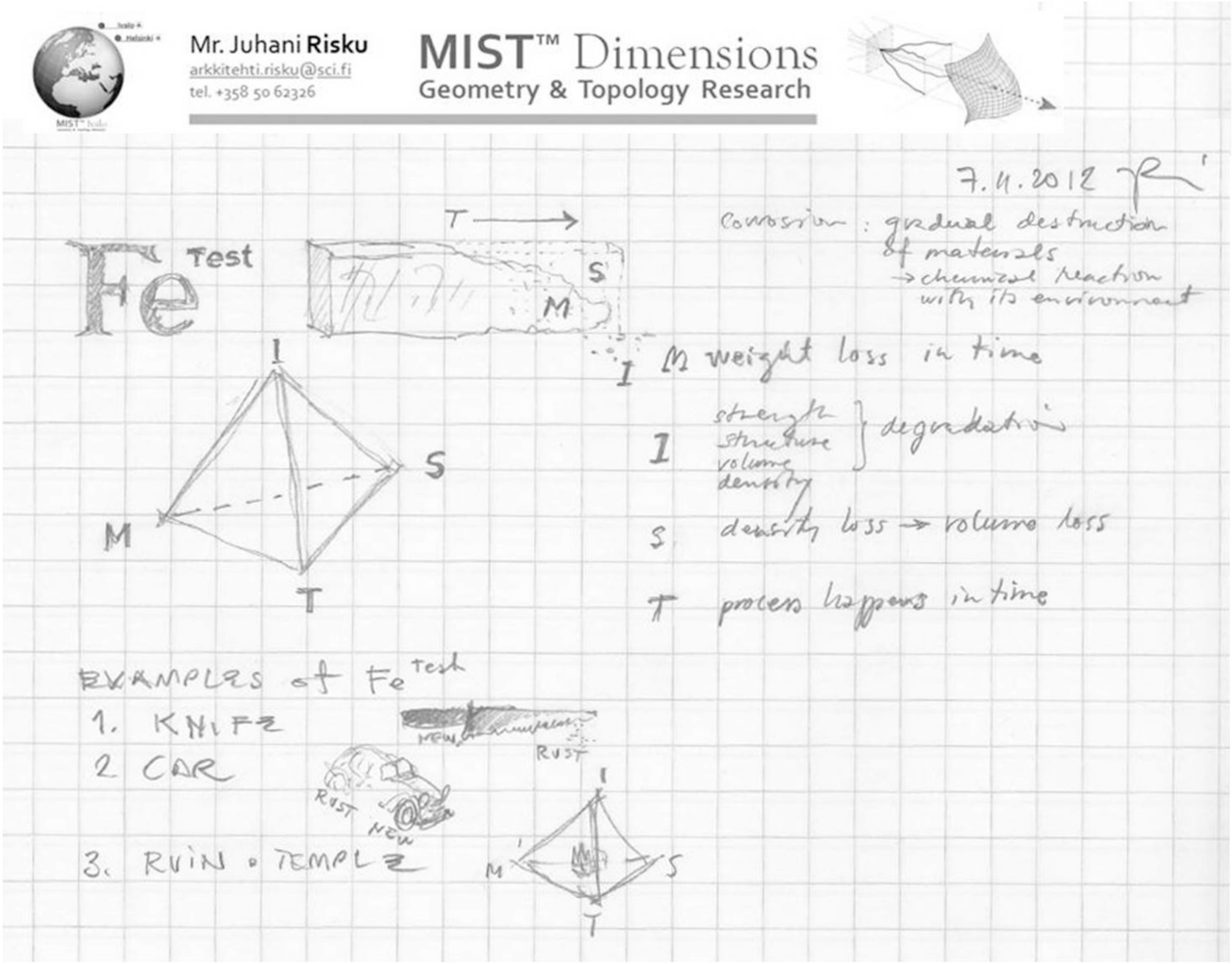

FIGURE 30. Sketch of the Fe-test for MIST dimensions: Corrosion of a knife

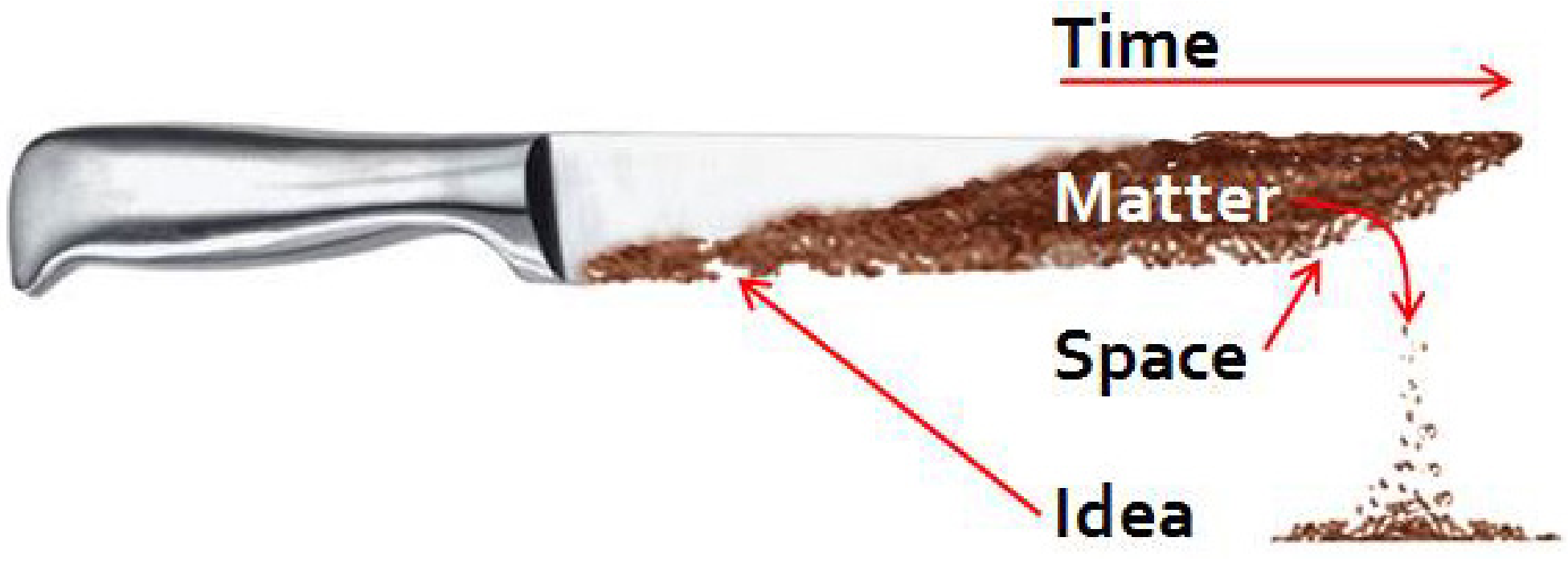

FIGURE 31. Corrosion of a knife with MIST dimensions

### City planning 1: NTNU campus concept

The new campus of the Norwegian University of Science and Technology (NTNU) in Gløshaugen was planned to be realized in 12 years (not in 25-35 years as in the official plan). This plan also brings byproducts as the Nansen Science Park, Trollscape amusement park, and Factory-In-Ship FISh concept of shipbuilding (Figure 32).

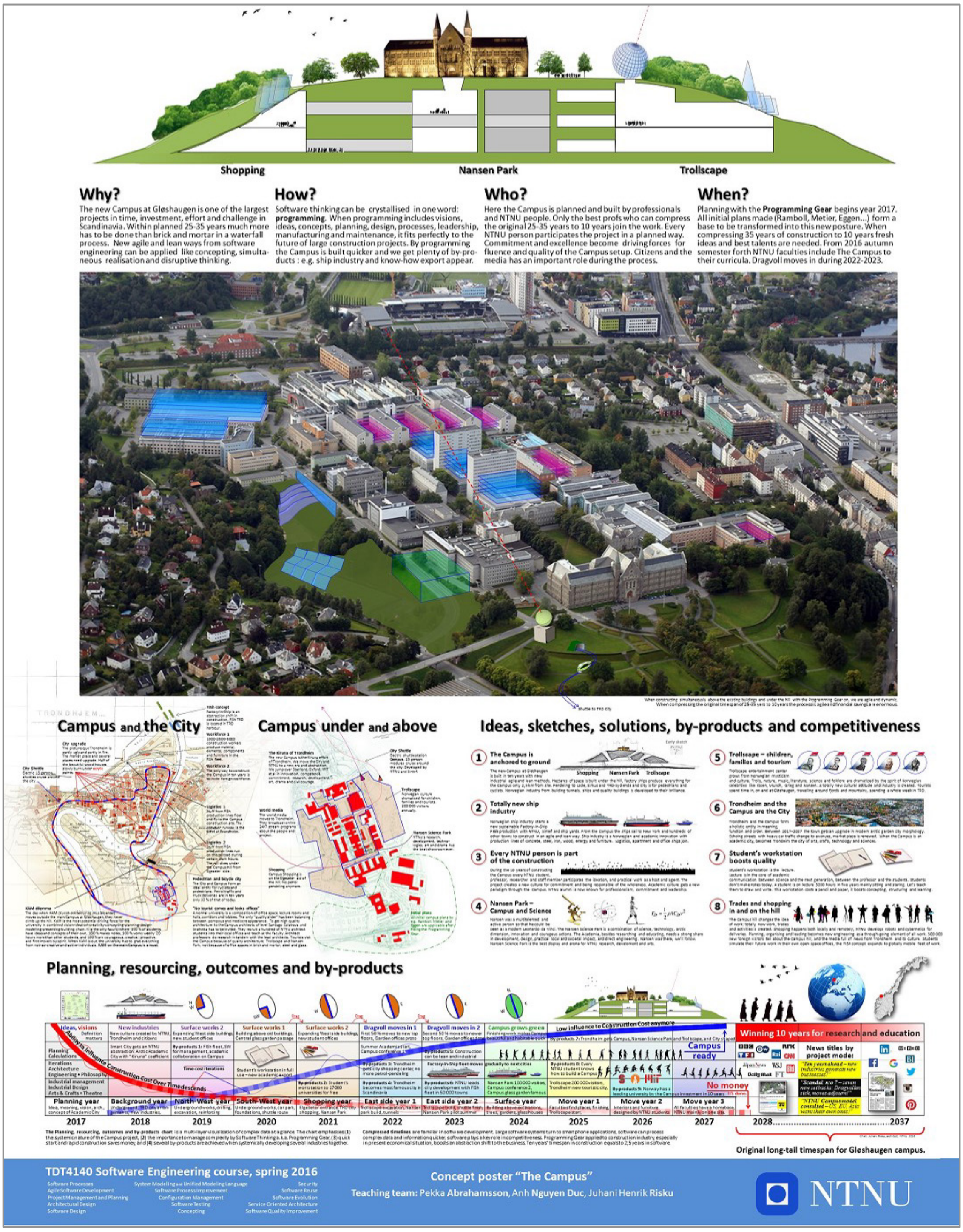

FIGURE 32. Campus at Gløshaugen, NTNU Trondheim (the author & Outi Alapekkala, 2016).

## City planning 2: Arctic Garden City

The Ledoux-Howard-Risku Garden City model is a hybrid version of a scalable and dynamic city planning platform (Figure 33). It is applicable in different cultures, geographical locations and interests of trades and activities. The Arctic version of the Garden City model is meant to be a prototype suitable in Canada, USA, Northern Europe, Russia and China. The ideal size of the city is between 10 000 and 100 000 inhabitants.

The city has a local geomorphological character with agriculture, industries and trades that follow the natural conditions of the site. The city gets its form following the geomorphology, which guides to place the operational and functional activities like residential areas, public buildings, industrial and market areas, parks, recreation grounds and agricultural areas.

The trades and services of an individual garden city is a function of natural resources, location and network of proximate garden cities in an agile and sound composition.

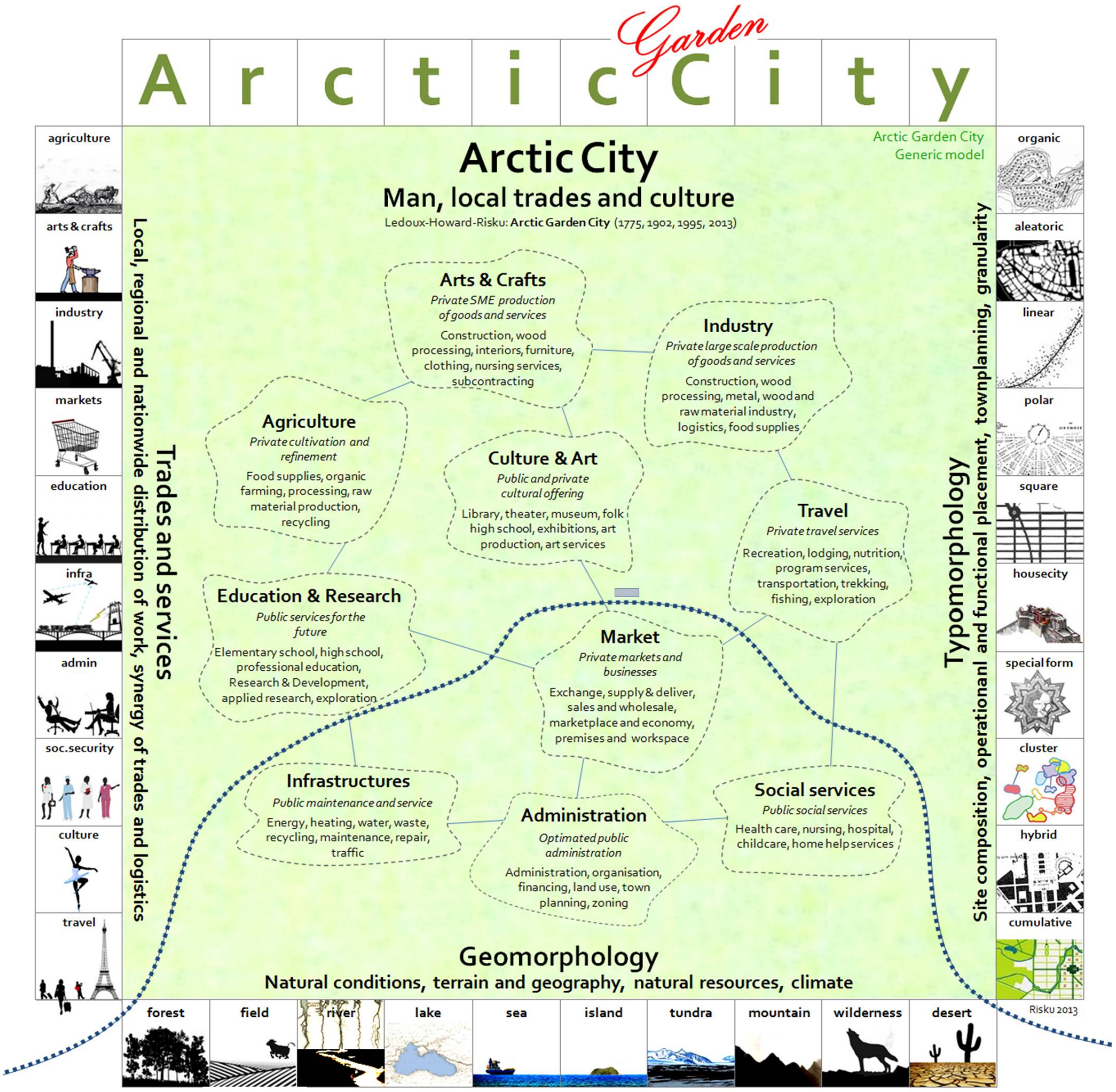

FIGURE 33. Ledoux-Howard-Risku Garden City model

### Educational concepts, student's workstation, FFE design

The academic lecture is the student's workstation (Figure 34). It enables taking notes and agile production of knowledge, innovation and critique in a professional way. The student can create several books on different topics for her/himself. The workstation prepares the student for further studies and jobs.

The "Fuzzy Front End Design" course is a hands-on design course that focuses on the early stages of a product development process and the design of software-intensive, physical products (Figure 35). The course is held at the JYU Startuplab in the University of Jyväskylä.

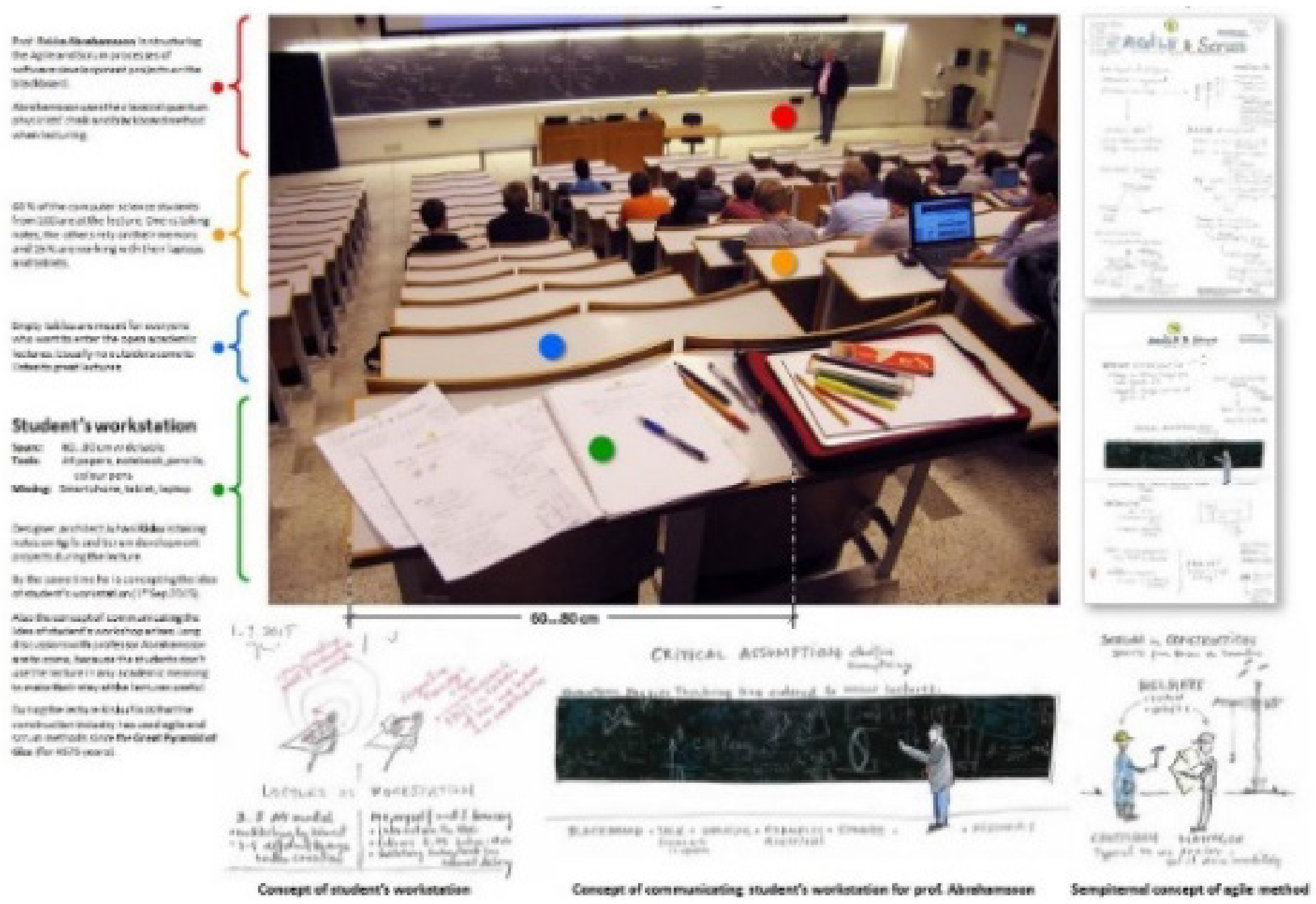

FIGURE 34. Academic lecture as the student's workstation

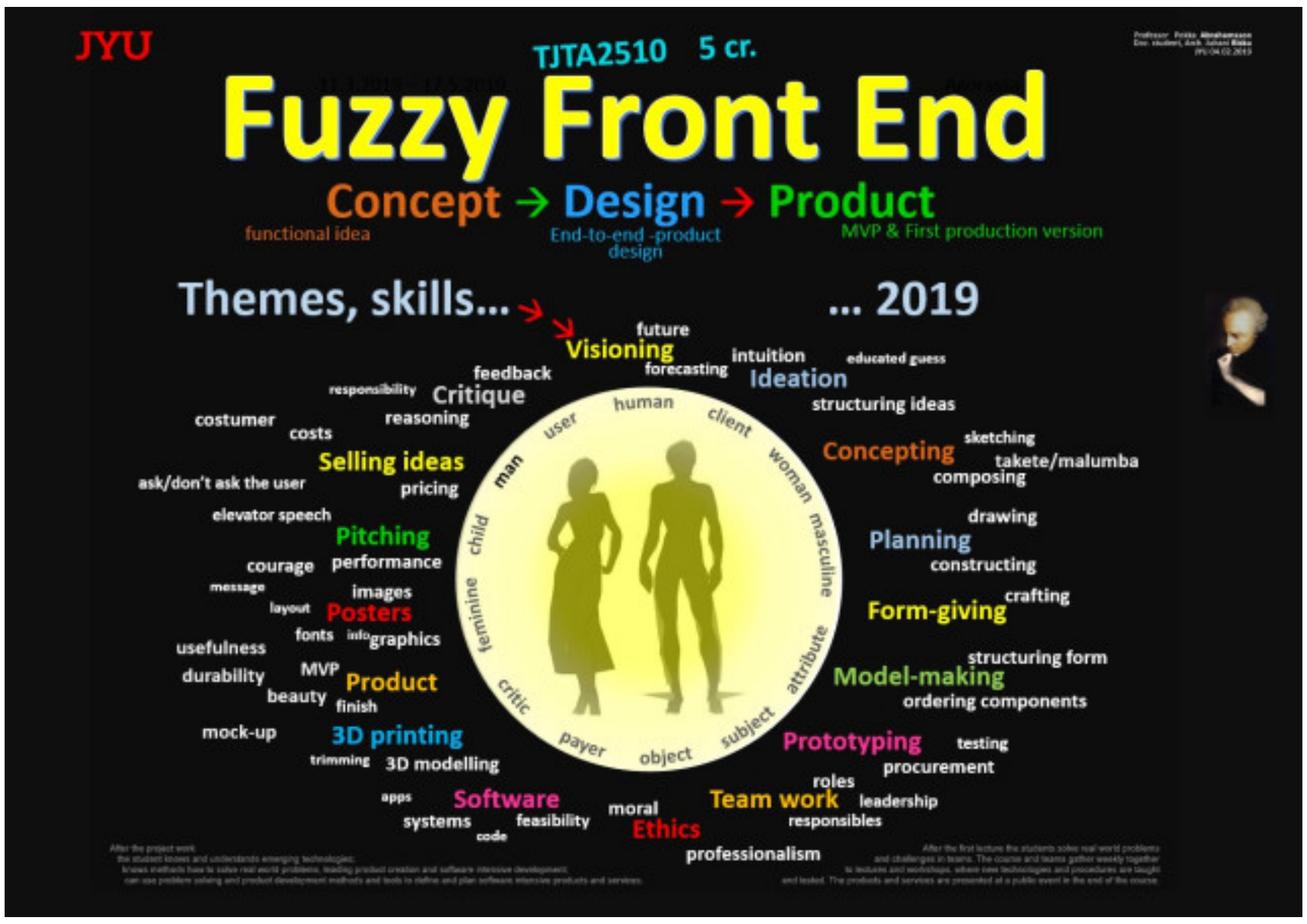

FIGURE 35. "Fuzzy Front End Design" course

**Music halls, room acoustics, structural acoustics**

The form of the chamber music hall is based on condensing acoustical sound field (Figure 36). The form strengthens the sound field the farther the sound advances. New ideas in architecture are usually gradual enhancements in style and technology. Seldom innovation happens in architecture. Special areas, like lighting, acoustics, structural engineering and digitalization, form a fruitful base for innovation in architectonic development.

The parameters of the intimate chamber music hall are (Figure 37)

- Condensing sound field with adjustable wall elements: tightened and tuned by the conductor
- Dynamic reverberation time RT (adjustable between 0.7-1.2 s)
- Dimensions: 25.1 x 20.3 = 510 $m^2$, h = 8.0 m
- Volume: 510 $m^2$ x 8 = 4080 $m^3$
- 330 seats, width 60 cm
- Max. 50-member chamber orchestra

Acoustic proximity and music tuning make the room a spatial player.

The form of the hall is based on condensing acoustic sound field. The form strengthens the sound field the farther the sound advances. The backward-narrowing music hall is the opposite of the fan-shaped widening hall. The tapered hall forms a condensing sound field that keeps the sound pressure stronger than the so-called shoe box hall. As the walls narrow the hall, the floor also rises backwards and the ceiling lowers. Acoustic research and architectural modeling can be used to determine the properties of music halls of different sizes using a condensing sound field model. The sensitivity of the chamber music hall's orchestra and instruments, and even the quiet piano pianissimo whispering sound are heard in the condensing sound field in the hall.

The walls of the hall are made of hard wood and the back wall behind the players is made of thick glass. The texture of the wood and glass walls is shaped with 3D cassettes. The walls are attached to the ceiling and floor, and can be tuned with tuning pegs along with the conductor and musicians affinity. Composers can also use hall tuning as part of their composition. The acoustic intimacy of the condensing sound field hall is equal: each seat is at the same time rich of its atmosphere, and special in detailed fine-tuning.

Acoustically, the best music halls in the world are often in the form of shoe boxes. These halls, such as the Wiener Musikverein and the Boston Symphony Hall, are also large, accommodating about 1,750 and 2,600 listeners. The chamber music halls are small compared to them, which means that the acoustic problems are also smaller. Therefore, the acoustics of chamber music halls can be developed to be more sensitive and more dynamic in structure. For example, tuning the rigidity of walls and the use of large glass surfaces are innovations that should be offered in new halls and music centers. In this way, the chamber music halls become one more component for composers and musicians, for the possibilities offered by the music and the instruments themselves.

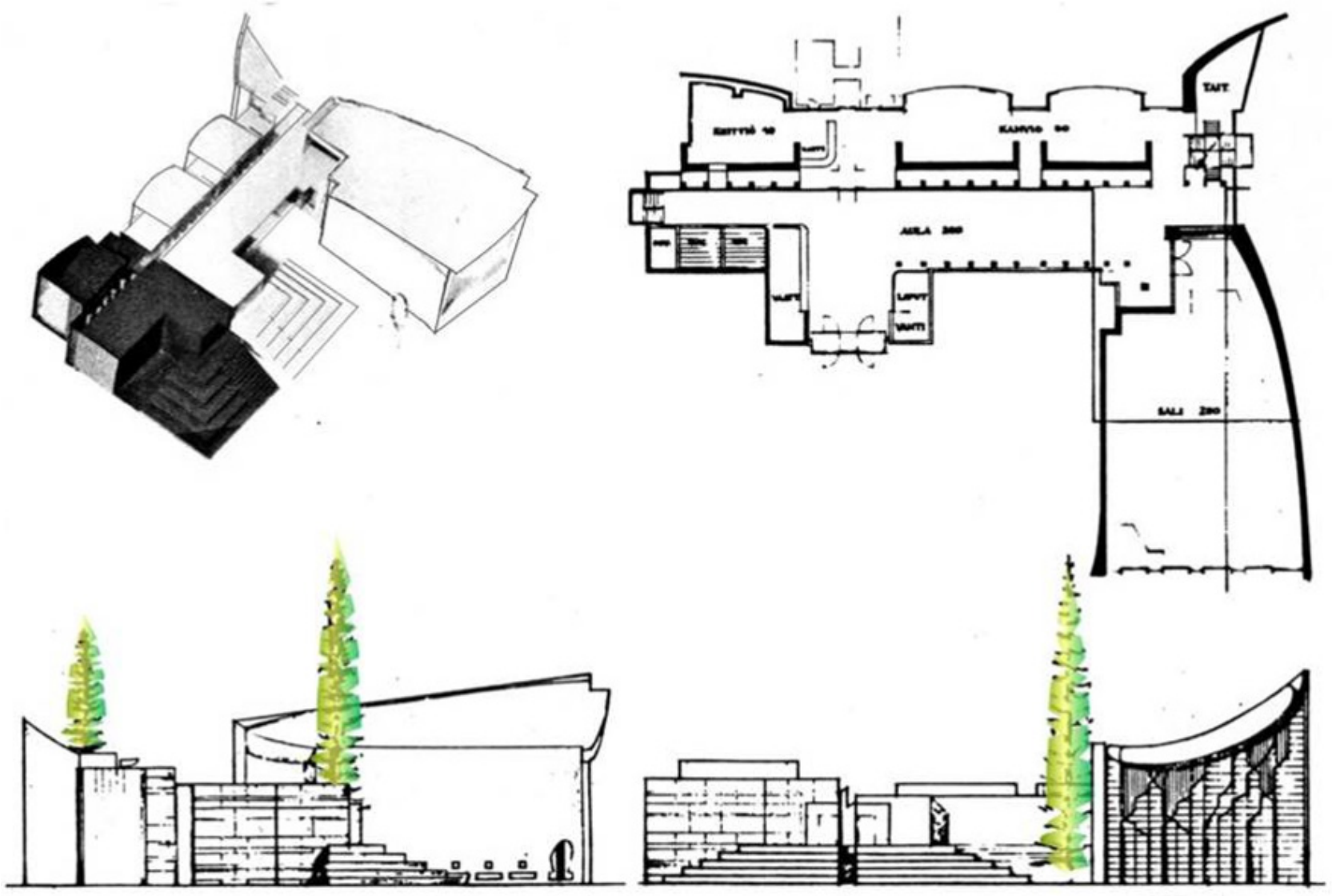

FIGURE 36. Chamber music hall; proposal to Besançon, France, 2013

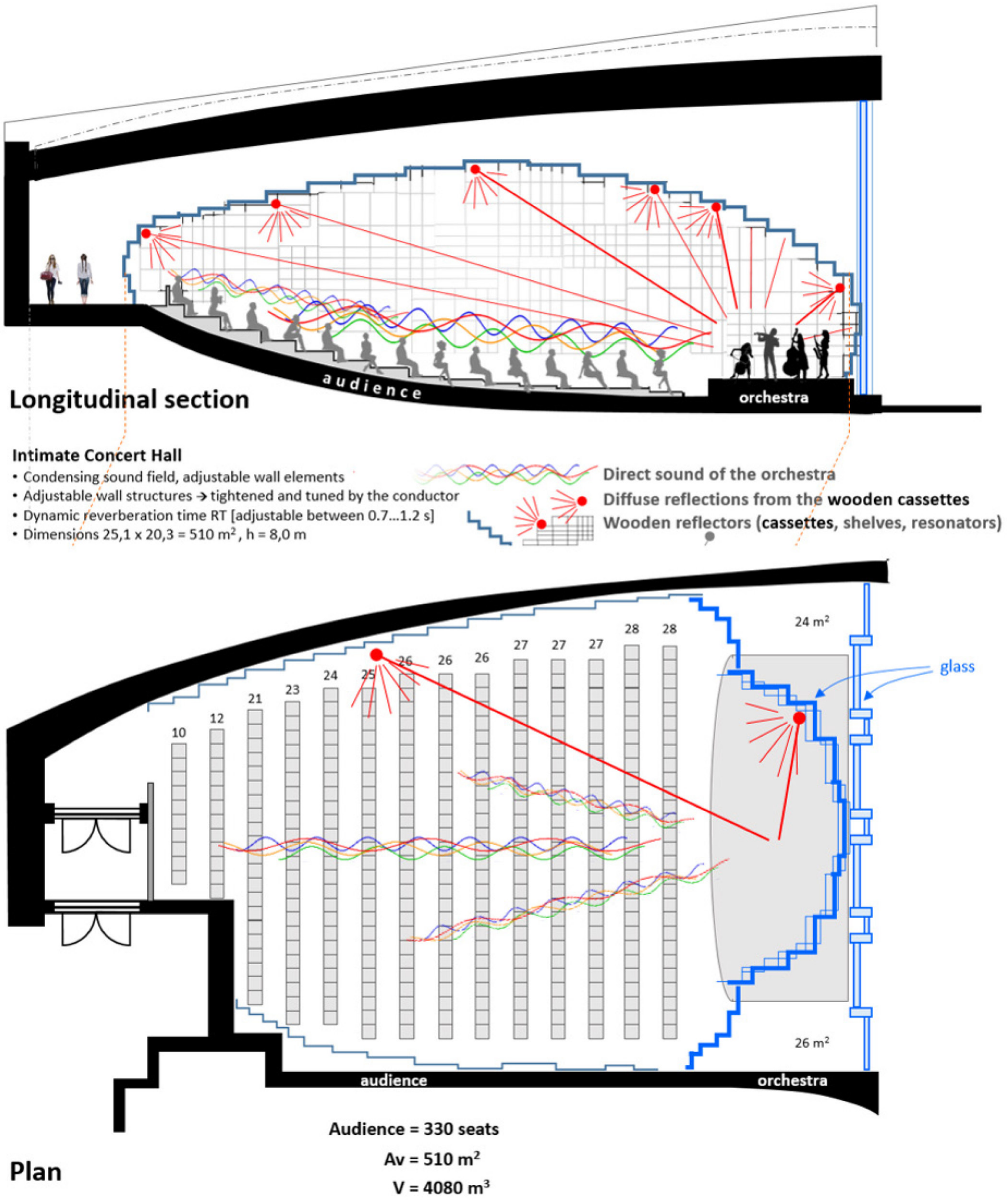

FIGURE 37. Condensing sound field in the chamber music hall

# ORIGINAL PAPERS

# I

# WHAT CAN SOFTWARE STARTUPPERS LEARN FROM THE ARTISTIC DESIGN FLOW? EXPERIENCES, REFLECTIONS AND FUTURE AVENUES

by

Juhani Risku & Pekka Abrahamsson, 2015

# What can software startuppers learn from the artistic design flow? Experiences, reflections and future avenues

Juhani Risku and Pekka Abrahamsson[1]

Norwegian University of Science and Technology, Department of Computer and Information Science, Trondheim, Norway
juhani.risku@idi.ntnu.no, pekkaa@ntnu.no

**Abstract.** **[Context and motivation]** Startups are one of the most important economic drivers in today's economy. The high failure rates have not discouraged communities, cities or universities from investing in startup ecosystems. We know that the great majority of startups today are software based. **[Problem]** It is equally well acknowledged that the early stage concept design and validation plays a key role in getting the first customers and ultimately attracting funding for the survival and possible success of a startup. Very little research, however, exists to support starttuppers in developing their early-stage ideas into concepts efficiently **[Results and contribution]** This paper aims contributing to this gap by studying the artistic design flow and the tools utilized by architects, industrial designers and artists, and as a result proposes concrete ways to improve the current state-of-practice. However, it is argued that borrowing techniques, approaches and ideas will not be sufficient, and a change in software education culture is required.

## 1 Introduction

Software startup is all about design: you need an idea to start with, you have to refine the idea into a crystallized concept which is then developed into a prototype. The prototype, whether it is on a paper, a simulation or a partly functioning application, is later realised and published as a software solution in the form of an application, a service or a large systemic solution. Design, as a historical entity and industry from early pyramids through Leonardo da Vinci´s art and machines, has their design origins in arts, architecture and craftsmanship.

During Nokia Corporation´s golden age 2004-2009, the money-maker was the so called Symbian based mobile phones. Phones became all-the-time software intensive products, in which usability and visualisation grew in importance. In Symbian core teams, the designers (artists, industrial designers and visual designers) had a twofold role: design individually holistic user interfaces so that several proposals were evaluated during one session. Then the proposals were merged into one version, which was developed both into a simulated version on a laptop and a functional version in an existing phone.

We observed then that a single designer can, at its best, design nearly everything from abstract to detail-level solution for a mobile application by herself. This means that a typical designer team of 5-7 designers can together cover up to 85-90 % of the application, in some days, to be put forward for coding. What is even more worth-noticing is that, designers are able to start any work initiatively alone as well as in a team. This may be because of their educational background in design and forward-looking attitude. If there were engineers or psychologists in the team they could participate right after the designers had produced the design. Engineers or psychologists seldom had a plan or design even in a raw format to be discussed. Now it seems that the idea of initiators and followers also scale to a single team level [1]. The initiators start and lead the creation and development process, and the followers participate mostly on technical implementation details. This detailed level is often divided according to educational and professional background to roles like user-interface designer, tester, coder, etc. The ideas of *first movers* and *second movers* emerge.

The notion of a first-mover can be found in Aristotle's idea of the prime mover. Aristotle's Metaphysics ("after the Physics"), develops his theology of the prime mover, as πρῶτον κινοῦν ἀκίνητον: an independent divine eternal unchanging immaterial substance [2]. Aristotle's Prime Mover causes the movement of other things. The Prime Mover is the purpose and the teleology of the movement when creating it. The Second Mover observes, waits and stays on the background. The Second Mover never starts the process and gives a meaning to it. But, as we know that both fast followers and late movers can profit from the first-movers´ mistakes.

A person, who takes the initiative in a product creation process, needs a wide skillset of the product related factors from idea conception, crafting, building and finishing the product. He (she) needs not only a deep understanding of the user needs and market demand, but also an ability to lead the creation and development process. In startup context, one person may have such responsibilities but may fall short in actual skills. Current research on software startupper competencies emphasizes software engineering skills but acknowledges a need for a broader skillset [3]. Seppänen et al. do not however indicate what these broader skills may be [3]. We hypothesize that first-movers are more likely to possess horizontal artistic and design skills, which means in practice having a deep knowledge in ideation, sketching, concepting, planning, design workflow, leadership, crafting and human understanding.

[1] The first author spent 9 years in Nokia Corporation in various design related roles where he also served as an innovation director. The second author has researched and taught software startups since 2011.

Universities play a major role in educating startuppers. Today's software startuppers often have a computer science education. When observing computer science master students of their fourth academic year at NTNU we noticed an interesting issue. Very few students if any take regularly notes during the lectures although it is well grounded that note-taking is essential in a learning process [4]. Based on our initial observation only one or two from 50 students are making notes, writing or drawing structures during the lectures. One worth-mentioned observation is that only small portion of students follow the lecture and something visible from it. In one lecture at NTNU, we have seen c.a. two thirds of the students appear to be sitting two hours listening or just eyeballing the wall behind the lecturer. When asked, other lecturers confirmed our observations. As a contrasting example, Figure 1 shows a picture of a designer´s notebook when following a computer science lecture.

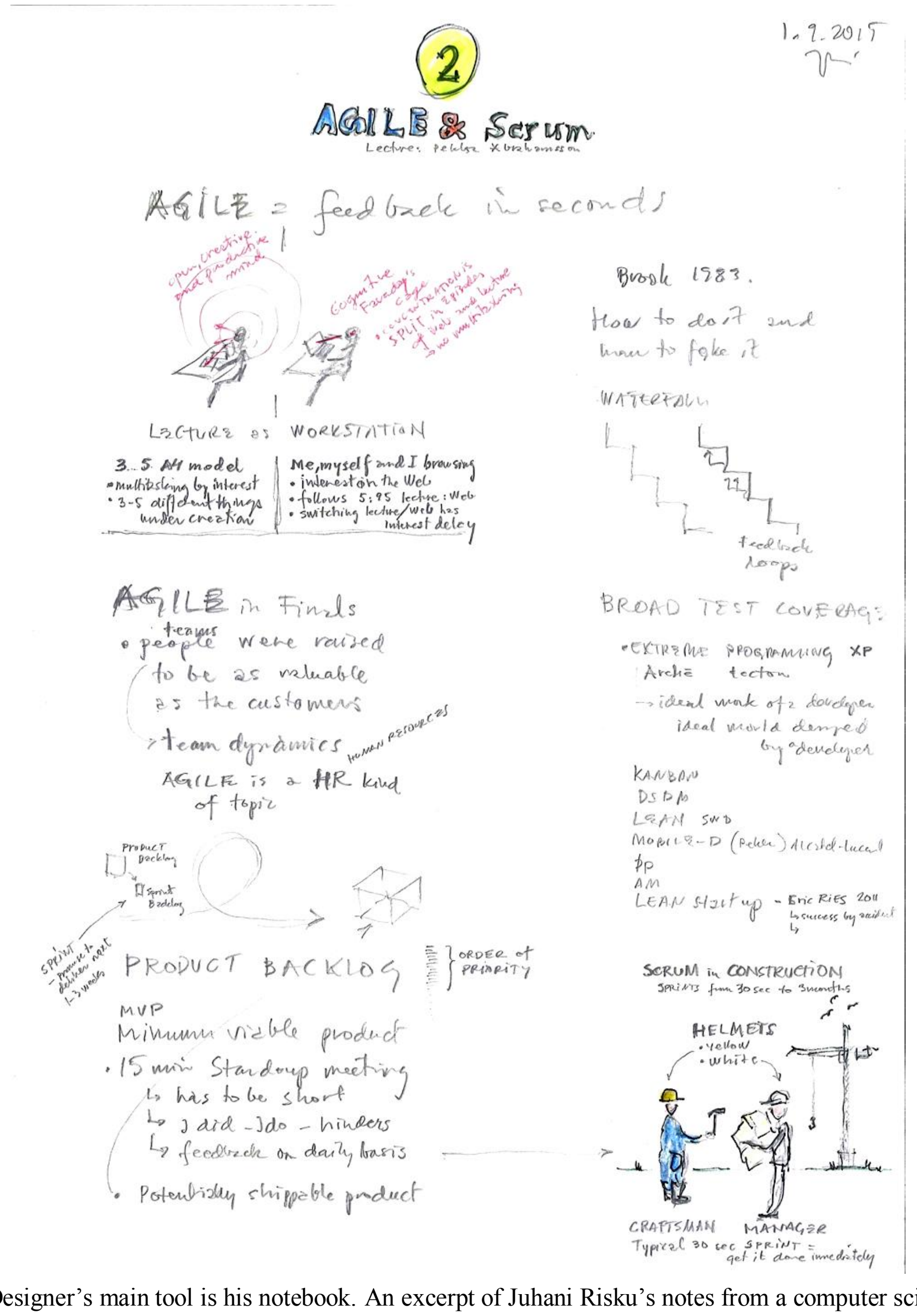

Figure 1. Designer's main tool is his notebook. An excerpt of Juhani Risku's notes from a computer science lecture.

When pulling together the Nokia and university experiences we formulated a hypothesis: It appears that designers have a proper skill-set to start to work with even the blue-sky ideas and that this skill-set is different from those of computer science students. Computer science students may be active, but they don´t take the initiative to start designing early on, which in this context means envisioning, planning or sketching as it is not part of their educational background.

We propose that the present day startup ideas and software solutions in general are based on ideas and solutions which have similar ideation and development processes like the arts or any design. Whereas smartphones as an example are certainly designed by industrial designers having artistic education, do the core elements like mobile applications, services and systems have an origin in the arts and artistic creation methods in their early phase of the development process? Are the art-based design methods applicable to software design? If we can find a working, perhaps a separate and independent design methodology in software development, does it scale to the arts?

The remainder of this paper is organized as below: Section 2 presents the background for artistic design, Section 3 reflects the current design process in modern software development considering from the startup viewpoint and finally Section 4 proposes alternative avenues to improve the current state of practice.

## 2. Considering an artistic design flow

In this section an artistic design flow, including the art creation principles and artistic creation processes are depicted. A comparative reflection on design thinking and designer thinking is also presented.

### 2.1. Art creation principles as basis for architecture and design

The artist, when starting to paint or to sculpt, has either an ideation session or just starts to craft. Crafting rarely starts from a scratch, however. The years of artist's experience have prepared the artist to the task at hand. The ideation session may consist of sketching and composing by drawing or with materials or items. Ideation and sketching before the practical crafting is a separate planning and design phase to structure the later work. Typically complex projects like house and car construction need to carefully plan and design in advance.

When giving a form to a car, a house, or a sculpture, artist needs several form-giving operations like moving, rotating, mirroring, breaking, cutting, drilling, joining, stretching, bending and resizing, as shown in Figure 2. These are operations to mold the structural form of the object, and the operations have their origin in sculpture. When the outward appearance, the surface, walls, priming coats and flat details on a car, house or a sculpture need color, coating or graphic figures—the work follows classical painting methods like sketching the structure of the figure, drawing details, mixing colors and coloring separate pictorial items. The outer form coating and decoration happens with 2D artistic methods of drawing, painting, graphic design and typography.

The creation process of the most noticeable buildings, cars and sculptures of the 20th and 21st century have been following the form-giving methods from classical arts and crafts, as described in Figure 2. When crafting, the artist may take drawing pauses to create or check details of the artefact. Iterative pauses and checking draft forms are similar to testing in software development. In both processes corrective actions and decisions for continuing the work are made by evaluating errors, need for changes and fine-tuning.

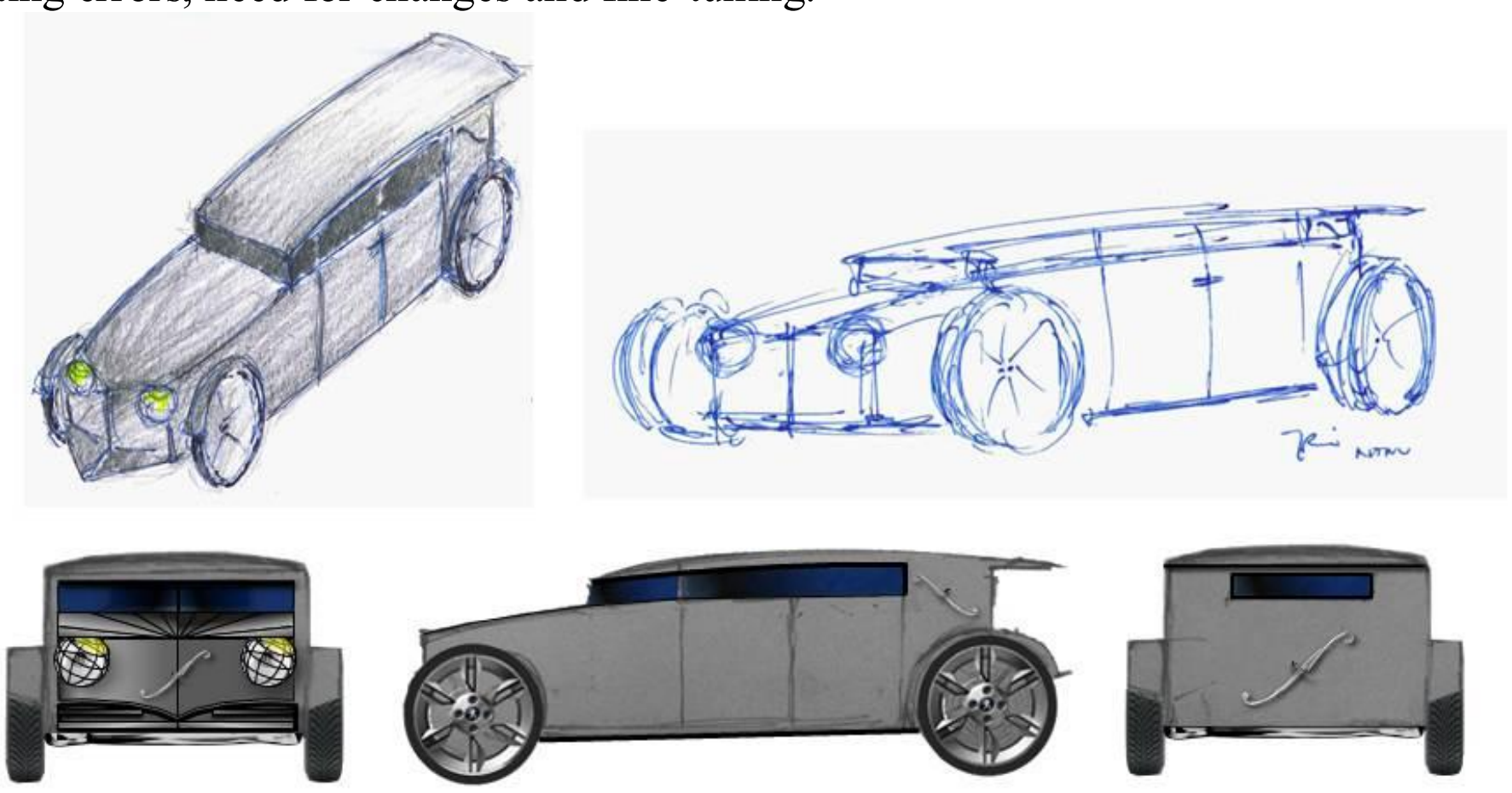

Figure 2. Form-giving with artistic methods: Juhani Risku, RodMobile 2015

Direct crafting is "sketching and freezing form" simultaneously. Direct crafting differs from art category and material to another e.g. giving form in clay allows additions, cut offs and reforming until the clay is dry or fired. Stonecutting and sculpting is all about cut off and giving form by diminishing the megalith.

Form-giving operations like moving, rotating, mirroring, breaking, cutting, drilling, joining, stretching, bending and resizing have been used by engineers and designers working with CAD (Computer Aided Design).

CAD allows, as direct crafting in sculpting, try-outs and experimental form-giving using solid object modification without preceding plans or sketches. When giving form to everyday items and complex systems, such a teapot or a car, drawing and sketching by hand ensures that the 3D modelling with CAD software is easier and more efficient.

When considering the automotive industry, the general appearance of the car is studied and developed and details created (Figure 2.). The car gets its forms during a process where the primary idea is in the form of a written synopsis. The birth of the car is given sketching by hand and crafting by clay, 3D models and prototypes. The car reaches the level of an entity when the industrial production line produces the first street legal versions. When the designers end the creation process of a car model they lose their grip on the car and it becomes an autonomous entity. The car manufacturer together with the design team create a brand philosophy, and the successful car becomes a part of the design history with a specific essence. This applies to other artistic domains as well. Figure 3. shows a result of a time consuming sketching period. During drawing a sketch turns to a stained glass window.

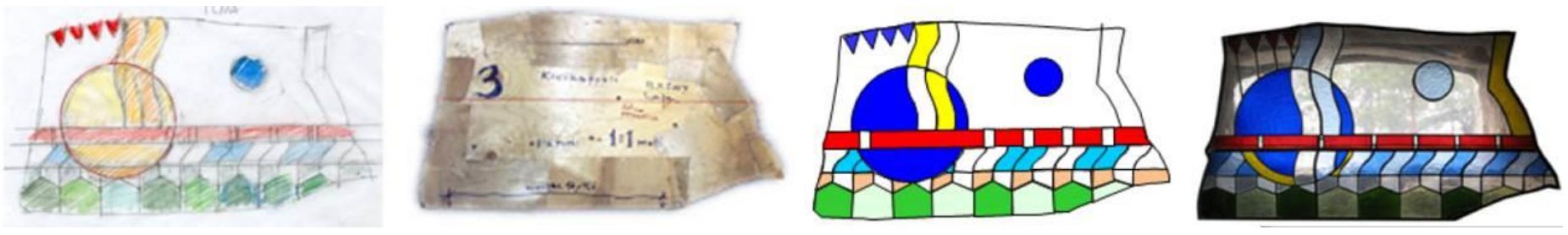

Figure 3. Form-giving from sketch to a stained glass: Juhani Risku 1995 Stone Chapel, Vivamo, Finland

Industrial design applies artistic form-giving principles and practices from the classic handicraft and its modern formats of digitalized 3D design methods. Industrial designers often have a thorough education in art, crafts and design with traditional Bauhaus methods [5]. Bauhaus's main objectives were to unify art, craft, and technology. This is still seen as an ideal for architecture and industrial design education.

### 2.2. Artistic creation process

Artistic creation process can be seen as an evolutionary pattern from an idea into a real world artefact. After initiating, the idea evolves either in a waterfall process or iteratively through several increments. When the artist is working alone (which is close to a standard), all feedback, iteration loops and test sessions are made in seconds inside the designer's brain. Similarly to writers, artists are sometimes faced with a "writer's block" in which they experience a creative slowdown. These are typically caused by external events and for example in Nokia some persons had encountered 12 lay-off warnings and being under subsequent co-operation negotiation processes diminishing their creative capacity to very low.

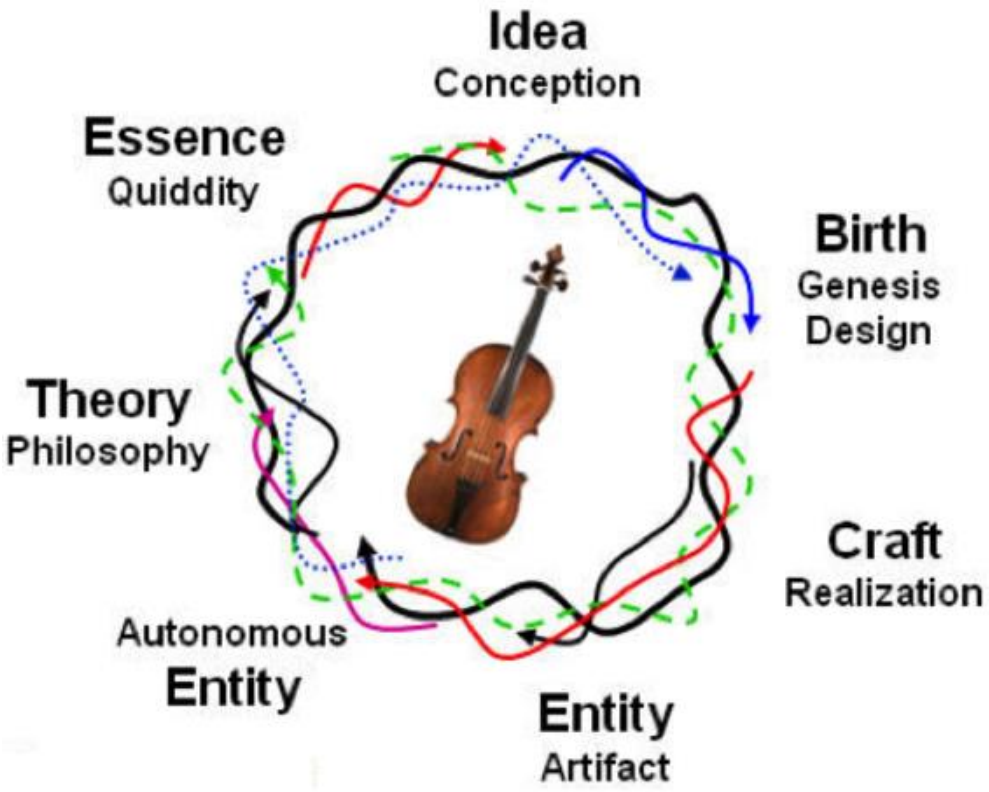

Figure 4. Artistic design circle: The creation process in art and architecture. [6]

Figure 4 depicts a typical iterative creation process from an idea to an independent product takes a full iteration round. The round is simultaneously a knowledge gathering session as well as a waterfall-like development process. The second round iterates the first round, the third the second and so on. This way we got e.g. the best violins by Antonio Stradivari. The violins differ from each other only one to two mm variation and the evolution in quality from the first Stradivarius to the 960$^{th}$ version of it.

Artistic design process is also closely connected to the concept of first-mover and second mover introduced in the introduction section. As proposed, the first-mover has the ability of pioneering firms to earn positive economic profits. The pioneering happens through advantages arisen from technological leadership, pre-emption of assets and buyer switching costs. Also advantages derived from a "learning" or "experience" curve, the success in patent or R&D races ensures first-mover advantages.

The second mover doesn't act in this way. The second mover rather observes, waits and stays on the background. A second mover never starts the process or gives meaning to it. But, as we know that fast followers and late movers can profit from the first-movers′ mistakes. Facebook overtook Myspace and Skype overtook the VoIP service providers. In both cases the second mover had time to conceptualize a more competitive solution, Facebook was a better fit for big audiences and Skype was free and it worked.

Artists, architects and designers act like first-movers when they are ideating and envisioning new things and objects. The artist's mind-set is based on originality, creation or even abiogenesis in the first place. Copying is not an option for a real artist. A second-movers role for an artist is a loser's role, despite you make money with the copies. Here, copies are usually edited and transformed to look different than the original ones, because the shame may be mortal.

An ideal software startup as a first-mover would be presenting an epoch-making application of an everyday action, which is overpriced and arrogant, and all this in a disruptive way. Do we have any examples of epoch-making startup incidents? Yes, Apple′s iPhone was made when the company still was in startup mode, Uber reorganizes the transportation business starting from the most conservative one, the cabs.

When the artistic creation process normally happens in silence and solitude (solitary artistic workflow) all decisions are being made without interaction with the outer world or interfering people. Quick and straight-forward decision-making is an important part of artistic quality and productivity. Quick decisions mean early "right or wrong" solutions. As a possible outcome can be an early sketch, illustration or a draft similar to the one presented in Figure 5, which is the result of a few minutes of ideation regarding lecture recording. This means either a fluent progress or early corrections to get back on track. This serves as a concrete example of how a design-driven startupper moves rapidly, facilitates communication, creates a solid plan from sketches and ideas thus enabling fast validation of the idea with real customers.

The artist has always been the leader of her own creation, design, execution, implementation and manufacturing process, alone. In case of Leonardo, Picasso and Dali, it is a question of detail and lifelong production, which scales into software design in the form of systemic solutions. Design output for Leonardo were ideas and drawn lines and systemic solutions in the form of helicopters, bridges and artwork to the creation of fundamentals of art by surgery.

Regarding the development and skill in artistic processes, seniority and mastery ensure quality. A senior, when being creative, active and diligent after getting the best education and criticism by peers and masters, is free from practical obstacles to produce individualistic art. Creativity is often connected to young artists but more equitable would be to see creativity as a function of novel ideas, full professional skill set, courage and diligence to work. If so, the youngster has years ahead to work hard.

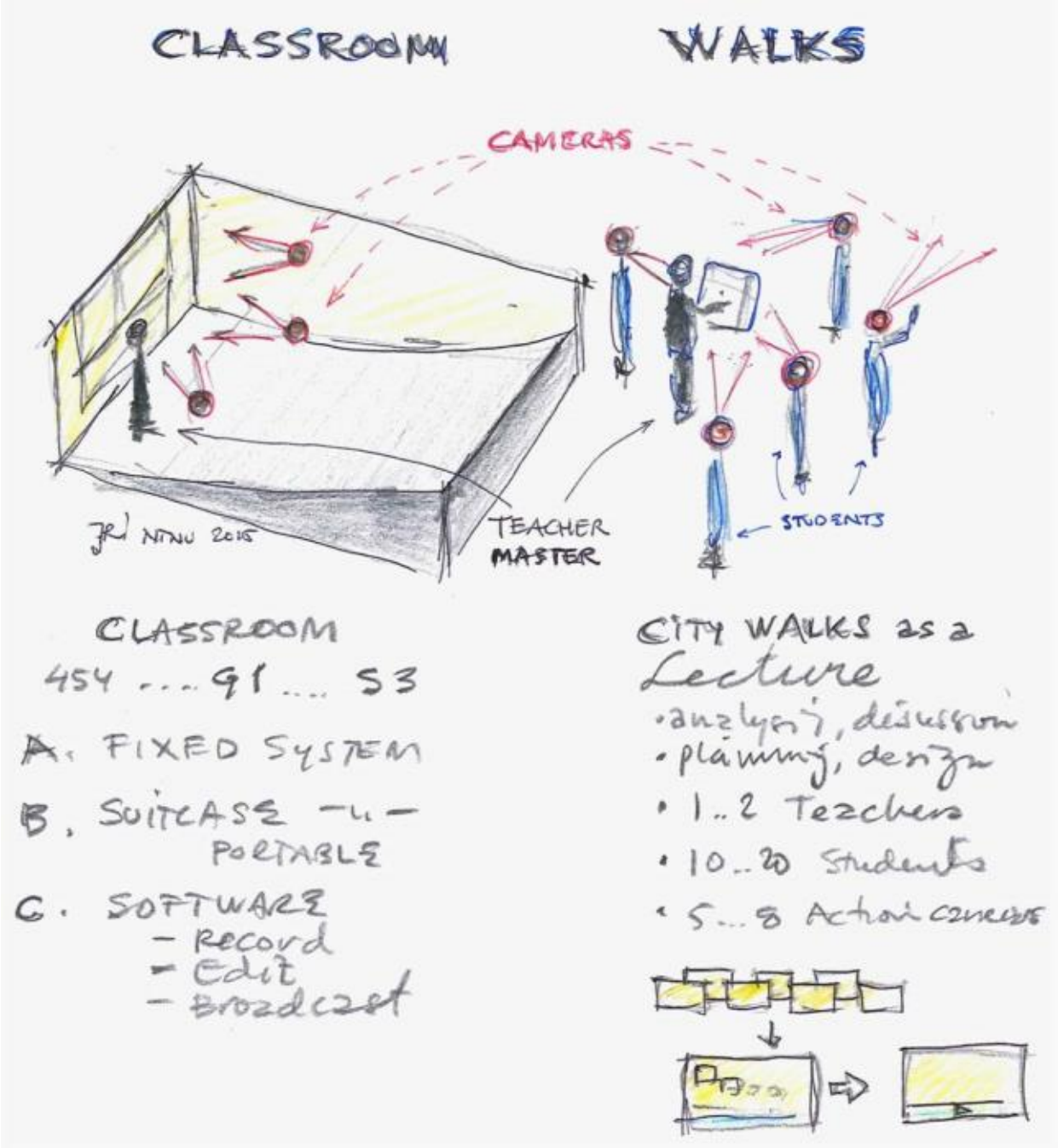

Figure 5. Early sketch of lecture recording scenarios

As a summary, designers mainly think and act according to design culture´s centuries-old ways of working and education. Design is based on artistic creation and craftsmanship. The young person studied as an apprentice under a master´s supervision learning all skills needed for the profession. When the apprentice was better than her master, she could continue her independent career. His target is to develop his design thinking capacity, which will be addressed in the following section.

### 2.3. Design thinking and designer thinking

Design thinking is a description of how designers think when creating. Design thinking refers to design-specific cognitive activities that designers apply during the process of designing [7].

Present design thinking has its origin in the psychology of science [8] and design engineering [9] Peter Rowe, in his book about Design Thinking [10], described methods and approaches used by architects and urban planners. Design thinking has two different goals: to help solving problems and to help creating solutions. In the problem-solving process, seven stages can be used: define, research, ideate, prototype, choose, implement, and learn. Creative solution-based thinking starts with a goal and strives to get alternative solutions to be developed. Solution-based thinking is problem-free and is commonly used by artists and architects.

Design thinking has become popular outside designers' own discipline. We may see it also as a movement (with positive goals, however) to smuggle design principles and habits in weeks as a superficial skill-set to persons outside the design culture. More importantly, we maintain that there is a danger when importing design thinking in a lightweight manner without professional designer's education or traditional apprenticeship work, the thinking is not the same as the designers practise.

An important question is that how far does design thinking lead without design skills? Does the thinking just remain as thoughts without further development? If not, who realises the results of the thinking? One answer is: A "designer thinking" is always a "designer crafting", but an "outsider thinking design thoughts" usually doesn´t craft the ideas further. In the outsider´s case design thinking is in a danger of remaining as a superficial and inefficient layer of knowledge and skills.

Seppänen et al. claim that startuppers competence needs are volatile and may change as the startup evolves [3]. We maintain that in the early phase of the startup, the competence needs are much more stable that we may think. The design skillset broadly considered represents those expertise areas required when acquiring funding for the startup. Design thinking is rooted in the arts and crafts traditions, and as such, offers a fruitful avenue to consider when planning for future computer science education curriculums.

The value of design thinking is at its best when proper professional design work is executed by the new-born thinkers. This is possible for the young startuppers because they haven´t forgotten art classes at the elementary school and senior high school. If the startuppers have a continuum in art and design education at university level they become more than design thinkers, they will become thinking designers.

## 3. Design in software development and startups today: Communicability should be the key

Modern software development has adopted several design principles from the traditional arts and design field. Usability, User Experience and User Centric Design have been in the core of industrial product and services design for decades. Graphic design of Web pages, icons and visuals is a centric component in creating attractive Web presence. These design dimensions are also crucial for startuppers when considering the quality of achieved applications, services and software.

Flow charts, paper prototypes, simulations and scale models are commonly used. Paper prototyping can be seen as a modern and simplistic way of developing even complex systems. This practice should not however be confused with ideation sessions with sticky notes or using them to monitor the progress of the project on a team room's wall for example. Paper prototyping (or prototyping on paper) should primarily be seen as a designer's planning tool for the product or service, not the project itself.

In paper prototyping non-artist people practice with artistic methods without necessarily knowing it. Generally there is no training to use these methods and developers learn from the experience. The underlying question here is naturally to what extent their capacity would improve if they would have better artistic and design skills.

The artists and designers, when creating their solitary artefacts, can use more profound tools for organizing complexity than those with paper prototypes. The "paper prototyping alike" methods of artists are structural sketching, visual versions of a product in drawings and scale models, storyboards, collages and iterative improvisation. Sketching allows the fail quickly method for the artists and designers to imagine the prototype, test immediately its structural entities and fine-tune, correct or restructure it in seconds.

Artistic methods are quicker, more informative and already close to 80 % of a functioning prototype, not just stickers on the wall. Those methods are also easier to communicate so that all parties can understand the goal better. Paper prototype needs always several design iteration rounds, several professionals and time to be taken seriously and being useful for the development process.

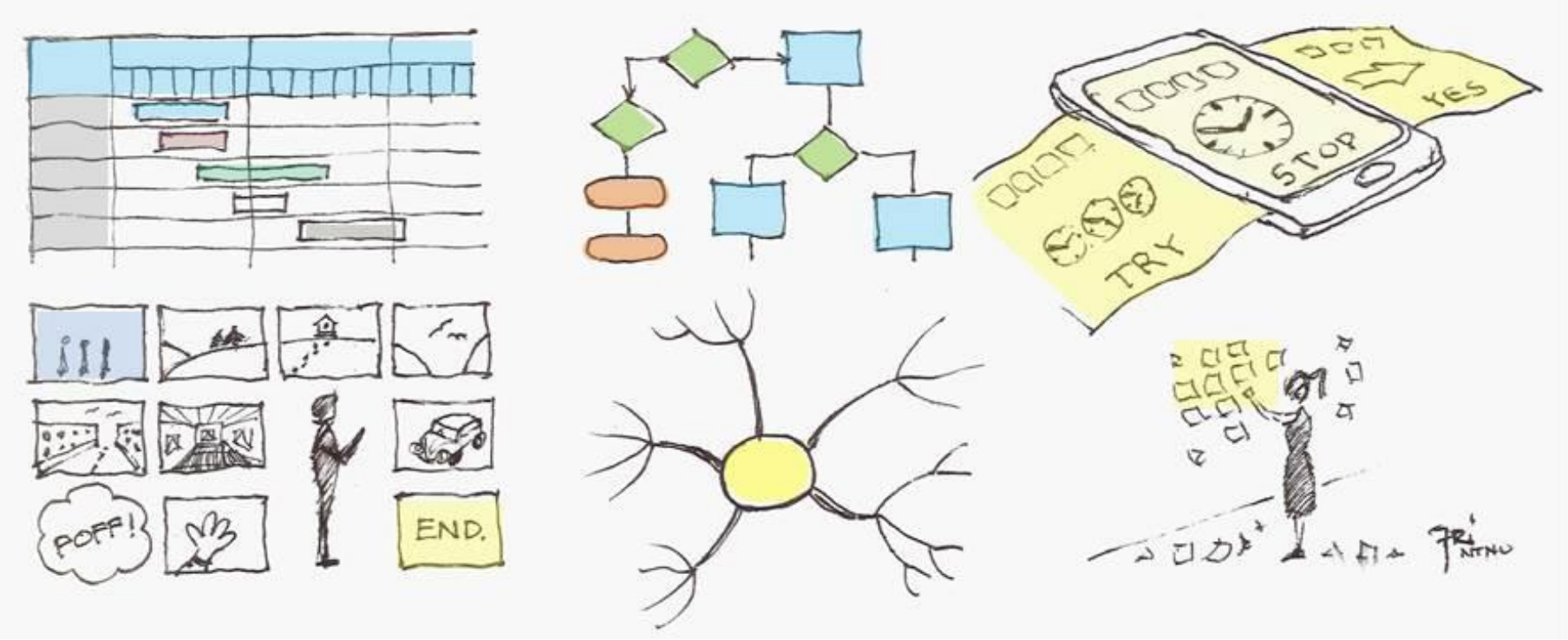

Figure 6. Planning and design tools employed by software developers.

Figure 6 presents planning and design tools employed by software developers including the Gantt-chart, flow-chart, paper prototyping, storyboard, mind-map and sticky note ideation. All tools can be useful for a specific purpose but they should employed with care. For example, sticky notes with relations, text and hierarchy form a complex system, which may be difficult to communicate. A drawing with real world look and feel is a prototype on a higher design abstraction and easier to communicate. We maintain that design skills are needed to ensure the efficient use of the selected tools.

When a startup team presents their proposal for a solution to the prospective funders, they face a test of efficiency in communication: Is a well-engineered structure good enough to get a positive funding decision or would a visible and touchable mock-up be better? We believe that the latter is more viable, tolerates more error, is producible faster and finally is more open for continuing the product development.

Major differences between the engineering thinking and the design thinking are: the engineering design is more difficult to communicate in two minutes, but for example a real world clay model of a smartphone engages the executives for several hours. The charts, diagrams and graphs of the same smartphone are too generic and dull. The model is touchable and highly crafted like a jewel. This is the difference between software and design thinking, design is sensual and charts are engineered. We believe that design thinking is actually closer to the startuppers' workflow than engineering thinking. Design thinking is, however, currently only superficially included in universities' curricula beyond designers' education.

## 4. Four alternative avenues for design and software

We propose that software development and startups have four different directions to go when dealing with the design opportunity:

1. (Radical) Embed designers' ideal as an integral part of software creation and development
2. (Conventional) Keep design as a separate area of the development process and use it when needed
3. (Arrogant) See design as an entity which already has an application in software development and merely needs improvement
4. (Ignorance) Ignore design as a systematic power player in software development.

The four future scenarios of software with or without design culture in depth are outlined below.

### 4.1. Embed designers' ideal as an integral part of software creation and development

Embedding designer's ideal as an integral part of the process means that computer science (CS) and software development (SWD) take part in the creation and evolution of design and apply all design principles that have their origins in design, as shown in Figure 7. However, it is proposed that software development would rather approach design discipline rather than vice-versa. This approach causes several needs and actions:

a) Need for fundamental program of research and development of design principles in CS and SWD. Here CS and SWD can be seen as prior inventors of design principles which can´t be found on traditional art and design areas because of their "non-software thinking" abilities.
b) Need for fundamental design education according to traditional design methods and newfound design methods from CS and SWD research
c) Apply the newfound CS and SWD design principles to traditional art and design areas to support their abilities to improve design quality and to ease CS/SWD and design professionals´ collaboration with lever and pulley effects.
d) Upgrade the researchers' and teaching staff´s skills in design to provide relevant and newfound design education
e) Need to gain street credibility in their overall SW development and design approach by delivering software applications and solutions related to research findings. CS and SW professors may have ideas, skills and brainpower to outperform and beat present operating systems and platforms like iOS and Android, and games like Angry Birds.

The first two points can be realised in weeks or for next season´s curriculum, especially at universities with design education programs (e.g. NTNU, Aalto, and Stanford). Point c is a game changer and future success story of CS´s and SWD´s impact in general design principle evolution. It is also a probable area of competition between universities and criteria for ranking their quality. The newfound CS and SWD design principles as an area of research and development is totally new and its value for diverse businesses is of millions and billions of euros. No university has leadership in this design approach or strategies to tackle the approach.

Point d is a simultaneous program to Point a and Point b. Point e is the hard one and gets possibly the most resistance. Practical output of research results in the form of applications and solutions can be seen as irrelevant effort of a free researcher´s role in science. Practical solutions may also be too demanding horizontal strives for highly vertical scientific targets. Or maybe the scientists merely get measured by "wrong" outputs. Alternatively, maybe the comfort zone reached is a direct hindrance for CS and SWD progress in general and an indirect barrier against advancement in university rankings.

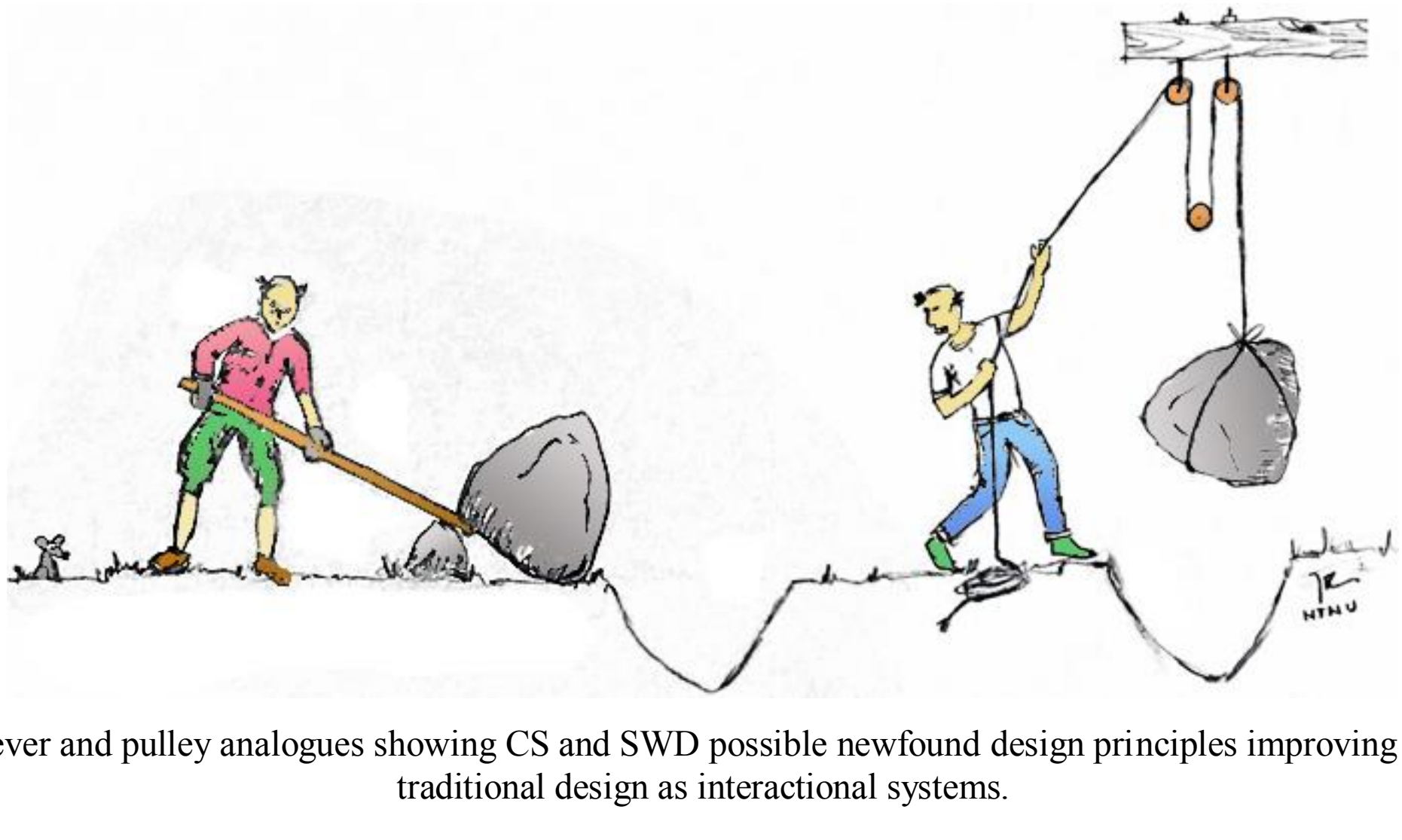

Figure 7. Lever and pulley analogues showing CS and SWD possible newfound design principles improving SW design and traditional design as interactional systems.

### 4.2. Keep design as a separate area of development process

We can remain conventional and continue treating design as a tool, and use it when needed. Indeed, this is the present day mind-set and practice at universities, software intensive companies and startuppers alike. Design in the form of graphics, visuals, usability, User Experience, and hardware like devices, accessories and mechanics is ordered separately from small-scale design consultant companies. Design is usually outsourced into part time and split sessions. Design services are also under high cost-reduction pressure. Those make design as a buyable entity for the host company and design becomes a subjugated and uncertain party. We however argue that this is in conflict with the idea that design could be a key competitive factor and a change maker for the host company. In other words, as subcontractors, design companies can´t give their best to their paying customer.

Continuing with design as a separate and subcontracted area for CS and SWD causes:

f) Inefficiency in getting best design solutions and details to CS and SWD.
g) Larger gap between CS/SWD and design collaborative interaction in agile and lean approaches, in human efforts and commitment, and holistic development and progress in quality.

Threats (or opportunities), if design would be embedded into CS and SWD:

h) CS and SWD as an area of discipline would grow stronger. One will not take over the other one but in either case, people would lose their "*governance by subjugation*" mentality, which is not necessarily a bad thing.
i) Trained designers could replace several "non-designers performing design duties" who have gained their positions as granted in a CS/SWD society or company. The latter type designers would be exposed but they would also learn novel ways of working.
j) On the background (when continuing to see design as an outsource opportunity) the designers´ justified and positive impact would grow outside the CS/SWD community.
k) Design would be a success factor for CS/SWD research and development and every CS/SWD student would get professional skills of a newfound CS/SWD designer.

### 4.3. Considering design as an entity, which already has an application in software development.

A basic set of design and design principles already are in CS and SWD. Design in its classical principles is adopted to Usability, User Experience Design (UXD) and User Centric Design[2] (UCD, HCI, cognition psychology). HCI and cognition psychology have separate CS/SWD formulations and conventions like Man-Machine-language MML, use cases and task flows as design. Certainly, these software related design areas are researched and developed in scientific circumstances, and taught for students. But the question is that are they researched and developed in a closed and hermetic vertical environment without influence from outside? Examples of outer influence are: reading design literature, practicing design form-giving techniques, taking design classes and having design people working, teaching and researching at CS/SWD institutions.

### 4.4. Ignoring design as a systematic power player

Ignoring design as a systematic power player in software development means that present day CS/SWD methods would be seen as self-sufficient and a hermetic attitude without external input. Interaction would be seen unnecessary or even hostile. This would cause degeneration and a slow withering would take over the CS/SWD community. Points presented in Section 4.3 and Section 4.4 are relatives with only a slight difference: In Section 4.3 the CS/SWD community is self-contained and self-contained with their superiority. Both in Section 4.3 and Section 4.4 it is a question of gained comfort zone and arrogance to not to renew and have constant progress.

## 5. Conclusions and concrete steps to take

When considering the possible futures and trends when combining software development and design principles we observe that there is an immense opportunity to improve the current-state of praxis and the existing design habits in software development.

Embedding design principles into software development is a future strength especially for the first-movers. A fundamental new strategy, however, is needed here. The strategy describes goals and provides a roadmap, organising patterns, participating organisations and people.

We propose that software development, startuppers and design can join their forces in combining design thinking and practices to software thinking and practices. To be successful this research and development approach requires the following:

[2] User Centric Design (UCD), Human Computer Interaction (HCI or CHI), and cognition psychology have a scientific approach to understand the human being and her needs, expectations and actions in contexts of systems, machines and daily Use Cases and task flows. UCD and HCI are typically developed by designers at e.g. mobile device and solution companies (Apple, Google, ex-Nokia).

a) Start: Get an overview of present day students´ awareness and skills to plan and design necessary entities in software development; envision the possible futures of new design skill-set need and form, set short and long term targets for the envisioned outcome, create a strategy to realise the development project. Find an owner for the project.
b) Research: Recruit eligible faculties, departments and people from the university to participate in structuring the project. Ownership at computer science departments sounds natural.
c) Development: Beside the research project establish a development project to pilot, experiment and test the research results and give further input for the researchers.
d) Piloting: Deliver the research and development results to participating faculties, departments and people to support their own education and training sessions.
e) Education: Start simultaneous classroom training for acquisition of information from all participating faculties and departments.
f) Joint projects: Find appropriate peer and partner courses to be in collaboration during the research, development and education phases.

We could call this novel type of software education by the Design Software Thinking and Praxis Program (DSTPP).

From the four alternative avenues for design and software the ideal choice would be the first, radical, way. When embedding designers' ideal as an integral part of software creation and development software startuppers would act as more independent creators and leaders during their products´ lifecycle. For a startupper a designer´s skill-set would give advantage in speed in product development, launch and go to market. The designer´s skill-set helps communication to the investor and customer, because the product is well structured and clearly organized with accurate design principles. And the most important outcome for a design-skilled startupper is that she gets a whole portfolio of sketches, organised ideas in structured format, interconnected product concepts to be scaled and further developed. As a by-product of design-thinking with design skills the startupper, very early, gets a holistic and forward-looking mindset, which helps her to succeed in the business.

The designer-developer-startupper, as the mindset and skill-set shows, is a systemic combination of wide and highly professional capabilities. Design, like planning, refers to broader contexts, more flexible and creative attitude [11, 12]

The nature of systemic design process to be an integrated, holistic, multidirectional approach to the design (of instruction). In a systemic approach the designer is frequently concious of the correlation of the total (instructional) system, and all its details [13]. Therefore design should be applied in its systemic form to startuppers´ skill-sets. This systemic and multidirectional design approach influences software development research and education in a very positive way.

## Acknowledgements

We want to thank the Software Startup Research Network and especially Dr. Anh Nguyen-Duc for his inspirational ideas on this work.

II

# SOFTWARE STARTUPPERS TOOK THE MEDIA'S PAYCHECK: MEDIA'S FIGHTBACK HAPPENS THROUGH STARTUP CULTURE AND ABSTRACTION SHIFTS

by

Juhani Risku & Outi Alapekkala, 2016

## Bibliographic information:

Outi **Alapekkala**, Juhani **Risku**

***Software startuppers took the media's paycheck. Media's fightback happens through startup culture and abstraction shifts***, ICE Conference, Trondheim Norway 2016.

This is an author-generated preprint version.

The final version is available at:

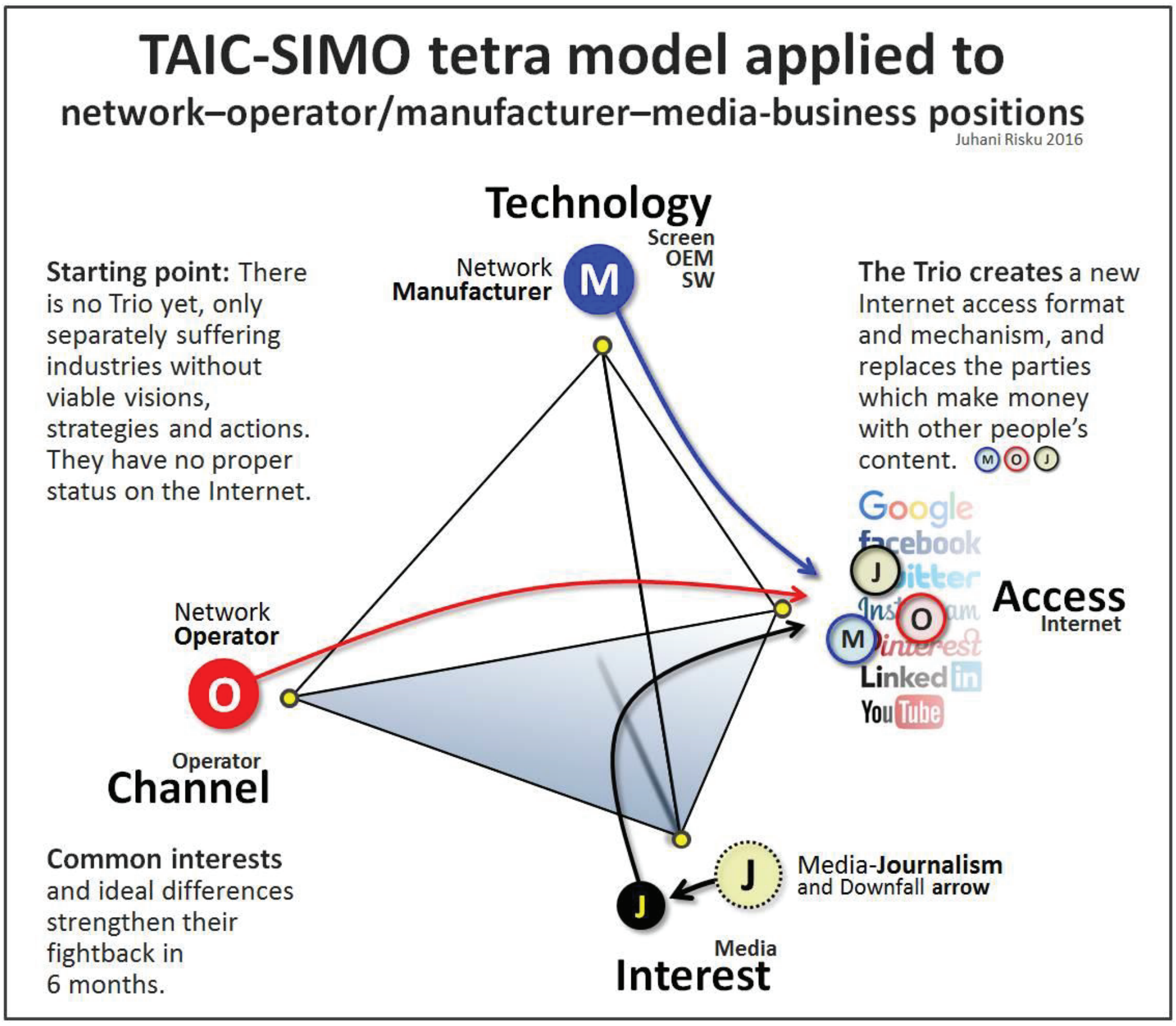

# Software startuppers took the media's paycheck

## Media's fightback happens through startup culture and abstraction shifts

Juhani Risku, Norwegian University of Science and Technology NTNU
Department of Computer and Information Science
Information Systems and Software Engineering (ISSE)
Trondheim, Norway
juhani.risku@idi.ntnu.no

Outi Alapekkala, In Action – Societal Innovation startup
Systemic functions
Media and Journalism
Tornio, Finland
outi_alapekkala@yahoo.fr

***Abstract*—The collapse of old print media and journalism happened when the Internet, its solutions, services and communities became mature and mobile devices reached the market. The reader abandoned printed dailies for free and mobile access to information. The business of core industries of the early Internet and mobile communication, the mobile network manufacturers and operators are also in stagnation and decline. Therefore these industries may have similar interests to improve or even restructure their own businesses as well as to establish totally new business models by going into media and journalism.**

**This paper analyses, first, the production flows and business models of the old and present media species. Second, it analyses the current market positioning of the network manufacturers and operators. Third, the paper suggests two avenues for media and journalism and the network manufacturers and operators, the Trio, to join their forces to update journalism and make all three stagnating industries great again. Last, we propose further research, development and discussion on the topic and envision possible futures for journalism, if the three would engage in cooperation. We see that the discussion should consist of ethical, societal and philosophical subjects because the development of the Internet solutions are based on "technology first" actions.**

**We find and outline a tremendous opportunity to create a new industry with new actors through combining the interests of the network manufacturers, network operators and journalism in a systemic solution through a strategic alliance and collaboration Fig. 1. Software startuppers with their applications and communities will be the drivers for this abstraction shift in media and journalism.**

**Our experiences in the media, journalism, mobile network, mobile phone manufacturing and startups provide the basis for our formulations on the future of those industries.**

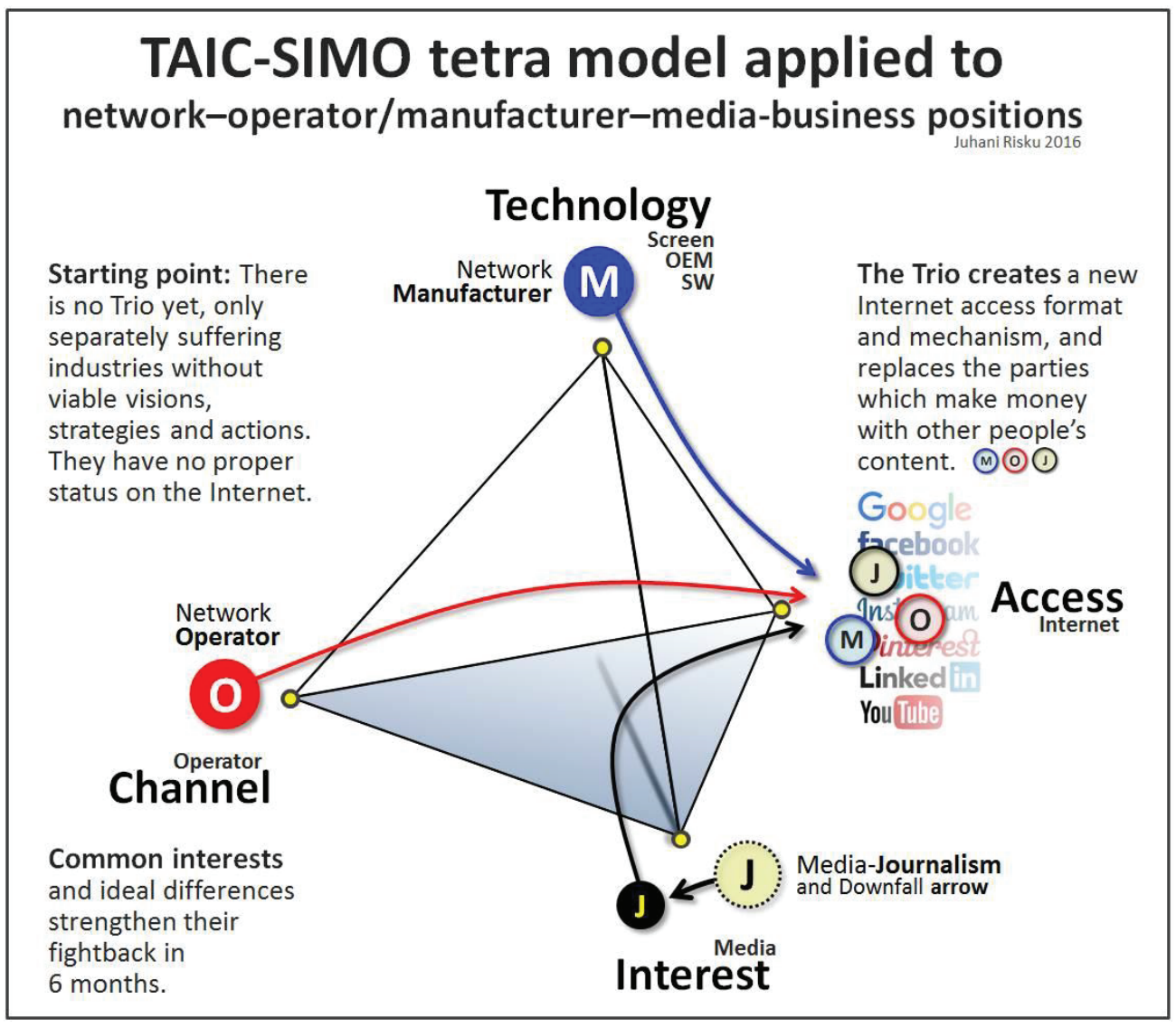

Fig. 1. TAIC-SIMO tetra model applied to network–operator/manufacturer–media-business' (the Trio's) positions. In the Trio's fightback they join their forces and build a new format and access to the Internet, and they override today's advertisement money hijackers. This requires abstraction shifts, startup culture, new leadership and rapid actions from the Trio. The tetra shows direct connections between edges (e.g. O–M collaboration), the tetra face (triangle) show a combined business area (e.g. M–O–J). The missing part of the tetra can be established on a new abstraction level (e.g. M–O–J Internet Access). This MOJ Access is a spin-off of a network–operator/manufacturer–media house consortium. The idea to establish an own Internet Access spin-off instead of acquiring an existing one is about creating a new actor in this business environment. This actor can disrupt the present business models and offer the users and consumers a combination of more interesting and fair services.

## I. Introduction

Media's and journalism's business models are today from the era of "owning the paper printing machine" and "owning the broadcast studios and channels". Owning the machine and channel was possible only for the tycoons and business moguls and they took the power through the printed newspaper and displays paid for by single copies, subscriptions and advertisements. The power came from owning the technology (expensive printing machine) and owning the channel. R.G.

Picard emphasizes that journalism's printed newspaper is extremely capital intensive business because of its high capital requirements, high fixed production, distribution, marketing and first-copy costs. [1]. Media is an umbrella term to combine business sectors and activities like broadcasting, TV, radio, Web publishing, print media and journalism into a one trade and business.

Picard underlines how journalism differs from media. Journalism is a cross-cutting discipline and a set of working methods and practices enabling content production for all publishing means from print, TV and radio to social media. He defines journalism as an activity with practices to gather, convey and process information and knowledge and insists on how journalism's functions stay prominent for society. Therefore, the practice of journalism is not a media, a distribution platform, or a business model. [2].

Having said this, journalism is in our focus to restructure the media sector business because its practices and ways of working are at the core of all media and publishing. While the media and ways to publish have evolved, no significant innovation in journalism has happened and merely "writing better articles" is clearly no longer an option. Through rethinking journalistic activity and its practices, new business can be developed that will allow media companies to overcome stagnation and to regain the customer. [3].

## II. Reasons for the lost paycheck

### *A. Startup culture replace old media's working patterns*

Software startuppers killed the traditional print journalism as they took both its readers and its income. Their algorithms have attracted the reader to access free-of-charge information and content, and advertisers have followed the customer. When software startuppers like Sergey Brin and Larry Page founded Google, it replaced older search engines like Alta Vista, Yahoo and Lycos, and made it difficult for newer engines like Ask Jeeves and Bing to gain market share. [4].

There are common factors of the media and journalism downfall: lack of innovation, comfort zone laziness, business focus failures, illusions of self-correcting actions like A. Taylor says, writing more and better articles as solution to journalism's failures [5], concentrating more on investigative journalism, improving their Web presence, publishing quality photos, visualisation. This is illusionary because journalists have certainly done good work all the time. New ideas appear include selling articles online by micro payments, and crowdsourcing as a collaborative action with the audience. A recent report commissioned to look for innovative media outlets and innovation in journalism startups found no single groundbreaking innovation. While many actions are considered as innovations in journalism, they are more of gradual improvements and tackling platforms, business models and processes, not journalistic content itself. Like Schiffrin (et.al) say, they did not find any revolutionary innovations by journalism professionals. [6].

Software startuppers' algorithms allow the theft and gathering of media outlet's public content, as well as lean and agile content production for bloggers and other citizen journalists: their articles, posts, videos and other content is published as it is ready and edited if needed. The content costs nothing to produce and nothing to publish, and it can be interactive. They write content iteratively so that articles gain readers, like Blank and Patenaude–Gaudet say. [7], [8]. In this sense, software startuppers have allowed just about anybody to become a one-man media house or a citizen journalist without having to follow some binding editorial guidelines. Membership of official professional bodies, the related fees, administrative red tape and other gate keeping activities have also become redundant.

Old media houses and journalists seek to be lean and agile through following the trend of providing part of their content for free and allowing some limited interactivity subject to moderation on their websites. They also seem to seek to attract free workforce and content through either hosting a blog platform or providing famous people from politics and business their own regular blog space.

Meanwhile software startuppers have developed their own Web presence solutions like search engines, social media chat services and short message services because they had to find business models to monetarise their companies. [9]. They harmfully put advertisement banners and later sensitive and contextual advertisement solutions on their Web pages and services.

The advertisers began to move from printed newspapers to Web services, because their visibility was nationwide or global, fees were priced by actual views of the advertisements, and the (automated) pay-per-click model (PPC) is cost efficient because of flat-rate agreements or bid-based systems. Contextual advertising programs with algorithms like Google AdWords and AdSense, and Microsoft AdCenter changed the Web advertising revolutionary from year 2006 onwards, according to Shatnawi [10]. At the same time investments in printing press, ink, paper and labour sunk. The decline of classified advertising in newspapers caused advertising revenue losses because of specialized digital online job recruitment, dating and real estate web services starting from Craigslist 1995, Leurdijk (et.al). [11].

The newspaper size format modification from broadsheet to tabloid is an indicator of print media's change: when paper consumption is halved, the number of articles has diminished by one-third. A change in the average size of articles has also occurred: there are fewer small and mid-size articles, but more large articles, like Andersson says [12]. Readers also move to superficial and sensational articles. Journalism, both in printed and investigative form is too slow in the middle of 24/7 Web publishing. Social media and new forms of instant messaging produce enormous amount of information, news and just-on-time text, while shared and combined editing offices partly cause generic offering so that the readers bump into exactly the same material on several newspapers and their Web sites. Free content becomes normality in journalism, and the original quality standards of journalism become outdated.

The rapid smartphone development started from Apple's iPhone year 2007 to accelerate the mobile revolution in mobile content creation, usage and mobile presence on news pages, knowledge search and social media. In 30 months Apple sold 42 million iPhones. As success factors for iPhone Laugesen (et.al.) mention market size, share and growth, average revenue per user (ARPU), usage of mobile data, content and services offered on AppStore, and consumer satisfaction, which in the first beginning was high. [13]. Advertisements became a part of Web and mobile content so that its share of ad turnover grew accordingly with the fall of ad turnover of the old media. It began to be difficult to make money with news and journalism, because less people were buying. [14].

*B. Startup culture redefines quality?*

As bloggers' and other citizen journalists' articles, posts and other content attract ever increasing numbers of readers and followers, old media seeks to, somewhat, undermine the quality and integrity of that content. It suggests that citizen journalism is not of as high quality, and thus not as credible as, the traditional media, which has an established profession, ethical guidelines and other general rules and norms for presenting things. For the old media quality is also a brand issue, it is a reputation built over the years on the assumption of being a trustworthy source of information.

Meanwhile, software startuppers, their algorithms, new communication platforms and applications allow the broadcasting of many more additional viewpoints, insights from professionals, experts and other stakeholders as well as for the expression of opinions that might otherwise be censored. Software startuppers have thus given platforms and potential visibility to far more views and opinions than the traditional media could have ever given. Far too strange opinions to the commonly accepted as well as stand-alone comments and insights have always been filtered out by the old media in the name of speedy production of news on all possible topics by a classic daily. A report by Johanna Vehkoo on quality journalism notes that editors and journalists basically have their own quality criteria and most publications have their own ethical code [15].

Old media no longer has the resources to do quality. Risto Uimonen suggests that editorial work has got a somewhat automated feel as the Internet values quantity and speed over quality, depth and analysis [16]. While facts and quotes may well be checked and certain ethical and journalistic standards respected, there is no time or space to voice all views. Therefore old media satisfies itself to repeat a standard explanation of events and often focuses on communicating political differences on the topics.

Citizen journalism allowed for by software startuppers upgrades the notion of freedom of speech and potential outreach of even singular opinions by allowing free publication and dissemination of one's content.

*C. Startup culture to claim the role of the Fourth Estate?*

Are press, media and journalists the Fourth Estate – the fourth power next to legislative, executive and judiciary powers, keeping a watchful eye on the three others running democratically?

Press and media affect decision-making and general opinion through deciding who and what gets visibility. In allocating this visibility, the old media sticks to old habits: reporting on and giving visibility to the views of established societal actors: governments, ministries, institutions, political parties and their people. New entrants seeking to get their voice heard are often simply dismissed or, worse, ridiculed, by the very same press and media that claim to value variety of opinions and views.

Old media is thus not the watchdog of the system as it claims, but rather the clue and visible network keeping it together and sealing the system from outsiders and new entrants.

The Internet, blogs and social media allow new entrants on the political and societal markets to voice and spread new ideas.

If the current three powers need a watchdog, the role could well be assumed by a large and democratic social media network of citizen journalists. Social media's business model is based on free speech and on the passion of professionals and experts to contribute and participate into a meaningful debate, whereas the old media cannot necessarily say and do everything because it has to keep advertisers and shareholders on board. Profoundly, journalists and media houses have different core interest: media house wants to make money for its shareholders, whereas a journalist is driven by the ideology of having a role of a societal watchdog. In the midst of these conflicting interests the weaker participant, the journalist, has to give up. This can be seen in mergers and close-downs of newspapers and magazines as well as in huge lay-offs of journalists, by Persefoni. [17]. This means that the journalists' professional ethics is diluted in the watchdog-Fourth-Estate and it is replaced by business interests of digital content services and other more popular channels for expression like chats on social media.

*D. Lean and agile reveal old media's imperfection*

Old media houses and their journalists want readers to pay for the content, events and decisions they decide to highlight or for advertisers to pay for their choices in exchange of visibility. But the reader no longer sees the point in paying for those highlights as she gets the equivalent information for free from Internet sources. Readers leave old media and advertisers follow the reader giving their money to software startuppers who have made available the multiple free platforms and services.

Software startuppers have revealed the old print media's and journalism's inadequacy, insufficiency and imperfection: old media's customer disappeared as soon as alternative sources of information become easily accessible.

In particular *four inadequacies* rise.

*One size fits all* is no longer an option. A reader does not want to buy a full paper that has information she's not interested in. She prefers going online and reading, eventually also paying for, only what she is interested in. New online platforms provide for an opportunity to establish communities defined by a common interest or topic. These social media communities and those who run them can also make money with their specialised platforms because advertisers find an extremely well targeted audience for their products and services on them. Individual bloggers who demonstrate a good number of followers are also subject to attract advertising and sponsoring money. These types of business models are of interesting value for advertisers' money and give readers direct access to topics of their interest.

*Too general and too neutral articles* of the old media are less attractive for the reader than some colourful and opinionated citizen journalist articles and a long list of readers' comments following it. Established journalists are mere observers reporting on what has been said, decided and done. They quote others sticking to what's being communicated in official press releases and by the various spokespersons and reprint information from press agencies without putting it into national and local context. Citizen journalists, bloggers and those creating content on social media platforms are themselves actors in society, doers and professionals with true insight on their topic. They are willing to contribute to societal discussion with their insights and sometimes even extremely opinionated views and other strong statements, that do not need to respect the political correctness traditional media does, in order to keep its advertisers, shareholders and access to official briefings.

*One-way communication* of the old print media is also no longer an option. The various online platforms and social media have the capacity to engage and sustain debate the old media has no capacity, resources or willingness to do. Engaging in a debate with your readers is simply not the old way of doing journalism. Readers' comments, additional information, views, opinions and corrections on content force old media to see that they don't know everything and may even have misunderstood something. Meanwhile, this agile, lean, humble and realistic attitude is fundamental for passionate citizen journalists.

*All words no deeds*. Old media is mere communication of observed events and quotes from actors. It presents the news, events and decisions as inevitable facts. "This is how it is and there's nothing you can do about it, but stay informed." Meanwhile, if it claims the role of a watchdog of other powers, it should be more engaging and suggest alternative or corrective paths and actions when it sees an injustice or an error in the system and become an actor that engages the reader. Online platforms and social media networks have a tremendous potential to initiate concrete, in particular collective, action and engage their readers in it through quick social media networking.

## III. Network Manufacturers and Operators are Strategic Relatives to the Media

Network manufacturers are few and their business is globally a mixture of expected strong growth and zero sum game. The sales of 2G and 3G networks markets will diminish steeply [18]. Growth comes from e.g. India and through 5G technologies after the year 2020. There are uncertainty factors from operators' investments in recent spectrum auctions and hard competition between network operators, says Kahn. [19]. In order to maintain their positions in the markets, mergers and acquisitions are a necessity.

The network operator industry is also under constant change Fig 2.

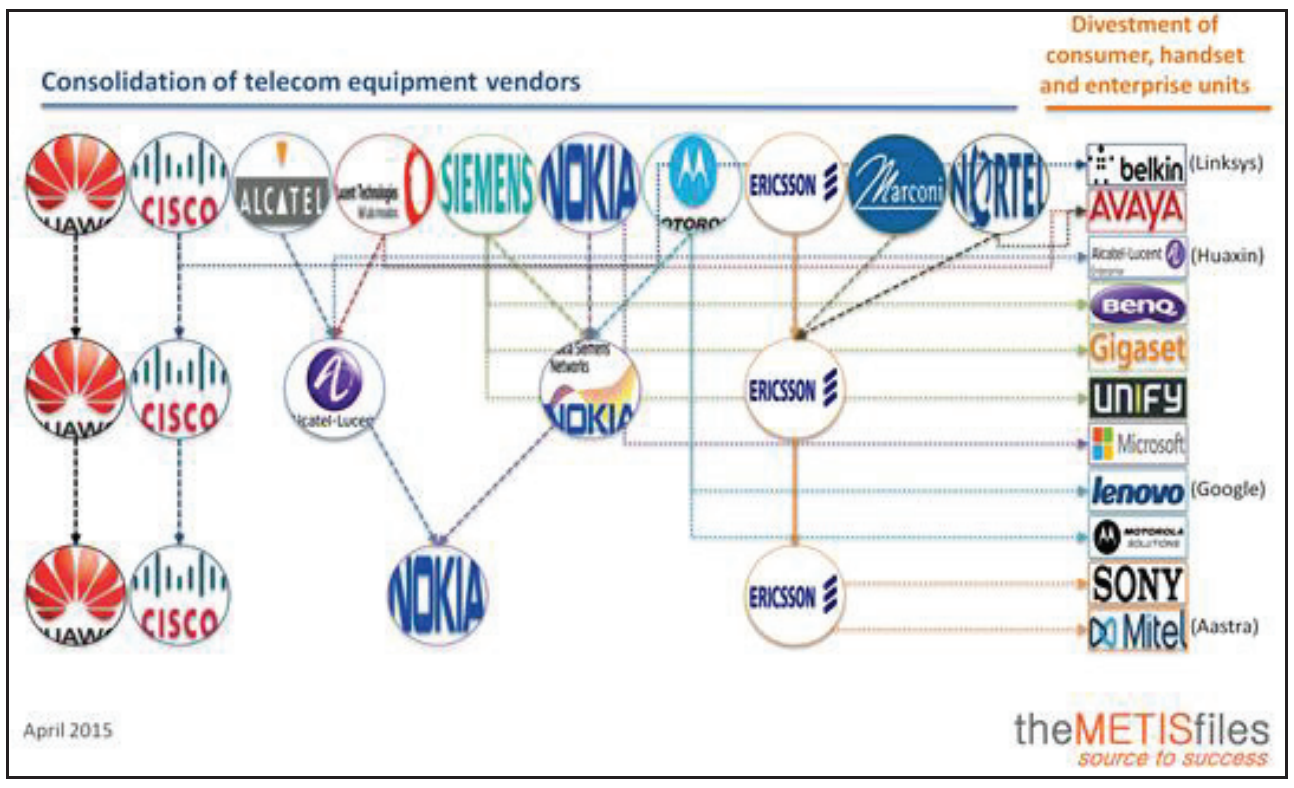

Fig. 2. How Apple, Cisco and Huawei disrupted the telecom equipment market, the Metisfiles, April 2015. [20]

The Verizon-Vodafone acquisition in 2014 was a transaction valued at approximately $130 billion. [21]. Good example of network operators' hard battle is the French companies Iliad-Free and Altice-Numericable-SFR to expand abroad and grow faster than the rival. Operators like T-Mobile US, Bouygues Télécom, Orange Suisse and Portugal Telecom have been on their shopping list. [22].

The changes in Network manufacturing and network operating businesses are as fundamental as that of journalism's. It is a battle of footprint, growth and existence.

## IV. Two Avenues and Some Winding Roads for Media-Journalism Through Alliances or Startups

When trying to give structure to operator-manufacturer-media houses', the Trio's, present position in their respective businesses, the Cynefin sense-making framework positions each of them clearly [23]. Cynefin also describes their common futures and strategic opportunities.

The Cynefin model of operator-manufacturer-media houses [Fig. 3] shows two different layers, the business awareness inside the industry (where we are today), and the ideal positioning for agile and creative future of the business (where we should be).

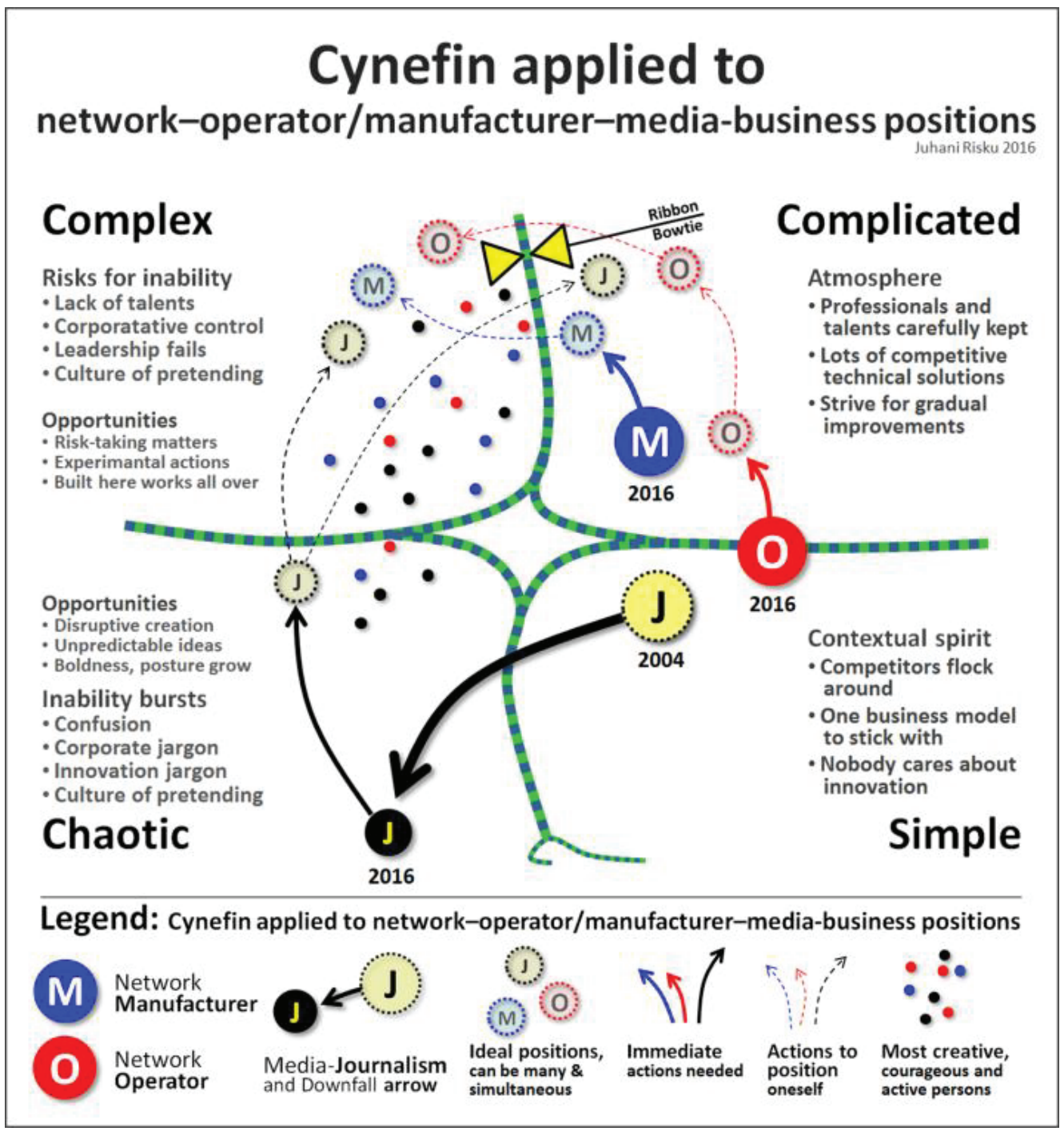

Fig. 3. Cynefin applied to network–operator/manufacturer–media-business positions 2016. Cynefin [ˈkʌnɨvɪn] [23] is framework with five domains: simple, complicated, complex, chaotic and disorder (in the intersection). Simple and complicated represent ordered world, complex and chaotic represent disordered world. When a business is in the simple section it is stable and predictable until disruptive changes happen. Companies with too high comfort zone face problems and may fall to chaotic section. The route clockwise from simple through chaos to complex and complicated is long and painstaking. Balancing in sections complex-complicated-simple allows creation, innovation and possibilities to develop stable simple businesses in a constant flow. Journalism fell from simple to chaotic, operators have stagnated more to simple, manufacturers are on the way to complex with 5G. They all need immediate actions to survive: journalism through complex to complicated, operators upwards between complicated and complex, manufacturers more to complex. Complex offers innovation and experiments, complicated offers stable business. The best place is the Ribbon-Bowtie, where good leadership guarantees innovation and successful business. The Ribbon-Bowtie positioning as a strategic endeavour requires balancing between creative and stable mindsets and giving more authority for creative and disruptive forces and talents. When M, O and J join their forces, it is a combination of creativity, making business from chaos, functioning experiments, and running solid and stable business. M, O and J have some overlap to understand each other, and they have extreme strengths to be integrated. The chaotic journey of journalism appers to be the disruptive strength of the Trio, because in its agony the industry is forced to find viable and competitive yet revolutionary solutions to survive. Now it seems that journalism has been separated from the successful media industry and it tries to survive alone.

Journalism and media houses have some ways to make steps or leaps for better business. Steps are gradual and mainly conservative, leaps are radical and innovative. We divide them to singular solutions and systemic solutions. In singular ones companies look for evolutionary steps and easy-to-communicate actions so that the stakeholders support them. Typically timid decisions like lay-offs, acquisitions, buying a startup or changing the CEO are more cosmetic actions than preparing winning businesses. In singular solutions only gradual growth and changes are allowed for understandable reasons: they are easy to communicate and to do for a corporation than turn the ship at once. It takes several years to rewrite the strategy, recruit and teach new talents, get production lines in order and get the customer to believe in the newborn brand promise. In singular solutions the superficial promise is as important as it is to look convincing. Gradual steps destroy startup culture: Skype was acquired by Microsoft, and it became part of a corporation and lost its startup culture. Nest was acquired by Google, and now Nest is losing its leaders and best people. [24]. Nest as a startup is gone.

Several corporations try to build startup culture inside the company as known as internal startups. For startuppers it may sound like adult caretakers want some entertainment from young radicals. It is impossible to think that a group of startuppers could make anything radical with limited authorisation and resources. Startuppers in a corporation are not given the role they should.

A corporation may also think that the top management and the staff can be surrounded by startup culture. A reasonable question is, that why would they become startup culture lovers by dictation and because it is fashionable? We don't have too many success stories of startup surgery in corporations.

Systemic solutions are more complex and fundamental than singular solutions. In a systemic solution a corporation or an industry envisions, plans and restructures its businesses into a new position with all available means and tools. In media, network manufacturing and network operating fields some companies could gain profits and footprint by letting an internal organization innovate and collaborate with similar organisations of other industries. This means that first steps in collaboration could be organised rather between larger teams or groups than to let only some individuals to from these companies to work together. Here these programs should be funded independently and led by someone else than the existing executives like an innovation officer or newly hired created person. Independence, freedom and changes in leadership are a crucial part of startup culture and internal startups.

Another systemic solution is to establish joint programs and strategic alliances inside one's own industry. This is lighter than a merger and acquisition, but it gives more critical mass. However, a joint program with a competitor in a stagnating business may look more like losers join their forces.

Instead, a radical joint program and a strategic alliance would consist of collaboration between different stagnating industries. Here network operators, network manufacturers and media houses, which are strategic relatives, could join their forces in a new way to build unforeseen models and mechanisms to monetarise content creation, online publishing, vlogging and blogging, developer communities, editing, visualisation, internet radio and TV, virtual and augmented reality, Fig. 1. The TAIC-SIMO tetra model shows the fundamental relation between four different industries in (e.g. network operator–network manufacturer–media–Internet). They are all dependent of each other so that the media serves content (Interest) to the Internet whereas the network operator (Channel) and manufacturer (Technology) enable the Internet services on the background. The Access is the user's favorite home entry page to the Internet [25].

A third avenue, to "Uberise" your corporation, is not explained but only mentioned in this paper. The idea of a completely disruptive model for an industry usually comes from startups e.g. Uber to offer taxi services, or a disruption to another industry as by-product e.g. Google was originally not a media company, but its search absorbed journalism's income, Google took journalism's paycheck, as we call it.

Like Bontemps says, uberisation of business is an increase in volume and expansion through several new and different fields. [26]. In large corporations uberisation could happen either through making a radical change in their business focus, products, services and leadership, or through trying to change the surrounding business environment and markets. Usually these actions are impossible because of the magnitude and investments of their present business and because of corporative ownership and management. The needed radical decisions to be made hardly get support amongst the shareholders, investors and board members. Only radical startups change the business environment on conventional and ordinary industries and change happens insidiously, like changing the media through search and chat.

## V. Conclusions and Aftermath

Without radical actions journalism does not survive in the pressure of technological development.

The circumstances of media and journalism are more uncertain and unpredictable. There are more questions, opinions and claims than innovation, radical actions and academic or engineered solutions for gaining back media's and journalism's grandeur. Software startuppers and their technological developments are driving tremendous change in media and journalism, which both have major societal roles. Journalists, political scientists and philosophers are merely trying to keep up the pace and observing the change, not leading it themselves. But shouldn't they be doing exactly that? Why don't they?

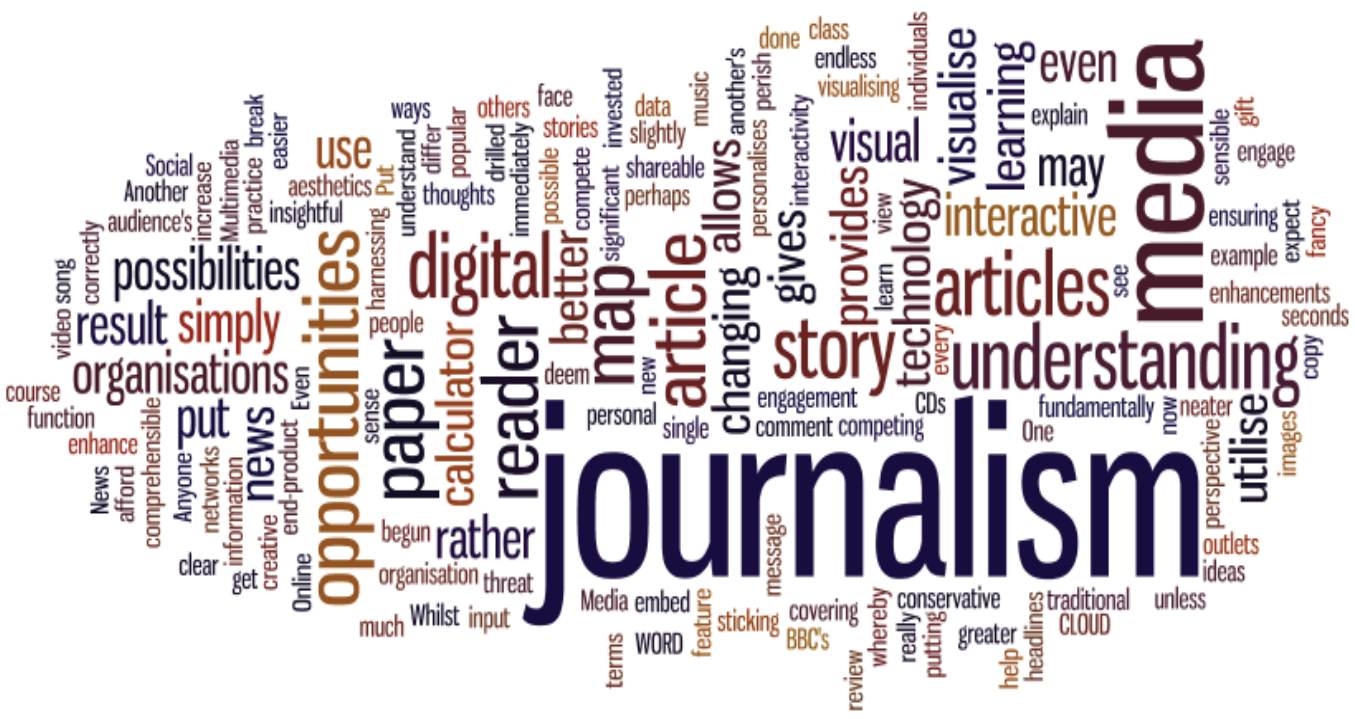

Fig. 4. Present day environment of journalism is filled with uncertainty, possibilies, jargon and gradual only actions. The picture above represents the status quo in many ways: there are 150 notions related to journalism, all the notions are either opportunities or reasons for further confusion. Perhaps more descriptive in the picture is that it is hard to find its origins: two different journalistic notions quote to each other. [27], [28].

If startuppers killed media and journalism – is the Fourth Estate dead? Who's the watchdog now? And should bloggers and other citizen journalists be granted access to press conferences and other sources of information traditionally reserved for those holding an official press card? What does journalism cost and why – and who should pay? These are few of the questions for the philosophers.

Traditional media relies on getting its information mostly from official sources (e.g. official press releases and conferences, press agencies) and through conducting personal interviews. It is thus merely repeating what an authority or a person said. Is that still the right way of doing it? Do traditional journalistic ways of working need redefinition? What is quality and how is political correctness linked to quality? These are few of the questions to the media professionals and academics.

From a systemic operator-manufacturer-media house collaboration totally new products, services, patents, formats and processes can be developed. The operator-manufacturer-media house Trio finds new startups and developers through establishing a global developer and creative community of hundreds of millions of people all equipped with smartphones, action cameras, electric bikes, editing software, bloggers' creativity and new creative culture.

Even a car with its in-car and car-to-car communication can be developed into a car reporter and journalist, which would be a radical abstraction shift. The car becomes anyway a scanner, camera and a total sensor, so a journalistic platform with technologies and formats for media houses could well be an equivalent of Google's Street View magnitude. This would be business as usual for the network manufacturer. This assumption suggests a window of opportunity for disruptive car-to-car innovation for informing, connecting and entertaining the drivers, passengers and citizens in cities and motorways as they read, listen and watch content on the road. The network manufacturer could easily build technology to manage all data flows, payments, videos, chats, studios, media centers and creative reporters' equipment. The network operator could easily build a local, national and continental delivery system on top of the manufacturer's and media house's innovations. Here three industries filled with uncertainty would work together with the ethics of journalism, manufacturer's engineering skills and operator's local customer base.

Operator-manufacturer-media house Trio finds easily strategic partners from areas where they don't have harsh competitors. The idea is that the Trio as an Internet veteran and trusted content creator takes the customer, user and creator as a partner, pays properly for the content, promotes the creative reporter community and acts like a peer startupper.

Industry evolution on the Trio's three specific, business areas need research, development and discussion, both singular and cross-sectoral. Research in the rapid evolution and changes in media industry, startup culture and technology has its inertias of research practices to reach applicable results in time. Therefore research should always be a counterpart of development. Development for its part happens either by the slow and at times pompous corporations or by novice startuppers with more eagerness than sense of professional execution. When Giardino (et.al.) [29] refer to Marmer (et.al.) [30] that more than 90 % of startups fail, no industry, trade or human activity can accept this enormous waste of startuppers' work. This is both a research and development question.

The Trio's fightback happens best through repairing journalism and making these industries become part of a highly ethical commitment. Here the participation of the faculties of social sciences and humanities is crucial. Software startuppers need partners from those areas to avoid a situation in which the societal development is allowed to happen "technology first". Ethical and societal startups need to stem from the faculties of drama, journalism, philosophy, social and political sciences to become part of and influence the rapid development in businesses that software startuppers have alone boosted.

## ACKNOWLEDGMENT

This paper has been a hard effort to formulate a proposal for several different industries to join their forces with startup culture. Pekka Abrahamsson has supported all hard initiatives which could be called complex systemic formulations. But, in his words, "write and produce text first, otherwise nothing happens." We want to thank Pekka for his encouraging words.

## REFERENCES

Beside scientific sources, media and journalistic articles and news have been taken as sources for references because the biggest concern about media's and journalism's economical downfall is expressed immediately on their news and articles. Media and journalism also try to find solutions for the fightback to get the advertisement money from Web services like Google, Facebook and Twitter.

As a source, media and journalism should be trustworthy because of their self-regulation, peer feedback, and especially because of their ethical rules and publishing policies. This applies, according to the so called quality media, at least to the quoted references. The reliability of the media and journalism is based on the idea of freedom of speech.

One important reason to have media and journalism as a source of data is that media and journalism trust on their own surveys, analyses and action models and they act according to these findings.

# III

# GAMIFYING THE ESCAPE FROM THE ENGINEERING METHOD PRISON

by

Kai-Kristian Kemell, Juhani Risku, Arthur Evensen, Pekka Abrahamsson, Aleksander Madsen Dahl, Lars Henrik Grytten, Agata Jcdryszek, Petter Rostrup & Anh Nguyen-Duc, 2018

# Gamifying the Escape from the Engineering Method Prison

## An Innovative Board Game to Teach the Essence Theory to Future Project Managers and Software Engineers

Kai-Kristian Kemell [https://orcid.org/0000-0002-0225-4560], Juhani Risku, Arthur Evensen, Pekka Abrahamsson [https://orcid.org/0000-0002-4360-2226]
Faculty of Information Technology
University of Jyväskylä (JYU)
Jyväskylä, Finland
{kai-kristian.o.kemell, juhani.risku, arthur.n.evensen, pekka.abrahamsson}@jyu.fi

Aleksander Madsen Dahl, Lars Henrik Grytten, Agata Jedryszek, Petter Rostrup
Faculty of Computer Science
Norwegian University of Science and Technology (NTNU)
Trondheim, Norway
{aleksamd, larshg, agataaj, petternr}@stud.ntnu.no

Anh Nguyen-Duc
Department of Business and IT
University of Southeast Norway
Bø I Telemark, Norway
Anh.Nguyen.duc@usn.no

***Abstract*—Software Engineering is an engineering discipline but lacks a solid theoretical foundation. One effort in remedying this situation has been the SEMAT Essence specification. Essence consists of a language for modeling Software Engineering (SE) practices and methods and a kernel containing what its authors describe as being elements that are present in every software development project. In practice, it is a method agnostic project management tool for SE Projects. Using the language of the specification, Essence can be used to model any software development method or practice. Thus, the specification can potentially be applied to any software development context, making it a powerful tool. However, due to the manual work and the learning process involved in modeling practices with Essence, its initial adoption can be tasking for development teams. Due to the importance of project management in SE projects, new project management tools such as Essence are valuable, and facilitating their adoption is consequently important. To tackle this issue in the case of Essence, we present a game-based approach to teaching the use Essence. In this paper, we gamify the learning process by means of an innovative board game. The game is empirically validated in a study involving students from the IT faculty of University of Jyväskylä (n=61). Based on the results, we report the effectiveness of the game-based approach to teaching both Essence and SE project work.**

### 1 Introduction

Software Engineering (SE) as a discipline is generally seen as lacking in general theories [6] [8]. Practitioners on the field employ a multitude of different SE methods and variations of the more common methods [8], while especially software startups commonly still work with purely ad hoc methods or various combination of mainly Lean and Agile practices [14]. While tackling the situation through the creation of a universal, context-independent software development methodology that suits every SE endeavor might be the ideal solution, this line of action has seen little success so far as is evident from the amount of various methods and practices being employed on the field. One recent effort to address this situation has been *the Essence Theory of Software Engineering* (Essence from here on out), proposed by the SEMAT initiative [8] [19]. Instead of aiming to be a one-size-fits-all SE method, the Essence specification is a modular framework that can instead be used to support the use of the various existing SE methods and practices [8].

Essence is built on the philosophy that methods are not supposed to be exclusive or monolithic by nature. Instead, it would be ideal if practitioners always sought to employ the methods and practices best suited for each SE context individually. In this context, [7] also refer to what they call *method prisons*. Method prison, they argue, is a situation where an organization is locked into using one or several specific method(s), regardless of whether they fit the current SE context of the organization. They consider this to be the normal state of an IT organization.

They posit that this is a result of methods being treated as being monolithic and exclusive, whereas there is actually nothing preventing practitioners from combining and modifying them as they wish. They have intended Essence to be a solution to method prisons by supporting the modification, combination, and tailoring of methods and practices to fit any possible SE context. This view on SE methods and practices proposed by Essence could potentially serve to improve the quality of SE work of practitioner organizations, and warrants studies looking into it. Acting in line with this view of SE methods and practices, however, requires lots of work, reflecting, and planning from the would-be users of Essence.

Being a new tool, Essence has yet to see widespread adoption among practitioners, although it has recently gained some more traction in the academia [20]. One reason for the relatively low practitioner interest is possibly the lack of tools to help implement it, as well as the failure of its would-be users to see its full potential [6]. Due to the modular nature of Essence, its full potential is not realized until it is tailored by its would-be users to suit their specific SE context. This may make it seem less attractive to potential users at a quick glance. Furthermore, learning Essence is not a quick process [15] and may necessitate the taking on a new perspective on the nature of SE methods and practices, which can deter potential users from exploring it.

Acknowledging the perceived difficulty of adopting Essence, the creators of the specification, as well as other individuals interested in it, have made efforts to facilitate the adoption and use of Essence. Some academic studies and other publications have proposed tools to aid in the implementation of the specification in practice (e.g. [6]). In this paper, we chose to tackle the adoption problem by means of gamifying SE project work and the use of Essence by means of a board game.

Although gamification as a concept is relatively new, the idea of using games for learning purposes, or the concept of *serious games* is not at all new [2]. In fact, the idea of using games for educational purposes by far predates digital games as a phenomenon, making gamification not at all limited to digital games specifically [2]. Reference [2] defines gamification to be "the use of game design elements in non-game contexts". In this particular case, we speak of gamification in the sense of gamifying the SE endeavor through means of simulation in the form of a board game, as well as the gamification of the adoption of Essence.

In this study, we develop and evaluate *The Essence of Software Development – The Board Game* through an empirical experiment. In the experiment, we observe groups of IT students play the board game and use mixed methods to gather data from the participants, as is discussed further in the fourth section. More specifically, the purpose of this study is to create an educational board game that fulfills the following objectives:

*1) First year SE students should learn the basic concepts of Essence and SE in a fun way*

*2) The board game should teach a method agnostic view of SE, and that methods are modular*

*3) The board game should teach the importance of teamwork and communication in SE project work*

The rest of this paper is structured in the following manner. Sections 2 and 3 discuss Essence and the board game respectively. We then go over the research methods of the study in section 4 and discuss the experiment in detail in section 5. The data from the experiment is analyzed in section 6. In section 7 we discuss our findings and their implications before concluding the article in the 8th and final section.

## 2 The Essence Theory of Software Engineering

As Essence has yet to become a widespread tool in the industry, and is still relatively new, having originally been proposed in 2012 [8], we will briefly describe the specification and its components in this chapter. The specification was proposed by the SEMAT (Software Engineering Method and Theory) community that consists of a number of different practitioner organizations and academic researchers [19]. The specification comprises both what the authors call a kernel, which they claim involves the elements that are present in every SE endeavor, and a language for extending the kernel as needed. The specification is therefore modular in nature and is intended to be modified as needed to fit any potential SE context. For example, extant literature has shown how to describe SCRUM with Essence [13].

The Essence kernel is split into three areas of concern: Customer, Solution, and Endeavor [8]. The core of the Essence kernel consists of seven alphas, which the authors refer to as “[the essential] things to work with” [8]. The seven alphas are elements the authors of the specification posit are present in every SE endeavor. The alphas are complemented by a number of Activity Spaces, or “[the essential] things to do” [8]. Each Activity space may contain one or more Activities, or no Activities at all [12]. Finally, the kernel also includes a third type of element: competencies [12]. The competencies underline the key capabilities required from the team in order to carry out the endeavor [8].

In practice, as the quoted descriptions above underline, the alphas of the specification are the trackable elements to be worked on. For example, one of the alphas in the kernel is simply ‘Software System’; the system that is being worked on [12]. The alphas are to be tracked to measure the progress being made on the SE endeavor at hand [12]. For the purpose of tracking the alphas, each alpha is assigned a set of states that are used to determine the progress on each alpha during the SE endeavor. Each state includes a brief, general description of the state, e.g. “Ready: the system (as a whole) has been accepted for deployment in a live environment”, as well as state checklists to help gauge whether the particular state has been reached [12].

Aside from the kernel, the Essence specification includes a language that is to be used in extending the kernel as needed [12]. The language contains the syntax for creating further alphas and other specification elements [12]. Akin to e.g. XML, it uses both natural and formal language to describe the specification elements. Most of the content in the kernel, and any context-specific versions of it, consists of context-dependent natural language while formal language is mainly used to structure the content written in natural language, as well as to guide users in writing it. Three levels of conformance are specified for descriptions written using the language, with level three descriptions being automatically trackable and actionable, and level one descriptions being rather freeform in nature. Lower level descriptions are easier to produce but offer less utility when used in conjunction with external tools for Essence.

In extant literature, Essence has been applied to student contexts before. Reference [16] conducted a field study on Essence by using student teams to assess the framework. The student teams were to use the framework in a real SE project undertaken as a part of their studies, and their utilization of the framework was monitored during the process. The authors concluded that, in comparison to the results of the same course from earlier years, the utilization of Essence seemed to make a difference in how well the project teams. Apart from academic literature, practitioner reports on the use of the framework are available online. For instance, the SEMAT community website features, among other things, experience reports from practitioners, e.g. [4].

## 3 The Essence of Software Development – The Board Game

The Essence of Software Development board game was developed by IT students from the Norwegian University of Science and Technology under the supervision of the more senior authors of this paper. We developed the board game in this fashion to ensure a student-oriented design approach, i.e. by having students develop a game they themselves would like to play. The game is intended to serve as a game-based learning tool for teaching the use of the Essence specification, as well as SE project work on a more general level.

In designing the game, we worked with several goals in mind. First, the game should be aimed at new SE students as an introduction to both SE project work and Essence. Secondly, the game should, in this vein, include some important elements of Essence. We decided to focus on the core philosophy of Essence: its method agnostic approach to SE project work, as well as the idea of methods being modular in the sense that they ought to be combined in a way that best suits each SE endeavor at hand. Additionally, we included the seven alphas of the Essence kernel into the game: opportunity, stakeholders, requirements, software system, work, team, and way of working are all present in the game under the surface, though just as in real life, they are not always visibly present as you play.

Thirdly, the game was to reflect the cooperative nature of SE project work by encouraging team work and communication rather than competition. Past research has established that team work and communication are two of the most important areas of SE project work [10]. Finally, the board game, despite being a game, was to be reasonably realistic in simulating an SE project. The resulting board game simulates in a simplified manner an SE endeavor and has the players assume the roles of the project team members, with one of the players acting as the team leader or, in other words, project manager. The goal of the game is to work as a team to complete an SE project. This is a rather novel design choice for a board game as most such games tend to focus on competition rather than cooperation, with players either winning or losing as individuals. In this board game, on the other hand, the players either win or lose as a team, much like in a real world SE project.

Each player controls a character in the game, each of which has a certain level of soft skills, hard skills, and energy. Soft skills are required to successfully cooperate on various project tasks, while hard skills are required to finish certain more difficult SE tasks at a high enough level. Energy, on the other hand, is the main resource in the game, spent on various actions and completing tasks in the project. These attributes of each character can be influenced by various events and items as the game goes on. For example, installing a coffee machine in the office results in everyone having a little bit more energy.

Each game starts with the players drawing a scenario card which dictates the nature of the project being worked on. For example, the players might work on a mobile game commissioned by an external client. The simulated SE endeavor then proceeds iteratively, with each iteration marking an arbitrary period of time. The amount of iterations each game takes is pre-determined by the scenario chosen for each game.

In order to finish the project, the players must work on various SE tasks. The number of tasks that are to be completed is denoted by the scenario drawn at the start of each game. The tasks in the game are split into front-end, back-end and architecture tasks. These are also departments physically present on the game board, along with the testing department. Each character works in one of the department, although players are free to switch departments as they wish during the game, but may only work on the tasks of the department their characters are currently located in. Each finished task, save for architecture tasks, is to be tested before deployment, and untested tasks may result in various risks manifesting.

During each iteration, the players are to cooperate in order to figure out how to best split their available resources between the tasks they must complete. There are no turns and each player is free to act as they wish at any given time during the iteration. While communication is encouraged, it is up to the team leader to make the final decision on what each team member is to work on during each iteration. Once the deadline for the scenario is reached after a certain amount of iterations, the team either wins if all tasks are finished, or loses if any tasks remain unfinished. Though the game is based on iterations, the iterations could just as well be called sprints or phases to account for e.g. a more waterfall-oriented development method.

Essence is present in the game in its method agnostic approach to SE. No method is imposed on the players and they may even choose to use an ad hoc approach to SE should they wish. In line with how Essence encourages combining and mixing various methods, the players are free to choose what methods and practices they employ during the project based on what they consider to be the most beneficial combination. Each practice affects the game in some way, and together the practices can heavily influence the way the game proceeds as they offer various beneficial and less beneficial combinations for the players to explore.

## 4 Research Methodology

This study was conducted as a mixed method study, with a focus on qualitative data. We chose a primarily qualitative approach to this study due to the nature of its research problem which is focused on the subjective experiences of the individuals playing the board game. The data were collected through three separate surveys, one multiple choice exam on SE project work, and written reports delivered by the participants. The underlying philosophical approach for this study is interpretivist, with the study explicitly focusing on the subjective perceptions and experiences of the participants [11]. In addition to contributing to the empirical body of knowledge on engineering in the area of Essence in educational use, drawing from the contribution typology that [14] adapted from [18], this study presents a contribution in the form of guidelines.

This study was carried out through an experiment that was conducted over the course of two successive evenings. The participants were to participate either only on the second evening, or on both evenings. All the participants of the experiment were students from the IT faculty of University of Jyväskylä. More specifically, some were Computer Science majors while others were Information Systems majors. Thus, all participants had some degree of knowledge of SE Engineering project work. On the other hand, all participants were unfamiliar with Essence.

The goal of the experiment was to evaluate whether the board game fulfilled the objectives presented in the introduction. For this purpose, we collected an extensive set of data, both qualitative and quantitative, on the learning experiences and game experiences of the participants involved in the experiment using multiple methods of data collection. The use of a pre-game and post-game survey was adapted from the gamification evaluation process used by [5] while the contents of the post-game survey were adapted from the evaluation criteria of [17]. Furthermore, we followed the general guidelines for planning experiments in SE of [21] in conducting the experiment and planning the data collection.

First, each of the participants filled out a pre-game survey which focused on demographic information, e.g. their age, the year course of the participants, as well as their previous work experience. Then, after the experiment on both days, the participants filled out a largely quantitative post-game survey. The survey was adapted from the evaluation criteria of [17], with some modifications made to the criteria in order for them to better fit into the context of a board game rather than a digital game. The detailed framework can be found in the results chapter of this paper in Table I. The post-game survey was conducted as a Likert five point scale survey, where the choices varied from “strongly disagree” (1) to “strongly agree” (5), with the statements focusing on the learning experience of the participants (e.g. “I learned something new about Software Engineering”), as well as their experience with the board game (e.g. “I had fun playing the Board Game”).

In addition to the pre-game and post-game surveys, the students were asked to complete a multiple-choice examination on Software Engineering projects adapted from several public online sources. Finally, all participants were to deliver a written report of two to four pages on their experiences with the board game after the experiment. For the purpose of the data analysis and reporting of the results, we employed the guidelines from [9].

## 5 The Experiment

The study was carried out on by conducting an experiment on two successive evenings, spanning five hours per evening. The participants were only given instructions to arrive at the location of the experiment at the given time and date, and that the experiment was for a scientific study. This was done to avoid having any of the participants familiarize themselves with Essence beforehand, i.e. to gather data as unbiased as possible about their learning. The participants were to either participate on both evenings or only the second evening. The participants were awarded one or two study credits for their participation based on whether they participated on one evening or both evenings. On the first evening, 37 students participated in the experiment, while 61 participated on the second evening, including the 37 that had also been present on the first evening. The protocol was largely the same for both evenings.

### *5.1 The First Day*

On the first day, by 16:00 (4 PM), all participants were to arrive at the scene of the experiment. Once all the participants had arrived at the scene, an introductory speech explaining the rules of the experiment was given. In short, they were to participate for the duration of the entire experiment while following any further instructions. While they were allowed to take short breaks to e.g. use the rest room, they were not allowed to leave for longer periods of time. They were then asked to fill out the pre-game survey

After the introduction and the pre-game survey, on the first evening two of the authors asked four students, eight in total, to join each of them in playing a round of the game to demonstrate it to the other participants. The purpose of this demo round was to make it easier for the participants to understand the game. After approximately thirty minutes of demonstration, the participants, save for those who participated in the demonstration, were split into seven groups.

The groups were formed randomly, decided by having the participants draw a piece of paper with a number between one and seven on it from a mug. Once the groups had been formed, each group was assigned one participant who had taken part in the demonstration round. The eighth demonstration round participant was assigned to one of the groups with five rather than six members in it. At approximately 17:00, the groups had been formed and the participants were instructed to play the game in the groups until the end of the experiment. The authors observed the process, as it is seen in Fig. 1, in a largely passive fashion. The purpose of the observation was primarily to ensure that each group was playing the game and following the rules.

Towards the end of the first experiment day, at 20:30, the participants were offered pizza and were asked to fill out the post-game survey while enjoying it. After completing the survey and jotting their names down on the list of participants, they were free to leave for the evening.

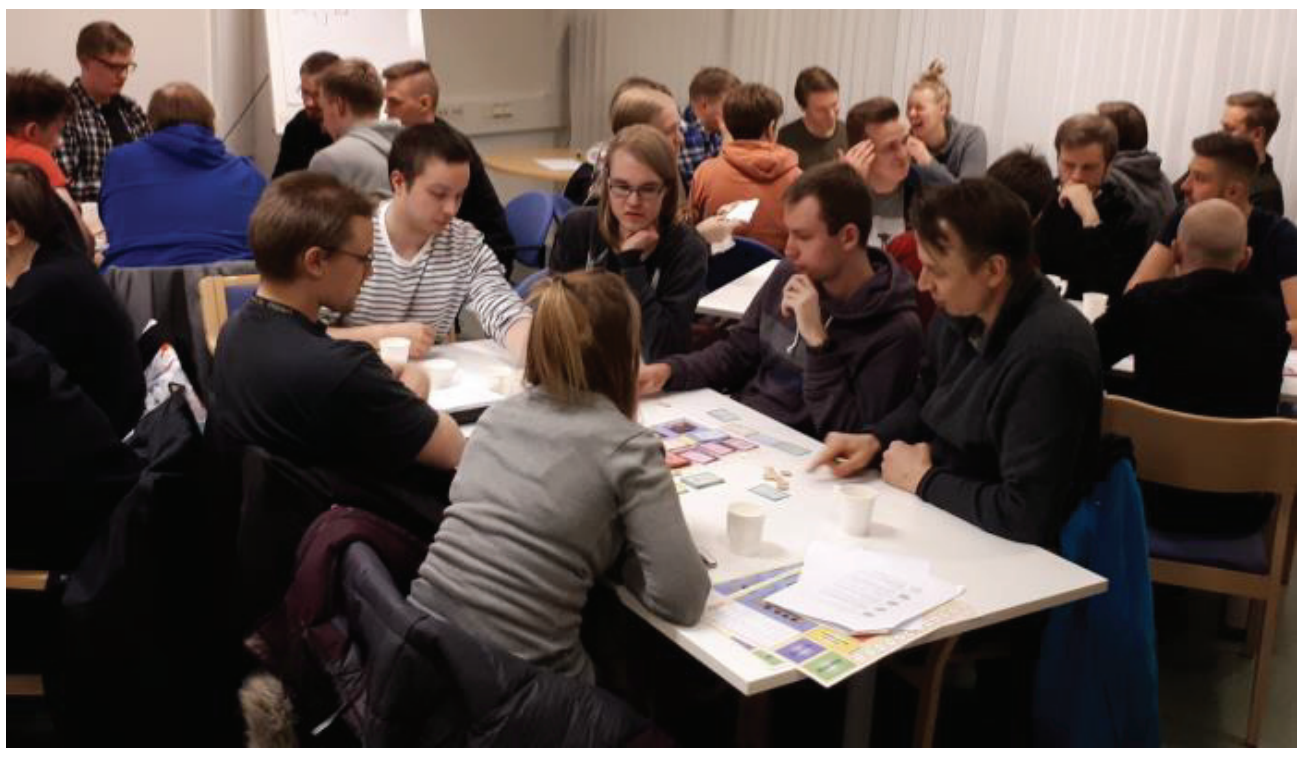

Figure 1 Groups of Participants Playing the Game

### *5.2 The Second Day*

The second experiment day was carried out largely in the same fashion. Shortly after 16:00, the participants were once again given an introduction to the experiment. Those that had not participated on the previous day were then asked to fill out the pre-game survey. As over half of the participants had been present on the previous day, no demonstration was given. Instead, the participants were directly split into ten groups in a random fashion, with one group consisting of seven participants and the rest of the groups consisting of six. At 16:30 the participants had been arranged into their respective groups and were asked to play the game until told otherwise.

At 20:10, the participants were asked to start filling out the data collection forms. All participants were asked to fill out the post-game survey, as well as to complete the multiple-choice examination on SE project work. In addition, those participants that had been present on both days were asked to fill out an open-ended survey on the game mechanics of the board game. The purpose of this survey was to collect data that could, in the future, be used to improve the board game, although it was not used in this particular study. At 20:30, the participants were once again offered pizza, and were asked to finish filling out the forms. Once finished with the forms, they were to confirm their attendance and were given instructions for writing their reflective report based on their experiences in the experiment.

## 6 Results

A diverse set of data was gathered from the experiment. The bulk of our findings is based on the quantitative Likert scale survey data from the post-game survey which was conducted following the evaluation criteria of [17], as stated earlier, as well as quantitative data from the multiple-choice examination on SE project work. In addition, these two sets of data are complimented by qualitative data from both the open-ended questions at the end of the post-game survey and the demographic data from the pre-game survey.

The results of the post-game survey are analyzed through the criteria we adapted from [17]. Modifications to the original evaluation criteria of [17] were made to make the framework more applicable to the context of a board game as opposed to a digital game. The main criteria categories of user experience and educational usability were also used to guide the analysis of the data. The criteria, seen in Table I below, were directly converted into statements for the Likert scale post-game survey, the results of which can also be found in the table. The survey results in the table are divided into four columns based on which group of participants the data were collected from. Group A participated in the experiment on both days, while Group B only participated on the second day. This was done to gain a better understanding of how the participants felt about playing the game for longer periods of time.

TABLE I. Evaluation Criteria and Post-Game Survey Results

| **User experience (UX)** | | | | |
|---|---|---|---|---|
| *1. Emotional issues* | *Day 1 (grp A)* | *Day 2 (grp B)* | *Day 2 (grp A)* | *Day 2 (all)* |
| 1.1. Playing the board game motivated me to learn more about Software Engineering | 2,43 | 2,8 | 2,39 | 2,56 |
| 1.2. Playing the board game was fun | 3,43 | 3,04 | 2,33 | 2,62 |
| 1.3. Playing the board game made me want to play more | 2,59 | 2,16 | 1,58 | 1,82 |
| 1.4. This way of learning about SE is exciting | 2,76 | 2,72 | 2,17 | 2,39 |
| 1.5. This way about learning SE is interesting | 2,92 | 3,12 | 2,39 | 2,69 |
| *2. User-centricity/engagement* | *Day 1 (grp A)* | *Day 2 (grp B)* | *Day 2 (grp A)* | *Day 2 (all)* |
| 2.1. The gamification elements enhanced my interest towards studying Software Engineering | 2,76 | 2,92 | 2,44 | 2,64 |
| 2.2. The visual representation of a Software Engineering project enhanced my engagement with the board game | 2,73 | 2,88 | 2,28 | 2,52 |
| 2.3. The interactive way of representing a Software Engineering project enhanced my engagement with the board game | 3,03 | 3,28 | 2,64 | 2,9 |
| 2.4. The textual information about Software Engineering enhanced my engagement with the board game | 2,62 | 2,68 | 2,28 | 2,44 |
| *3. Appeal* | *Day 1 (grp A)* | *Day 2 (grp B)* | *Day 2 (grp A)* | *Day 2 (all)* |
| 3.1. I was interested in playing the board game | 3,57 | 3,64 | 2,25 | 2,82 |
| 3.2. The board game was visually appealing | 2,59 | 2,92 | 2,19 | 2,49 |
| *4. Satisfaction* | *Day 1 (grp A)* | *Day 2 (grp B)* | *Day 2 (grp A)* | *Day 2 (all)* |
| 4.1. The board game experience added fun to the learning opportunity | 3,54 | 3,56 | 2,67 | 3,03 |
| 4.2. This way of learning about Software Engineering is motivating | 3,11 | 2,92 | 2,42 | 2,62 |
| 4.3. I felt a satisfying sense of achievement at some point during the game session | 3,62 | 3,12 | 2,72 | 2,89 |
| 4.4. The board game made me interested in its contents (SE) | 2,95 | 2,96 | 2,42 | 2,64 |
| **Educational usability** | | | | |
| *1. Error recognition, diagnosis and recovery* | *Day 1 (grp A)* | *Day 2 (grp B)* | *Day 2 (grp A)* | *Day 2 (all)* |
| 1.1 The player(s) can make mistakes while playing the board game. I felt like the mistakes I (or we as a team) made were useful learning experiences | 3,16 | 3,08 | 2,56 | 2,77 |
| 1.2 After playing the board game, I feel like I can avoid making similar errors in the future | 2,92 | 2,84 | 2,28 | 2,51 |
| *2. General learning experiences* | *Day 1 (grp A)* | *Day 2 (grp B)* | *Day 2 (grp A)* | *Day 2 (all)* |
| 2.1 Playing the board game resulted in useful learning experiences about Software Engineering | 2,35 | 2,6 | 2,08 | 2,3 |
| 2.2 The contents of the board game (e.g. the vocabulary used) was related to other things I have learned about Software Engineering during my university studies | 3,22 | 3,56 | 2,92 | 3,18 |
| 2.3 The board game taught me new things about Software Engineering | 2,05 | 2,4 | 2,03 | 2,18 |
| 2.4 I feel like the board game was a successful representation of a Software Engineering project | 2,3 | 2,72 | 2,19 | 2,41 |

*6.1 User Experience*

The board game was generally considered to be a positive experience by the participants. The large majority of the participants felt they had both had fun playing the board game and had been interested in doing so. Similarly, the participants generally thought that the board game had added fun to the learning opportunity, and considered a board game to be a motivating way of learning SE. In particular, the participants enjoyed working as a team to win in the game, and some of the participants noted that the social aspect of the gameplay was what they had liked the most about the experience.

Despite having considered the board game experience both fun and interesting, the participants would not have liked to keep playing the game after the duration of the experiment, or even until the very end of it. In their reports and in the open-ended closing questions of the post-game survey, the common sentiment among the participants was that the game was fun for a few rounds, but slowly became less and less interesting as they kept playing. This, many of them added, was a result of the game having little replay value. This can also be seen in Table I when comparing the answers of the participants who participated on both days, i.e. when comparing the responses of group A from the first day to their responses from the second day. Those who participated on both days enjoyed the game less and felt it was less useful on the second day, as evidenced by the averages of almost every survey question. Even the participants who felt the most negative about the game towards the end of the experiment nonetheless typically reported that they had enjoyed the game during the first game round or two.

The participants generally felt that the game became too predictable due to the lack of competitive elements in the board game, and due to the game in general having relatively few random elements in it for a board game. Even more importantly, most participants felt the game was in fact too easy with more than four or five players. This was especially noticeable in the data gathered from the second day of the experiment when the participants were playing in groups of six or seven as opposed to the groups of five on the first experiment day. As the game difficulty did not scale based on the number of players involved in a round, having more players playing the game simply added more resources for the team to use, indeed resulting in the game becoming easier with more players.

As the participants were instructed to keep playing the game until the end of the experiment, some of the groups tackled the problems they felt the game had in terms of game mechanics by establishing house rule. For example, to add an element of competition into the game, one group of participants had one of their members play the role of the “son of the boss”. The son of the boss would seemingly be a part of the project team in the game but would seek to sabotage the project from within for his own gain. Some other groups simply lowered the number of players playing the game or imposed restrictions on the amount of resources they had in the game to make the game more difficult and therefore more interesting.

Aside from these game design issues the participants felt the game had, the participants generally reported positive experiences. It is hardly surprising that the participants would not have liked to keep playing the game after already playing it for over four hours in one go, or eight hours on two successive evenings. Given the educational nature of the game, it was not intended to be played for lengthened periods of time for entertainment purposes. After all, once the intended pedagogical goals of the game have been reached, it has served its purpose.

*6.2 Educational Usability*

In evaluating the educational value of the game, we consider teaching both Essence and SE project work as its pedagogical objectives. Though the game is primarily meant to serve as a brief introduction to Essence, the game simulates the process of carrying out an SE project, and consequently is also meant to teach SE project work to students.

The participants largely felt that they had not learned much new about SE while playing the board game, underlining in their qualitative responses that they felt like the game primarily served as a way of revising what they had already learned. Only three respondents agreed with the statement “the board game taught me new things about Software Engineering” in the post-game survey. This sentiment could also be observed through the responses to the post-game survey: 6 participants out of 62 agreed or strongly agreed with the statement “playing the board game taught me new things About Software Engineering.” Furthermore, 12 participants out of 62 agreed or strongly agreed with the statement “playing the board game resulted in useful learning experiences about Software Engineering.”

While new learning experiences among the participants were seldom reported, 34 out of the 62 participants agreed or strongly agreed with the statement “the contents of the board game (e.g. the vocabulary used) was related to other things I have learned about Software Engineering during my university studies,” in addition to 12 participants neither disagreeing nor agreeing with the statement. This suggests that the game does nonetheless successfully teach SE project work in a relevant manner. The participants of the experiment were not limited to first year students, and as a result, largely already had a fair understanding of SE project work. Taking this into account, the lack of new learning experiences is not surprising. It is likely that the game would result in more new learning experiences when played exclusively between first year SE students.

When going into specifics about what they had learned or what they thought the game mainly taught, the responses indicated that the participants felt the game had reinforced their idea of the importance of teamwork in SE project work. Many participants also added that the game emphasized soft skills that they felt are seldom discussed in relation to SE.

Apart from SE project work in general, the board game did not directly teach much about Essence. When asked what they considered the most important in an SE endeavor, based on their experiences with the board game, none of the participants mentioned the kernel or the practices present in the board game. In their written report on the experiment, the participants were also asked to describe Essence in their own words. They were asked to do so without consulting online sources, while at the same time being reminded that the report is not graded and that e.g. "I don't know" is as such a fair answer as well. All of the participants simply wrote that they had no clue as to what Essence was based on the board game. It can nonetheless be argued that the board game did in fact teach the players Essence by conveying the idea of SE methods being modular, along with involving the seven alphas of the Essence kernel, as we will discuss later in this chapter.

### *6.3 Objectives of the Board Game*

In the introduction, we defined three objectives for the board game that were evaluated through the experiment. We will now analyze the data directly in relation to these objectives.

*4) First year SE students should learn the basic concepts of Essence and SE in a fun way*

As established in the User Experience subchapter A., the participants nearly universally reported having had fun playing the board game at least for the first one or two rounds, with most of the participants agreeing with the statement "I had fun playing the board game" towards the end of the experiment as well after hours of playing the game. In addition, most participants agreed with the statement "The contents of the board game (e.g. the vocabulary used) was related to other things I have learned about SE during my university studies", which points to the board game successfully capturing the basics of SE project work.

To further gauge whether this goal was reached, we had the participants complete a multiple-choice examination on SE project work after playing the game. The examination was mostly compiled from multiple public online sources, though we added a few additional questions at the end of the survey that were directly related to the contents of the game. However, as we did not have the participants take this examination both before and after the experiment, its results cannot be used make conclusive statements.

The main observation to be made from the multiple-choice examination data is that the majority of the participants passed the examination, as can be seen in Fig. 2. Out of the 61 responses we received in total, 17 were discarded on the basis of being incomplete or otherwise not properly answered, resulting in 45 complete responses. Out of these 45 participants, 34 (75%) would have passed the examination had it been graded, having received more than 50% of the maximum score. The median score was 16 out of 29.

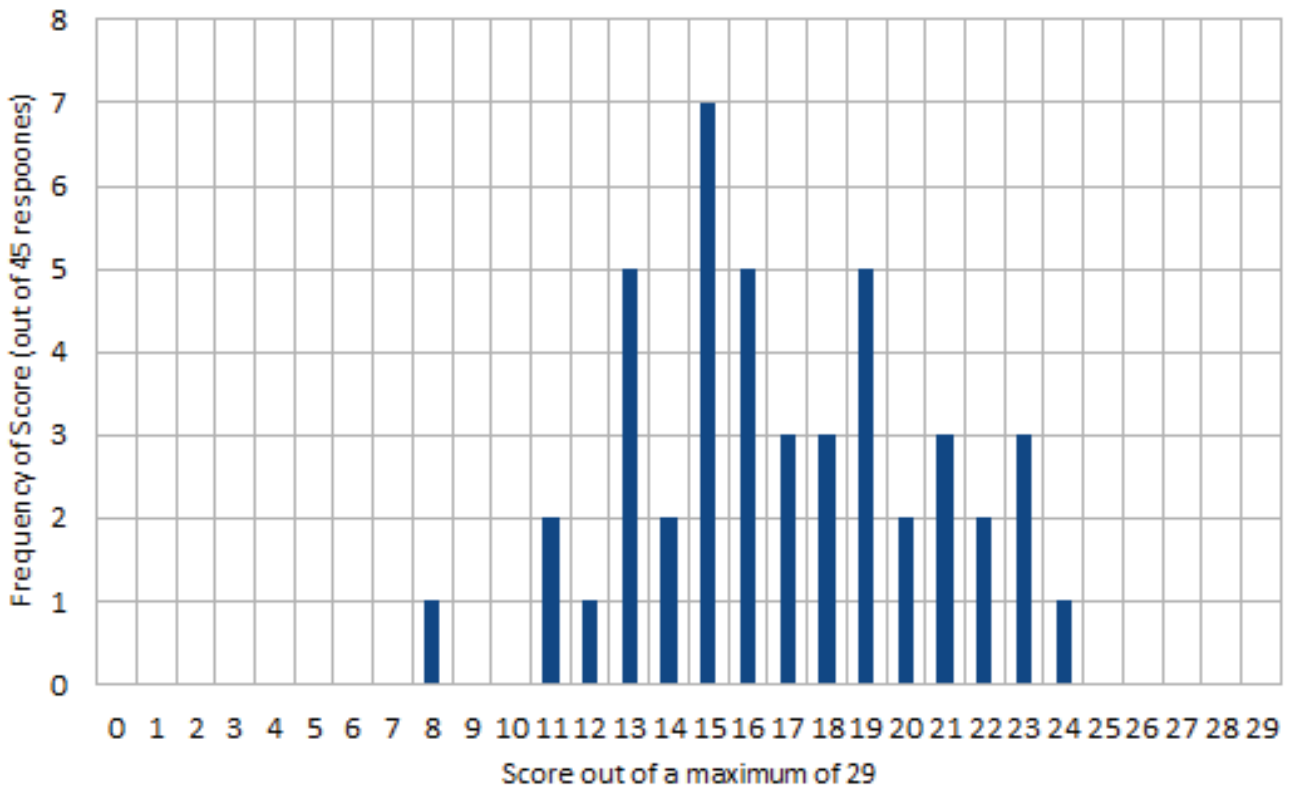

Figure 2 Multiple Choice Examination Results by Score Totals

It is worth noting that there was a possibility of adverse learning while playing the game as well, based on the results of the examination. Being a board game, the game mechanics do result in some generalizations and simplifications of the nature of SE project work, which may be misleading to those with little prior knowledge on the topic. For example, when asked if "the only reason for testing during software development is to mitigate risk at that point in time", 10 respondents out of 45 falsely responded "true". In the context of the board game, that is indeed the only reason to test the software. Furthermore, when asked whether "it's always beneficial to add more developers to a project", in line with how the game became easier the more players (developers) were present, five participants falsely answered "true".

While it is not possible to accurately gauge what effect playing the game may have had on the results of the multiple-choice examination as far as the participant scores go, we nonetheless argue based on our data that this objective was reached. In combination with the multiple-choice examination results, the results of the post-game survey indicate a positive overall result in the context of this objective.

*5) The board game should teach a method agnostic view of SE, and that methods are modular*

This was one of the key principles we followed in designing the game, as was discussed in the third chapter. The participants played the game following the rules as far as the modular use of methods went, and in doing so were introduced to this view on SE methods. More explicit learning experiences in relation to this view on SE methods could certainly be achieved by introducing the players to Essence beforehand, though in this case we chose to not do so to gather as neutral as possible data on what exactly the game taught without outside guidance. Though the participants largely considered Essence to have remained unknown to them after playing the game, we nonetheless argue that this objective was fulfilled through the game mechanics of the game, which pave way for future adoption of Essence among participants.

*6) The board game should teach the importance of teamwork and communication in SE project work*

In response to being asked what they considered important in SE project work based on their experiences with the game, the single most common theme in the responses of the participants was communication and teamwork. One participant, going into more detail, responded that the most important in SE project work was, in their opinion, "an atmosphere that encourages discussion and where one does not have to regret mistakes, as well as communication [in general]". Furthermore, when asked what they had considered to be positive in the game as an open-ended question, a large number of participants mentioned getting to work as a team to have been fun, as well as having enjoyed the social aspect of the game in general. We therefore argue that the third and final objective set for the game was also fulfilled.

## 7 Discussion

Through the experiment, we studied the game-based learning of the Essence specification. Our data indicate that the game-based approach was an enjoyable experience for the participants, and that the board game fulfilled the objectives we outlined in the introduction. In this section, we discuss our findings in relation to teaching Essence, as well as using a board game for educational purposes in the area SE project work.

### *7.1 Implications of the Findings*

Extant literature, as well as official SEMAT statements, have suggested that Essence still suffers from a lack of interest among practitioners (e.g. [6] [20]), likely stemming from its resource-intensive adoption and the lack of tools to aid practitioners in adopting it [6]. Past studies in various fields (e.g. [3]) have also shown that game-based learning is a suitable approach. As with any form of teaching, however, the teaching, and in this case the instrument used in it, needs to fit the context and the intended learning goals. We therefore posit that teaching Essence by game-based means is a proposal worth pursuing, serving as a motivation this study. A game-based approach is particularly suitable for this context as the instrument can then be used by other parties to teach Essence and SE in the future.

Analyzing the feedback gathered from the participants on the board game and its game mechanics, the major shortcomings of the game are related to the core game loop which the participants considered to have become too predictable after some rounds, as well as the lack of scaling in the game mechanics. This was an adverse effect of our decision to focus on cooperation and teamwork in designing the game. While the participants enjoyed the social aspect of the game and the cooperation, many of them noted that the lack of competitive elements also made the game less interesting after some time spent playing. To what extent this is to be considered a downside is debatable as the game was not intended to be played for lengthened periods of time. Being an educational game, the game will have already reached its educational objectives after a few rounds. Nonetheless, we did also discover a clear problem we with the game mechanics: the difficulty of the board game presently does not scale based on the number of players. This can make the game too easy, and thus less interesting, when played with a larger group of players.

Aside from these problems the participants reported having had with the game mechanics, the pedagogical side of the game in relation to Essence can also be seen as lacking to some extent based on the data. While the game involves the seven alphas of the Essence kernel, they largely remain under the surface, as discussed in section three. Similarly, though the game is built around the method agnostic nature of Essence that posits that methods and practices should be combined as is seen beneficial in each unique SE context, this is not the focus of the game. Unless the players reflect on this philosophy on their own, they may simply end up playing the game without paying any mind to it. It may thus be beneficial to heighten the role of Essence in the game by e.g. involving the use of the Essence specification language into the gameplay to make the learning experience more purposeful. In its current form, the board game does not directly teach the use of Essence in practice.

Presently, the game is well-suited as a first touch SE project work and project management for new SE students. It is best played for small amounts of time due to the major design decisions behind it which encouraged teamwork and communication at the cost of competitive, replayability-enhancing elements. Our findings indicate that the game successfully: (1) teaches first year Software Engineering students the basic concepts of Essence and Software Engineering in a fun way, (2) teaches a method agnostic view of Software Engineering, and that SE methods are modular, and (3) teaches the importance of teamwork and communication is SE project work.

Putting our findings into a broader perspective, we encourage the use of board games for educational purposes, especially in the context of SE project work. The participants of our experiment reported that they had particularly enjoyed the social aspect of the learning experience and regarded working as a team to beat the game to be an enjoyable activity. Board games offer a chance for students to either learn in a social F2F setting while competing against each other or while collaborating as a team, which is something that students seemed to enjoy based on our results.

### *7.2 Limitations of the Study*

The reported results of this study are based on a varied set of data which has some shortcomings. In evaluating what the participants had learned while playing the game, we conducted a multiple-choice exam on SE. However, the data gathered through this exam lacks a point of comparison as it was only gathered after the experiment. It is therefore not possible to accurately determine what exactly was learned from the game and what the participants may have known beforehand. Additionally, though the use of students as subjects for empirical experiments is at times questioned [1], in this case the students were the intended target group of the board game being studied, and thus their use as subjects was well justified.

### *7.3 Recommendations for Future Research*

In this paper, we have highlighted some points of improvement in the board game employed in the study. Those interested in developing the game further, or using the game for educational or other purposes, are encouraged to do so as the board game is, as of this publication, available as open source through FigShare. We also have plans to take this board game further so any interested parties are encouraged to contact the authors for possible future cooperation. Though the game examined in this study does succeed in conveying the general philosophy on SE methods behind Essence, it does not concretely teach the use of Essence. This makes it consequently more useful for SE students than practitioners looking to start using Essence. We thus urge those interested in Essence to continue working on tools to help facilitate its adoption. Especially such tools aimed at practitioners are still needed.

## 8 Conclusions

In this study, we built *The Essence of Software Engineering – The Board Game* to teach the Essence specification and SE project work and demonstrated its effectiveness by means of an empirical experiment. We invited IT students (n=61) to play the board game in an experimental setting and gathered a diverse set of data from the experiment. Based on our findings, we conclude that the board game fulfills the goals set for it. I.e. the board game (1) teaches first year SE students the basic concepts of Essence and SE in a fun way, (2) teaches a method agnostic view of Software Engineering, and that SE methods are modular, and (3) teaches the importance of teamwork and communication is SE project work. On the negative end, our findings indicate that the game has a low replay value and some issues related to game mechanics. Furthermore, the game presently does not teach the use of Essence in practice. To this end, we also discuss possible future improvements to the game and plan on working on it further based on our data. Though the board game is fit to be played as is and is available as such, we will continue to work on the game further and plan to introduce a version with improved replayability through e.g. competitive elements, as well as a heightened role of Essence.

Whereas gamification and serious games are typically discussed primarily in relation to digital games [2], we recommend that board games are also considered for game-based learning purposes in the field of SE. We suggest that future research could investigate the possibility of introducing other board games for teaching SE topics. We also posit that there is still a further need for tools to aid in the adoption of Essence. Due to the central role of project management in the success of SE projects, facilitating the adoption of project management tools is important as well.

# IV

# WHAT MAKES A DIGITAL GAME ADDICTIVE? A PLAYER VIEW-POINT ON PLAYER RETENTION

by

Juhani Risku, Kai-Kristian Kemell, Sandra Schweizer, Anh Nguyen-Duc, Mari Suoranta & Pekka Abrahamsson, 2020

# What Makes a Digital Game Addictive? A Player Viewpoint on Player Retention

Juhani Risku
Faculty of Information Technology
University of Jyväskylä
Jyväskylä, Finland
0000-0002-0587-4431

Kai-Kristian Kemell
Faculty of Information Technology
University of Jyväskylä
Jyväskylä, Finland
0000-0002-0225-4560

Sandra Schweizer
Department of Computer Science
Norwegian University of Science and Technology, NTNU

Anh Nguyen-Duc
Department of Business and IT
University of Southeast Norway
Bø, Norway
0000-0002-7063-9200

Mari Suoranta
School of Business and Economics
University of Jyväskylä
Jyväskylä, Finland
0000-0002-3849-4902

Xiaofeng Wang
Faculty of Computer Science
Free University of Bozen-Bolzano
Bozen-Bolzano, Italy
0000-0001-8424-419X

Pekka Abrahamsson
Faculty of Information Technology
University of Jyväskylä
Jyväskylä, Finland
0000-0002-4360-2226

*Abstract*— Acquiring and retaining users is a persistent issue for any digital innovation. In the case of digital games, as they move increasingly towards free-to-play models, which rely on various in-game transactions to generate profit, retaining active players becomes increasingly vital for game developers. This has especially been the trend in online games in the past decade, as well as mobile games specifically. Player retention as a phenomenon, however, has been the focus of few academic studies. While related topics such as what makes games enjoyable or interesting, as well as digital game addiction, have been studied, these are not entirely synonymous to player retention. Moreover, existing studies on player retention have taken a game-focused, quantitative approach. Thus, to further our understanding of this emerging practitioner topic, we conduct a qualitative multiple case study focusing on the point of view of the players. Based on our findings, we highlight the importance of some existing good practices in the area, while also discussing potential negative impacts of some. We also propose a hypothesis, that higher self-efficacy has an impact on player retention.

## I. Introduction

As digital games move increasingly towards free-to-play models which rely on various in-game transactions to generate profit, retaining active players becomes increasingly vital for game developers. This has especially been the trend in online games in the past decade, as well as mobile games specifically. Following this change in business models, it has become vital for online games, like various SAAS services, to retain users, especially ones that spend money on the game. Despite this paper focusing on digital games, attracting new users and retaining them is a topic pertaining to digital innovations at large, Nambisan et.al. [21], rather than simply digital games and innovations in that context.

Various past studies have focused on what makes a game enjoyable, or immersive. Studies have been conducted, e.g., on the effects of virtual avatar customization on immersion (e.g., Turkay et.al. [27]), on social interaction in online games (e.g. Liu et.al., [18]), as well as what is referred to as the balance principle in relation to game difficulty (e.g. Ducheneaut et.al., [12]). However, studies on what keeps the players coming back to the same game are scarce.

Player retention is an emerging, practitioner-oriented topic. Few academic studies on the topic currently exist, with the few existing ones having taken a game-specific and quantitative approach to the topic. Given the novelty of the research area, qualitative studies could arguably help us to better understand the phenomenon. Moreover, as games are typically developed in an agile manner, this approach is consistent with the user-focused approach of agile itself.

In this paper, we take on a player-focused view on retention. By means of a multiple case study focusing on a group of players (n=14), we seek to understand what impact various online game elements have on their players. We focus on game elements from the point of view of the players, seeking to understand what makes an individual keep playing the same game over time. Specifically, we approach this through two sub research questions which are formulated as follows:

RQ1: What online game elements keep players playing the game? and

RQ2: How do rewards affect the player's motivation to continue playing video games?

The rest of this paper is structured as follows. Section 2 provides context for the topic by examining existing, relevant literature. In section 3 we discuss the research framework for the study that was used to guide the data collection. In section 4 we discuss the study design and thus methodology of the study. Section 5 provides the results of the analysis, which are then discussed in section 6. Section 7 concludes the paper..

## II. Background: User Retention

### A. User Retention

Player retention refers to the idea of retaining players, i.e. how many players come back to the service after initially trying it out. It is a practitioner-oriented construct primarily used by companies in analyzing their business. A practitioner expert on the topic, Luton [19], defines retention as the number of users retained over a given period of time.

Retention is typically viewed as a percentage-based number in B2C contexts.

Retention is most important for video games utilizing free-to-play business models where the game itself is free and the revenue comes from various transactions within the game. As users can install and try the game without having to pay, they can simply choose to quit after mere minutes of playing without any sunk costs involved. This is often the case as well, as mobile games tend to lose most of their new players after they have installed and tested the game for the first time, Drachen et.al. [11]. Keeping the game viable for a long period of time requires focus on user retention.

Academic research on player retention is still lacking, especially in terms of qualitative studies. Primarily quantitative studies on player retention have been conducted [11], utilized machine learning on a large quantitative data set comprising multiple games. Weber and Mateas [28] conducted another study utilizing quantitative data and AI utilizing data from a single game, providing a tool for developers to utilize in analyzing player retention. Compared to the existing studies on the topic, we take on a different approach both in terms of using quantitative data instead of qualitative, and in terms of studying the phenomenon from the point of view of the players.

Past the construct of player retention, similar phenomena have been studied in the field of technology acceptance and technology use (e.g., Davis [10]), and more specifically continuance of use (e.g., Battacherjee et.al. [5], Soliman et.al. [24]). These are all long-standing areas of research especially in Information Systems (IS) research. However, these studies have seldom focused on games specifically.

Extant studies explaining players' participation in games have focused primarily on immersion. Based on a literature review on the topic, Liu et.al. [18] identified six sub-dimensions of immersion that have been studied: sensory immersion, spatial immersion, tactical immersion, strategic immersion, narrative immersion, and social immersion. These, they continue, are not particularly well-established, however [18]. To this end, we argue that a qualitative study on the topic could help us further understand what contributes to player retention, possibly past these immersion factors as well.

## III. Research Framework

A research framework was formulated to support the data collection and analysis in this study. The framework is composed of four elements, as shown in Table 1.

TABLE I. Research Framework

| | |
|---|---|
| Social | Contact with other players, with or without own contribution |
| Rewards | Positive feelings; Motivates |
| Game Mechanics | Game difficulty increases in relation to the progress of the player |
| Self-efficacy | The perception one has of their own abilities in performing a task |

### *A. Social*

Social dimension is the player's relation to teamwork, relationship between other players, and socializing.

### *B. Rewards*

Rewards relates to the player in receiving in-game rewards, and the feelings when getting of rewards.

### *C. Game Mechanics*

Game Mechanics describes how the game is built, how the game changes when progressing during the play.

### *D. Self-efficacy*

Self-efficacy refers to the perception one has of their own abilities in performing a task. In the context of gaming, self-efficacy relates to the belief to succeed during the progress of playing.

Self-efficacy refers to the perception one has of their own abilities in performing a task, Carberry et.al. [7]. E.g., one's own perception of whether their C++ programming skills are adequate accomplish an assignment. Self-efficacy is considered important in performing successfully diverse tasks. If a person believes in her own abilities, she is more likely to survive when facing challenges, more likely to strive for related tasks in the first place and more intrinsically motivated, Bandura [3]. Similarly, one may feel less inclined to begin carrying out a task if their self-efficacy is low, even if, objectively, they would have the required skills or knowledge to carry out the task. Self-efficacy can thus result in productivity, Gist et.al. [14]. In this study, we compared the eight self-efficacy categories' and 29 activities' (Appendices) correlation to the Primary Empirical Conclusions 1-8.

According to Bandura [3], self-efficacy is influenced by four factors: (1) mastery experiences, (2) vicarious experiences, (3) social persuasion, and (4) physiological states. Mastery experiences are past task completions (or failures) and hold a notable impact on one's self-efficacy. In the absence of, and in addition to, mastery experiences, vicarious experiences can weigh on one's self-efficacy. Vicarious experiences are gained by observing others with similar ability perform tasks. Social persuasion, then, refers to support from prestigious individuals or individuals we respect, and can positively affect self-efficacy. Finally, our current physiological states, such as stress or simply being tired, can influence our self-efficacy.

Whereas self-esteem refers to one's sentiments about oneself, with low self-esteem relating to feeling negatively about oneself, self-efficacy is simply related to tasks. Despite having low self-efficacy in relation to a task such as winning a baseball match, one may consider this to not matter, continuing to feel confident in them-selves [14]. Moreover, self-efficacy is related to specific tasks [7]. Though we have our perception of e.g. our own ability to program, it is typically considered in relation to a specific task at hand. We may consider ourselves to be good at programming, and yet know that developing an autonomous vehicle from scratch is well beyond our means.

Self-efficacy changes over time, and training or studying a skill is likely to positively influence our self-efficacy in relation to tasks related to that skill [7]. We look at the self-efficacy of students before and after taking a design course intended to support self-efficacy in relation to various skills required in design work. However, we also wish to understand the potential link between self-efficacy in these skills and creativity in practice. We thus further discuss self-efficacy in the specific context of creativity in the following subsection.

### 1. *Self-Efficacy in the Context of Creativity*

Studies in relation to creativity also include self-efficacy, and specifically, creative self-efficacy, which Tierney and Farmer [25] define as being “the belief one has the ability to produce creative outcomes”. Past studies have established a link between creative self-efficacy and creativity, Mathisen and Bronnick [20]. I.e. one’s confidence in one’s own creativity makes one more creative. Furthermore, Mathisen et.al. argued based on their data that creative self-efficacy can be improved by training, while going on to note that further studies on whether doing so also improved creative performance were required. This is something we seek to further address in this study, although extant studies on the matter also exist.

Schack [23] studied self-efficacy and creative productivity in gifted children. Their results pointed towards (creative) self-efficacy training not necessarily producing creativity in the students and having little bearing on their subsequent initiation of independent (creative) projects. Self-efficacy training may in fact lead to lower self-efficacy e.g. if the participants find the tasks challenging, resulting in negative mastery and/or vicarious experiences [23]. This presents challenges to those seeking to devise training courses for self-efficacy.

While much effort has gone towards studying the relationship between creativity and creative self-efficacy, studies looking into self-efficacy in other skills in relation to creativity are fewer in number. Ahlin et.al. [1] studied entrepreneurial self-efficacy in relation to creativity and innovation in the context of entrepreneurship, while Khedhaouria et.al. [16] studied general self-efficacy in the same context. Beeftink et al. [4] studied design self-efficacy in relation to performance in creative profession, and also argued that self-regulation strongly supported self-efficacy in the context of design work.

## IV. Study Design

This study was split into two parts: a pre-study and the main study. Data collected in both stages was qualitative, focusing on in-depth data.

### A. *Pre-Study Phase*

In the pre-study phase, from the participants (n=14) nine were male of age range 16-29 years, and five were female of age range 20-24 years. The average hours of daily play was 1,5 hours (males) and 1,3 hours (females). Average hours of play during weekends was 4 hours (males) and 6,2 hours (females). All of the 14 players were of expert rank.

The participants were to first fill out a questionnaire, either individually or with the help of one of the authors. The questionnaire contained five questions related to the game playing habits of the respondents. The first three questions were general in nature, focusing on how much time they spent playing games on average and what their favorite games were. The last two questions contained lists of factors, which they were asked to sort by priority, that would make the participants either start or stop playing a video game.

Afterwards, a structured interview was conducted to gain further insights into their answers in the questionnaire. The structured interview consisted of 13 questions related to the mood, experiences and feelings of the players while playing digital games. These questions focused on the constructs in the research framework.

### B. *Main Study*

In the main study, the participants of the pre-study who volunteered for a more personal, in-depth case study (n=5) were studied further. Data were collected by means of 1) observation, and 2) semi-structured interviews. The more in-depth data gathered from the main study was intended to provide richer data to support the findings of the pre-study, as well as to provide more insight as to why some of the game elements influenced retention.

The participants were observed playing an online video game of their choice for an hour. The observation of the players happened in their “natural habitat”, letting the players play as they usually did in own familiar contexts.

During the observation, the think-aloud protocol was utilized to collect data. The participants were to narrate their experiences, and typically did so quite elaborately, creating a dialogue as they explained what they were doing and why. After the observation session, a semi-structured interview was conducted with each participant.

Similar to the pre-study, the questions prepared beforehand focused on the constructs of the research framework. Past the pre-planned questions, each participant was asked questions based on both the observation and their responses to the pre-planned questions.

Both the interview and observation audio were recorded and later transcribed. The transcripts were then analyzed for this study. This was the case for both the pre-study and main study data.

## V. Results

In our analysis, we highlight our findings as Primary Empirical Conclusions (PECs), which are numbered. The analysis is structured into subsections according to the central themes in the research framework constructed based on extant research. The first subsections discusses the social aspects of online games, while the second subsections discussed rewards specifically. The third subsection then contains findings related other, less specific game elements such as difficulty, which is highly game-specific in terms of game mechanics.

### A. *Social*

In terms of the social aspect of video games, the participants all enjoyed social interaction, citing it as one of the main reasons for them to play an online game. However, the participants preferred playing with friends, making the digital game simply another way to spend time with their existing friends.

PEC1: Though the social aspect of online games is important, players seem to enjoy playing with people they already knew beforehand.

Notably, the primary observation made in terms of social interaction was that the participants played digital games differently with strangers than they did while playing with friends. When playing with strangers, the participants were likely to take on a more passive role in the game, unless they felt particularly confident in their abilities. When playing with friends, on the other hand, they took on a more active role regardless of their skill level or confidence. Moreover, some of the participants noted they did not enjoy talking to strangers online at all, unless they had to in order to progress in the game.

*"I don't kn- it's scary to just talk to people you don't know? But like, these last people that I played a few games with I would be comfortable with... talking with them by like the second or third game. (...) Because then we would be FRIENDS! [laughs]"* (Interview)

The participants discussed this behavior themselves, noting that they did not want to embarrass themselves by playing too brashly if they were not confident in what they were doing. Unless they felt confident in their abilities, they felt afraid of being shamed by strangers inside the game, should they fail to perform some task. The participants were more relaxed in this sense while playing with friends.

PEC2: Players take on a more passive role when new to the game, while taking on a more active role as they become more experienced. This behavior is negated if playing with a familiar group of people.

PEC3: Players take on a more active role when playing with people they are familiar with, while taking on a more passive role while playing with people they are unfamiliar with, unless they feel particularly confident in their abilities in the context of the game.

Some of the participants also discussed the importance of in-game communities from the point of view of their social lives. Some considered online interaction simply a substitute for face-to-face interaction when the latter was not available, while others considered online relationships as meaningful as any other.

Though the social aspect of online games was considered positive by all participants, many also noted that they preferred playing alone or simply playing offline games when tired. While playing alone, they felt no peer pressure to perform and thus found it more relaxing when not feeling competitive. This ties with the aforementioned PEC2 to some extent. Even if one felt confident in their abilities on average, being tired would make one feel unsure about their performance, potentially resulting in some pressure to nonetheless perform.

### *B. Rewards*

Rewards were considered important by all participants, though some more than others. Rewards in general were considered pleasurable, as opposed to only certain, game-specific ones. Often, the more the participants felt they had had to work for a particular reward, the more satisfying it felt. In fact, the participants generally did not care much for rewards they felt like they had not worked for. Moreover, the participants generally considered achieving goals they had set themselves more satisfying than simply pursuing tasks the game directly rewarded them for, although their personal goals sometimes aligned with ones set by the game.

*“Because then it seems like, we worked hard, and we overcome all-, or overcame all the difficulties, and then we got a reward. And it's not fun when we just "breeze" though everything.”* (Interview)

PEC4: The balance of increasing challenges and in-game progression is important for all players.

The sense of achievement from overcoming difficulties was considered desirable by all participants in terms of motivating them to keep playing. Some noted that it was empowering in general, making them feel more confident in their abilities outside digital games as well.

Randomness in rewards consisting of digital items was considered to make them more exciting. The chance to e.g. receive a rare digital item kept the participants interested in pursuing the rewards. Similarly, discovering various rewards while exploring the game world, not knowing for example what could be in the chest over yonder, was also considered exciting.

Occasionally, achievements were also linked to the social aspect of the game. Possessing rare items and other digital signs of status resulted in gaining positive attention from other players of the game. Sometimes this type of sense of accomplishment also resulted from simply belonging to a particularly esteemed group (guild) of players inside a game. In this fashion, the participants felt even more satisfied with their achievements if they were considered impressive by other people as well.

PEC5: Achieving something in a game increases player motivation greatly to continue playing.

### *C. Other Game Elements Associated with Retention by the Participants*

On a similar note, in terms of game mechanics, the participants discussed the importance of challenge. Overcoming challenges was considered to be something that made the game fun, although some easier parts in-between would make the experience overall more enjoyable as opposed to a continuous need to perform. As long as the challenges were perceived as something the players could see themselves beating, they considered scaling difficulty a positive element. Without a challenge, a game would quickly become boring. However, on the contrary to challenge being considered positive, losing was also cited as a common reason for stopping a gaming session. Losing streaks were demotivating, even though occasionally the participants would wish to continue until they won or completed the task at hand.

PEC6: Difficulty spikes are generally considered particularly rewarding as long as they are perceived possible to overcome.

In seeking to understand retention, we also discussed addictive behavior with the participants by asking them to describe times when they felt they wanted to stop playing a game but for some reason had also felt they could not do so yet. Typically, these types of situations were a result of the participants wanting to finish some specific task in the game before stopping. E.g., defeating a particularly difficult enemy. Another reason in online games was social pressure: though the participant felt like stopping, their group in the game egged them on to continue.

The participants also discussed feeling an obligation to play in relation to frequency rewards. I.e., typically, being rewarded for playing the game every day. These types of rewards were sometimes considered negative, as the participants felt they were missing out on something if they did not do the daily tasks, as opposed to simply feeling rewarded when they did complete the tasks. For the most part, these frequency rewards were considered irrelevant or positive in terms of motivation, though.

*"They could have "dailies" that you do for -hours-, that you -had- to do EVERY day to get the reputation and you'd have to do it every day for like, a month or two months, I don't know. And then it was very frustrating, cause then you would be like, if you have been out, having fun, you come home and it's like eleven in the evening, you're like "Oh no, I have to do my "dailies" today...". You know then it becomes a "have-to""* (Interview)

PEC7: Common game design decisions aiming at player retention may have the opposite effect if the players consider them excessive.

Finally, in the interviews, multiple participants occasionally discussed the importance of avatars either directly or in-directly. Playing as a character that they felt they could identify with in some way was important to them. They either liked being able to customize their own character or being able to pick from a wide variety of pre-made ones in hopes that some of them would be to their liking.

*"Oh, I really like characters, and the setting, and it's not like your standard 'Call of Duty' things, because that's just military, very boring, they're studs, middle-aged white guys and everyone is a middle-aged white guy and that's no fun."* (Interview)

PEC8: Relating to the character one plays as contributes to feeling immersed.

### *D. Self-efficacy*

PEC9: The role of self-efficacy in player retention.

Self-efficacy relates to the player retention on several standpoints of Primary Empirical Conclusions. The self-efficacy activities are classified into eight categories: creativity, exploration, iteration, implementation, communication, resourcefulness, synthesis, and vision [7]. According to this classification, the social aspect (PEC1) relates to exploration of self-efficacy activities like understand the needs of people and consider the viewpoints of others. Exploration is also a part of PEC8 (relating to the character), where the player dynamically adjusts with corresponding self-efficacy activities. These relations between self-efficacy activities are in alignment with the player´s retention on playing of multiple reasons.

Achievement relates to the importance of challenges (PEC5) and difficulty spikes are considered rewarding (PEC6), which both increase the player's motivation to continue playing. On self-efficacy scale these factors relate to implementation (make risky choices, suggest new ways to achieve goals). PEC6 relates also to resourcefulness when finding new uses for exiting methods or tools, like in PEC4, where increasing challenges and in-game progression are in balance.

PEC2 and PEC3 (balancing between active or passive role) relate to exploration, where observing people, consider viewpoints and understand people's needs dynamically guides the player's attitude in his/her role.

Self-efficacy creativity cluster (come up with imaginative solutions, do things in original way, think of new and creative ideas) is related to the player's attitude when facing challenges, reaching the sense of achievement and surviving with random elements when playing.

Player retention refers to high self-efficacy of players by descriptions of the found Primary Empirical Conclusions (PECs). Especially social aspect and relating to the character are related to exploration of self-efficacy activities. Achieving something increases motivation (PEC5), and difficulty spikes are rewarding (PEC6), that relate to implementation category of self-efficacy activities. The sense of achievement (PEC6) is one of the most significant emotions of player retention, which relates to self-efficacy's resourcefulness by terms troubleshoot problems, solve most problems, and handling unforeseen situations.

When self-efficacy describes a person's ability to manage in particular missions or perform an assignment, it expresses the person's belief to treat with objective,, tasks, and challenges.

According to juxtaposition of player retention and self-efficacy activities, high self-efficacy reflects simultaneously the player´s enjoyment, motivation, achievement, and success when gaining rewards. As a hypothesis, higher self-efficacy has an impact on player retention in its positive characteristics.

## VI. Discussion

The Primary Empirical Conclusions (PECs) of the study are summarized in the table below (Table 2).

TABLE I. PRIMARY EMPIRICAL CONCLUSIONS OF THE STUDY

| # | Description |
|---|---|
| 1 | Though the social aspect of online games is important, players seem to enjoy playing with people they already knew beforehand. |
| 2 | Players take on a more passive role when new to the game, while taking on a more active role as they become more experienced. |
| 3 | Players take on a more active role when playing with people they are familiar with, while taking on a more passive role while playing with people they are unfamiliar with. |
| 4 | The balance of increasing challenges and in-game progression is important for all players. |
| 5 | Achieving something in a game increases player motivation greatly to continue playing. |
| 6 | Difficulty spikes are generally considered particularly rewarding as long as they are perceived possible to overcome. |
| 7 | Common game design decisions aiming at player retention may have the opposite effect if the players consider them excessive. |
| 8 | Relating to the character one plays as contributes to feeling immersed. |
| 9 | Player retention corresponds to players' high self-efficacy when comparing the descriptions found in Primary Empirical Conclusions 1-8. |

PEC1. Past the implication that people prefer to spend time with their friends, this poses some practical implications for game developers. While in competitive online games automatic matchmaking systems are currently widely used, one downside to their use is that the players seldom end up playing with the same people again. This can be seen as a negative as it makes it more difficult for the players to meet new people through the game, which otherwise might make them more inclined to continue playing the game [18]. Thus, game developers should consider trying to make it easier for people to bond with each other in the game by making it easier for players to meet the same people again even in match-based games without a persistent game world.

Furthermore, studies on hedonistic information system use and the usage of (dis)continuance have argued that peer influence and the loss of value stemming from lost connections with others using the current system can discourage one from switching from one virtual world social networking system to another, Berger et.al. [6]. Helping players establish relationships within the digital game and in this fashion make them more likely to keep playing.

PECs2-3. One practical implication to be drawn from these observations is that players tend to be more reserved when new to the game. This can make them feel more inhibited around strangers, making them enjoy their game sessions less, as social interaction has been argued by past studies to be important in making a game more immersive and thus more enjoyable [18]. Game mechanics that encourage other players to mentor and act friendly towards newer players could be utilized to alleviate this issue.

The importance of the balance principle, i.e. scaling difficulty, in keeping a digital game interesting for longer periods of time has been highlighted in past studies, Hunicke et.al. [15], Richter et.al. [22]). Our findings support this notion (PEC4). The players of a game need to face challenges, lest the game become boring, which is what typically leads to the players moving on to another game (PEC7).

Emotions are important in keeping the gameplay interesting. Even typically negative feelings such as frustration can be motivating in helping overcome challenges [1]. When the players feel that they have experienced everything the game has to offer or simply no longer find it challenging, they end up feeling bored and are likely to stop playing that particular game, Baker et.al. [2], Chumbley et.al. [9].

However, the challenges should also be perceived as surmountable by the players. A player is likely to keep trying, regardless of feelings of frustration, as long as the challenge at hand is not considered impossible. The feeling of accomplishment upon overcoming challenges is particularly rewarding (PEC5). Yet, if the game continuously maintains a high sense of challenge, it may become tiring for the players. If the challenges simply keep getting tougher and tougher, one may eventually lose motivation to continue.

The sense of achievement from overcoming challengers seems to be one of the most important sensations in terms of player retention (PEC6). Players long for the feeling of accomplishment that stems from overcoming challenges with their own skills. This can serve to even boost one's confidence outside the realm of digital games. This ties closely to the notion of difficulty, as overcoming a challenge is likely to result in a sense of accomplishment. This is in line with the findings of Richter et al. [22] and Lafrenière et al. [17] who also concluded that difficult situations are considered more rewarding than constantly smooth gameplay.

A sense of achievement can also stem from social interaction. Some of the participants remarked that they felt even better about their achievements if they could somehow boast with them to other players, either directly or with the help of various game elements such as leaderboards. Giving players opportunities to showcase their progress or skills even in games that are not team-based or particularly social in nature can help in player retention. Indeed, outside challenges posed directly by the game in various forms (e.g. unlockable achievements for completing certain tasks), players tend to set themselves goals as well, and achieving these personal goals can be even more rewarding than achieving goals set or suggested by the game.

Related to challenges and PEC6 is the importance of rewards. While rewards alone can be motivating, they are most effective when tied to a sense of achievement. Receiving a reward for completing a difficult task was considered more rewarding by the participants than simply receiving a reward they felt they had not worked for. Rewarding new players frequently can assist in player retention. Giving a new player a feeling of gaining an advantage by frequently playing the game as a new player can motivate them to come back after the initial session, improving day one retention as suggested by Cheung et al. [8].

Random elements can help keep rewards interesting. We have seen various practitioner organizations leverage this by utilizing mechanics such as loot boxes where the rewards for spending real money on the game are random items, cosmetic or functional. Much of the time, in-game rewards for defeating enemies also contain a notable degree of randomness.

In relation to PEC8, the participants discussed a common player retention strategy: activity rewards. The games they played would reward them for participating in a certain activity every day. Sometimes this activity was simply starting up the game every day, but in most cases, receiving the rewards required the participants to complete various tasks in the game. In some cases, the participants felt these tasks were simply extra rewards for doing what they would usually do.

On the other hand, in some cases, the participants felt compelled to complete the tasks, lest they feel that they are missing out on rewards they considered highly important. This resulted in negative feelings, as the participants felt they had to devote some time to the game even if they did not feel like doing so for any other reason than to gain these activity rewards. This in turn could potentially lead to frustration, which could then lead to the players simply giving up upon feeling left behind upon missing the rewards.

Avatars, in terms of PEC9, have been noted to be important in generating a feeling of immersion, Turkay et.al. [26]. Being able to customize one's avatar makes it easier for the players to relate to it [26]. Alternatively, as was the case in our data, providing the players with various choices to pick from in choosing an avatar can also work if the developers opt for premade avatars. Having one predetermined main character to play as it can make it harder for the players to become immersed, although it may arguably be preferable for story-telling purposes in single player games.

### *A. Limitations of the Study*

Due to the chosen qualitative case study approach, we are unable to make any claims related to demographic factors. E.g., does a certain demographic of players simply enjoy rewards without a sense of accomplishment attached? To this end, we cannot study the effects of personality traits or other individual factors. This can be an interesting avenue to explore as future work.

The generalizability of the findings in this study is limited by the chosen case study approach, even if the number of cases in this study also falls within the recommended number of cases in case study research of Eisenhardt [13]. Following a theory-building logic, we proposed 9 PECs that should be further studied using a larger sample size and more quantitative research methods, to achieve a higher level of generalizability.

Finally, it should be noted that no silver bullet for player retention exists. Player retention mechanics cannot make up for a lack of an interesting core gameplay loop. Player retention mechanics can help keep the players interested for a longer time, but if they are not interested in the game in the slightest, no combination of player retention mechanics will make them enjoy it for long.

## VII. Conclusions

In this study, we have taken on a player-focused view on player retention by means of a qualitative multiple case study. We utilized multiple data collection methods to better understand player retention, conducting a questionnaire, semi-structured interviews, and by observing a number of participants playing a game of their choice.

Our findings were primarily practical, providing a better understanding of what could aid game developers achieve better player retention. Returning to the two research questions posed in the introduction, we answer them as follows.

RQ1. What online game elements keep players playing the game? The balance principle was considered the most important one by the participants. I.e. scaling difficulty that keeps the game challenging, even as the player becomes more skilled and/or their character becomes stronger. To this end, feelings of achievement, stemming from various sources, were considered highly motivating.

Social interactions are also important in online games. While players prefer playing with their existing friends, fostering player interaction inside the game world also contributes to creating new social interactions between previously unfamiliar people.

RQ2. How do rewards affect the player's motivation to continue playing video games? Rewards on their own are satisfying. However, players prefer being rewarded for overcoming challenges. A sense of achievement is important, and the resulting positive feelings can be further strengthened through in-game rewards.

Rewarding different kinds of activities to appeal to different playstyles, depending on the type of the game, can be helpful. As opposed to simply rewarding the players for winning, rewards for activity, exploration and completing various extra tasks can be motivating.

Finally, future research on the topic would benefit from a less general approach. Rather than studying player retention on a general level across games, more focus should be put on the effect of specific game elements on player retention. This could be done inside a game genre or by focusing on a specific game. In addition, a specifically customized questionnaire for gaming and player retention, with a large player base, would explain more in detail the players' attitude to self-efficacy categories. Simultaneously the game developers' would get ideas to design games more creatively, and rewarding for the players.

# APPENDICES

The full list of 29 activities of Self-efficacy questionnaire. [7]

**Appendix: Innovation Self-Efficacy (ISE) Survey**

*Directions:* Rate your degree of confidence that you can do each of the activities listed below on a scale from 0 (not at all confident) to 100 (extremely confident).

1. Understand the needs of people by listening to their stories.
2. Find connections between different fields of knowledge.
3. Seek out information from other disciplines to inform my own.
4. Identify opportunities for new products and/or processes.
5. Question practices that others think are satisfactory.
6. Come up with imaginative solutions.
7. Make risky choices to explore a new idea.
8. Consider the viewpoints of others/stakeholders.
9. Evaluate the success of a new idea.
10. Apply lessons from similar situations to a current problem of interest.
11. Envision how things can be better.
12. Do things in an original way.
13. Set clear goals for a project.
14. Troubleshoot problems.
15. Keep informed about new ideas (products, services, processes, etc.) in my field.
16. Communicate ideas clearly to others.
17. Provide compelling stories to share ideas.
18. Learn by observing how things in the world work.
19. Solve most problems if I invest the necessary effort.
20. Be resourceful when handling an unforeseen situation.
21. Suggest new ways to achieve goals or objectives.
22. Test new ideas and approaches to a problem.
23. Share what I have learned in an engaging and realistic way.
24. Make a decision based on available evidence and opinions.
25. Relate seemingly unrelated ideas to each other.
26. Think of new and creative ideas.
27. Model a new idea or solution.
28. Find new uses for existing methods or tools.
29. Explore and visualize how things work.

Eight innovation-related categories of self-efficacy 29 activities of Carberry et al. 2018 [7]:

creativity (questions 6, 12, 26)
exploration (1, 8, 18, 29)
iteration (9, 22, 27)
implementation (7, 13, 21, 24)
communication (16, 17, 23)
resourcefulness (3, 14, 15, 19, 20, 28)
synthesis (2, 10, 25)
vision (4, 5, 11)

# V

# EXPLORING THE RELATIONSHIP BETWEEN SELF-EFFICACY AND CREATIVITY: CASE IT & BUSINESS EDUCATION

by

Juhani Risku, Kai-Kristian Kemell, Joni Kultanen, Polina Feshchenko, Jeroen Carelse, Krista Korpikoski, Mari Suoranta & Pekka Abrahamsson, 2020

# Exploring the Relationship Between Self-Efficacy and Creativity: Case IT & Business education[1]

Juhani Risku[1] [0000-0002-0587-4431], Kai-Kristian Kemell[1] [0000-0002-0225-4560], Joni Kultanen[1], Polina Feschenko[1], Jeroen Carelse[2], Krista Korpikoski[3], Mari Suoranta[1] and Pekka Abrahamsson[1]

**Abstract.** Self-efficacy belief affects humans in life, action and work. The higher self-efficacy, the stronger contribution in fulfilling tasks, helping others in a team, and survive when facing obstacles and failures. Based on the literature we hypothesize that creativity correlates to higher self-efficacy. 49 students' self-efficacy was recorded by a questionnaire based on Innovation Self-Efficacy Measure (ISE) method. In the experiment, the students acted as designers and completed the design task. Results were analyzed and reflected against the self-efficacy questionnaire results to evaluate the impact of the course in the progress in design skills and creativity. In the paper, creativity is evaluated against the standard definition of creativity, which explains, that originality and effectiveness are required factors of creativity. The results show that creativity and self-efficacy are not strongly correlated. As a conclusion, we develop a vision of design program vision according to our findings of the results to enhance IT students' self-efficacy through design skills.

## 1 Introduction

Creativity, and generally innovation potential, scale to almost any use. Creativity belongs as a skill for anyone, not only for those working in professions that are conventionally considered creative ones, such as architecture. For example, creativity and innovativeness have been extensively studied in relation to entrepreneurship, e.g. [1], [2], where innovative ideas can shape and create markets. In the context of Information Technology (IT), Graziotin et al. (2014) underlined the importance of creativity in relation to problem solving among programmers [3], and Carberry et al. (2018) stress the importance of innovativeness among engineering students [4].

Creativity is considered as an important factor for the students at universities, because they need to address themselves to new and unfamiliar tasks and missions and defer to their ever-changing future work, vocation and circumstances they are to undergo [5]. The problem of creativity at the academia is not that it would be missing

[1]An early version of this paper was published at 10th International Conference on Software Business (ICSOB 2019) [40]

from there; it is that analytic culture penetrates the academic domain [6]. Jackson presents five factors influencing the education at universities. First, creativity is essential for the human existence, creativity is built into human proficiency, and it effects our self-knowledge and how we carry out our objectives and life. Secondly, both creative and academic progress are essential, if we consider that higher education is crucial for students' potential in affirming their progress favorable life. Thirdly, academic teachers can through demanding and fascinating courses push significant stimulus on the creative growth of the students. Fourthly, the taught subjects and the upcoming profession itself can generate insights to creative advancement. Creativity, mastery, and crafts and abilities of a profession are linked to students' creative progress. As fifth factor, arisen creativity of a student scales and applies to all other themes in life than to academic context only. Here the higher education could have a remarkable influence to advance the students' lifelong aspiration of creativity in studies and improvement [7], [8], [9].

In relation to creativity at the academia, several disciplines and curricula emphasizes note-making: In art, design and teaching diaries, research logs and fieldbooks are used as tools for note-taking, but also for producing new when acting in the contexts of in the real world [10], [11]. Architects use drawing and sketchbooks to improve their action and operating planning [12].

These examples of the importance of note-making and the use note-making as a medium for creation, highlights the need of both note-taking and its intensified continuation, note-making, to be a self-evident tool in the design classes, if not at the whole academia. We wanted to evaluate the students' notebooks and find correlations between measured self-efficacy rates.

In this study, we utilized the ISE Measure instrument by Carberry 2018 to measure the innovation self-efficacy of students. This was done before and after a university course intended to teach design and to support innovation self-efficacy. Additionally, we evaluated, using a panel of judges, the creativity of the students based on materials they produced during the course. Taking on an explorative approach, we then sought correlations between creative output and innovation self-efficacy.

**RQ: Does the students' creativity correlate to self-efficacy questionnaire results and design work evaluations?**

The rest of this paper is structured as follows. Section 2 presents the theoretical background of this study was we discuss self-efficacy, creativity, and creativity in the context of self-efficacy. In section 3 we declare our research methodology from course design to data collection and analysis. We then explain the analysis of our data in section 4, and discuss the implications of the results in section 5. In section 6 we present discussion and implications. Finally, section 7 concludes the paper.

At the University of Jyväskylä there is no faculty of engineering, art and design, but creativity related courses are included to the curricula. Art is taught at the Faculty of Humanities and Social Sciences, at the Department of Music, Art and Culture Studies and Department of Teacher Education. Art and design is covered according to the curricula in Art and Teaching, Education and Early Years pedagogy, and during the master program of museology. At the Faculty of Information Technology, design is taught at courses Digital Service Innovation and Design, Service Design Project, Interface of Things, Interaction Design, Web and Usability, Life-Based Design, Designing for Life, and Fuzzy Front End Design. These courses do not form a designer's profession, but they are deepening the student's understanding in information technology. The target course at hand for this study, was originally meant for students at the Faculty of Information Technology. When allowing access for students from all faculties, half of the students came from other than the IT Faculty. So, creativity and self-efficacy was evaluated among students with various disciplines.

## 2 Theoretical Background

In this section, we discuss the theoretical background of the study. The first subsection discusses self-efficacy in general, while the second one discusses creativity in general. The third subsection then connects these two as we discuss creativity and its relationship with (different types of) self-efficacy.

### 2.1 Self-Efficacy

Self-efficacy refers to the perception one has of their own abilities in performing a task (Carberry 2018). E.g., one's own perception of whether their C++ programming skills are adequate accomplish an assignment. Self-efficacy is considered important in performing successfully diverse tasks. If a person believes in her own abilities, she is more likely to survive when facing challenges, more likely to strive for related tasks in the first place and more intrinsically motivated [13]. Similarly, one may feel less inclined to begin carrying out a task if their self-efficacy is low, even if, objectively, they would have the required skills or knowledge to carry out the task. Self-efficacy can thus result in productivity [14].

According to Bandura (1994), self-efficacy is influenced by four factors: (1) mastery experiences, (2) vicarious experiences, (3) social persuasion, and (4) physiological states. Mastery experiences are past task completions (or failures) and hold a notable impact on one's self-efficacy. In the absence of, and in addition to, mastery experiences, vicarious experiences can weigh on one's self-efficacy. Vicarious experiences are gained by observing others with similar ability perform tasks. Social persuasion, then, refers to support from prestigious individuals or individuals we respect, and can positively affect self-efficacy. Finally, our current physiological states, such as stress or simply being tired, can influence our self-efficacy.

Whereas self-esteem refers to one's sentiments about oneself, with low self-esteem relating to feeling negatively about oneself, self-efficacy is simply related to tasks. Despite having low self-efficacy in relation to a task such as winning a baseball match, one may consider this to not matter, continuing to feel confident in themselves (Gist & Mitchell, 1992). Moreover, self-efficacy is related to specific tasks (Carberry et al., 2018). Though we have our perception of e.g. our own ability to program, it is typically considered in relation to a specific task at hand. We may consider ourselves to be good at programming, and yet know that developing an autonomous vehicle from scratch is well beyond our means.

Self-efficacy changes over time, and training or studying a skill is likely to positively influence our self-efficacy in relation to tasks related to that skill (Carberry et al. 2018). This is something we study in this paper as well as we look at the self-efficacy of students before and after taking a design course intended to support self-efficacy in relation to various skills required in design work. However, we also wish to understand the potential link between self-efficacy in these skills and creativity in practice. We thus further discuss self-efficacy in the specific context of creativity in the following subsection.

### 2.2 Creativity and Measuring Creativity

Creativity has been studied in a plentiful contexts and through all fields of science. It can be seen as the production of novel and useful ideas in any domain [15]. While such a general definition for creativity can largely be agreed-upon, creativity often has to be further defined when seeking to measure it for the purpose of a study.

Creativity is typically measured by examining outcomes of the process that leads to the creation of creative results [16], [17]. In practice, this often means having participants generate creative solutions for uncommon problems [18], [19]. The problems should be uncommon to ensure that the participants are not familiar with the problem at hand, which would enable them to use solutions they know are well suited for solving it in a creative manner. These solutions are then scored by judges, e.g. (some of) the authors of the study, in order to assess the creativity of the solutions and simultaneously the individuals (Graziotin et al. 2014).

In measuring creativity, much effort has also gone towards understanding what factors affect creativity. To this end, factors such as personality, Wolfradt and Pretz, 2001 [20], cognitive style [which Hayes and Allison (1998) [21] define as “the way people perceive stimuli and how they use this information to guide their behavior”] Beeftink et al., 2011 [22], and psychosocial work environment (Shalley et al., 2000) [23], have been linked with creative performance in different contexts. In the context of IT and software development, Graziotin et al. (2014) linked developer happiness with their creativity, providing further support for the findings of Forgeard (2011) who suggested that one’s current mood affected one’s creative thinking.

### 2.3 Self-Efficacy in the Context of Creativity

Studies in relation to creativity also include *self-efficacy*, and specifically, *creative self-efficacy*, which Tierney and Farmer (2002) [24] define as being “the belief one has the ability to produce creative outcomes”. Past studies have established a link between creative self-efficacy and creativity (Mathisen & Bronnick, 2009) [25]. I.e. one’s confidence in one’s own creativity makes one more creative. Furthermore, Mathisen and Bronnick (2009) argued based on their data that creative self-efficacy can be improved by training, while going on to note that further studies on whether doing so also improved creative performance were required. This is something we seek to further address in this study, although extant studies on the matter also exist.

Schack (1989) [26] studied self-efficacy and creative productivity in gifted children. Their results pointed towards (creative) self-efficacy training not necessarily producing creativity in the students and having little bearing on their subsequent initiation of independent (creative) projects (Schack 1989). Self-efficacy training may in fact lead to lower self-efficacy (Schack 1989) e.g. if the participants find the tasks challenging, resulting in negative mastery and/or vicarious experiences. This presents challenges to those seeking to devise training courses for self-efficacy.

While much effort has gone towards studying the relationship between creativity and *creative* self-efficacy, studies looking into self-efficacy in other skills in relation to creativity are fewer in number. Ahlin et al. (2014) studied entrepreneurial self-efficacy in relation to creativity and innovation in the context of entrepreneurship, while Khedhaouria et al. (2015) studied *general* self-efficacy in the same context. Beeftink et al. (2012) studied design self-efficacy in relation to performance in creative profession, and also argued that self-regulation strongly supported self-efficacy in the context of design work.

In this study, we take on an explorative approach in studying self-efficacy in a large number of skills, using the instrument developed by Carberry et al. (2018) to measure self-efficacy. In doing so, we seek to further our understanding of how self-efficacy, and specifically, how various different types of self-efficacy, affect (or do not affect) creativity, as we discuss in detail next while presenting our research methodology.

### 2.4 Design assignments using note-making

During the courses, the design assignments required different skills to make notes, visualize, communicate and create structured plans. Therefore note-making was taught briefly to show its power when designing.

Note making is considered as an active and versatile method of ideation, creation, getting intentions, and note-taking a passive action capturing information more by dictation. As Neville 2014 expresses, that note-taking is the start of note recording, which leads to more fundamental note-making. Linear note-taking is of recording the lecturer's with summaries, sentences, abbreviations, central points after listening or seeing. When moving from passive and linear note-taking process to an active, reviewing, synthesizing, and generating process, where ideas are presented in a readable and creative way, the information is easier to remember and recall [27].

Buzan & Buzan (2010) explain in the context of mind mapping that "It is vital for note-making – explaining your own thoughts, planning, organizing, thinking creatively, making relationships and attaining views - and for note-taking – recording information in meetings, debate, lectures etc. or summarizing books and other written material" [28]. At the course, the students were explained that all means and skills like writing, sketching, drawing, concepting, making graphs, tables, charts, maps and other explanatory jottings and visualizations connote wider perspectives and abstractions during the lectures. On the contrary, note-taking was described as a tenuous and insufficient method when aiming to ideate and create new ideas and concepts. During the course, note-making was seen as a designer's tool for collecting own ideas and designs in a visual and structured order for later use. The five factors and proposals by Jackson were built into the class as instructional advice for the students to start their notebook of personal designs and innovations. The students also new, that the notebooks were a mandatory outcome of the course.

Note-making can be considered a crucially important method for students on the academic lectures, because they are important when transferring knowledge and academic qualities to new generations. Longman and Atkinson 1999 found, that students could not recall 95 % of the information from the lecture, which they had not captured to their notes [29]. Another alarming observation by Ruiz et. al. (2010) is that all students do not take notes. Females more frequently take notes, and males take notes more occasionally [30]. Of same opinion is Powell and Wimmer 2014 when expressing that at present fewer students take notes during lectures. They propose a screencast method for students in hands-on computer programming class. When creating an own screencast supported with a hands-on instruction, it is a practical method in increasing the students learning results [31].

Several universities guide students in note-making on their Web pages, because "students frequently do not realize the importance of note taking and listening" (Darmouth College, 2013) [32]. In the same way, University of Leicester, University of New Brunswick, University of York, Cardiff Metropolitan University and University of the Witwatersrand among several other. Note-making is considered by universities in their preliminary guidelines as an important matter in learning.

## 3 Research Methodology

### 3.1 Course Design and Participant Characteristics

The data for this study were collect from a design course (YRIS2410 Venture Lab & TJTS5791 Lean Startup) and from the course (TJTS1000 Design Practices in Contemporary World) at the University of Jyväskylä. The courses in question were open to IT and economy students and the first course also involved international students.

Data for this study was collected from 49 (23 in the preliminary study published in [40]) participants. Both, the ISE questionnaire and the course assignment were required for the student to be included in the study. The collected data were anonymized after collection and the data from each participant were denoted by unique IDs instead.

Out of the 49 participants included in this study, 27 were female and 22 were male. Eight of the participants were between 18-24 years of age, 30 were between 25-34 years, Five between 35-44 years. The degree of the participants' current studies divided as 25 master's and 1 bachelor's students. The major subjects of the participants were Business (17, International Business, Marketing, and Entrepreneurship), Information Systems Science (8), Information Technology (1), Sports (1), and Sports (1). At the University of Jyväskylä, Information Systems studies is combined with Business and Economics studies.

### 3.2 Data Collection Methodology

Data were collected from the twenty-six participants who completed the previously described course. Two sets of data were collected for the study during the activities of the experimental procedure: (1) Innovation Self-Efficacy questionnaire, and (2) course assignment (design related task).

The data measuring the level of participants' innovation self-efficacy were collected using the Innovation Self-Efficacy (ISE) Measure instrument developed by Carberry et al. (2018). The ISE Measure was developed to measure the innovation self-efficacy of engineering students. It can be used to evaluate the positive or negative impacts of an intervention, such as training, on the participants' judgment of their own innovation ability (Carberry et al. 2018), as we do in this paper.

The ISE measure is utilized by having an individual give a numeric rating to their confidence in an activity. There are 29 activities related to innovation and creativity in the survey (e.g. "Identify opportunities for new products and/or processes"). The activities are categorized into eight categories: creativity, exploration, iteration, implementation, communication, resourcefulness, synthesis, and vision. (Carberry et al. 2018) The full list of 29 activities can be found in detail in our analysis of our data in the Appendices.

Participants completed the Innovation Self-Efficacy (ISE) Measure instrument based questionnaire (Carberry et al., 2018) twice during the course, first during the first lecture of the course and the second time during the final lecture. The questionnaire was introduced to the students in the class as a class activity that would be used for further research and the students had 15 minutes to complete each questionnaire. The questionnaire was completed on a Web Survey service. For this research, background information of the participants was collected from the first questionnaire and the innovation self-efficacy results from the questionnaire on the final lecture.

In addition to the ISE Measure data, for the creativity evaluation, students created two design related tasks during the course. The first was to design appealing services and entertainment for new visitors to Neste Rally motorsport event in Jyväskylä. This was a visual one-pager concept with drawings, text and event related factors. The second assignment was to design new city centric attractions to invite young people to settle and enjoy in Hämeenlinna city in Finland. This was a two-pager with a rough ideation page and a conceptual design page. A brief training session preceded the task, including presentations on writing, drawing, sketching or cutting and pasting pictures from a source of their own choice.

## 4 Data Analysis

The data set included answers on 29 questions of 49 participants. The questions were aimed at finding the respondents view (each graded 0-100) on his/her own self-efficacy. Thus, in order to calculate the self-efficacy variable for each participant, the mean of all 29 responses has been calculated and used for checking the assumption of its correlation with the evaluated creativity variable.

To evaluate creativity, we followed the model Graziotin et al. (2014) used to measure creativity in their study. We utilized a panel of judges to evaluate the creative outputs of the student participants in the form of a notebook, based on the judges' own definitions of creativity. Each judge individually rated each notebook on a 7 point Likert scale, where 1 equaled to the output not being creative and 7 equaled to high creativity.

The panel of judges consisted of three experienced designers, one of whom was an author of this paper as well. One of the designers is a doctoral student, industrial and service designer with six years of experience. The other one is an architect, acoustician, scenographer, carpenter, UI designer and doctoral student, with over 30 years of design practice. The third person is an industrial designer, artist, art and design teacher and carpenter, with over 25 years of design practice.

In this study measurements of quality were utilized for the assessment of creativity, which is where we diverted from Graziotin et al. (2014), who in addition used measurements of quantity. This diversion was due to our study only having one object to evaluate the creativity on, whereas Graziotin et al. (2014) had several outputs from each participant to measure. Graziotin et al. (2014) measured quality by two scores: the average score based on all the outputs of a participant evaluated for creativity (ACR) and the best score among the outputs of a participant (BCR).

During the evaluation of the student's assignments according to their creativity, the three judges used also a combined version of the standard definition of creativity added with creative productivity. According to the standard definition of creativity Runco et.al. 2012 explain, that originality and effectiveness are required factors of creativity [34].

Effectiveness included notions useful, solid, adequate and suitable; commodity, adaptiveness and firmness. Originality included notions authenticity, genuine, real and original, beauty and delight. Creative productivity was evaluated both by quantity in generation of ideas that are original and adaptive, meaning the quality dimension. The judges also reflected the evaluation against their own definition of creativity as a comprehensive reason.

The evaluation was judged on Likert scale 1…7, and brief a description was written as reasoning for the judgement.

## 5 Results

In this section, the findings revealed during the statistical analysis for finding correlation between creativity and self-efficacy are discussed. The dataset contains data from 49 respondents.

Both self-efficacy and evaluated creativity variables data have been tested for normality before testing the assumption of their correlation. Both variables' skewness and kurtosis values were within their accepted ranges (1 and -1 for skewness, 2 and -2 for kurtosis). Besides, the normality test results for both variables did not show any level of significance ($p > 0.05$), so we can conclude that they are normally distributed. That can be also observed from their histograms.

Boxplots of the data showed an outlier for one of the variables and it is possible to observe that no clear linear relationship between them is present either (Figure 1). Thus, Spearman's correlation coefficient was computed in order to check the dependence between two variables. To do that, IBM SPSS software was used.

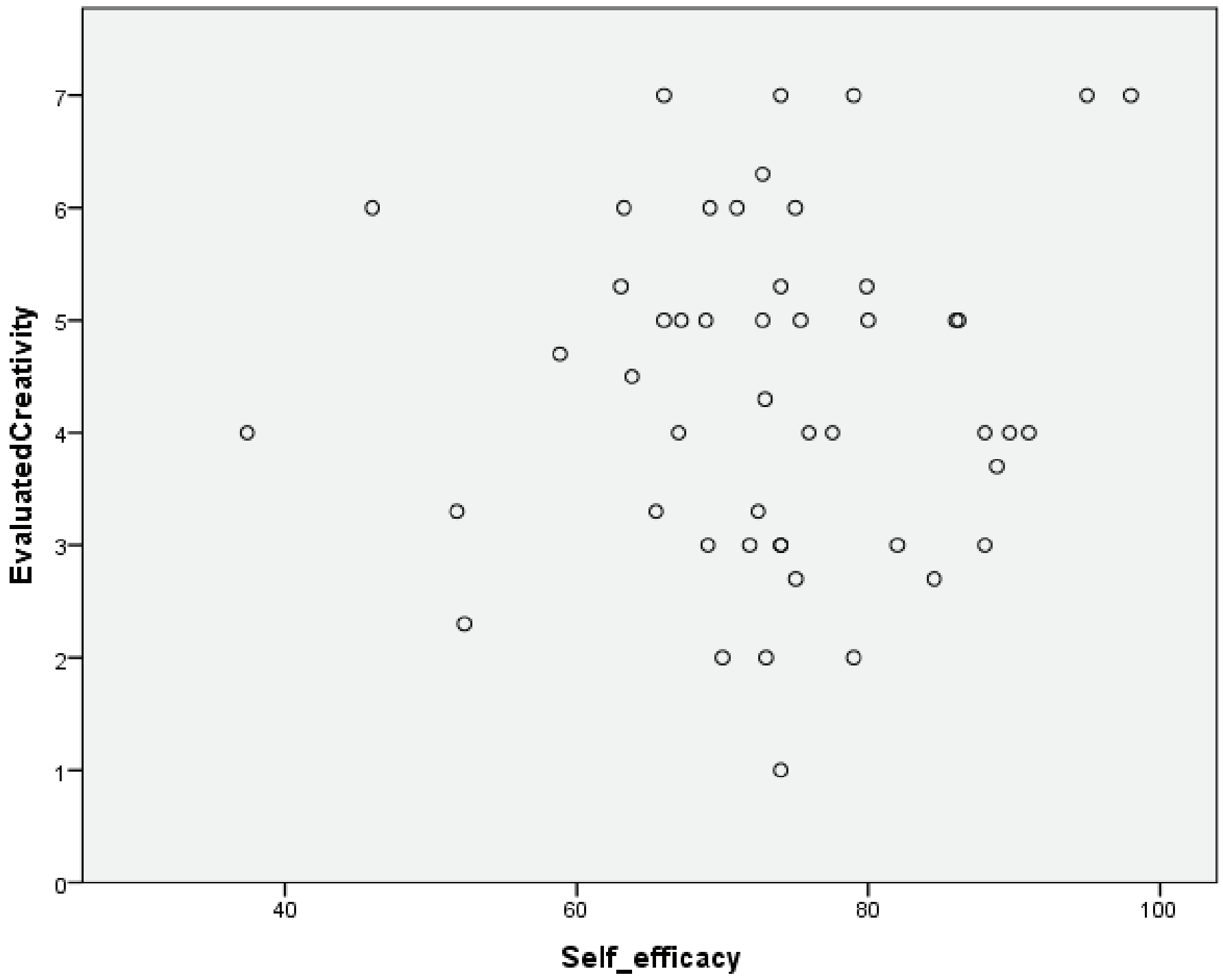

**Fig. 1.** Scatter plot showing no linear relationship between the studied variables.

As shown in the Table 1, the correlation coefficient turned out to be -0.20, what indicates the negative correlation between the variables (Freedman et. al. 1978) [33]. However, since the significance is equal to 0.924, it is possible to conclude that there is not enough evidence to say that the correlation exists, even though in the sample a small negative correlation has been observed (Weinberg & Knapp 2002) [35]. Thus, the new data analysis has shown the same results as discussed in the previous study.

**Correlations**

| | | | Evaluated Creativity | Self_efficacy |
|---|---|---|---|---|
| Spearman's rho | EvaluatedCreativity | Correlation Coefficient | 1,000 | -,020 |
| | | Sig. (2-tailed) | . | ,894 |
| | | N | 49 | 49 |
| Self_efficacy | Correlation Coefficient | -,020 | 1,000 |
| | Sig. (2-tailed) | ,894 | . |
| | N | 49 | 49 |

Table 1. Spearman's rho values for studied variables.

From the above analysis, we can say that with the current number of observants (49 individuals), presence of correlation between self-efficacy and creativity can not be confirmed, even though a small negative correlation was noticed. For further research, the bigger study should be conducted in order to prove the assumptions of this study and be able to draw conclusions regarding the relationship between the two constructs.

## 6 Discussion and implications

Universities have a significant role in finding and developing methods and curricula to improve students' self-efficacy. It advances innovation, academic motivation, remembering, learning and achievement. As proposals, Risku and Abrahamsson [36], present a whole set of embedding designer's ideal to computer science (CS) and software development (SWD). A thorough program of research and development of design principles is needed, as well as design education according to traditional design methods and newfound design methods from CS and SWD research. Also design principles from CS and SWD can be transferred to traditional art and design to refresh the cooperation between SW startups and design professionals. Researchers and teaching staff with their upgraded and new design skills can create a new design culture to the academia [35].

Companies, especially software startups, are typically first-movers in the sense of Aristotle's prime mover. The startuppers need a whole set of skills to create products and services starting from idea generation, constructing, and perfecting the solutions. Four ways are proposed for startuppers to move forward: a Radical, Conventional, Arrogant and Ignorant. The radical alternative, merge designers' ideal to startuppers' creation process, coincides the message of the importance of note-making, wide skill-set in design and street credibility in applying design and research findings [36]. These notions are in line to gain higher self-efficacy so, that the efficacious persons endure among barriers and debacles, are highly motivated and engaged, until they reach prosperity (Carberry 2018). Software startups are in the center of high creativity and self-efficacy on account of note-making and design, because practical skills in note-making is part of creation and design, and success in executing own ideas to the markets relates to self-efficacy source of mastery experiences. It means that the startupper performs the tasks in her own control. A positive vicarious experience can be in startuppers' context interpret to be as working in and for the team [13].

## 7 Conclusions

We collected altogether data of 49 students and it shows that self-efficacy beliefs do not describe the creativity grades of design class students. This leads to two conclusions: either run the research in a designer student class and in a non-designer student class, and compare the results. Then the wider experience of design students may show differences with the other group. On the other hand, a comparative research made between specially design-trained non-designers versus ordinary non-designers may lead to differences, after which it is possible to know, if the training made the difference.

From individual details in the study can be found, that students with long experience in design got high grades in creativity by the judges' evaluations. These students also had strong self-efficacy belief, and they were of age 25-34.

Self-efficacy theory has faced methodological critique [37]. In our cases, the students performed in making practical designs, which were evaluated by senior design professionals. As an assumption, a skillful student's self-efficacy could be either high or realistic, depending on the self-criticism and awareness of own professional capabilities. This was not investigated in our studies. In individual results, we found negative self-efficacy and performance relationships, when the student got high grade in creativity evaluated by the judges, but low self-efficacy belief according to the questionnaire. Here we a similar finding as Taylor (2014), when she argues, that "self-efficacy levels and their class placement was not discovered" during a study amongst Middle School Students [37]. This indicates, that students understand the notion of self-efficacy in a different way considering their experience, skills and motivation.

Future research on self-efficacy and its relation to creativity and innovation is important for present day companies and universities. A wide range of methods, tools, practices, curricula and research can help students to encourage themselves, inspire oneself to create and innovate privately and in a team. This development often can happen in companies on the fly, but the academia can develop the means in advance and supported by research. By the same time with a theoretical, practical and educational procedure, new ways of working motivates students to meaningful studies and exploration. According to the design course investigated, the spirit and eagerness amongst the students from diverse faculties and backgrounds form an ideal setup for teams.

When following Carberry et. al. [4], self-efficacy belief effects the students and teaching personnel, people, in life, work and play, it is important to enable all factors strengthening people's skills and motivation. Here Risku et.al, proposes in the Discussion and implications section a wide set of objectives and means. When outlining these objectives and means, education that offers diversely theory and practice, handicrafts and design training, the comprehensive skill-set of the student cannot hinder the growth of her self-efficacy. On the contrary, the encompassing skill-set strengthens her courage to start projects, act in a multifaceted team, find sensible roles in work, and take responsibility in leadership.

Further research is needed in comparison of creative professions and domains. Like Rolling lists, Leonardo da Vinci was an architect, architect, engineer, artist, sculptor, painter, musician, inventor, mathematician and anatomist [38]. The question is, does creativity stem from the same origin. Is the creativity different between his various professions, or did he use the same creativity base but emphasizing it case by case? Rolling continues by saying that STEAM education, a combination of Science, Technology, Engineering, Arts, and Mathematics, offers diversity of knowledge from all these domains by joining them together. STEAM education is immediately applicable to the academia as a meaningful medium to advance creativity and self-efficacy. Still research and pilot programs are needed.

The described objectives require an appropriate context, which in this case sounds like a combination of a Humboldtian university, Arts & Crafts & Design faculty, and faculty of Technology. Creativity and design as forward looking and planning acts belong to any faculty, besides the conventional arts, design and engineering schools and faculties. Therefore self-efficacy belief as an indicator in these contexts could be an instrument to develop a model for multifaceted education environment.

Limitations were also found in this study. Self-efficacy and creativity was evaluated within one week and a definite group, but still on a sharp testbed. Now when started, a wider research and evaluation program at relevant academic course could be organized. After the results, the interpretation of the results can lead to actions, and to be included to the university's strategic program. This iterative process, when consisting of practical actions and research, forms a positive self-perpetuating circle.

We believe that the new generations, the Y and Z ones, grown with the Internet, social media, technology disruption and Virtual Reality, even Artificial Intelligence, want to see the university with a magic touch. The Z generation, as called as the Millenials, are starting their adult life at the universities and work. We pioneers can enable the young people's learning and self-efficacy through cultivating the Academia.

## Appendices

The full list of 29 activities of Self-efficacy questionnaire. [4]

**Appendix: Innovation Self-Efficacy (ISE) Survey**

*Directions:* Rate your degree of confidence that you can do each of the activities listed below on a scale from 0 (not at all confident) to 100 (extremely confident).

1. Understand the needs of people by listening to their stories.
2. Find connections between different fields of knowledge.
3. Seek out information from other disciplines to inform my own.
4. Identify opportunities for new products and/or processes.
5. Question practices that others think are satisfactory.
6. Come up with imaginative solutions.
7. Make risky choices to explore a new idea.
8. Consider the viewpoints of others/stakeholders.
9. Evaluate the success of a new idea.
10. Apply lessons from similar situations to a current problem of interest.
11. Envision how things can be better.
12. Do things in an original way.
13. Set clear goals for a project.
14. Troubleshoot problems.
15. Keep informed about new ideas (products, services, processes, etc.) in my field.
16. Communicate ideas clearly to others.
17. Provide compelling stories to share ideas.
18. Learn by observing how things in the world work.
19. Solve most problems if I invest the necessary effort.
20. Be resourceful when handling an unforeseen situation.
21. Suggest new ways to achieve goals or objectives.
22. Test new ideas and approaches to a problem.
23. Share what I have learned in an engaging and realistic way.
24. Make a decision based on available evidence and opinions.
25. Relate seemingly unrelated ideas to each other.
26. Think of new and creative ideas.
27. Model a new idea or solution.
28. Find new uses for existing methods or tools.
29. Explore and visualize how things work.

APPENDIX 1. Descriptive statistics of the studied variables (Dataset 1).

**Descriptives**

| | | | Statistic | Std. Error |
|---|---|---|---|---|
| Evaluated Creativity | Mean | | 4,35 | ,214 |
| | 95% Confidence Interval for Mean | Lower Bound | 3,90 | |
| | | Upper Bound | 4,79 | |
| | 5% Trimmed Mean | | 4,35 | |
| | Median | | 4,40 | |
| | Variance | | 1,196 | |
| | Std. Deviation | | 1,094 | |
| | Minimum | | 2 | |
| | Maximum | | 6 | |
| | Range | | 4 | |
| | Interquartile Range | | 2 | |
| | Skewness | | -,107 | ,456 |
| | Kurtosis | | -,827 | ,887 |
| Self_efficacy | Mean | | 70,42 | 2,320 |
| | 95% Confidence Interval for Mean | Lower Bound | 65,64 | |
| | | Upper Bound | 75,19 | |
| | 5% Trimmed Mean | | 71,01 | |
| | Median | | 72,60 | |
| | Variance | | 139,928 | |
| | Std. Deviation | | 11,829 | |
| | Minimum | | 37 | |
| | Maximum | | 90 | |
| | Range | | 52 | |
| | Interquartile Range | | 13 | |
| | Skewness | | -,766 | ,456 |
| | Kurtosis | | 1,300 | ,887 |

APPENDIX 2. Normality tests (Dataset 1).

**Tests of Normality**

| | Kolmogorov-Smirnov[a] | | | Shapiro-Wilk | | |
|---|---|---|---|---|---|---|
| | Statistic | df | Sig. | Statistic | df | Sig. |
| Grade | ,148 | 26 | ,147 | ,965 | 26 | ,494 |
| Self_efficacy | ,126 | 26 | ,200* | ,950 | 26 | ,238 |

*. This is a lower bound of the true significance.
a. Lilliefors Significance Correction

APPENDIX 3. Histograms showing normal distribution of the two studied variables (Dataset 1).

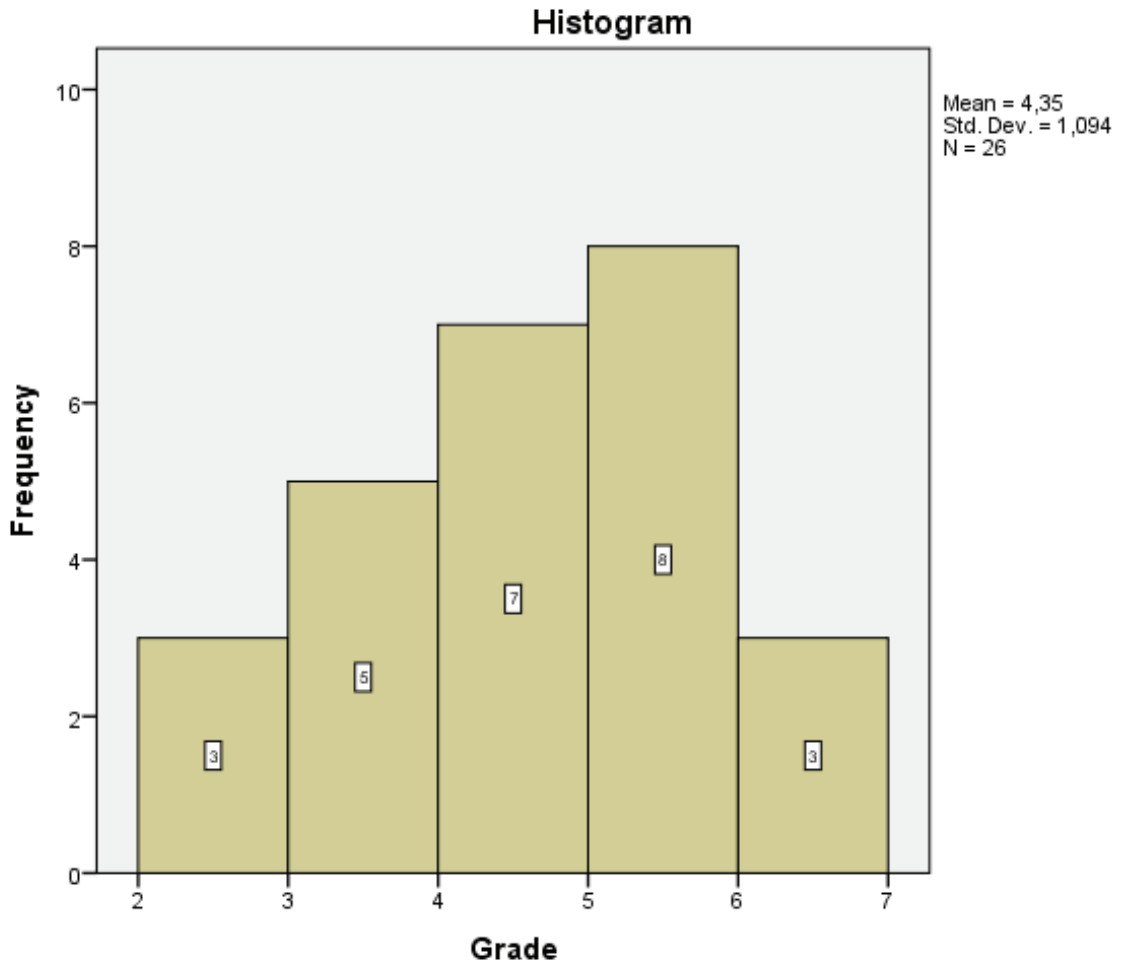

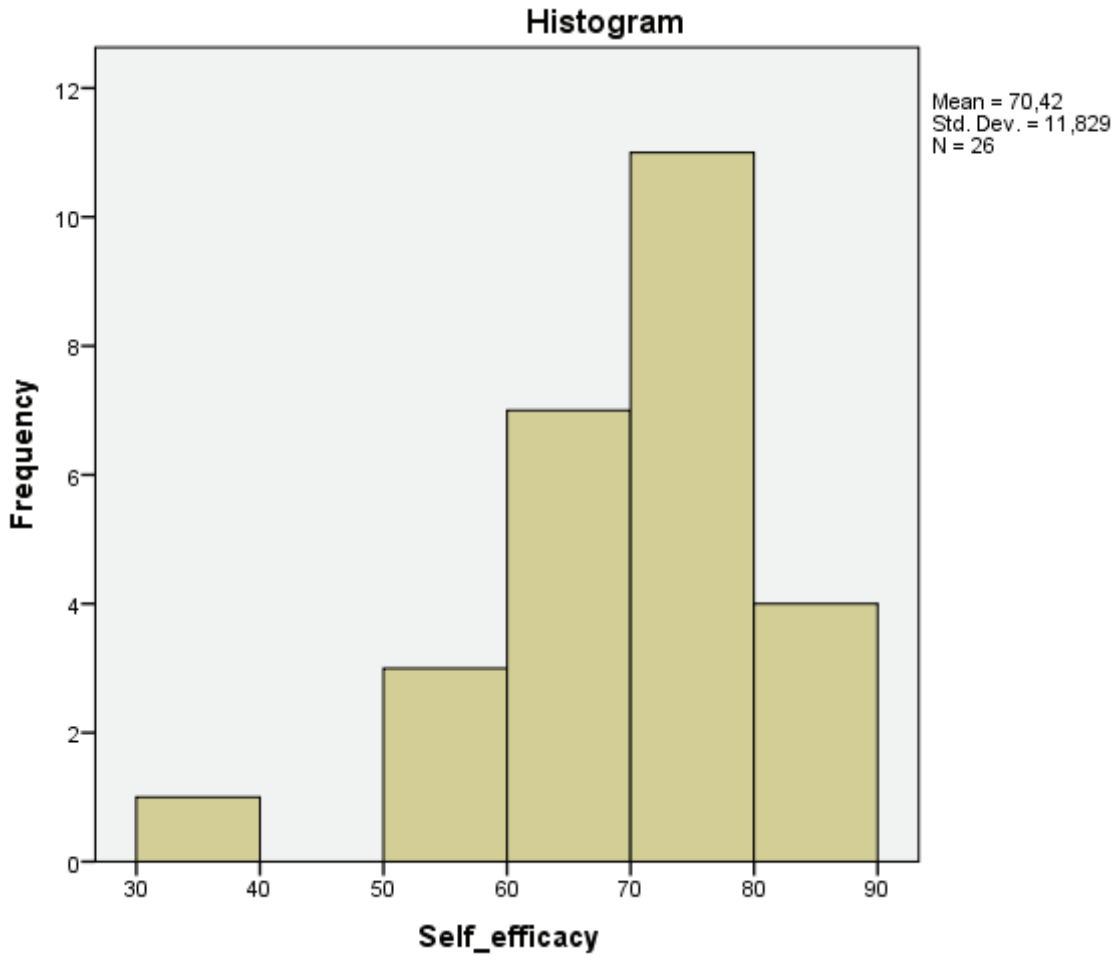

APPENDIX 4. Boxplots of two variables (Dataset 1).

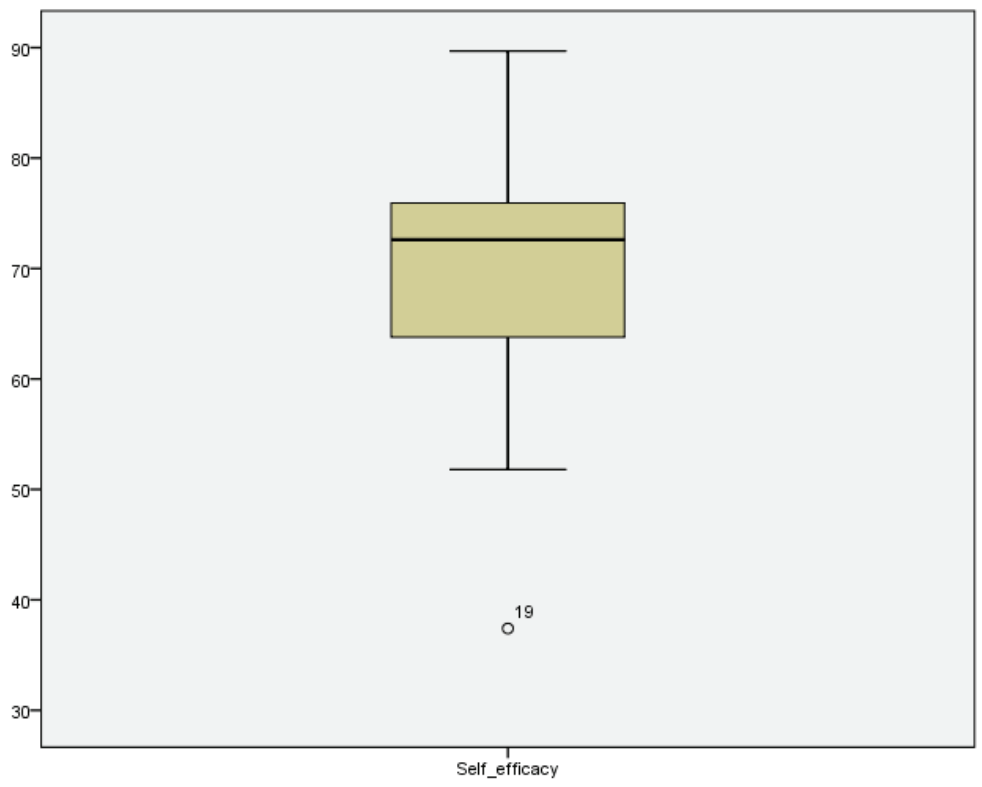

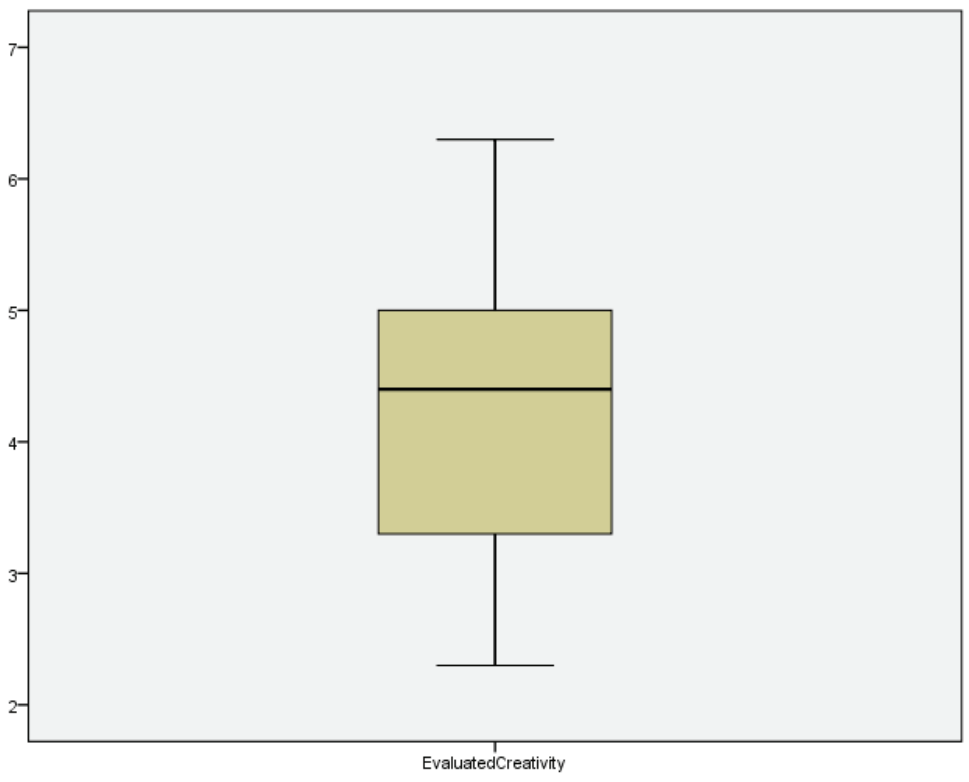